\documentclass[longauth,traditabstract]{aa}
\usepackage{graphicx}
\usepackage{txfonts}
\usepackage[breaklinks,colorlinks,citecolor=blue]{hyperref}
\usepackage{fixltx2e}
\usepackage{natbib}
\usepackage{ifthen}

\bibpunct{(}{)}{;}{a}{}{,}

\def\setsymbol#1#2{\expandafter\def\csname #1\endcsname{#2}}
\def\getsymbol#1{\csname #1\endcsname}

\def\Planck{\textit{Planck}}

\def\alltwentythirteenresultspapers{\nocite{planck2013-p01, planck2013-p02, planck2013-p02a, planck2013-p02d, planck2013-p02b, planck2013-p03, planck2013-p03c, planck2013-p03f, planck2013-p03d, planck2013-p03e, planck2013-p01a, planck2013-p06, planck2013-p03a, planck2013-pip88, planck2013-p08, planck2013-p11, planck2013-p12, planck2013-p13, planck2013-p14, planck2013-p15, planck2013-p05b, planck2013-p17, planck2013-p09, planck2013-p09a, planck2013-p20, planck2013-p19, planck2013-pipaberration, planck2013-p05, planck2013-p05a, planck2013-pip56, planck2013-p06b, planck2013-p01a}}

\def\alltwentyfifteenresultspapers{\nocite{planck2014-a01, planck2014-a03, planck2014-a04, planck2014-a05, planck2014-a06, planck2014-a07, planck2014-a08, planck2014-a09, planck2014-a11, planck2014-a12, planck2014-a13, planck2014-a14, planck2014-a15, planck2014-a16, planck2014-a17, planck2014-a18, planck2014-a19, planck2014-a20, planck2014-a22, planck2014-a24, planck2014-a26, planck2014-a28, planck2014-a29, planck2014-a30, planck2014-a31, planck2014-a35, planck2014-a36, planck2014-a37, planck2014-ES}}

\newbox\tablebox    \newdimen\tablewidth
\def\leaderfil{\leaders\hbox to 5pt{\hss.\hss}\hfil}
\def\endPlancktable{\tablewidth=\columnwidth 
    $$\hss\copy\tablebox\hss$$
    \vskip-\lastskip\vskip -2pt}
\def\endPlancktablewide{\tablewidth=\textwidth 
    $$\hss\copy\tablebox\hss$$
    \vskip-\lastskip\vskip -2pt}
\def\tablenote#1 #2\par{\begingroup \parindent=0.8em
    \abovedisplayshortskip=0pt\belowdisplayshortskip=0pt
    \noindent
    $$\hss\vbox{\hsize\tablewidth \hangindent=\parindent \hangafter=1 \noindent
    \hbox to \parindent{$^#1$\hss}\strut#2\strut\par}\hss$$
    \endgroup}
\def\doubleline{\vskip 3pt\hrule \vskip 1.5pt \hrule \vskip 5pt}

\def\L2{\ifmmode L_2\else $L_2$\fi}

\def\DeltaT{\ifmmode \Delta T\else $\Delta T$\fi}
\def\deltat{\ifmmode \Delta t\else $\Delta t$\fi}
\def\fknee{\ifmmode f_{\rm knee}\else $f_{\rm knee}$\fi}
\def\Fmax{\ifmmode F_{\rm max}\else $F_{\rm max}$\fi}
\def\solar{\ifmmode{\rm M}_{\mathord\odot}\else${\rm M}_{\mathord\odot}$\fi}
\def\Msolar{\ifmmode{\rm M}_{\mathord\odot}\else${\rm M}_{\mathord\odot}$\fi}
\def\Lsolar{\ifmmode{\rm L}_{\mathord\odot}\else${\rm L}_{\mathord\odot}$\fi}
\def\inv{\ifmmode^{-1}\else$^{-1}$\fi}
\def\mo{\ifmmode^{-1}\else$^{-1}$\fi}
\def\sup#1{\ifmmode ^{\rm #1}\else $^{\rm #1}$\fi}
\def\expo#1{\ifmmode \times 10^{#1}\else $\times 10^{#1}$\fi}
\def\,{\thinspace}
\def\lsim{\mathrel{\raise .4ex\hbox{\rlap{$<$}\lower 1.2ex\hbox{$\sim$}}}}
\def\gsim{\mathrel{\raise .4ex\hbox{\rlap{$>$}\lower 1.2ex\hbox{$\sim$}}}}

\def\simprop{\mathrel{\raise .4ex\hbox{\rlap{$\propto$}\lower 1.2ex\hbox{$\sim$}}}}
\def\deg{\ifmmode^\circ\else$^\circ$\fi}
\def\pdeg{\ifmmode $\setbox0=\hbox{$^{\circ}$}\rlap{\hskip.11\wd0 .}$^{\circ}
          \else \setbox0=\hbox{$^{\circ}$}\rlap{\hskip.11\wd0 .}$^{\circ}$\fi}
\def\arcs{\ifmmode {^{\scriptstyle\prime\prime}}
          \else $^{\scriptstyle\prime\prime}$\fi}
\def\arcm{\ifmmode {^{\scriptstyle\prime}}
          \else $^{\scriptstyle\prime}$\fi}
\newdimen\sa  \newdimen\sb
\def\parcs{\sa=.07em \sb=.03em
     \ifmmode \hbox{\rlap{.}}^{\scriptstyle\prime\kern -\sb\prime}\hbox{\kern -\sa}
     \else \rlap{.}$^{\scriptstyle\prime\kern -\sb\prime}$\kern -\sa\fi}
\def\parcm{\sa=.08em \sb=.03em
     \ifmmode \hbox{\rlap{.}\kern\sa}^{\scriptstyle\prime}\hbox{\kern-\sb}
     \else \rlap{.}\kern\sa$^{\scriptstyle\prime}$\kern-\sb\fi}
\def\ra[#1 #2 #3.#4]{#1\sup{h}#2\sup{m}#3\sup{s}\llap.#4}
\def\dec[#1 #2 #3.#4]{#1\deg#2\arcm#3\arcs\llap.#4}
\def\deco[#1 #2 #3]{#1\deg#2\arcm#3\arcs}
\def\rra[#1 #2]{#1\sup{h}#2\sup{m}}

\def\dots{\relax\ifmmode \ldots\else $\ldots$\fi}
\def\WHzsr{\ifmmode $W\,Hz\mo\,sr\mo$\else W\,Hz\mo\,sr\mo\fi}
\def\mHz{\ifmmode $\,mHz$\else \,mHz\fi}
\def\GHz{\ifmmode $\,GHz$\else \,GHz\fi}
\def\mKs{\ifmmode $\,mK\,s$^{1/2}\else \,mK\,s$^{1/2}$\fi}
\def\muKs{\ifmmode \,\mu$K\,s$^{1/2}\else \,$\mu$K\,s$^{1/2}$\fi}
\def\muKRJs{\ifmmode \,\mu$K$_{\rm RJ}$\,s$^{1/2}\else \,$\mu$K$_{\rm RJ}$\,s$^{1/2}$\fi}
\def\muKHz{\ifmmode \,\mu$K\,Hz$^{-1/2}\else \,$\mu$K\,Hz$^{-1/2}$\fi}
\def\MJysr{\ifmmode \,$MJy\,sr\mo$\else \,MJy\,sr\mo\fi}
\def\MJysrmK{\ifmmode \,$MJy\,sr\mo$\,mK$_{\rm CMB}\mo\else \,MJy\,sr\mo\,mK$_{\rm CMB}\mo$\fi}
\def\microns{\ifmmode \,\mu$m$\else \,$\mu$m\fi}

\def\muK{\ifmmode \,\mu$K$\else \,$\mu$\hbox{K}\fi}
\def\microK{\ifmmode \,\mu$K$\else \,$\mu$\hbox{K}\fi}
\def\muW{\ifmmode \,\mu$W$\else \,$\mu$\hbox{W}\fi}
\def\kms{\ifmmode $\,km\,s$^{-1}\else \,km\,s$^{-1}$\fi}
\def\kmsMpc{\ifmmode $\,\kms\,Mpc\mo$\else \,\kms\,Mpc\mo\fi}
\providecommand{\sorthelp}[1]{}

\def\WMAP{WMAP}
\def\LCDM{$\Lambda$CDM}
\def\nside{N_{\mathrm{side}}}
\def\npix{N_{\mathrm{pix}}}
\def\fsky{f_{\mathrm{sky}}}

\def\healpix{\texttt{HEALPix}}
\def\commander{\texttt{Commander}}
\def\ruler{\texttt{Ruler}}
\def\nilc{\texttt{NILC}}
\def\sevem{\texttt{SEVEM}}
\def\smica{\texttt{SMICA}}

\def\XFaster{\texttt{XFaster}}

\def\MHz{\ifmmode $\,MHz$\else \,MHz\fi}

\title{\Planck\ 2015 results.\ IX.\\Diffuse component separation: CMB maps}
\author{\small
Planck Collaboration: R.~Adam\inst{76}
\and
P.~A.~R.~Ade\inst{87}
\and
N.~Aghanim\inst{61}
\and
M.~Arnaud\inst{74}
\and
M.~Ashdown\inst{70, 5}
\and
J.~Aumont\inst{61}
\and
C.~Baccigalupi\inst{86}\thanks{\hspace{-4pt}Corresponding author: C.~Baccigalupi, \url{bacci@sissa.it}.}
\and
A.~J.~Banday\inst{96, 9}
\and
R.~B.~Barreiro\inst{66}
% ...
}
\institute{\small
APC, AstroParticule et Cosmologie, Universit\'{e} Paris Diderot, CNRS/IN2P3, CEA/lrfu, Observatoire de Paris, Sorbonne Paris Cit\'{e}, 10, rue Alice Domon et L\'{e}onie Duquet, 75205 Paris Cedex 13, France\goodbreak
\and
African Institute for Mathematical Sciences, 6-8 Melrose Road, Muizenberg, Cape Town, South Africa\goodbreak
\and
Agenzia Spaziale Italiana Science Data Center, Via del Politecnico snc, 00133, Roma, Italy\goodbreak
\and
Aix Marseille Universit\'{e}, CNRS, LAM (Laboratoire d'Astrophysique de Marseille) UMR 7326, 13388, Marseille, France\goodbreak
\and
Astrophysics Group, Cavendish Laboratory, University of Cambridge, J J Thomson Avenue, Cambridge CB3 0HE, U.K.\goodbreak
\and
Astrophysics \& Cosmology Research Unit, School of Mathematics, Statistics \& Computer Science, University of KwaZulu-Natal, Westville Campus, Private Bag X54001, Durban 4000, South Africa\goodbreak
\and
Atacama Large Millimeter/submillimeter Array, ALMA Santiago Central Offices, Alonso de Cordova 3107, Vitacura, Casilla 763 0355, Santiago, Chile\goodbreak
\and
CITA, University of Toronto, 60 St. George St., Toronto, ON M5S 3H8, Canada\goodbreak
\and
CNRS, IRAP, 9 Av. colonel Roche, BP 44346, F-31028 Toulouse cedex 4, France\goodbreak
\and
CRANN, Trinity College, Dublin, Ireland\goodbreak
\and
California Institute of Technology, Pasadena, California, U.S.A.\goodbreak
\and
Centre for Theoretical Cosmology, DAMTP, University of Cambridge, Wilberforce Road, Cambridge CB3 0WA, U.K.\goodbreak
\and
Centro de Estudios de F\'{i}sica del Cosmos de Arag\'{o}n (CEFCA), Plaza San Juan, 1, planta 2, E-44001, Teruel, Spain\goodbreak
\and
Computational Cosmology Center, Lawrence Berkeley National Laboratory, Berkeley, California, U.S.A.\goodbreak
\and
Consejo Superior de Investigaciones Cient\'{\i}ficas (CSIC), Madrid, Spain\goodbreak
\and
DSM/Irfu/SPP, CEA-Saclay, F-91191 Gif-sur-Yvette Cedex, France\goodbreak
\and
DTU Space, National Space Institute, Technical University of Denmark, Elektrovej 327, DK-2800 Kgs. Lyngby, Denmark\goodbreak
\and
D\'{e}partement de Physique Th\'{e}orique, Universit\'{e} de Gen\`{e}ve, 24, Quai E. Ansermet,1211 Gen\`{e}ve 4, Switzerland\goodbreak
\and
Departamento de F\'{\i}sica, Universidad de Oviedo, Avda. Calvo Sotelo s/n, Oviedo, Spain\goodbreak
\and
Department of Astronomy and Astrophysics, University of Toronto, 50 Saint George Street, Toronto, Ontario, Canada\goodbreak
\and
Department of Astrophysics/IMAPP, Radboud University Nijmegen, P.O. Box 9010, 6500 GL Nijmegen, The Netherlands\goodbreak
\and
Department of Physics \& Astronomy, University of British Columbia, 6224 Agricultural Road, Vancouver, British Columbia, Canada\goodbreak
\and
Department of Physics and Astronomy, Dana and David Dornsife College of Letter, Arts and Sciences, University of Southern California, Los Angeles, CA 90089, U.S.A.\goodbreak
\and
Department of Physics and Astronomy, University College London, London WC1E 6BT, U.K.\goodbreak
\and
Department of Physics, Florida State University, Keen Physics Building, 77 Chieftan Way, Tallahassee, Florida, U.S.A.\goodbreak
\and
Department of Physics, Gustaf H\"{a}llstr\"{o}min katu 2a, University of Helsinki, Helsinki, Finland\goodbreak
\and
Department of Physics, Princeton University, Princeton, New Jersey, U.S.A.\goodbreak
\and
Department of Physics, University of California, Santa Barbara, California, U.S.A.\goodbreak
\and
Department of Physics, University of Illinois at Urbana-Champaign, 1110 West Green Street, Urbana, Illinois, U.S.A.\goodbreak
\and
Dipartimento di Fisica e Astronomia G. Galilei, Universit\`{a} degli Studi di Padova, via Marzolo 8, 35131 Padova, Italy\goodbreak
\and
Dipartimento di Fisica e Scienze della Terra, Universit\`{a} di Ferrara, Via Saragat 1, 44122 Ferrara, Italy\goodbreak
\and
Dipartimento di Fisica, Universit\`{a} La Sapienza, P. le A. Moro 2, Roma, Italy\goodbreak
\and
Dipartimento di Fisica, Universit\`{a} degli Studi di Milano, Via Celoria, 16, Milano, Italy\goodbreak
\and
Dipartimento di Fisica, Universit\`{a} degli Studi di Trieste, via A. Valerio 2, Trieste, Italy\goodbreak
\and
Dipartimento di Fisica, Universit\`{a} di Roma Tor Vergata, Via della Ricerca Scientifica, 1, Roma, Italy\goodbreak
\and
Dipartimento di Matematica, Universit\`{a} di Roma Tor Vergata, Via della Ricerca Scientifica, 1, Roma, Italy\goodbreak
\and
Discovery Center, Niels Bohr Institute, Blegdamsvej 17, Copenhagen, Denmark\goodbreak
\and
Dpto. Astrof\'{i}sica, Universidad de La Laguna (ULL), E-38206 La Laguna, Tenerife, Spain\goodbreak
\and
European Southern Observatory, ESO Vitacura, Alonso de Cordova 3107, Vitacura, Casilla 19001, Santiago, Chile\goodbreak
\and
European Space Agency, ESAC, Planck Science Office, Camino bajo del Castillo, s/n, Urbanizaci\'{o}n Villafranca del Castillo, Villanueva de la Ca\~{n}ada, Madrid, Spain\goodbreak
\and
European Space Agency, ESTEC, Keplerlaan 1, 2201 AZ Noordwijk, The Netherlands\goodbreak
\and
Facolt\`{a} di Ingegneria, Universit\`{a} degli Studi e-Campus, Via Isimbardi 10, Novedrate (CO), 22060, Italy\goodbreak
\and
Gran Sasso Science Institute, INFN, viale F. Crispi 7, 67100 L'Aquila, Italy\goodbreak
\and
HGSFP and University of Heidelberg, Theoretical Physics Department, Philosophenweg 16, 69120, Heidelberg, Germany\goodbreak
\and
Helsinki Institute of Physics, Gustaf H\"{a}llstr\"{o}min katu 2, University of Helsinki, Helsinki, Finland\goodbreak
\and
INAF - Osservatorio Astronomico di Padova, Vicolo dell'Osservatorio 5, Padova, Italy\goodbreak
\and
INAF - Osservatorio Astronomico di Roma, via di Frascati 33, Monte Porzio Catone, Italy\goodbreak
\and
INAF - Osservatorio Astronomico di Trieste, Via G.B. Tiepolo 11, Trieste, Italy\goodbreak
\and
INAF/IASF Bologna, Via Gobetti 101, Bologna, Italy\goodbreak
\and
INAF/IASF Milano, Via E. Bassini 15, Milano, Italy\goodbreak
\and
INFN, Sezione di Bologna, Via Irnerio 46, I-40126, Bologna, Italy\goodbreak
\and
INFN, Sezione di Roma 1, Universit\`{a} di Roma Sapienza, Piazzale Aldo Moro 2, 00185, Roma, Italy\goodbreak
\and
INFN, Sezione di Roma 2, Universit\`{a} di Roma Tor Vergata, Via della Ricerca Scientifica, 1, Roma, Italy\goodbreak
\and
INFN/National Institute for Nuclear Physics, Via Valerio 2, I-34127 Trieste, Italy\goodbreak
\and
IPAG: Institut de Plan\'{e}tologie et d'Astrophysique de Grenoble, Universit\'{e} Grenoble Alpes, IPAG, F-38000 Grenoble, France, CNRS, IPAG, F-38000 Grenoble, France\goodbreak
\and
IUCAA, Post Bag 4, Ganeshkhind, Pune University Campus, Pune 411 007, India\goodbreak
\and
Imperial College London, Astrophysics group, Blackett Laboratory, Prince Consort Road, London, SW7 2AZ, U.K.\goodbreak
\and
Infrared Processing and Analysis Center, California Institute of Technology, Pasadena, CA 91125, U.S.A.\goodbreak
\and
Institut N\'{e}el, CNRS, Universit\'{e} Joseph Fourier Grenoble I, 25 rue des Martyrs, Grenoble, France\goodbreak
\and
Institut Universitaire de France, 103, bd Saint-Michel, 75005, Paris, France\goodbreak
\and
Institut d'Astrophysique Spatiale, CNRS (UMR8617) Universit\'{e} Paris-Sud 11, B\^{a}timent 121, Orsay, France\goodbreak
\and
Institut d'Astrophysique de Paris, CNRS (UMR7095), 98 bis Boulevard Arago, F-75014, Paris, France\goodbreak
\and
Institute of Astronomy, University of Cambridge, Madingley Road, Cambridge CB3 0HA, U.K.\goodbreak
\and
Institute of Theoretical Astrophysics, University of Oslo, Blindern, Oslo, Norway\goodbreak
\and
Instituto de Astrof\'{\i}sica de Canarias, C/V\'{\i}a L\'{a}ctea s/n, La Laguna, Tenerife, Spain\goodbreak
\and
Instituto de F\'{\i}sica de Cantabria (CSIC-Universidad de Cantabria), Avda. de los Castros s/n, Santander, Spain\goodbreak
\and
Istituto Nazionale di Fisica Nucleare, Sezione di Padova, via Marzolo 8, I-35131 Padova, Italy\goodbreak
\and
Jet Propulsion Laboratory, California Institute of Technology, 4800 Oak Grove Drive, Pasadena, California, U.S.A.\goodbreak
\and
Jodrell Bank Centre for Astrophysics, Alan Turing Building, School of Physics and Astronomy, The University of Manchester, Oxford Road, Manchester, M13 9PL, U.K.\goodbreak
\and
Kavli Institute for Cosmology Cambridge, Madingley Road, Cambridge, CB3 0HA, U.K.\goodbreak
\and
LAL, Universit\'{e} Paris-Sud, CNRS/IN2P3, Orsay, France\goodbreak
\and
LAPTh, Univ. de Savoie, CNRS, B.P.110, Annecy-le-Vieux F-74941, France\goodbreak
\and
LERMA, CNRS, Observatoire de Paris, 61 Avenue de l'Observatoire, Paris, France\goodbreak
\and
Laboratoire AIM, IRFU/Service d'Astrophysique - CEA/DSM - CNRS - Universit\'{e} Paris Diderot, B\^{a}t. 709, CEA-Saclay, F-91191 Gif-sur-Yvette Cedex, France\goodbreak
\and
Laboratoire Traitement et Communication de l'Information, CNRS (UMR 5141) and T\'{e}l\'{e}com ParisTech, 46 rue Barrault F-75634 Paris Cedex 13, France\goodbreak
\and
Laboratoire de Physique Subatomique et Cosmologie, Universit\'{e} Grenoble-Alpes, CNRS/IN2P3, 53, rue des Martyrs, 38026 Grenoble Cedex, France\goodbreak
\and
Laboratoire de Physique Th\'{e}orique, Universit\'{e} Paris-Sud 11 \& CNRS, B\^{a}timent 210, 91405 Orsay, France\goodbreak
\and
Lawrence Berkeley National Laboratory, Berkeley, California, U.S.A.\goodbreak
\and
Lebedev Physical Institute of the Russian Academy of Sciences, Astro Space Centre, 84/32 Profsoyuznaya st., Moscow, GSP-7, 117997, Russia\goodbreak
\and
Max-Planck-Institut f\"{u}r Astrophysik, Karl-Schwarzschild-Str. 1, 85741 Garching, Germany\goodbreak
\and
McGill Physics, Ernest Rutherford Physics Building, McGill University, 3600 rue University, Montr\'{e}al, QC, H3A 2T8, Canada\goodbreak
\and
National University of Ireland, Department of Experimental Physics, Maynooth, Co. Kildare, Ireland\goodbreak
\and
Niels Bohr Institute, Blegdamsvej 17, Copenhagen, Denmark\goodbreak
\and
Optical Science Laboratory, University College London, Gower Street, London, U.K.\goodbreak
\and
SB-ITP-LPPC, EPFL, CH-1015, Lausanne, Switzerland\goodbreak
\and
SISSA, Astrophysics Sector, via Bonomea 265, 34136, Trieste, Italy\goodbreak
\and
School of Physics and Astronomy, Cardiff University, Queens Buildings, The Parade, Cardiff, CF24 3AA, U.K.\goodbreak
\and
School of Physics and Astronomy, University of Nottingham, Nottingham NG7 2RD, U.K.\goodbreak
\and
Sorbonne Universit\'{e}-UPMC, UMR7095, Institut d'Astrophysique de Paris, 98 bis Boulevard Arago, F-75014, Paris, France\goodbreak
\and
Space Research Institute (IKI), Russian Academy of Sciences, Profsoyuznaya Str, 84/32, Moscow, 117997, Russia\goodbreak
\and
Space Sciences Laboratory, University of California, Berkeley, California, U.S.A.\goodbreak
\and
Special Astrophysical Observatory, Russian Academy of Sciences, Nizhnij Arkhyz, Zelenchukskiy region, Karachai-Cherkessian Republic, 369167, Russia\goodbreak
\and
Sub-Department of Astrophysics, University of Oxford, Keble Road, Oxford OX1 3RH, U.K.\goodbreak
\and
Theory Division, PH-TH, CERN, CH-1211, Geneva 23, Switzerland\goodbreak
\and
UPMC Univ Paris 06, UMR7095, 98 bis Boulevard Arago, F-75014, Paris, France\goodbreak
\and
Universit\'{e} de Toulouse, UPS-OMP, IRAP, F-31028 Toulouse cedex 4, France\goodbreak
% ...
}

\authorrunning{Planck Collaboration}
\titlerunning{Diffuse component separation: CMB maps}

\begin{document}

\abstract{We present foreground-reduced CMB maps derived from the full
  \Planck\ data set in both temperature and polarization. Compared to
  the corresponding \Planck\ 2013 temperature sky maps, the total data
  volume is larger by a factor of 3.2 for frequencies between 30 and
  70\GHz, and by 1.9 for frequencies between 100 and 857\GHz.  In
  addition, systematic errors in the forms of
  temperature-to-polarization leakage, analogue-to-digital conversion
  uncertainties, and very long time constant errors have been
  dramatically reduced, to the extent that the cosmological
  polarization signal may now be robustly recovered on angular scales
  $\ell\gtrsim40$. On the very largest scales, instrumental systematic
  residuals are still non-negligible compared to the expected
  cosmological signal, and modes with $\ell < 20$ are accordingly
  suppressed in the current polarization maps by high-pass
  filtering. As in 2013, four different CMB component separation
  algorithms are applied to these observations, providing a measure of
  stability with respect to algorithmic and modelling choices. The
  resulting polarization maps have rms instrumental noise ranging
  between 0.21 and 0.27$\,\mu\textrm{K}$ averaged over $55\arcm$
  pixels, and between 4.5 and 6.1$\,\mu\textrm{K}$ averaged over
  $3\parcm4$ pixels.  The cosmological parameters derived from the
  analysis of temperature power spectra are in agreement at the
  $1\sigma$ level with the \Planck\ 2015 likelihood.  Unresolved
  mismatches between the noise properties of the data and simulations
  prevent a satisfactory description of the higher-order statistical
  properties of the polarization maps. Thus, the primary applications
  of these polarization maps are those that do not require massive
  simulations for accurate estimation of uncertainties, for instance
  estimation of cross-spectra and cross-correlations, or stacking
  analyses.  However, the amplitude of primordial non-Gaussianity is
  consistent with zero within $2\sigma$ for all local, equilateral,
  and orthogonal configurations of the bispectrum, including for
  polarization $E$-modes. Moreover, excellent agreement is found
  regarding the lensing $B$-mode power spectrum, both internally among
  the various component separation codes and with the best-fit
  \Planck\ 2015 \LCDM\ model.}


\maketitle

\alltwentyfifteenresultspapers
\alltwentythirteenresultspapers

% main text 
\section{Introduction}
\label{sec:introduction}

This paper, one of a set associated with the 2015 release of data from
the \Planck\footnote{\Planck\ (\url{http://www.esa.int/Planck}) is a
  project of the European Space Agency (ESA) with instruments provided
  by two scientific consortia funded by ESA member states and led by
  Principal Investigators from France and Italy, telescope reflectors
  provided through a collaboration between ESA and a scientific
  consortium led and funded by Denmark, and additional contributions
  from NASA (USA).} satellite, presents maps of the Cosmic Microwave
Background (CMB) anisotropies derived from the full \Planck\ data set,
for a total of 50 months of observations from the Low Frequency
Instrument (LFI) and 29 for the High Frequency Instrument (HFI)
\citep{planck2014-a01}. This analysis updates the temperature-only
analysis of the first 15.5 months of \Planck\ observations discussed
in \citet{planck2013-p06}, and presents the first CMB polarization
maps derived from \Planck\ observations.

Much of the \Planck\ analysis effort since the 2013 data release has
revolved around understanding and reducing instrumental systematic
uncertainties. As summarized in \citet{planck2014-a01}, this work has
been highly successful, reducing the net power from systematic errors
in the HFI CMB channels by almost two orders of magnitude on large
angular scales \citep{planck2014-a13}. The main contributions to these
improvements have come from improved temperature-to-polarization
leakage modelling, reduced Analogue-to-Digital Conversion (ADC)
errors, and improved modelling of Very Long Time Constants (VLTCs;
\citealp{planck2014-a07,planck2014-a09}). With these improvements, the
\Planck\ observations are now sufficiently free of instrumental
artifacts to allow a robust determination of the CMB polarization
anisotropies on intermediate and small angular scales, covering
multipoles $\ell \gtrsim 20$. However, as described both in this paper
and in \citet{planck2014-a09,planck2014-a13}, residual systematics are
still not negligible compared to the CMB signal on the very largest
scales ($\ell \lesssim 20$), and these modes are therefore removed by
a high-pass filter from the current maps. The rate of progress is still
excellent, though, and updated all-scale maps with low large-scale
systematics are expected to be released in the near future.

For temperature, the most significant improvement in the \Planck\ 2015
analysis pipeline is absolute calibration based on the orbital CMB
dipole rather than the Solar dipole \citep{planck2014-a01}. This
change, combined with a better understanding of both the
\Planck\ beams and transfer functions
\citep{planck2014-a07,planck2014-a09}, has reduced the uncertainties
in absolute calibration to a few tenths of a percent, and the
agreement between LFI, HFI, and the Wilkinson Microwave Anisotropy Probe (WMAP) 
has improved to the level of the uncertainties \citep{planck2014-a01}.

The component separation efforts of the \Planck\ 2015 release are
summarized in three papers. The current paper is dedicated to CMB
extraction, and presents the main \Planck\ 2015 CMB maps in both
temperature and polarization. \citet{planck2014-a12} addresses
astrophysical component separation as implemented by
\texttt{Commander} \citep{eriksen2004,eriksen2008}, a Bayesian
component separation algorithm, and presents a global model of the
microwave temperature sky ranging from 408\MHz\ to 857\GHz, including
detailed maps of synchrotron, free-free, spinning dust, thermal dust,
and CO emission.  In addition, a few minor components (line emission
around 90\GHz\ and the thermal Sunyaev-Zeldovich effect near the Coma
and Virgo clusters) are included in the model, as are instrumental
parameters in the form of calibration, bandpasses, monopoles, and
dipoles. The corresponding polarization model includes only
synchrotron and thermal dust emission. \citet{planck2014-a31}
presents a detailed analysis of the foregrounds below
$\sim$100\GHz\ in both temperature and polarization. For detailed
descriptions of the various foreground components relevant for
microwave component separation, see either Sect.~2 of
\citet{planck2013-p06} or Sect.~4 of \citet{planck2014-a12}.

The foreground amplitude relative to CMB polarization is such that
effective foreground suppression is required for almost any
cosmological analysis, but the optimal approach depends sensitively on
the topic in question. For instance, because the fluctuation power of
diffuse polarized foregrounds decays as a power-law in multipole
moment $\ell$ \citep{page2007,gold2010,planck2014-a12,
  planck2014-XXX}, it is of greater importance for high-$\ell$ CMB
power spectrum and likelihood estimation to minimize noise sensitivity
and to marginalize over unresolved point sources, than it is to model
diffuse foregrounds with high accuracy. In this case, it is more
convenient to parametrize the residual foregrounds in terms of power
spectrum models, and to marginalize over these in terms of a few
global parameters, than to marginalize over a large number of
per-pixel foreground parameters.  The \Planck\ 2015 CMB likelihood
therefore employs cross-spectra coupled to simple harmonic space
foreground modelling \citep{planck2014-a13}, rather than the detailed
foreground modelling described in this paper.

Similarly, because of the low --- but non-negligible --- level of
residual instrumental systematics on large angular scales in the
\Planck\ 2015 data, the low-$\ell$ likelihood also implements a
special purpose cleaning algorithm, in terms of a simple template fit
including only the cleanest 30, 70, and 353\GHz\ channels
\citep{planck2014-a07,planck2014-a09,planck2014-a13}. This approach is
similar to that adopted by the 9-year WMAP likelihood
\citep{bennett2012}, and allows for easy propagation of uncertainties
from correlated noise in terms of full pixel-pixel noise covariance
matrices.

Other applications, however, such as gravitational lensing and
Integrated Sachs-Wolfe (ISW) reconstructions
\citep{planck2014-a17,planck2014-a26}, constraints on isotropy and
statistics \citep{planck2014-a18}, searches for primordial
non-Gaussianity \citep{planck2014-a19}, and constraints on global
geometry and topological defects \citep{planck2014-a20}, require
actual CMB maps, and these are all based on the products described in
this paper.

As in 2013, we apply four complementary CMB component separation
algorithms to the \Planck\ 2015 sky maps. In alphabetical order, these
are 1)~\texttt{Commander}, a parametric pixel-based Bayesian CMB Gibbs
sampler \citep{eriksen2004,eriksen2008}; 2)~\texttt{NILC}, a
needlet-based internal linear combination method
\citep{2012MNRAS.419.1163B, 2013MNRAS.435...18B}; 3)~\texttt{SEVEM},
which implements linear template fitting based on internal templates
in pixel space \citep{2008A&A...491..597L,2012MNRAS.420.2162F}; and
4)~\texttt{SMICA}, a semi-blind spectral-matching algorithm fully
defined in harmonic space \citep{cardoso2008, planck2013-p06}. These
codes were all applied to CMB temperature reconstruction in
\citet{planck2013-p06}, and have now been extended to polarization, as
described in Appendices~\ref{sec:commander}--\ref{sec:smica}. In
addition, each algorithm is applied to several subsets of the full
data set, including half-ring, half-mission, and yearly data splits
\citep{planck2014-a01}. Comparing the resulting maps, both between
algorithms and data splits, provides a good understanding of both
instrumental and algorithmic uncertainties.

The paper is organized as follows. In Sect.~\ref{sec:inputs} we
describe the \Planck\ 2015 data selection and pre-processing. In
Sect.~\ref{sec:pipelines} we briefly review the component separation
methods, deferring mathematical details to
Appendices~\ref{sec:commander}--\ref{sec:smica}. In
Sect.~\ref{sec:maps} we present the derived CMB maps. In
Sect.~\ref{sec:crosscorrelation} we quantify the residual foreground
emission present in the maps by cross-correlation with foreground
templates. In Sect.~\ref{sec:powspec} we present angular power spectra
and corresponding cosmological parameters.  In
Sects.~\ref{sec:higherorder} and \ref{sec:lensing} we consider
higher-order statistics and gravitational lensing. In
Sect.~\ref{sec:discussion} we summarize the main features and
limitations of these maps, and provide recommendations on their
applications. We conclude in Sect.~\ref{sec:conclusions}.

\section{Data selection and pre-processing}
\label{sec:inputs}

In this paper we use the full-mission \Planck\ data
\citep{planck2014-a07,planck2014-a09} and accompanying
simulations~\citep{planck2014-a14}.  The CMB temperature maps are
derived using all nine frequency channels, from 30 to 857\,GHz.  One
of our component separation methods, \texttt{Commander}, additionally
uses the 9-year WMAP temperature sky maps~\citep{bennett2012} and a
408$\,$MHz survey map~\citep{haslam1982}.  The CMB polarization maps
are derived from the seven frequency channels sensitive to
polarization, from 30 to 353\,GHz. In most cases, the component
separation methods use frequency channel maps as input, with the
exception of the temperature analysis performed by \commander, which
uses maps from subsets of detectors in each frequency channel, as
specified in Table~1 of~\citet{planck2014-a12}.

The primary foreground-reduced CMB maps are derived from full-mission
maps, maximizing signal-to-noise ratio and minimizing destriping
errors.  In addition, a number of data splits are analysed to enable
internal consistency checks and to make estimates of the properties of
the CMB maps.  These data splits can also be used in analyses where
more than one map is required as input.  For the purposes of component
separation, each subset of the data must contain the same combination
of frequency channels as the full-mission data set.  The following
data splits have been analysed:

\begin{itemize}

\item maps from the first and second half of each pointing period
  (``half ring''; HR1 and HR2);
  
\vskip 3pt
  
\item maps from odd and even years, consisting of year 1+3 maps for
  the LFI channels plus year 1 maps for the HFI channels, and year 2+4
  maps for the LFI channels plus year 2 maps for the HFI channels (YR1
  and YR2);
  
\vskip 3pt
  
\item maps from the first and second half of the mission, consisting of year
  1+2 maps for the LFI channels plus half-mission 1 maps for the HFI
  channels, and year 3+4 maps for the LFI channels plus half-mission 2
  for the HFI channels (HM1 and HM2).
  
\end{itemize}

The HFI maps in the half-ring data split have spurious correlations
between them~\citep{planck2014-a09}.  If these maps are used to
estimate the noise level in the power spectrum, the correlations cause
the estimate to be biased low.  This was already seen in the analysis
of the nominal-mission CMB maps in the 2013
release~\citep{planck2013-p06}, and the same effect is seen in the
full-mission maps.  The odd- and even-year data split does not include
all of the data; it omits HFI survey 5.  As a consequence, the HFI
maps have more missing pixels than in the half-mission data split.
For these reasons, the half-mission maps have been used as the primary
data split for assessing the properties of the CMB maps and for
further analysis.

\subsection{Data}

The frequency maps are described in detail in \citet{planck2014-a03}
for LFI and \citet{planck2014-a08} for HFI.  The maps have several
features that are relevant to component separation.  We summarize
them here, and the reader is recommended to consult the references for
further details:

\begin{itemize}

\item Monopole and dipole contributions from the CMB, CIB, and other
  astrophysical components are estimated and removed during 
  mapmaking \citep{planck2014-a07,planck2014-a09}.  This has
  an effect on component separation, and each method treats monopoles 
  and dipoles in a different way, as described in
  Appendices~\ref{sec:commander}--\ref{sec:smica}.
                                                                            
\vskip 3pt
  
\item The HFI maps are corrected for zodiacal light emission (ZLE) by
  subtracting a model of the emission at the ring level during
  mapmaking~\citep{planck2014-a09}.  This differs from the treatment
  in the 2013 component separation~\citep{planck2013-p06}, where the
  ZLE model was not subtracted.

\vskip 3pt
  
\item Leakage from intensity to polarization due to bandpass
  mismatches between detectors is estimated based on the ground
  measurements of the 
  bandpasses~\citep{planck2014-a05,planck2014-a07,planck2014-a09}. These
  estimates are subtracted from the polarized maps.

\vskip 3pt
  
\item The HFI maps at 100, 143, and 217\GHz\ are renormalized in
  order to correct for far-sidelobe effects in the
  calibration~\citep{planck2014-a08}.

\vskip 3pt
  
\item Missing pixels are filled with the average values of
  pixels in the surrounding area. This area is defined as being within
  a radius of 1\degr\ for LFI maps and HFI frequency maps, and within
  a radius of 1\fdg5 for HFI detector subset maps.

\end{itemize}

Point source catalogues and masks have been provided as input to the
component separation.  Construction of the PCCS2 catalogue is
described in~\citet{planck2014-a35}.  Masks have been constructed
based on these catalogues.  For each frequency channel, the intensity
source mask removes all sources detected down to a S/N threshold of 4
for the LFI channels and 5 for the HFI channels.  For each polarized
frequency channel, the polarization source mask removes all sources
detected at 99.99\,\% significance or greater.  The details of the
masking procedures used by each component separation method are
described in Appendices~\ref{sec:commander}--\ref{sec:smica}.

\subsection{Simulations}
\label{sec:ffp8_simulations}

To validate our results, we analyse realistic simulations of the
\Planck\ data set called full focal plane 8 (FFP8).  They are based on
detailed models of the instrument and sky, and are described in full
in \citet{planck2014-a14}.  We summarize their contents here.

\subsubsection{CMB}

The CMB was simulated using an input \LCDM\ model based on the 2013
cosmological parameter results~\citep{planck2013-p11}.  The fiducial
simulation contains no primordial tensor modes or primordial
non-Gaussianity.  However, four variants of the same CMB realization
have been produced that include non-zero values of the
tensor-to-scalar ratio and non-Gaussianity of a local type.

\subsubsection{Foregrounds} 

The \Planck\ sky model (PSM) has been used to simulate the foreground
components.  The intensity part of the simulation includes all
astrophysical components that were identified in the 2013 release.
The diffuse components that are relevant at low frequencies consist of
synchrotron, free-free, and anomalous dust.  At high frequencies, CO,
thermal dust, and CIB are included.  The foreground modelling has been
improved in this version of the simulations.  In particular,
\Planck\ 353\,GHz data were used to improve the frequency scaling of
the dust emission.  In polarization, the diffuse foregrounds are
synchrotron and thermal dust.  The extragalactic emission from radio
and infrared sources has been simulated in intensity and polarization,
and the SZ effect from clusters of galaxies has been included in
intensity.

\subsubsection{Simulated observations} 

Time-ordered data (TOD) for each detector were simulated using the
satellite pointing, and the individual detector beams, bandpasses,
noise properties, and data flags.  The same mapmaking used for the
data is used to generate maps from the simulated TOD.  All of the maps
from subsets of the data have also been generated from the
simulations.

Two versions of the maps are available, with and without bandpass
mismatch leakage.  The latter is simulated using the average bandpass
for all detectors in a frequency channel, eliminating the leakage
effect.  The version of the maps without bandpass leakage is
considered in this paper.

In addition to the fiducial maps, a set of 10\,000 Monte Carlo
realizations of CMB and noise has been generated.  These realizations
are intended to be used to assess the uncertainties on the results.

\subsubsection{Mismatch between simulations and data}

In analysing the simulations, a number of deficiencies have become
evident.  First, the amplitude of the CMB component does not match
that of the data.  This is the (expected) consequence of the fact that
the CMB model for the simulations was specified before the 
recalibration of the \Planck\ data between the 2013 release and the
present one was completed.  This mismatch can be mitigated by increasing the
amplitude of the CMB simulations by 1.3\,\% when comparing them to the
data.  Second, the noise properties of the simulated maps do not
precisely match those of the data.  This does not significantly affect
the analysis of the CMB temperature maps, since they are
signal-dominated.  However, it does affect the analysis of CMB
polarization maps, because they are more noisy.  The noise mismatch
appears to be scale-dependent, since the adjustment of the amplitude
of the noise simulations to match the data depends on the resolution
of the maps.  This is explored in Sect.~\ref{sec:higherorder}.

Any analysis that relies on simulations to estimate the uncertainties
of a result from the CMB polarization will be limited by these
mismatches.  Despite this, many analyses are possible, including those
using cross-spectrum, cross-correlation, or stacking techniques.

\begin{table*}[tb]
\begingroup
\newdimen\tblskip \tblskip=5pt
\caption{Masks and statistics of component-separated CMB maps from
  data and FFP8 simulations.\label{tab:solution_parameters}}
%\nointerlineskip
\vskip -4mm
\footnotesize
\setbox\tablebox=\vbox{
\newdimen\digitwidth
\setbox0=\hbox{\rm 0}
\digitwidth=\wd0
\catcode`*=\active
\def*{\kern\digitwidth}
\newdimen\signwidth
\setbox0=\hbox{+}
\signwidth=\wd0
\catcode`!=\active
\def!{\kern\signwidth}
\newdimen\decimalwidth
\setbox0=\hbox{.}
\decimalwidth=\wd0
\catcode`@=\active
\def@{\kern\signwidth}
\halign{ \hbox to 1.9in{#\leaderfil}\tabskip=2.0em&
    \hfil#\hfil\tabskip=2em&
    \hfil#\hfil\tabskip=4em&
    \hfil#\hfil\tabskip=3.5em&
    \hfil#\hfil\tabskip=0em\cr
\noalign{\doubleline}
\omit&\multispan4\hfil\sc Method\hfil\cr
\noalign{\vskip -3pt}
\omit&\multispan4\hrulefill\cr
\noalign{\vskip 3pt}
\omit\hfil\sc Parameter\hfil& \commander& \nilc& \sevem& \smica\cr 
\noalign{\vskip 5pt\hrule\vskip 5pt}
\noalign{\vskip 6pt}
\multispan5{\bf Sky fraction, $\fsky$ [\%]}$^a$\hfil\cr 
\noalign{\vskip 6pt}
\hglue 1em Data **confidence mask **$T$&    81.9& 96.4& 84.5& 85.0\cr
\hglue 1em \phantom{Data} **\phantom{confidence mask} **$Q,U$&  83.1& 96.5& 79.4& 85.0\cr
\noalign{\vskip 2pt}
\hglue 1em \phantom{Data} **preferred mask ***$T$&      \multispan4 \hfil77.6\hfil\cr
\hglue 1em \phantom{Data} **\phantom{preferred mask} ***$Q,U$&    \multispan4 \hfil77.4\hfil\cr
\noalign{\vskip 4pt}
\hglue 1em FFP8 **confidence mask *$T$&   75.3& 96.9& 82.8& 86.5\cr
\hglue 1em \phantom{FFP8} **\phantom{confidence mask} *$Q,U$& 87.5& 96.1& 79.3& 83.4\cr
\noalign{\vskip 2pt}
\hglue 1em \phantom{FFP8} **preferred mask **$T$&     \multispan4 \hfil73.5\hfil\cr
\hglue 1em \phantom{FFP8} **\phantom{preferred mask} **$Q,U$&    \multispan4 \hfil75.7\hfil\cr
\noalign{\vskip 6pt}
\multispan5{\bf Standard deviation at FWHM 10\arcm, N$_{\mathbf{side}}$ = 1024 [$\mu$K]}$^b$\hfil\cr 
\noalign{\vskip 6pt}
\hglue 1em Data **confidence mask **$T$& 101.8 (4.4)& 101.6 (5.4)& 101.4 (3.2)& 101.1 (4.2)\cr
\hglue 1em \phantom{Data} **\phantom{confidence mask}**$Q$& **6.3 (5.8)& **5.3 (4.8)& **6.3 (6.3)& **5.2 (4.7)\cr
\hglue 1em \phantom{Data} **\phantom{confidence mask}**$U$& **6.3 (5.8)& **5.3 (4.7)& **6.3 (5.9)& **5.2 (4.5)\cr
\hglue 1em \phantom{Data} **preferred mask ***$T$& 101.3 (4.4)& 100.9 (5.3)& 101.3 (3.2) & 101.0 (4.1)\cr
\hglue 1em \phantom{Data} **\phantom{preferred mask}***$Q$& **6.3 (5.8)& **5.2 (4.7)& **6.3 (6.3)& **5.2 (4.7)\cr
\hglue 1em \phantom{Data} **\phantom{preferred mask}***$U$& **6.3 (5.8)& **5.2 (4.5)& **6.3 (5.9)& **5.2 (4.5)\cr
\noalign{\vskip 4pt}
\hglue 1em FFP8 **confidence mask *$T$& 104.5 (3.5)& 106.5 (4.6)& 104.0 (4.5)& 104.3 (3.5)\cr
\hglue 1em \phantom{FFP8} **\phantom{confidence mask}*$Q$& **5.6 (5.0)& **5.1 (4.4)& **6.1 (5.6)& **5.0 (4.2)\cr
\hglue 1em \phantom{FFP8} **\phantom{confidence mask}*$U$& **5.7 (5.1)& **5.2 (4.4)& **6.1 (5.6)& **5.0 (4.3)\cr
\hglue 1em \phantom{FFP8} **preferred mask **$T$&    104.3 (3.5)& 107.5 (4.2)& 104.3 (4.4)& 104.5 (3.5)\cr
\hglue 1em \phantom{FFP8} **\phantom{preferred mask}**$Q$&    **5.6 (5.0)& **5.0 (4.3)& **6.1 (5.6)& **4.9 (4.2)\cr
\hglue 1em \phantom{FFP8} **\phantom{preferred mask}**$U$&    **5.7 (5.1)& **5.0 (4.3)& **6.1 (5.6)& **5.0 (4.2)\cr 
\noalign{\vskip 6pt}
\multispan5{\bf Standard deviation at FWHM 160\arcm, N$_{\mathbf{side}}$ = 64 [$\mu$K]}$^b$\hfil\cr 
\noalign{\vskip 6pt}
\hglue 1em Data **confidence mask **$T$& 48.0* (1.30)& 48.7* (1.00)& 47.5* (0.37)& 47.5* (0.79)\cr
\hglue 1em \phantom{Data} **\phantom{confidence mask}**$Q$& *0.25 (0.21)& *0.31 (0.27)& *0.29 (0.26)& *0.29 (0.26)\cr
\hglue 1em \phantom{Data} **\phantom{confidence mask}**$U$& *0.25 (0.21)& *0.31 (0.26)& *0.29 (0.25)& *0.28 (0.25)\cr
\hglue 1em \phantom{Data} **preferred mask ***$T$& 47.4* (1.29)& 47.1* (1.01)& 47.3* (0.37)& 47.1* (0.78)\cr
\hglue 1em \phantom{Data} **\phantom{preferred mask}***$Q$& *0.25 (0.21)& *0.30 (0.27)& *0.29 (0.27)& *0.28 (0.26)\cr
\hglue 1em \phantom{Data} **\phantom{preferred mask}***$U$& *0.25 (0.21)& *0.30 (0.26)& *0.29 (0.25)& *0.28 (0.25)\cr
\noalign{\vskip 4pt}
\hglue 1em FFP8 **confidence mask *$T$& 55.3* (0.34)& 59.8* (0.83)& 55.4* (0.48)& 55.6* (0.50)\cr
\hglue 1em \phantom{FFP8} **\phantom{confidence mask}*$Q$& *0.23 (0.18)& *0.27 (0.22)& *0.27 (0.22)& *0.24 (0.19)\cr
\hglue 1em \phantom{FFP8} **\phantom{confidence mask}*$U$& *0.23 (0.18)& *0.27 (0.23)& *0.26 (0.23)& *0.34 (0.20)\cr
\hglue 1em \phantom{FFP8} **preferred mask **$T$& 55.4* (0.35)& 61.7* (0.79)& 55.6* (0.48)& 56.0* (0.49)\cr
\hglue 1em \phantom{FFP8} **\phantom{preferred mask}**$Q$& *0.23 (0.18)& *0.26 (0.22)& *0.27 (0.22)& *0.24 (0.19)\cr
\hglue 1em \phantom{FFP8} **\phantom{preferred mask}**$U$& *0.23 (0.18)& *0.26 (0.22)& *0.26 (0.23)& *0.24 (0.20)\cr 
\noalign{\vskip 5pt\hrule\vskip 2pt}
}}
\endPlancktablewide 
\tablenote {{\rm a}} Sky fractions are given at $\nside = 2048$ for
$T$ and $\nside = 1024$ for $Q$ and $U$.\par
\tablenote {{\rm b}} Values in brackets are standard deviations of
half-mission half-difference maps, giving an indication of the level
of residual noise and systematic effects.  Standard deviations of $Q$
and $U$ have been computed from high-pass filtered maps.  For details
of the downgrading and high-pass filtering procedures, see text.\par
\endgroup
\end{table*}

\begin{figure*}[t]
\begin{center}
  \includegraphics[width=\columnwidth]{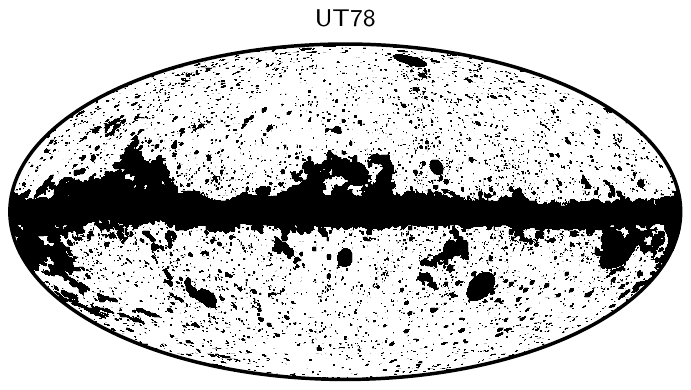}
  \includegraphics[width=\columnwidth]{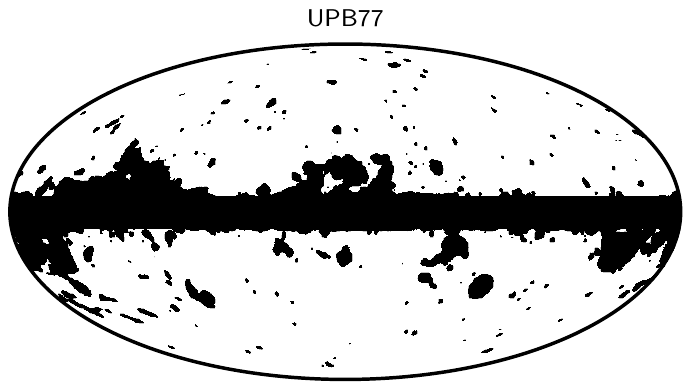}
\end{center}
\caption{Preferred masks for analysing component-separated CMB maps in
  temperature (\emph{left}) and polarization (\emph{right}).}
\label{fig:dx11_masks}
\end{figure*}

\section{Component separation methods}
\label{sec:pipelines}

The four methods used by \Planck\ to separate the CMB from diffuse
foreground emission were described in detail
in~\citet{planck2013-p06}.  They are representative of the main
approaches to component separation developed in recent years.  The
methods can be divided into two types.  The first type assumes only
knowledge of the blackbody spectrum of the CMB, and the foregrounds
are removed by combining the multi-frequency data to minimize the
variance of the CMB component.  The second type constructs an explicit
parameterized model of the CMB and foregrounds with an associated
likelihood, and the CMB component is obtained by maximizing or
sampling from the posterior distribution of the parameters.  Either
type of method may be implemented in the map domain or in the harmonic
domain.  We recall briefly their main features and comment on their
application to polarization data.  Descriptions of the changes in each
algorithm since 2013 are given in the appendices.

\begin{itemize}

\item \commander~\citep{eriksen2006,eriksen2008} is a Bayesian
  parametric method that works in the map domain.  Both the CMB and
  foregrounds are modelled using a physical parameterization in terms
  of amplitudes and frequency spectra, so the method is
  well suited to perform astrophysical component separation in
  addition to CMB extraction \citep{planck2014-a12}.  The joint
  solution for all components is obtained by sampling from the
  posterior distribution of the parameters given the likelihood and a
  set of priors.  To produce a high-resolution CMB map, the separation
  is performed at multiple resolutions with different combinations of
  input channels.  The final CMB map is obtained by combining these
  solutions in the spherical harmonic domain.  This obviates the need
  for the \ruler\ step that was used in 2013 to extend the
  \commander\ solution to high resolution.  A low-resolution version
  of the separation is used to construct the temperature power
  spectrum likelihood for large angular scales, as described
  in~\citet{planck2014-a13}. Note that \commander\ employs detector
  and detector set maps rather than full frequency maps, and excludes
  some specific detector maps judged to have significant systematic
  errors, in addition to incorporating the 9-year WMAP temperature sky
  maps and a 408$\,$MHz survey map. Thus, the selection of data is not
  identical between \commander\ and the other three methods.
  
\vskip 3pt

\item \nilc~\citep{delabrouille2009} is an implementation of internal
  linear combination (ILC) that works in the needlet (wavelet) domain.
  The input maps are decomposed into needlets at a number of different
  angular scales.  The ILC solution for the CMB is produced by
  minimizing the variance at each scale.  This has the advantage that
  the weights used to combine the data can vary with position on the
  sky and also with angular scale.  The solutions are then combined to
  produce the final CMB map.

\vskip 3pt

\item \sevem~\citep{2012MNRAS.420.2162F} is an implementation of the
  template cleaning approach to component separation that works in the
  map domain.  Foreground templates are typically constructed by
  differencing pairs of maps from the low- and high-frequency
  channels.  The differencing is done in order to null the CMB
  contribution to the templates.  These templates are then used to
  clean each CMB-dominated frequency channel by finding a set of
  coefficients to minimize the variance of the map outside of a mask.
  Thus \sevem\ produces multiple foreground-cleaned frequency channel
  maps.  The final CMB map is produced by combining a number of the
  cleaned maps in harmonic space.

\vskip 3pt

\item \smica~\citep{cardoso2008} is a non-parametric method that works
  in the spherical harmonic domain.  Foregrounds are modelled as a
  small number of templates with arbitrary frequency spectra,
  arbitrary power spectra and arbitrary correlation between the
  components.  The solution is obtained by minimizing the mismatch of
  the model to the auto- and cross-power spectra of the frequency
  channel maps.  From the solution, a set of weights is derived to
  combine the frequency maps in the spherical harmonic domain to
  produce the final CMB map.  Maps of the total foreground emission in
  each frequency channel can also be produced.  In the analysis
  performed for the 2013 release \citep{planck2013-p06}, \smica\ was
  the method that performed best on the simulated temperature data.

\end{itemize}

\subsection{Extension to polarization}
\label{sec:polarization}

The methods described above were applied to \Planck\ temperature data
for the 2013 release~\citep{planck2013-p06}, and they have been
extended to operate on polarization data in the present work.  A key
distinction between the methods is the choice of operating domain.
Two of the methods, \commander\ and \sevem, operate in the map domain,
so it is most natural for them to do the polarized component
separation on the $Q$ and $U$ maps.  The other two methods, \nilc\ and
\smica, operate in the harmonic or needlet domain.  An intrinsic part
of the transform of polarized maps to these domains is the
decomposition of $Q$ and $U$ into $E$ and $B$ modes, which is
accomplished by using spherical harmonic transforms on the full sky.
Thus these two methods perform their separation directly on $E$ and
$B$.

\subsection{Outputs}
\label{sec:outputs}

In addition to producing CMB maps, each method provides ``confidence''
masks to define the region of the sky in which the CMB solution is
trusted in temperature and polarization.  The procedure each method
uses to define the mask is described in
Appendices~\ref{sec:commander}--\ref{sec:smica}.  The confidence masks
are used to define masks for further analysis of the data.  Two of the
pipelines, \commander\ and \smica\, also produce foreground products,
described in \citet{planck2014-a12}.

The first 1000 Monte Carlo realizations of CMB and noise have been
propagated through the four pipelines.  This has been done twice, once
using the parameters derived from the data, and once using the
parameters derived from the fiducial FFP8 maps, to provide a set of
simulations to accompany each data set.

The methods produce maps at different resolutions, as described in
Appendices~\ref{sec:commander}--\ref{sec:smica}.  The products have
been brought to a standard resolution to compare them and for
distribution.  Standard resolution temperature maps have a Gaussian
beam of 5\arcm\ FWHM and
\healpix\footnote{\url{http://healpix.sourceforge.net}} resolution
$\nside = 2048$.  Standard resolution polarization maps have a
Gaussian beam of 10\arcm\ FWHM and \healpix\ resolution $\nside =
1024$.  If maps are produced at higher resolution then they are
downgraded to these standard resolutions.  The downgrading procedure
for maps is to decompose them into spherical harmonics ($T$, or $E$
and $B$, as appropriate) on the full sky at the input
\healpix\ resolution.  The spherical harmonic coefficients, $a_{\ell
  m}$, are convolved to the new resolution using
\begin{equation}
a_{\ell m}^{\rm out} = \frac{b_{\ell}^{\rm out}p_{\ell}^{\rm
    out}}{b_{\ell}^{\rm in}p_{\ell}^{\rm in}} a_{\ell m}^{\rm in}
\end{equation}
where $b_{\ell}$ is the beam transfer function, $p_{\ell}$ is the
\healpix\ pixel window function, and the ``in'' and ``out''
superscripts denote the input and output resolutions.  They are then
synthesized into a map directly at the output \healpix\ resolution.
Masks are downgraded in a similar way.  The binary mask at the
starting resolution is first downgraded like a temperature map.  The
smooth downgraded mask is then thresholded by setting pixels where the
value is less than $0.9$ to zero and all others to unity to make a
binary mask.  This has the effect of enlarging the mask to account for
the smoothing of the signal. In addition to the standard resolution
products, the maps, masks, and Monte Carlo realizations have been
downgraded to lower resolutions for analyses that need them, using the
above procedure.

The polarization maps and Monte Carlo realizations have been
decomposed into $E$ and $B$ mode maps, and downgraded to lower
resolutions too, for analyses that work on $E$ and $B$ directly.  The
decomposition is done on the full-sky maps using spherical harmonic
transforms.  The CMB maps from the real-space methods, \commander\ and
\sevem, are inpainted before doing the decomposition.  Both, the standard
\commander\ and \sevem\ CMB maps are inpainted inside their corresponding confidence masks using a
constrained realization (see Appendix~\ref{sec:commander} for
details). For both methods, only the CMB
maps are inpainted, not the Monte Carlo realizations as this would be
too computationally expensive.  The other two methods, \nilc\ and
\smica, work on $E$ and $B$ modes, so it is possible to make $E$ and
$B$ maps directly from their outputs in addition the the standard $Q$
and $U$ maps.

\section{CMB maps}
\label{sec:maps}

\begin{figure*}
\begin{center}
\begin{tabular}{cc}
\includegraphics[width=0.97\columnwidth]{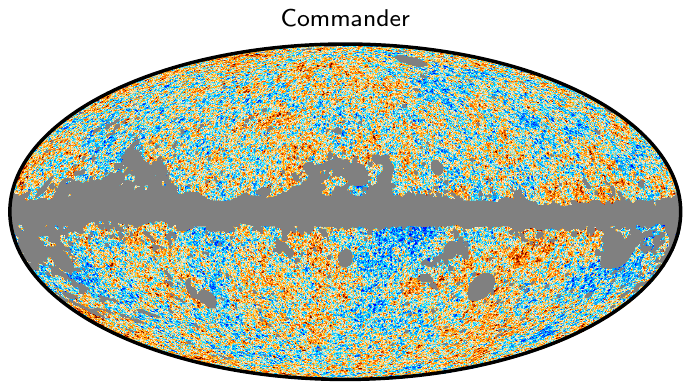}&
\includegraphics[width=0.97\columnwidth]{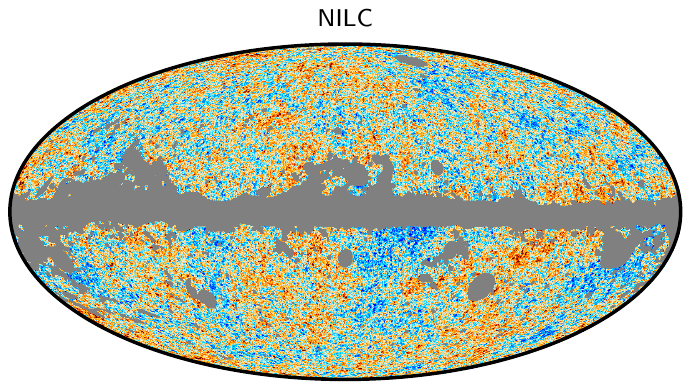}\\
\includegraphics[width=0.97\columnwidth]{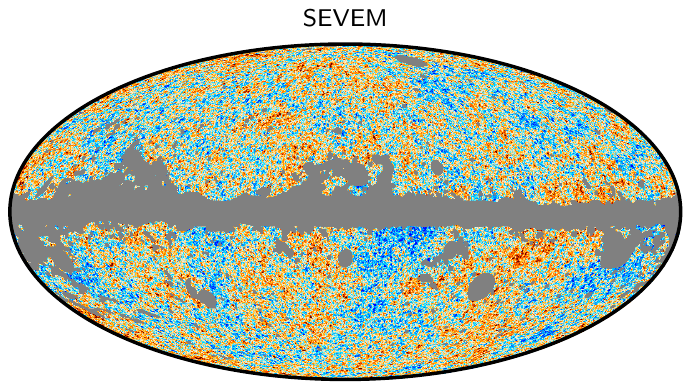}&
\includegraphics[width=0.97\columnwidth]{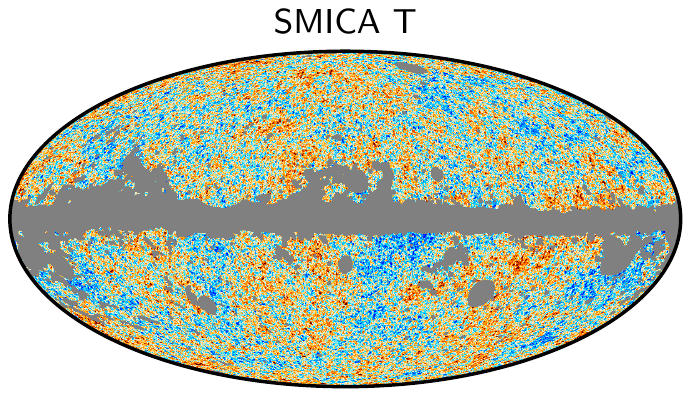}\\
\multicolumn{2}{c}{\includegraphics[height=1cm]{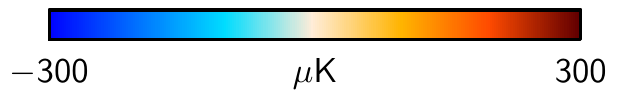}}
\end{tabular}
\end{center}
\caption{Component-separated CMB temperature maps at full resolution,
  FWHM 5\arcm, $\nside = 2048$.}
\label{fig:dx11_map_I}
%\end{figure*}

%\begin{figure*}
\begin{center}
\begin{tabular}{cc}
\includegraphics[width=0.97\columnwidth]{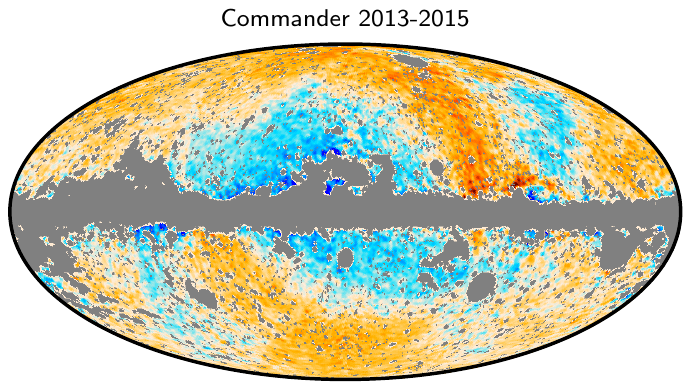}&
\includegraphics[width=0.97\columnwidth]{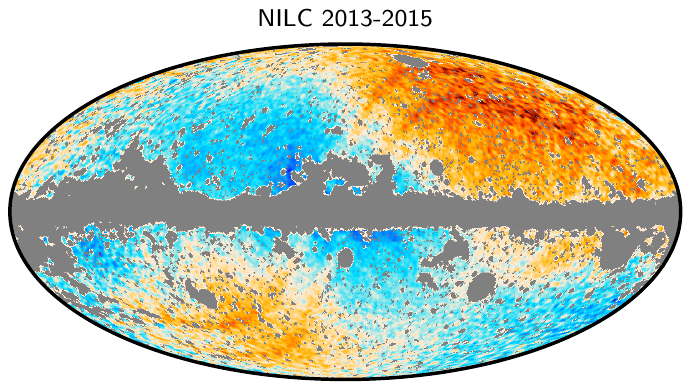}\\
\includegraphics[width=0.97\columnwidth]{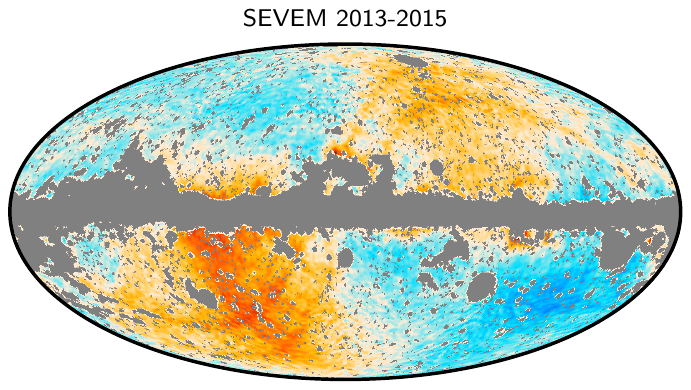}&
\includegraphics[width=0.97\columnwidth]{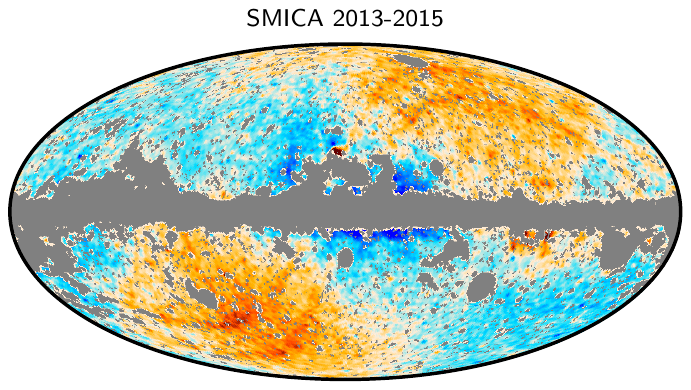}\\
\multicolumn{2}{c}{\includegraphics[height=1cm]{figs/colourbar_15uK}}
\end{tabular}
\end{center}
\caption{Differences between the component-separated CMB temperature
  maps from the 2013 and the 2015 releases.  The maps have been
  smoothed to FWHM 80\arcm\ and downgrading to $\nside = 128$.}
\label{fig:2015_2013_diff_I}
\end{figure*}

\begin{figure*}
\begin{center}
\begin{tabular}{cc}
\includegraphics[width=0.77\columnwidth]{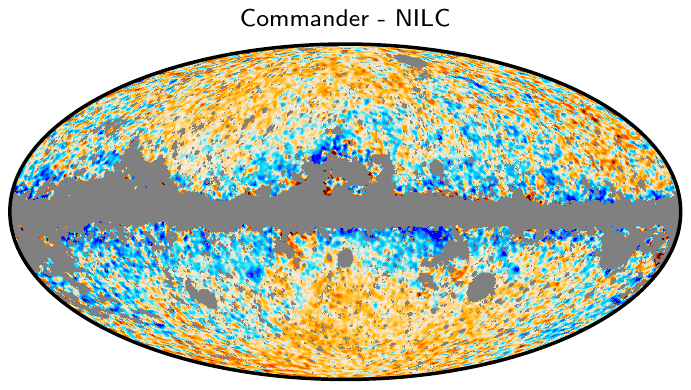}&
\includegraphics[width=0.77\columnwidth]{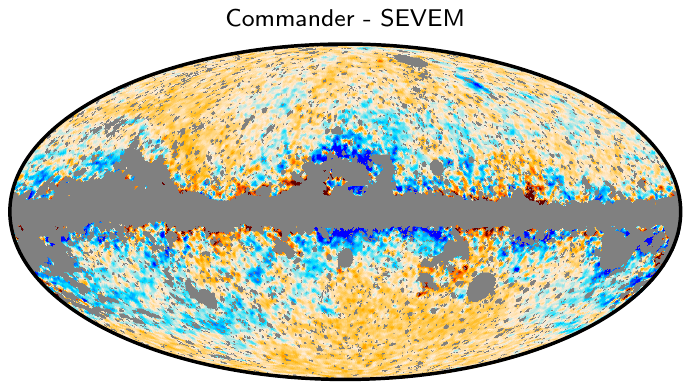}\\
\includegraphics[width=0.77\columnwidth]{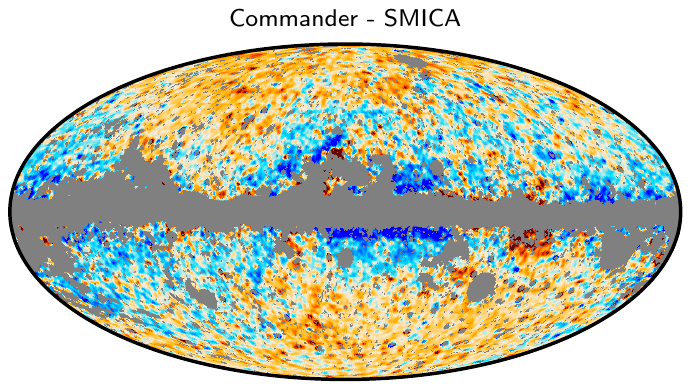}&
\includegraphics[width=0.77\columnwidth]{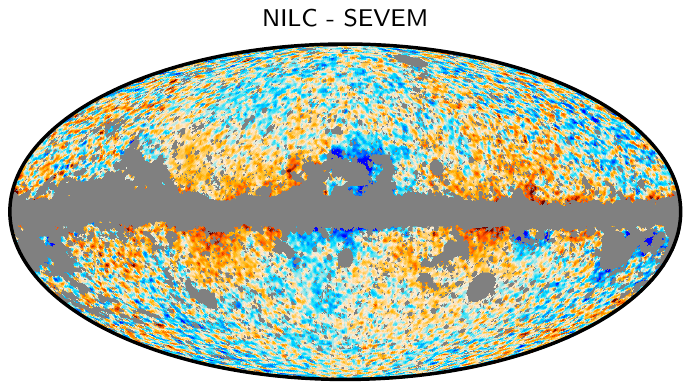}\\
\includegraphics[width=0.77\columnwidth]{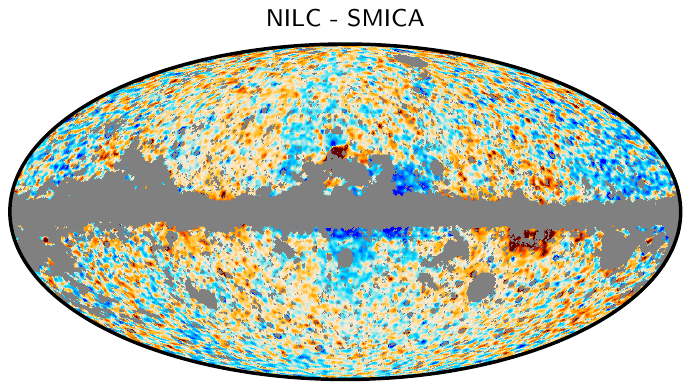}&
\includegraphics[width=0.77\columnwidth]{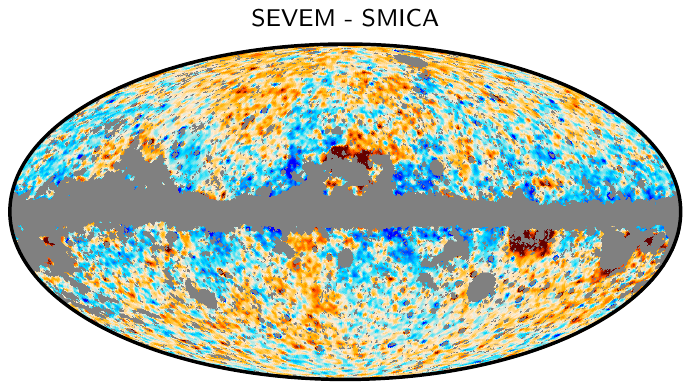}\\
\multicolumn{2}{c}{\includegraphics[height=1cm]{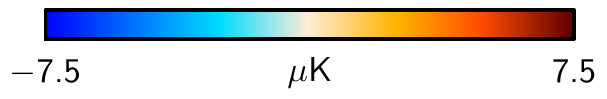}}
\end{tabular}
\end{center}
\caption{Pairwise difference maps between CMB temperature maps.  As in
  the previous Fig.~\ref{fig:2015_2013_diff_I}, the maps have been
  smoothed to FWHM 80\arcm\ and downgraded to $\nside =128$.}
\label{fig:dx11_diff_I}
%\end{figure*}

%\begin{figure*}[b]
\begin{center}
\begin{tabular}{cc}
\includegraphics[width=0.77\columnwidth]{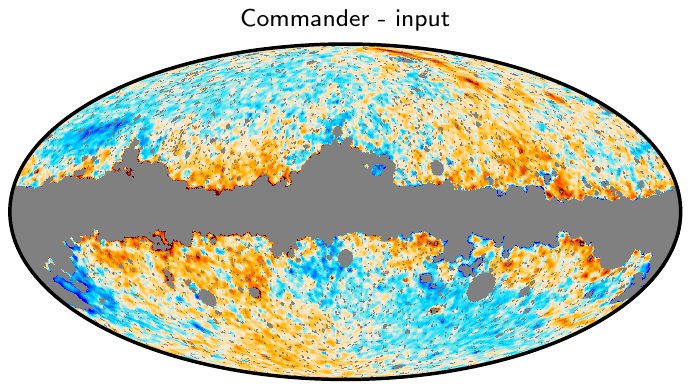}&
\includegraphics[width=0.77\columnwidth]{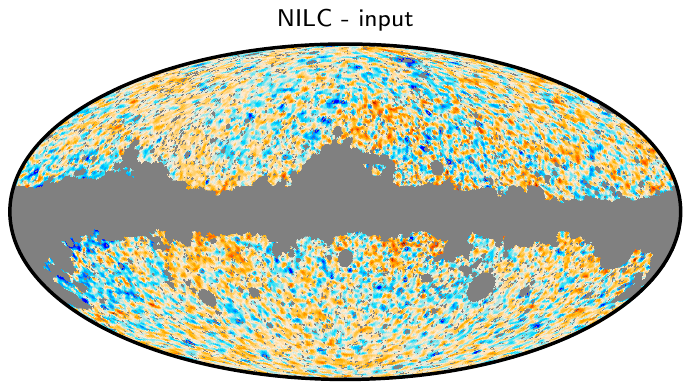}\\
\includegraphics[width=0.77\columnwidth]{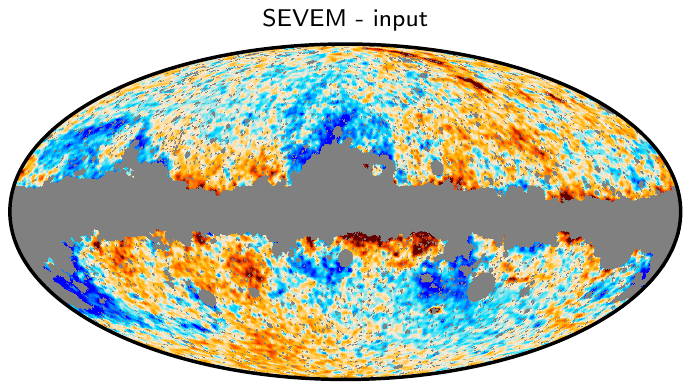}&
\includegraphics[width=0.77\columnwidth]{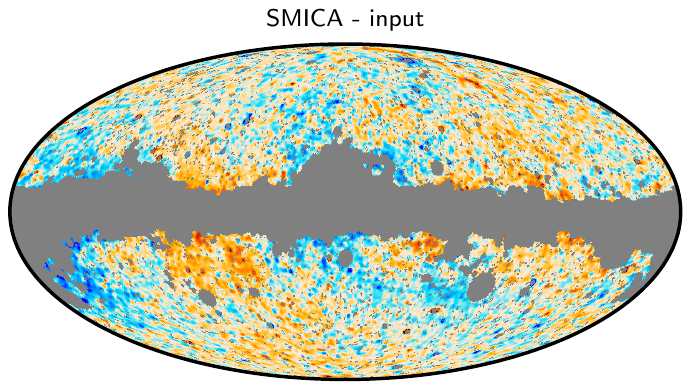}\\
\multicolumn{2}{c}{\includegraphics[height=1cm]{figs/colourbar_7_5uK}}
\end{tabular}
\end{center}
\caption{Difference between output and input CMB temperature maps from
  FFP8 simulations.  Smoothing and downgrading as in
  Fig.~\ref{fig:dx11_diff_I}.}
\label{fig:ffp8_res_I}
\end{figure*}

In this section we present and discuss the component separated CMB
maps in temperature and polarization.  For temperature, we compare the
maps to those extracted from the nominal mission data in 2013.  We
compare the maps from different methods to assess their consistency, and
we use the FFP8 simulations to assess the accuracy of the methods and the 
robustness of the solutions.  Throughout the discussion, we make use
of appropriate masks in order to highlight differences at intermediate
and high Galactic latitudes, ignoring the plane of the Galaxy where
differences are much higher due to the complexity of the foreground
signal and its dominance over the CMB.

\subsection{Temperature maps}
\label{sec:maps_temperature}

Temperature confidence masks produced by the methods have been used to
make combined masks for further analysis of the data.  The first mask
is constructed as the union of the \commander, \sevem, and
\smica\ confidence masks.  The \nilc\ mask is not included in the
union because it removes a significantly smaller fraction of the sky.
This union mask has $\fsky = 77.6\,\%$. We refer to it as
\texttt{UT78} and adopt it as the preferred mask for analysing the
temperature maps.  It is shown on the left in
Fig.~\ref{fig:dx11_masks}.  An extended version of the mask has been
constructed by adding to the \texttt{UT78} mask those pixels where the
the standard deviation between the temperature maps is greater than
10$\,\mu\mathrm{K}$.  This mask has $\fsky = 76.1\,\%$, and we refer
to it as \texttt{UTA76}.  A union mask has been created for the FFP8
simulations in the same way as for the data.  It has $\fsky =
73.5\,\%$, and we refer to it as \texttt{FFP8-UT74}.

The CMB temperature maps produced by the four methods are shown in
Fig.~\ref{fig:dx11_map_I}. No obvious differences are seen in these
maps, and they appear visually consistent outside the mask. An
important assessment of the robustness and consistency of the CMB $T$
component separation solutions is provided in
Fig.~\ref{fig:2015_2013_diff_I}, which shows the differences between
the \Planck\ 2013 and 2015 maps for each method. Several interesting
features may be seen in these differences, most of which correspond
directly to a better understanding of the systematic uncertainties in
the new maps. Starting with \commander, the most striking features are
large-scale swaths tracing the \Planck\ scanning strategy with a
peak-to-peak amplitude of $\sim 10\muK$. This pattern is very similar
to that originally pointed out by \citet{larson2014}, who found this
by subtracting the 9-year WMAP ILC map \citep{bennett2012} from a
template-cleaned version of the \Planck\ 2013 100\GHz\ map. Similar
patterns are also seen in the \commander\ residual maps shown in
Fig.~2 of \citet{planck2014-a12}, corresponding to detector maps that
are rejected from the new 2015 analysis. These structures are
primarily due to two effects, namely destriping errors from bandpass
mismatch between detectors and far sidelobe contamination
\citep{planck2014-a09,planck2014-a12}. By rejecting particularly
susceptible channels in the updated \commander\ analysis, these errors
are greatly reduced in the new map.

Turning to the other three difference maps in
Fig.~\ref{fig:2015_2013_diff_I}, we see that the residuals are
internally very similar, but quite different from the
\commander\ residuals. In these cases, the two most striking features
are, first, a $\sim 5\muK$ quadrupole roughly aligned with the CMB
solar dipole (Galactic coordinates $(l,b)=(264\deg,48\deg)$;
\citealp{planck2014-a01}), and, second, clear traces of the zodiacal
light emission. The former is explained by the fact that the HFI data
processing in 2013 did not subtract the relativistic Doppler
quadrupole of $\sim 6\muK$, which is now subtracted in the 2015
maps. Similarly, the latter is explained by the fact that ZLE was not
subtracted in the mapmaking in 2013, but is subtracted in the updated
processing~\citep{planck2013-p03,planck2014-a09}. \commander, on the
other hand, is less sensitive to residual ZLE, for the following two
reasons. First, in 2013 it used channels only up to 353\,GHz, which
are less affected by the ZLE than the higher frequencies.  Second, by
virtue of fitting independent thermal dust spectral parameters (index
and dust temperature) per pixel, it can efficiently absorb the ZLE in
the thermal dust component. However, some of the ZLE may still be
observed in the \commander\ differences: remnants of the ``red arcs''
typical in the second and fourth quadrant of the sky are just visible
in the difference of the \commander\ solutions, while being very
evident in all other cases.

The residuals seen in Fig.~\ref{fig:2015_2013_diff_I} are small
compared to the typical CMB anisotropies, with features mostly
constrained to $\lesssim 5$--10\,\muK, with a distinct large-scale
pattern. In particular, these small differences are completely
negligible for power spectrum and cosmological parameter
estimation. The only cosmological application for which some care is
needed is the study of large scale isotropy \citep{planck2013-p09},
for instance with respect to CMB quadrupole--octopole alignment. An
updated isotropy analysis of the new sky maps will be presented in
\citet{planck2014-a18}; no significant differences are found compared
to the 2013 results.

Table~\ref{tab:solution_parameters} summarizes main properties of the
CMB maps derived both from the data and from the fiducial set of FFP8
simulations.  We evaluate standard deviations in two cases,
corresponding to high (FWHM 10\arcm, $\nside = 1024$) and intermediate
resolution (FWHM 160\arcm, $\nside = 64$).  The values in parentheses
are standard deviations of half-mission half-difference (HMHD) maps,
and they give an estimate of the level of uncertainties due to
instrumental noise and systematic effects.  The same quantities are
given for the FFP8 simulations.  At this level, results show good
consistency for both data and simulations.  The \sevem\ maps have the
lowest standard deviation, as measured by the HMHD maps, at small and
intermediate angular scales.

Pairwise differences between all four maps are shown in
Fig.~\ref{fig:dx11_diff_I}, after smoothing to 80\arcm\ FWHM and
downgrading to $\nside =128$. As expected, differences are largest
close the edge of the mask, where the absolute foreground level is the
highest.  Comparing these with the corresponding maps from 2013 shown
in Fig.~6 of \citet{planck2013-p06}, and noting that the new color bar
spans a range that is four times narrower than the previous one, we
see that the internal agreement between the four methods is
substantially better in the new maps, typically by about a factor of
two.

Figure~\ref{fig:ffp8_res_I} shows the differences between the FFP8
outputs and the input CMB map. The residuals are smallest for
\nilc\ and largest for \sevem. However, we note that the foreground
model adopted for the simulations was chosen to be more complex than
the real sky, in order to explicitly probe modelling errors. In
particular, the simulated thermal dust frequency spectrum exhibits a
strong positive (and spatially dependent) curvature at low frequency
that is neither captured in the parametric models adopted by
\commander, nor easily modelled by the spatial templates adopted by
\sevem. This additional complexity makes it hard to draw a strong
conclusion about the performance on the real data, for which the
thermal dust spectrum may be very well approximated by a simple
one-component greybody component with a nearly spatially constant
spectral index \citep{planck2014-a12}.

\begin{figure*}
\begin{center}
\begin{tabular}{cc}
\includegraphics[width=0.97\columnwidth]{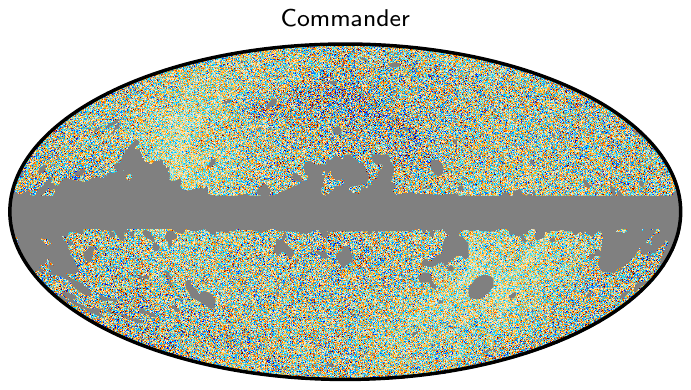}&
\includegraphics[width=0.97\columnwidth]{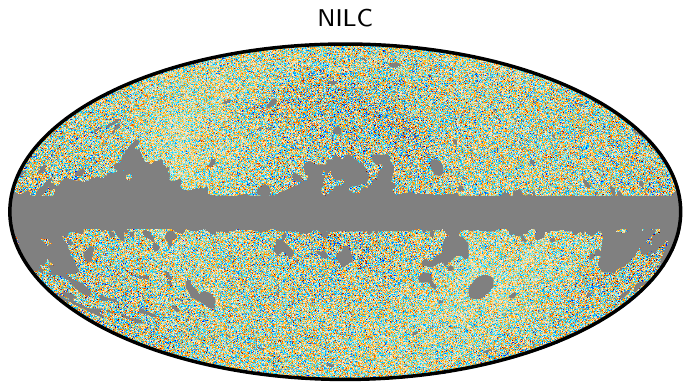}\\
\includegraphics[width=0.97\columnwidth]{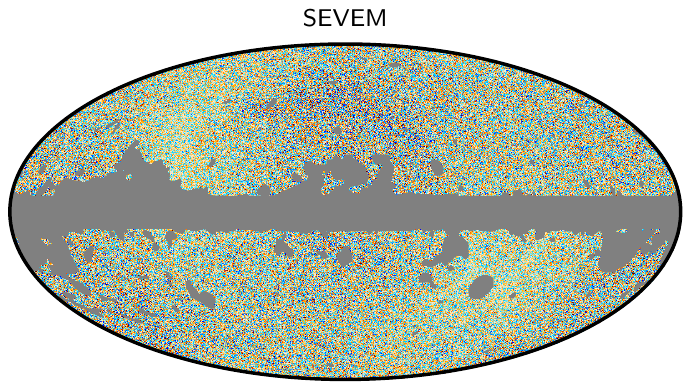}&
\includegraphics[width=0.97\columnwidth]{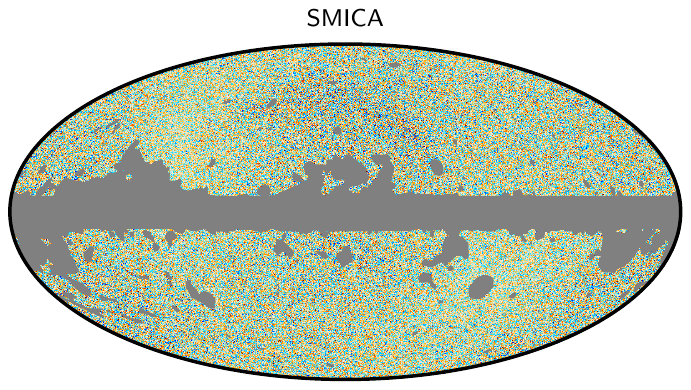}\\
\multicolumn{2}{c}{\includegraphics[height=1cm]{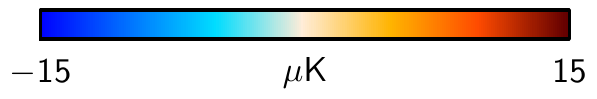}}
\end{tabular}
\end{center}
\caption{Component-separated CMB $Q$ maps at resolution FWHM 10\arcm,
  $\nside = 1024$.}
\label{fig:dx11_map_Q}
%\end{figure*}

%\begin{figure*}
\begin{center}
\begin{tabular}{cc}
\includegraphics[width=0.97\columnwidth]{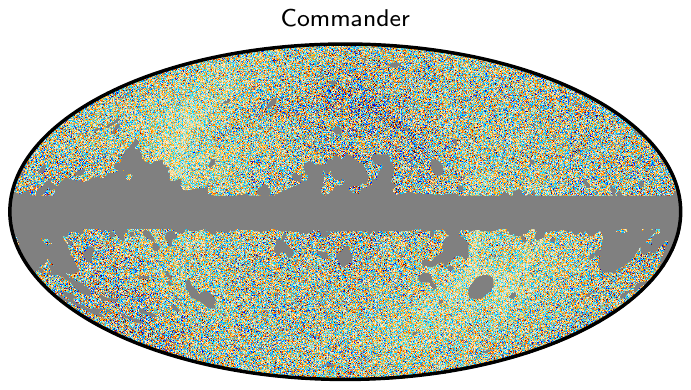}&
\includegraphics[width=0.97\columnwidth]{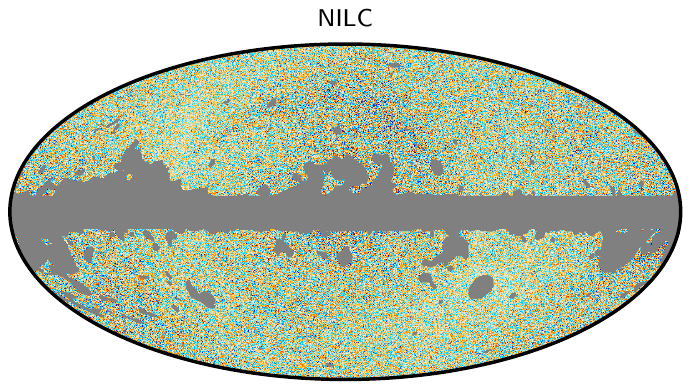}\\
\includegraphics[width=0.97\columnwidth]{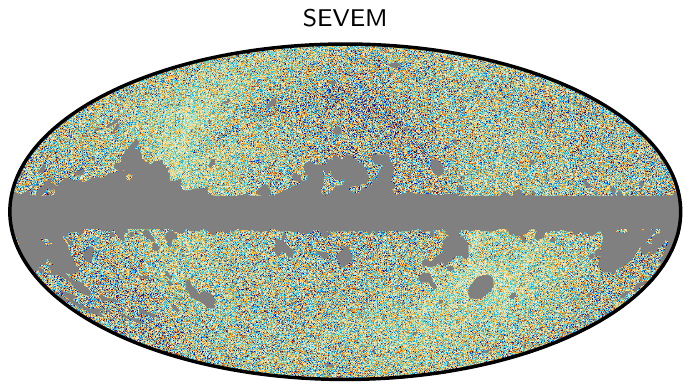}&
\includegraphics[width=0.97\columnwidth]{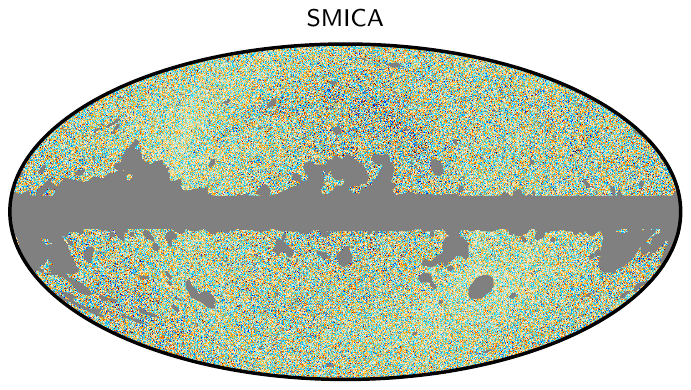}\\
\multicolumn{2}{c}{\includegraphics[height=1cm]{figs/colourbar_15uK.pdf}}
\end{tabular}
\end{center}
\caption{Component-separated CMB $U$ maps at resolution FWHM 10\arcm,
  $\nside = 1024$.}
\label{fig:dx11_map_U}
\end{figure*}

\begin{figure*}
\begin{center}
\begin{tabular}{cc}
\includegraphics[width=0.75\columnwidth]{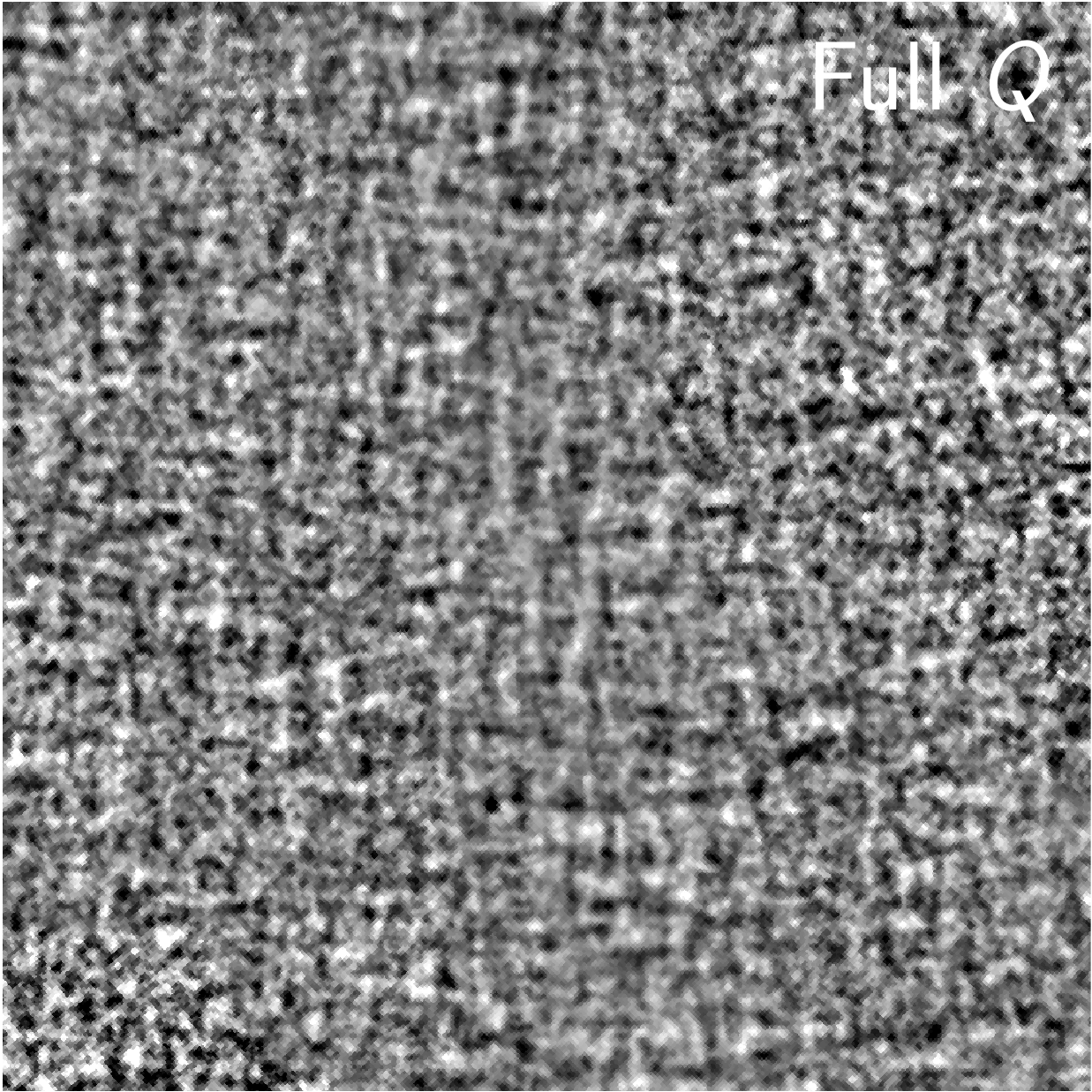}&
\includegraphics[width=0.75\columnwidth]{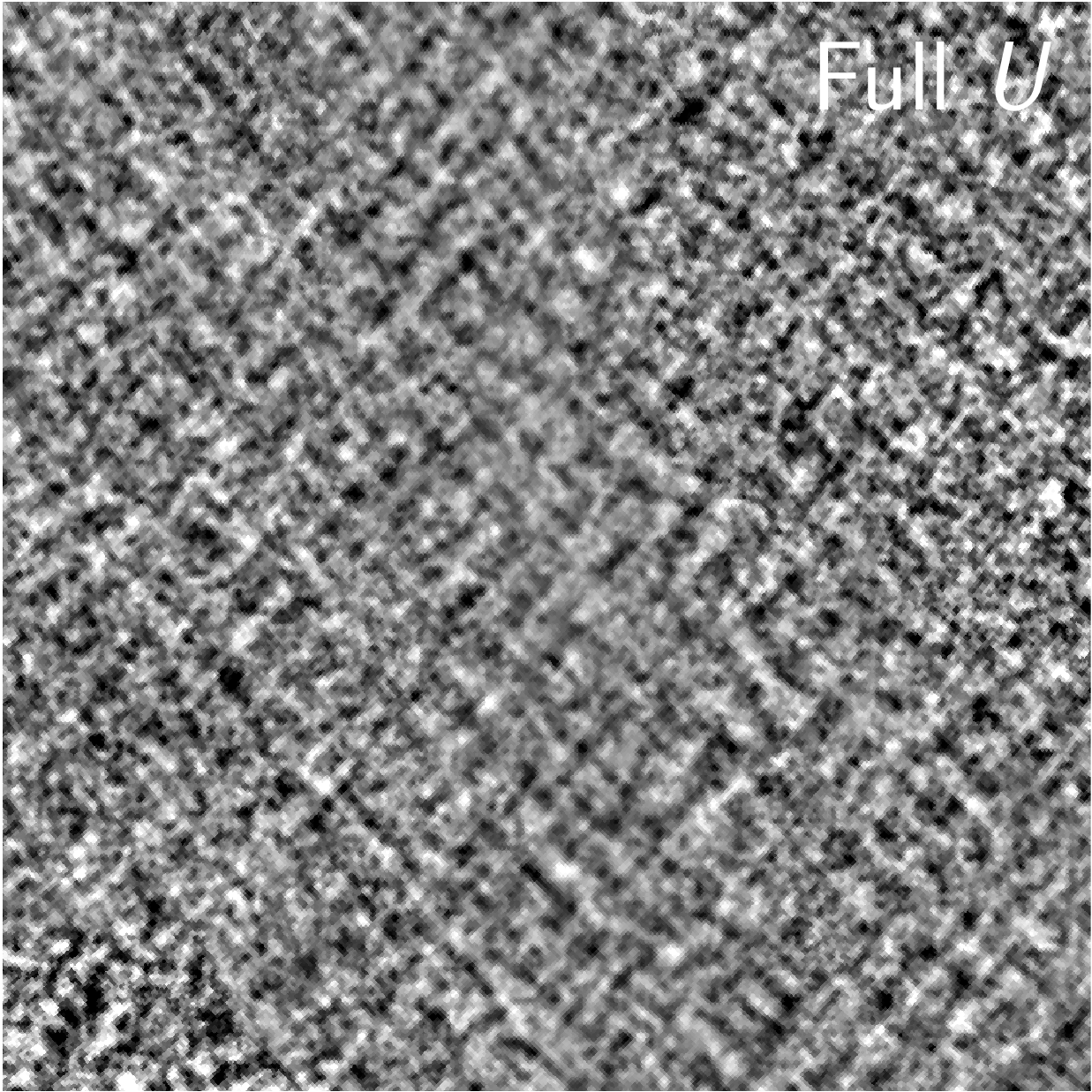}\\
\includegraphics[width=0.75\columnwidth]{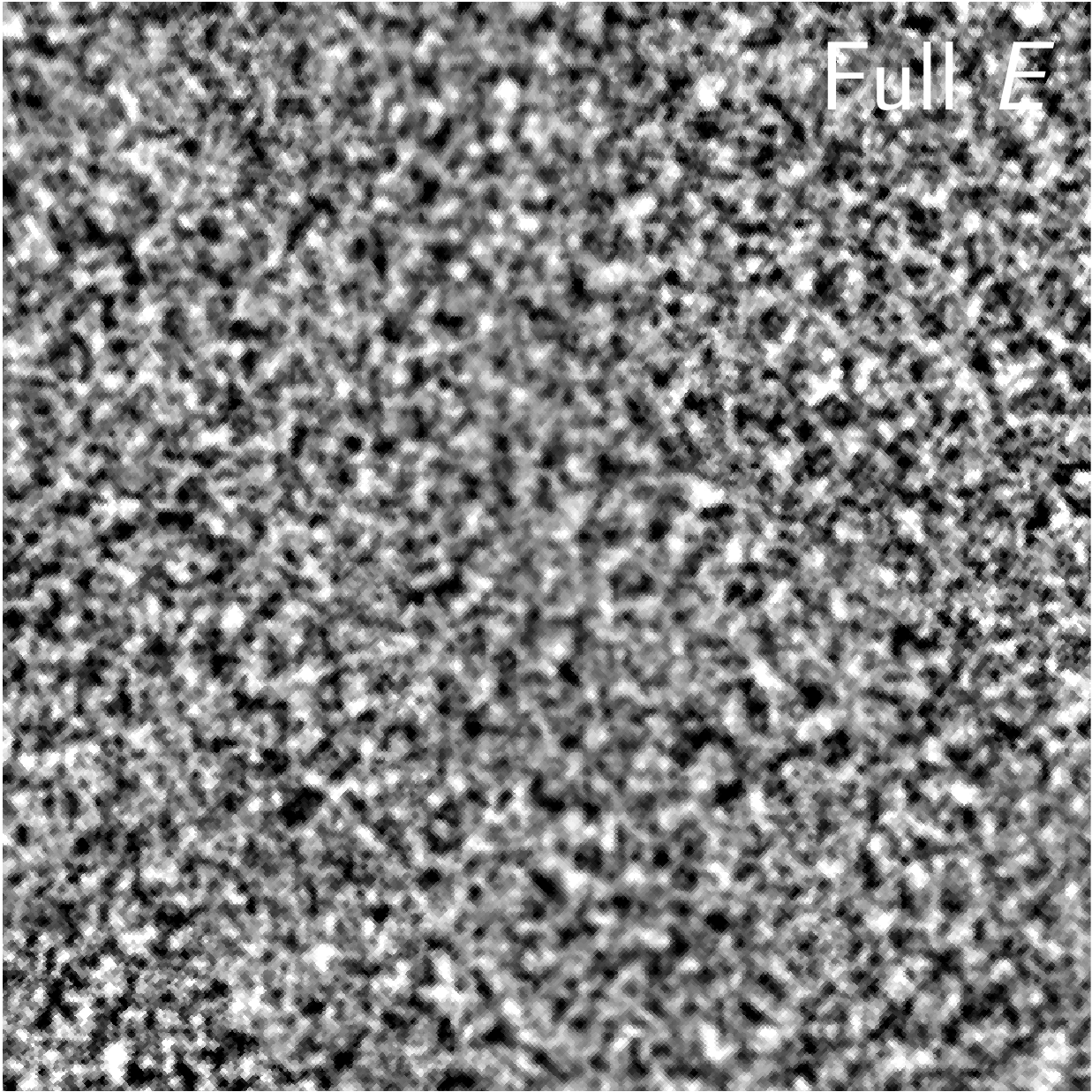}&
\includegraphics[width=0.75\columnwidth]{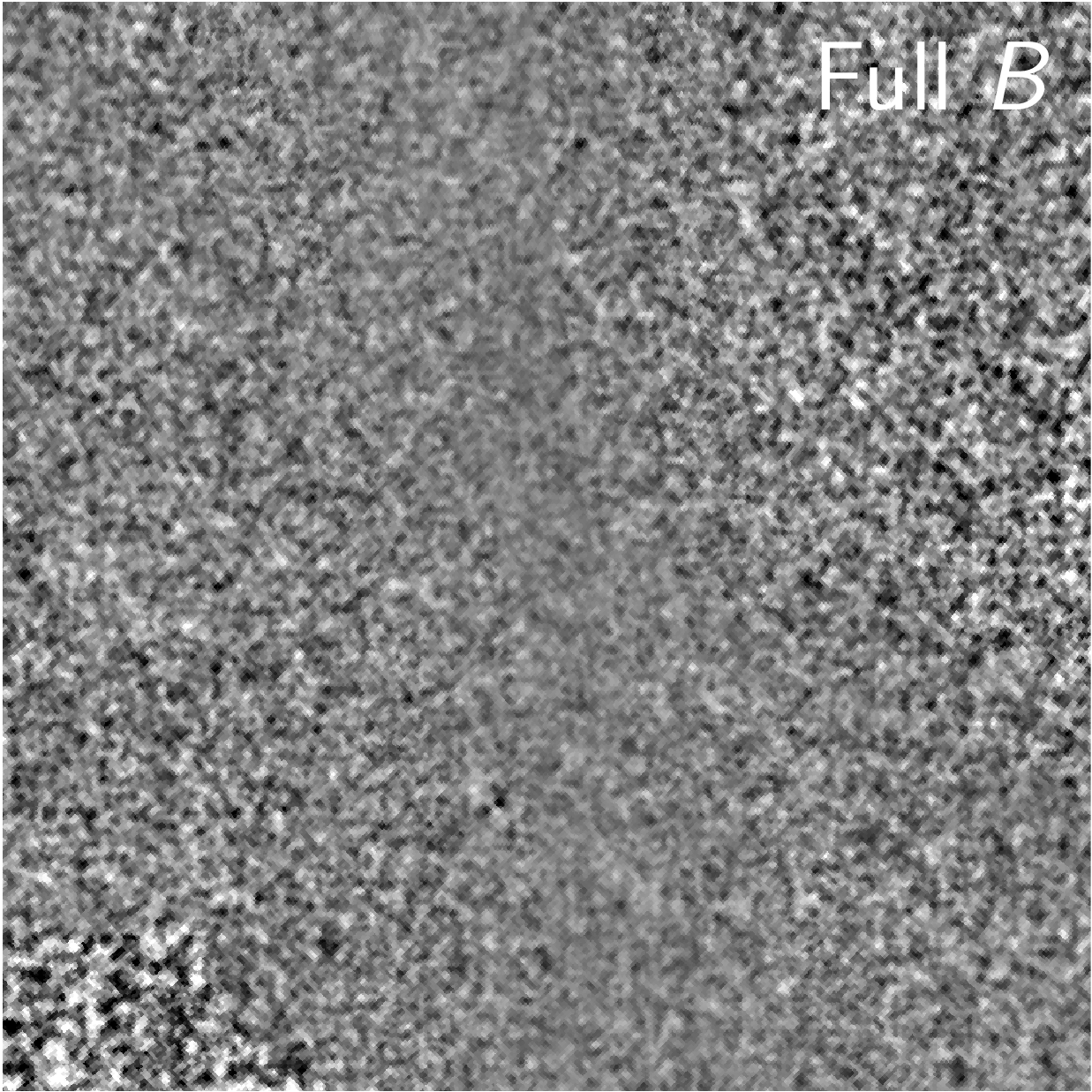}\\
\includegraphics[width=0.75\columnwidth]{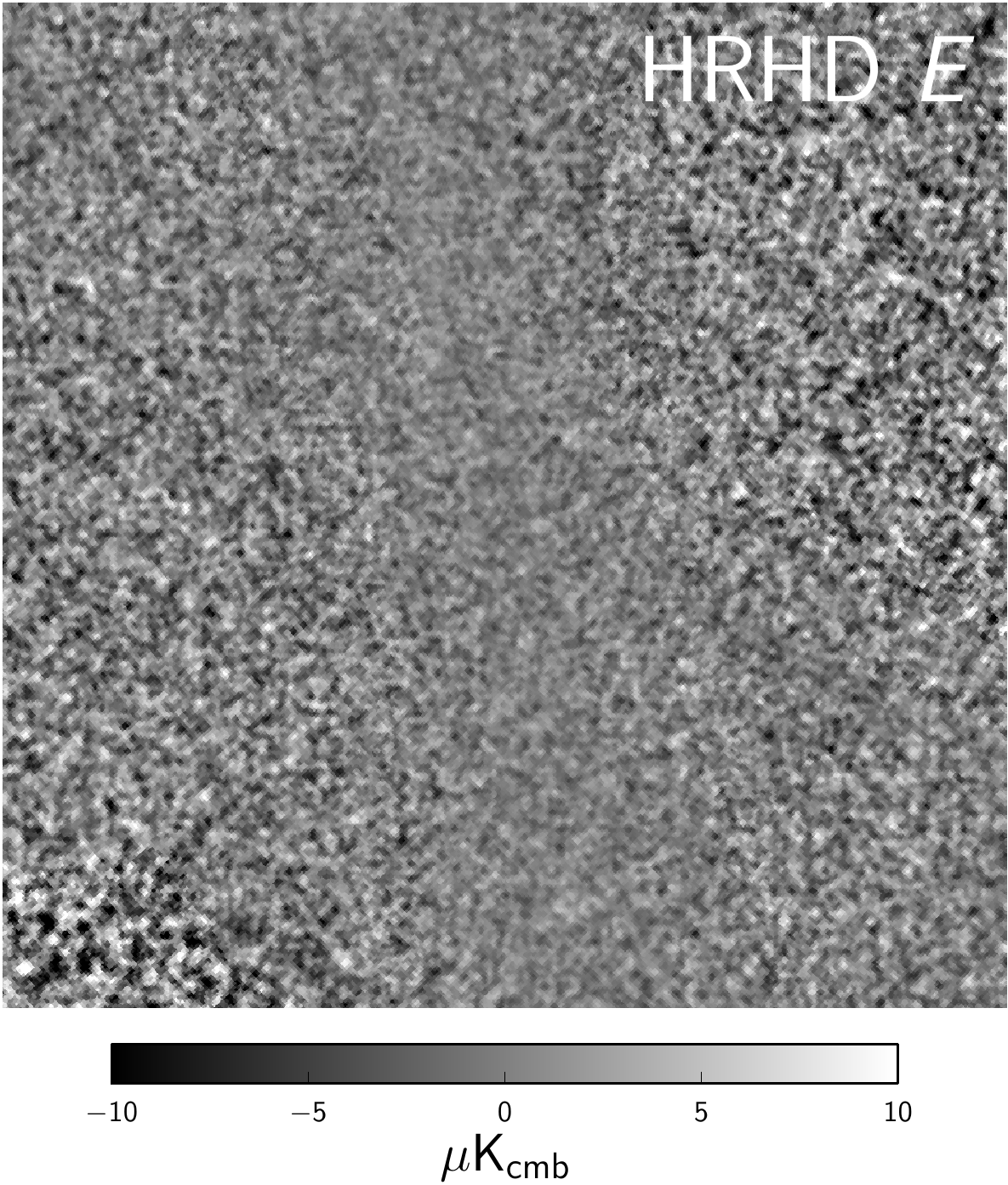}&
\includegraphics[width=0.75\columnwidth]{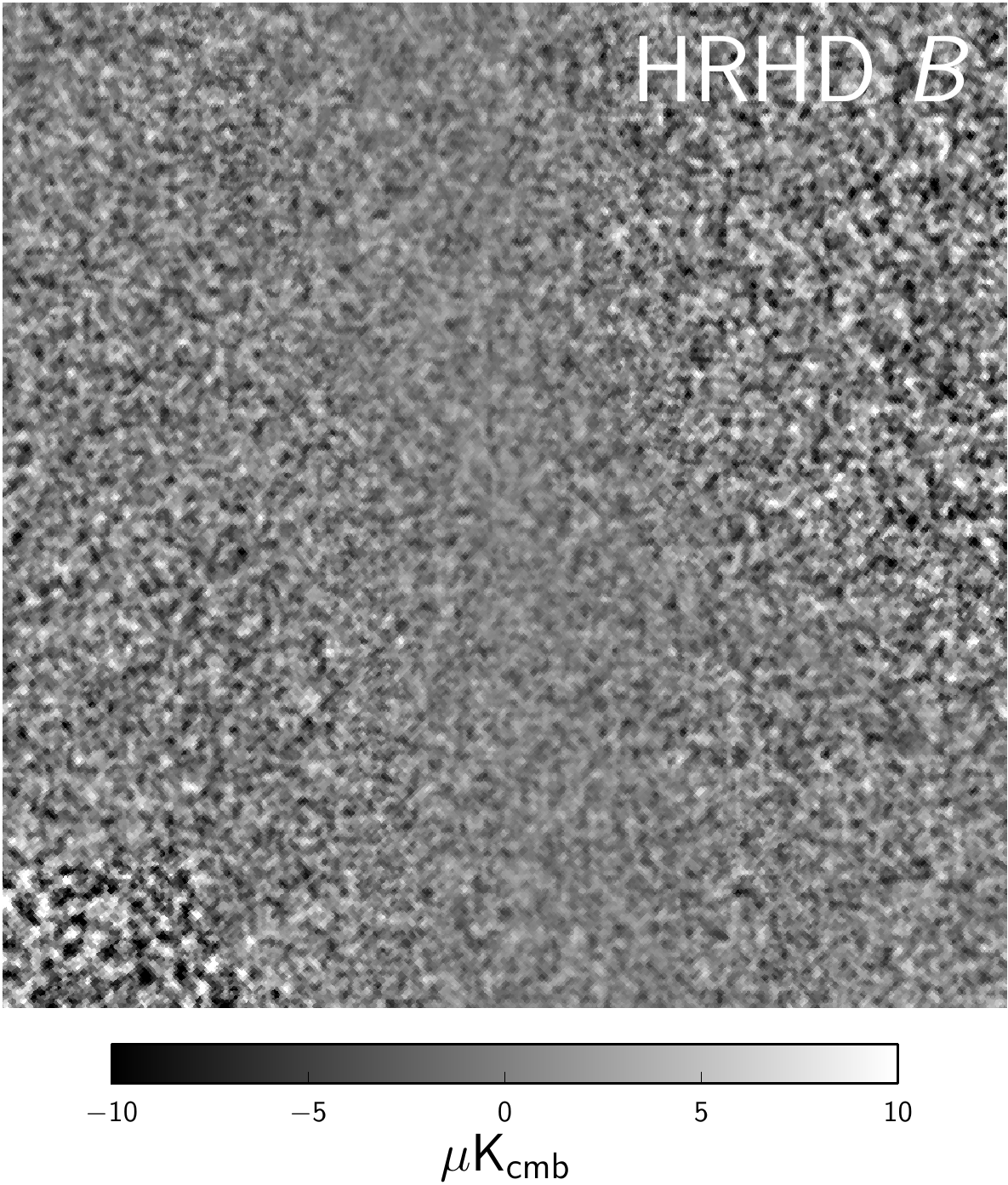}
\end{tabular}
\end{center}
\caption{$20\deg\times20\deg$ patch of the high-pass filtered
  \commander\ CMB polarization map, centered on the North Ecliptic
  Pole, $(l,b)=(96\deg,30\deg)$.  Each map is pixelized with a
  \healpix\ resolution of $\nside=1024$, and has an angular resolution
  of $10\arcm$ FWHM.  The top row shows $Q$ and $U$ maps derived from
  the full-mission data set, the middle row shows the corresponding
  $E$ and $B$ maps, and the bottom row shows the $E$ and $B$ maps of
  the half-ring half-difference (HRHD) map.  Note the characteristic
  $+$ and $\times$ patterns in the $Q$ and $U$ maps, and the clear
  asymmetry between $E$ and $B$ in the full data set.  Also note that
  the HRHD $E$ map is consistent with both the full and HRHD $B$
  maps.}
\label{fig:dx11_map_zoom_P}
\end{figure*}

\begin{figure*}
\begin{center}
\begin{tabular}{cc}
\includegraphics[width=1.00\columnwidth]{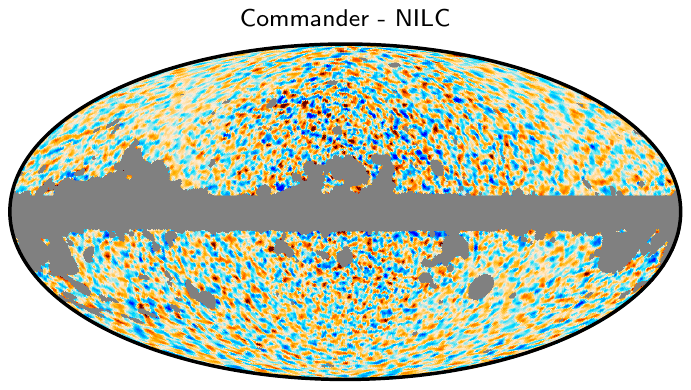}&
\includegraphics[width=1.00\columnwidth]{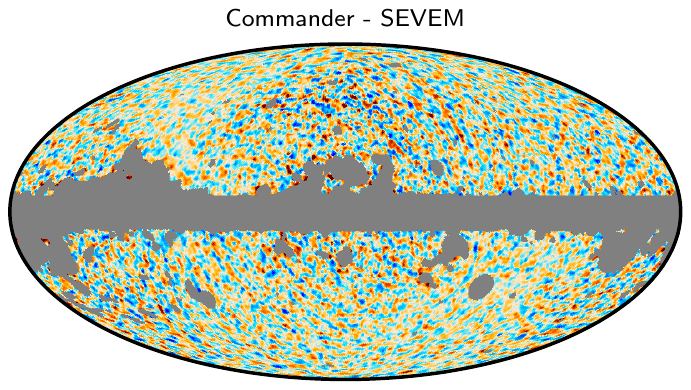}\\
\includegraphics[width=1.00\columnwidth]{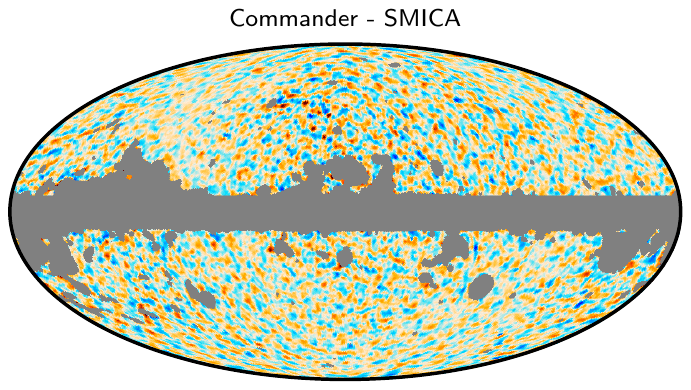}&
\includegraphics[width=1.00\columnwidth]{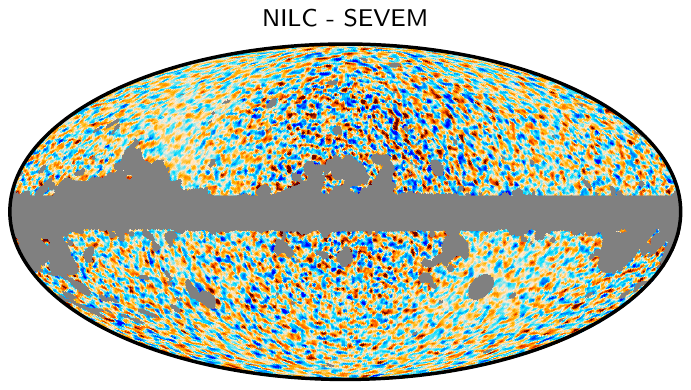}\\
\includegraphics[width=1.00\columnwidth]{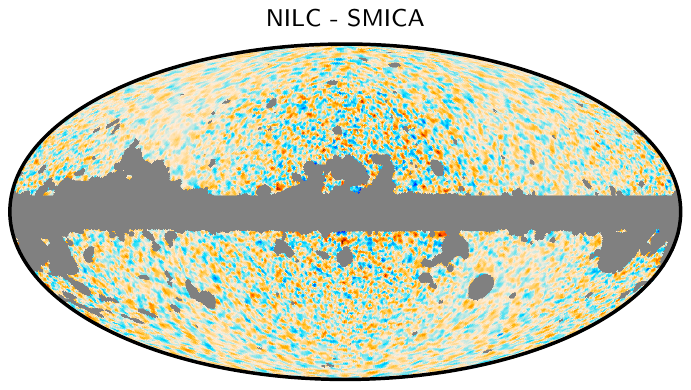}&
\includegraphics[width=1.00\columnwidth]{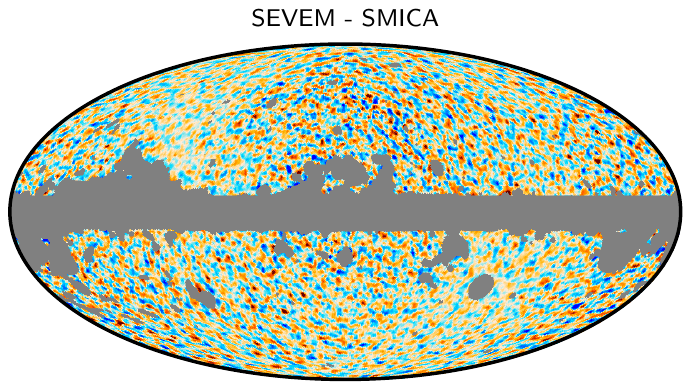}\\
\multicolumn{2}{c}{\includegraphics[height=1cm]{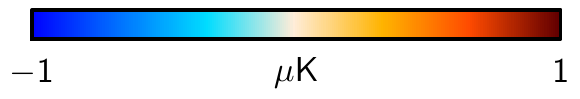}}
\end{tabular}
\end{center}
\caption{Pairwise differences between CMB $Q$ maps, after smoothing to
  FWHM 80\arcm\ and downgrading to $\nside = 128$.}
\label{fig:dx11_diff_Q}
\end{figure*}

\begin{figure*}
\begin{center}
\begin{tabular}{cc}
\includegraphics[width=1.00\columnwidth]{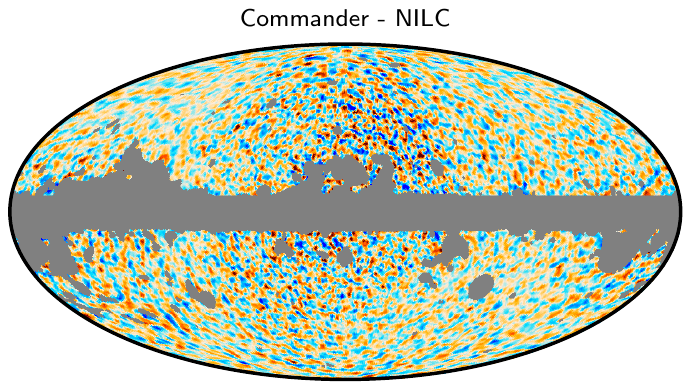}&
\includegraphics[width=1.00\columnwidth]{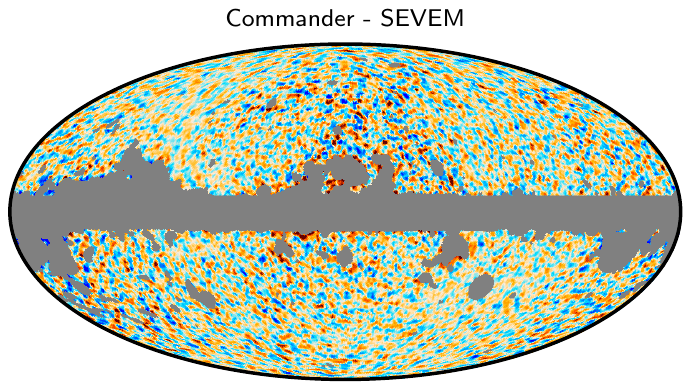}\\
\includegraphics[width=1.00\columnwidth]{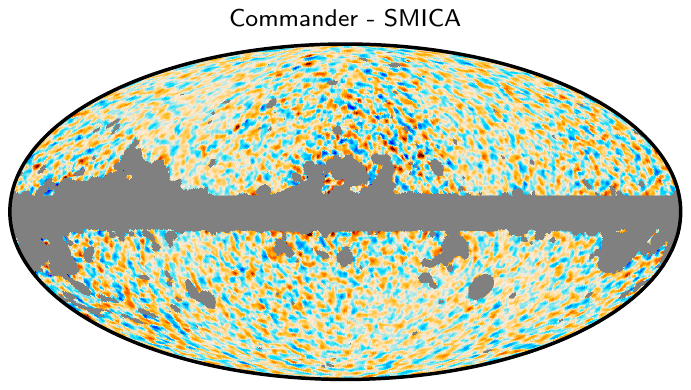}&
\includegraphics[width=1.00\columnwidth]{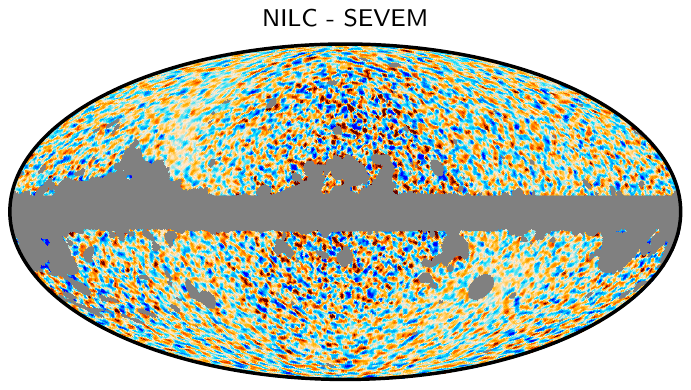}\\
\includegraphics[width=1.00\columnwidth]{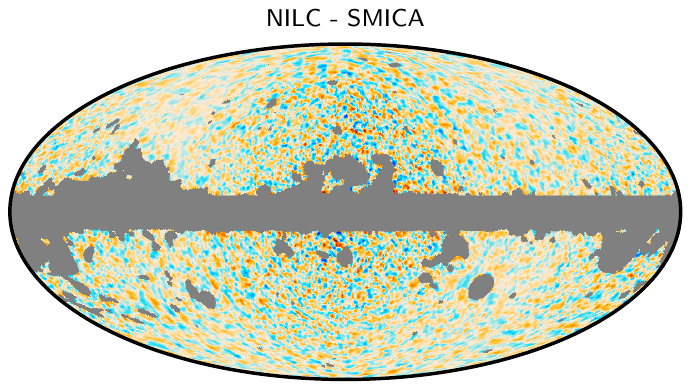}&
\includegraphics[width=1.00\columnwidth]{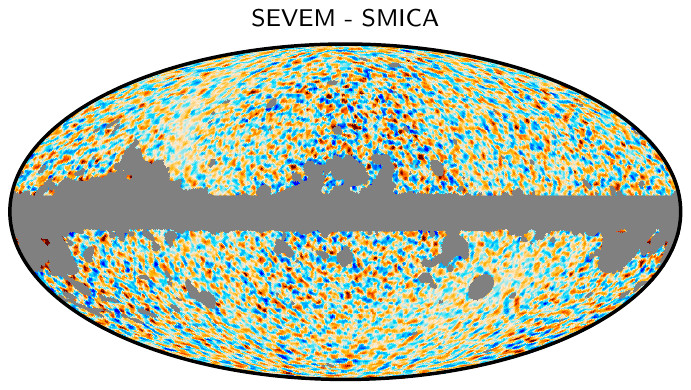}\\
\multicolumn{2}{c}{\includegraphics[height=1cm]{figs/colourbar_1uK.pdf}}
\end{tabular}
\end{center}
\caption{Pairwise differences between CMB $U$ maps, after smoothing
  and downgrading as in Fig.~\ref{fig:dx11_diff_Q}.}
\label{fig:dx11_diff_U}
\end{figure*}

\begin{figure*}
\begin{center}
\begin{tabular}{cc}
\includegraphics[width=0.97\columnwidth]{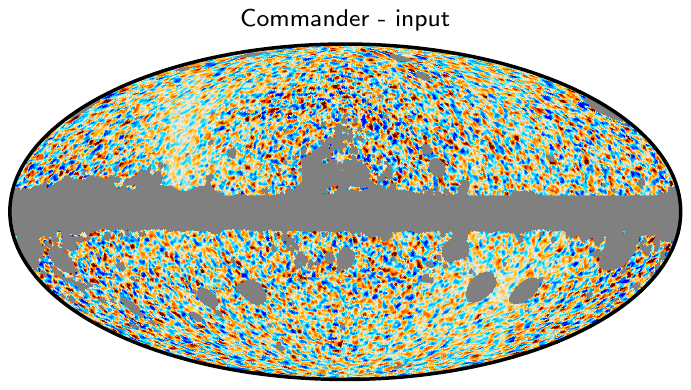}&
\includegraphics[width=0.97\columnwidth]{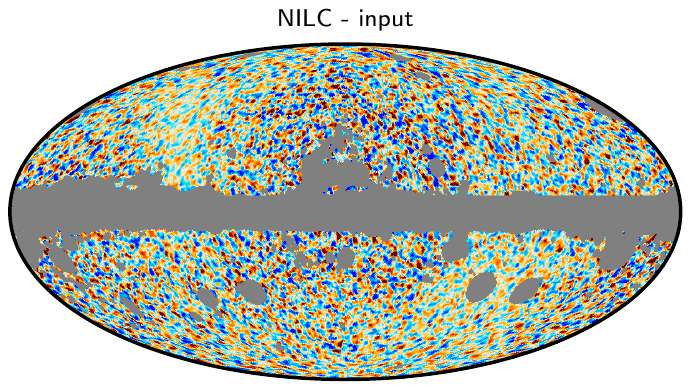}\\
\includegraphics[width=0.97\columnwidth]{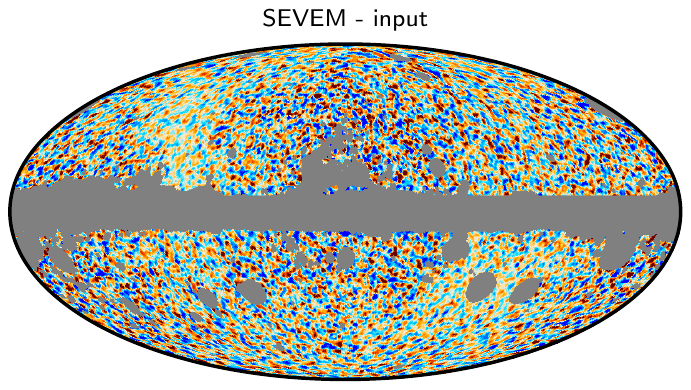}&
\includegraphics[width=0.97\columnwidth]{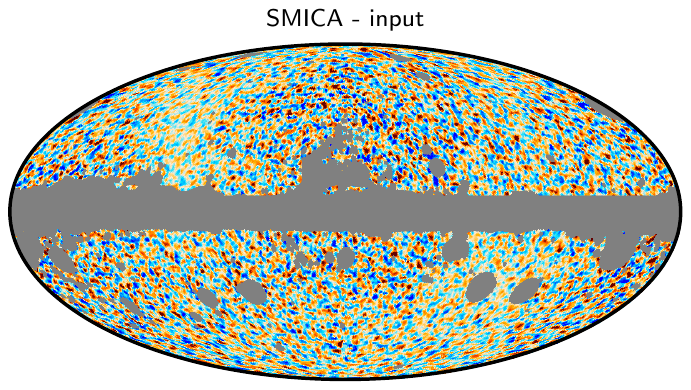}\\
\multicolumn{2}{c}{\includegraphics[height=1cm]{figs/colourbar_1uK}}
\end{tabular}
\end{center}
\caption{Difference between output and input CMB $Q$ maps from FFP8
  simulations.  Smoothing and downgrading as in
  Figs.~\ref{fig:dx11_diff_Q} and \ref{fig:dx11_diff_U}.}
\label{fig:ffp8_res_Q}
%\end{figure*}

%\begin{figure*}
\begin{center}
\begin{tabular}{cc}
\includegraphics[width=0.97\columnwidth]{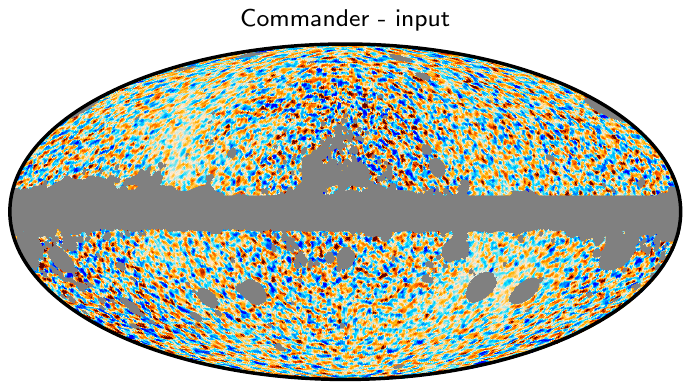}&
\includegraphics[width=0.97\columnwidth]{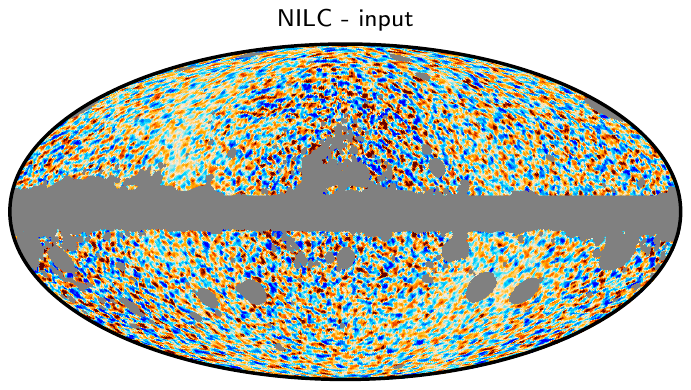}\\
\includegraphics[width=0.97\columnwidth]{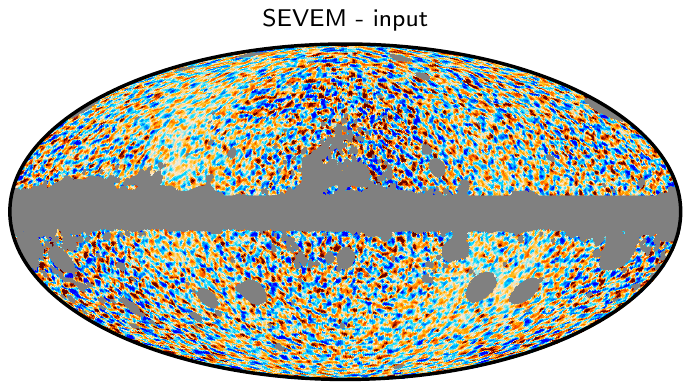}&
\includegraphics[width=0.97\columnwidth]{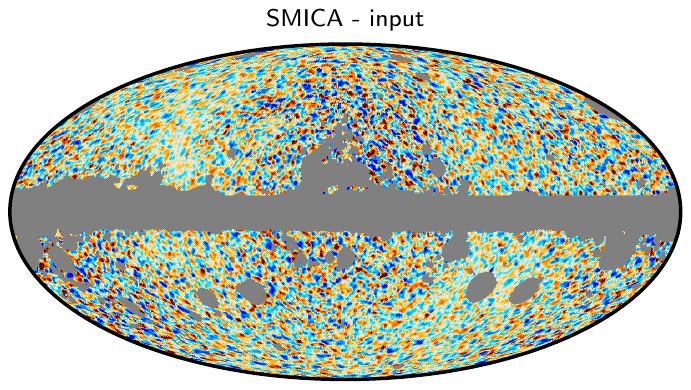}\\
\multicolumn{2}{c}{\includegraphics[height=1cm]{figs/colourbar_1uK}}
\end{tabular}
\end{center}
\caption{Difference between output and input CMB $U$ maps from FFP8
  simulations.  Smoothing and downgrading as in
  Fig.~\ref{fig:ffp8_res_Q}.}
\label{fig:ffp8_res_U}
\end{figure*}

\subsection{Polarization maps}
\label{sec:map_polarization}

We now turn our attention to the foreground-reduced CMB polarization
maps.  As discussed extensively in \citet{planck2014-a01},
\citet{planck2014-a07}, and \citet{planck2014-a09}, the residual
systematics in the \Planck\ polarization maps have been dramatically
reduced compared to 2013, by as much as two orders of magnitude on
large angular scales.  Nevertheless, on angular scales greater than
10\deg, correponding to $\ell \lesssim 20$, systematics are still
non-negligible compared to the expected cosmological signal.
Different combinations of input frequency channels for the component
separation have been explored in order to mitigate the polarization
residuals. However, it was not possible, for this data release, to
fully characterize the large-scale residuals from the data or from
simulations.  Therefore the CMB polarization maps provided in the
current release have been high-pass filtered to remove the large
angular scales.  This has been implemented by applying a cosine filter
to the $E$ and $B$ spherical harmonic coefficients of the maps. This
filter is defined as
\begin{equation}
w_{\ell}=\left\{ \begin{array}{lc}
0 & \ell < \ell_{1} \\
\frac{1}{2}\left[1-\cos\left(\pi\frac{\ell-\ell_{1}}{\ell_{2}-\ell_{1}}\right)\right] & \ell_{1}\le\ell\le\ell_{2}\\
1 & \ell_{2} < \ell\end{array}\right.,
\label{eq:high_pass_filter}
\end{equation}
and we have used $\ell_{1} = 20$ and $\ell_{2}=40$. The same filtering
has been applied to the FFP8 fiducial maps and to the Monte Carlo
simulations.

As for temperature, individual polarization confidence masks derived
for each method have been used to make combined masks for further
analysis. Our first polarization mask is simply the union of the
\commander, \sevem, and \smica\ confidence masks, which has $\fsky =
77.6\,\%$ and we refer to it as \texttt{UP78}.  However, the 1-point
statistics analysis summarized in Sect.~\ref{sec:onepointstatistics},
revealed significant point source contamination in the \commander,
\nilc, and \smica\ maps using this mask. \sevem\ was not affected by
this problem because it applies an inpainting technique to remove the
brightest point sources (see Appendix~\ref{sec:sevem} for further
details). For this reason, two extended versions of the mask were
created, the first by excluding in addition the pixels where the
standard deviation between the CMB maps, averaged in $Q$ and $U$,
exceeds 4\muK. This mask has $\fsky = 76.7\,\%$, and we refer to it
as \texttt{UPA77}.  The second was made by additionally excluding from
the union mask the polarized point sources detected at each frequency
channel.  It has $\fsky = 77.4\,\%$, and we refer to it as
\texttt{UPB77}. This mask is shown on the right of side of
Fig.~\ref{fig:dx11_masks}, and we adopt this as the preferred
polarization mask, since it is physically better motivated than
\texttt{UPA77}. Also, although it keeps a larger fraction of the sky
than \texttt{UPA77}, it is sufficient to alleviate the point source
contamination.

Masks have been made in a similar way for the FFP8 simulations.  The
union mask has $\fsky = 76.3\,\%$, and we refer to as
\texttt{FFP8-UP76}.  An extended mask that also excludes polarized
point sources has $\fsky = 75.7\,\%$, and we refer to it as
\texttt{FFP8-UPA76}.  \texttt{FFP8-UPA76} is the preferred mask for
FFP8 polarization analysis.

Figures~\ref{fig:dx11_map_Q} and \ref{fig:dx11_map_U} show the
high-pass filtered $Q$ and $U$ Stokes parameters of the CMB maps after
applying the \texttt{UPB77} mask. The maps are shown at full
resolution, and are thus dominated by instrumental noise except in the
regions at the ecliptic poles where integration time is greatest.
Visually, the methods operating in the harmonic (\nilc\ and \smica)
and spatial (\commander\ and \sevem) domains are more similar to each
other than the other methods.

\begin{table*}[tb]
\begingroup
\newdimen\tblskip \tblskip=5pt
\caption{Correlation coefficients of CMB maps with foreground
  templates for temperature and polarization.}
\label{tab:cross_corr}
\vskip -4mm
\footnotesize
\setbox\tablebox=\vbox{
\newdimen\digitwidth
\setbox0=\hbox{\rm 0}
\digitwidth=\wd0
\catcode`*=\active
\def*{\kern\digitwidth}
\newdimen\signwidth
\setbox0=\hbox{+}
\signwidth=\wd0
\catcode`!=\active
\def!{\kern\signwidth}
\newdimen\decimalwidth
\setbox0=\hbox{.}
\decimalwidth=\wd0
\catcode`@=\active
\def@{\kern\signwidth}
\halign{ \hbox to 1.5in{#\leaderfil}\tabskip=2em&
    \hfil$#$\hfil\tabskip=2em&
    \hfil$#$\hfil\tabskip=2em&
    \hfil$#$\hfil\tabskip=2em&
    \hfil$#$\hfil\tabskip=0pt\cr
\noalign{\doubleline}
\omit&\multispan4\hfil\sc Correlation Coefficient\hfil\cr 
\noalign{\vskip -3pt}
\omit&\multispan4\hrulefill\cr
\noalign{\vskip 3pt}
\omit\hfil\sc Foreground Template\hfil&\omit\hfil\commander\hfil&\omit\hfil\nilc\hfil&\omit\hfil\sevem\hfil&\omit\hfil\smica\hfil\cr
\noalign{\vskip 4pt\hrule\vskip 8pt} 
\omit\bf TEMPERATURE\hfil\cr
\noalign{\vskip 6pt}
\hglue 1em $H\alpha$&!0.010\pm0.071&  !0.011\pm0.071& !0.019\pm0.071&  !0.003\pm0.057\cr
\hglue 1em CO      &  -0.004\pm0.027&  -0.003\pm0.027& -0.003\pm0.027&  -0.007\pm0.022\cr
\hglue 1em 857\,GHz &  -0.043\pm0.084&  -0.032\pm0.084& -0.037\pm0.084&  -0.029\pm0.083\cr
\hglue 1em Haslam  &  -0.062\pm0.115&  -0.051\pm0.116& -0.065\pm0.115&  -0.023\pm0.069\cr
\noalign{\vskip 10pt}
\omit\bf POLARIZATION\cr
\noalign{\vskip 6pt}
\hglue 1em WMAP K-Ka&          -0.057\pm0.026& -0.116 \pm0.024& -0.026\pm0.025& -0.027\pm0.026\cr
\hglue 1em WMAP K-Ka (HPF) & !0.0042\pm0.0036& !0.0054\pm0.0037& !0.0147\pm0.0037& !0.0092\pm0.0036\cr
\noalign{\vskip 5pt\hrule\vskip 5pt}
}}
\endPlancktablewide                                                                                                                                         
\endgroup
\end{table*}

In order to have better visual insight at the map level, in
Figure~\ref{fig:dx11_map_zoom_P} we show a $20\degr\times 20\degr$
patch of the high-pass-filtered \commander\ polarization maps centred
on the North ecliptic pole. The top row shows the full-mission $Q$ and
$U$ maps. Note the characteristic ``$+$'' pattern in $Q$ and ``$\times$''
pattern in $U$; this is the expected signal for a pure $E$ mode
signal. To make this point more explicit, the middle row shows the
same map decomposed into $E$ and $B$ components. There is a clear
asymmetry between them, with $E$ having visibly more
coherent power than $B$, again as expected for an $E$-dominated
signal. Finally, the third row shows the half-ring half-difference
(HRHD) $E$ and $B$ maps, illustrating the noise level in the
full-mission maps. Comparing the middle and bottom rows, there is
clearly an $E$-mode excess in the full-mission map, whereas the
corresponding full-mission $B$-mode map is consistent with the HRHD
$B$-mode map. In addition, the HRHD $E$-mode map is also consistent
with both the full-mission and HRHD $B$-mode maps, suggesting that all
are consistent with instrumental noise.

Pairwise differences between the four polarization maps are shown in
Figs.~\ref{fig:dx11_diff_Q} and \ref{fig:dx11_diff_U}. The \nilc\ and
\smica\ solutions appear closest to each other.  The regions of the
sky that are most affected by differences appear to be those with a
higher noise level, as may be seen by comparing to
Figs.~\ref{fig:dx11_map_Q} and \ref{fig:dx11_map_U}.  For
completeness, Figs.~\ref{fig:ffp8_res_Q} and \ref{fig:ffp8_res_U} show
the differences between the FFP8 outputs and the input CMB map. The
differences show a pattern similar to that of the noise, though with
higher amplitude with respect to pairwise differences of solutions
from data, possibly reflecting again the enhanced complexity of the
simulated sky with respect to real data.

The combination of high-pass filtering and noise makes visual
comparison of these maps difficult. The rms summary provided in
Table~\ref{tab:solution_parameters} is more informative in this
respect. Comparing the HMHD rms values listed in parentheses, we see
that \nilc\ and \smica\ have the lowest effective polarization noise
levels at high angular resolution, with rms values that are roughly
20\,\% lower than those observed for \commander\ and \sevem. One
plausible explanation for this difference is the angular resolution
adopted for the fitting process by the four methods; whereas
\commander\ and \sevem\ perform the polarization analysis at $10\arcm$
FWHM resolution, \nilc\ and \smica\ adopt a $5\arcm$ FWHM
resolution. When comparing the rms values at an angular resolution of
$10\arcm$, as presented in Table~\ref{tab:solution_parameters}, the
maps from the latter two methods are smoothed by post-processing to a
lower resolution, whereas the maps from the two former codes are not.

This effect is not relevant at intermediate angular scales, for
instance at $160\arcm$ FWHM, as shown in the bottom section of
Table~\ref{tab:solution_parameters}. On these angular scales, we see
that the situation among the codes is reversed, and
\commander\ provides a 20\,\% lower effective noise than the other
three codes.

\section{Correlation with external templates}
\label{sec:crosscorrelation}

Correlation of the CMB maps with templates of foreground emission
provides a first diagnostic of residual contamination in the maps.  We
compute the correlation coefficient $r$ between two maps $\mathbf{x}$
and $\mathbf{y}$ as
\begin{equation}
\label{eq:cross_corr}
r = \frac{1}{\npix-1} \sum_i \,
\frac{\mathbf{x}_i -\langle\mathbf{x}\rangle}{\sigma_\mathbf{x}}\,
\frac{\mathbf{y}_i -\langle\mathbf{y}\rangle}{\sigma_\mathbf{y}}\,,
\end{equation}
where the index $i$ runs over the $\npix$ pixels observed with the
common mask, $\langle\mathbf{x}\rangle = \sum_i \mathbf{x}_i /\npix$,
$\sigma_{\mathbf{x}} = [\sum_i (\mathbf{x}_i
  -\langle\mathbf{x}\rangle)^2 /(\npix-1)]^{1/2}$, and similarly for
$\mathbf{y}$.  Both maps and templates are smoothed to FWHM 
1\degr\ and downgraded to $\nside = 256$ before computing the
correlation, and the analysis is performed separately on temperature
and polarization maps.

The foreground templates considered for temperature are: the 408\,MHz
radio survey of \cite{haslam1982}; the velocity-integrated CO map of
\cite{dame2001}; the full-sky H${\alpha}$ template of
\cite{finkbeiner2003}; and the \Planck\ 857\,GHz channel map.  The
uncertainty in the value of $r$ due to chance correlations between
foregrounds and the cleaned maps is estimated by computing the
correlations between the templates and the 1000 simulations of CMB and
noise provided by each method.  For polarization, the only template we
consider is one constructed from the \WMAP\ 9-year maps.  The template
is made by smoothing the K and Ka band maps to FWHM 1\degr\ and
differencing them to remove the CMB contribution.  This produces a
template containing the low-frequency polarized foreground emission
and noise.  The correlation analysis is done twice, once with the
original maps and templates, and once with a high-pass filtered
version.  The resulting coefficient factors and $1\,\sigma$
uncertainties are shown in Table~\ref{tab:cross_corr}.

For temperature, the results for all methods are compatible with no
correlation within $1\sigma$.  From this, we conclude that there are
no significant temperature foreground residuals with the same
morphology as the templates in the map, to a precision set by cosmic
variance.  For polarization, the analysis of unfiltered maps show that
\sevem\ and \smica\ are compatible with no correlation at the
$1\,\sigma$ level, \commander\ has a moderate level of correlation at
the $2\,\sigma$ level, and for \nilc\ we find a correlation around
$4.5\,\sigma$.  For high-pass filtered maps (labelled by HPF), the
level of correlation is reduced for \commander\ and \nilc\ to about
$1.5\,\sigma$, while it increases for \sevem\ and \smica\ to about 4
and $2.5\,\sigma$, respectively.  These detections may be due to
accidental correlations of the map with the template which are not
taken into account when computing the uncertainty.  Neither the noise
in the template nor the systematics in the maps at large angular
scales are modelled, the latter being important for the unfiltered
case.  In the filtered case, the signal is reduced relative to the
noise in the template, which could exacerbate spurious correlations.
From this analysis, it is difficult to draw strong conclusions about
the residual contamination in the polarized CMB maps.

\section{Power spectra and cosmological parameters}
\label{sec:powspec}

In this section we evaluate the foreground-cleaned maps in terms of
CMB power spectra and cosmological parameters.  We emphasize that the
\Planck\ 2015 results parameters are not based on the high-resolution
foreground-cleaned CMB maps presented in this paper, but are instead
derived from the likelihood described in \citet{planck2014-a15}.  That
likelihood combines the low-resolution \commander\ temperature map
derived from \Planck, WMAP, and 408$\,$MHz with a template-cleaned LFI
70\GHz\ polarization map in a pixel-based low-$\ell$ likelihood, and
adopts a cross-spectrum-based estimator for the high-$\ell$
temperature and polarization likelihood.  The high-$\ell$ likelihood
relies on a careful choice of masks along with templates and
modelling, all in the power spectrum domain, to reduce the
contribution from diffuse Galactic and discrete Galactic and
extragalactic foreground emission.  The templates and source models
are marginalized over when estimating cosmological parameters.

The parameter estimation methodology that we use here is primarily a
tool to evaluate the quality of the high-resolution CMB maps and to
assess overall consistency with the \Planck\ 2015 likelihood.  We
start with the foreground-cleaned CMB maps and masks described in
Sect.~\ref{sec:maps}.  The maps have been cleaned from diffuse
foregrounds and, to a varying extent, from extragalactic foregrounds.
We construct simplified templates for the residual extragalactic
foregrounds that are marginalized over when estimating cosmological
parameters.

While the \Planck\ 2015 likelihood takes into account calibration and
beam uncertainties, we have not done so in this analysis.  Two of the
methods, \commander\ and \smica, fit for the relative calibration
between frequency channels, but the uncertainties from this process
are not propagated into the maps.  The other two methods, \nilc\ and
\sevem, assume that the frequency channels are pefectly calibrated.
None of the four methods propagate the uncertainties in the beam
transfer functions into the CMB maps.

We are interested in assessing the relative quality of the CMB maps,
for which it is more important to assess the spread of parameters
between methods and as a function of angular scale, rather than to
provide absolute numerical values.  However, since the CMB maps we
describe here are the basis for the analysis of the statistical
isotropy of the CMB, primordial non-Gaussianity, and gravitational
lensing by large scale structure, it is both important and reassuring
that the parameter values that we find are reasonably close to those
of \citet{planck2014-a15}.

\begin{figure*}
  \begin{center}
    \includegraphics[width=\columnwidth]{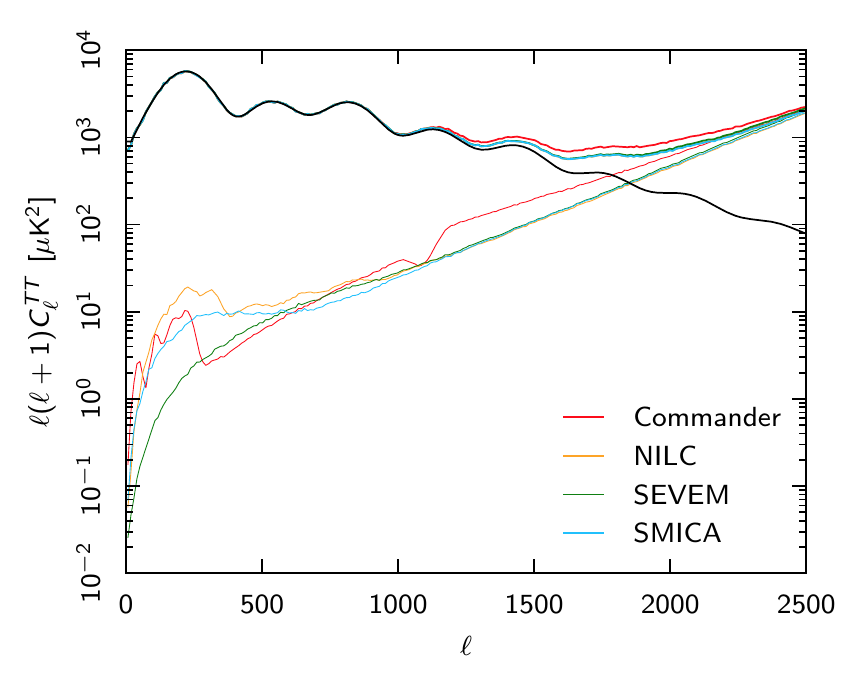}
    \includegraphics[width=\columnwidth]{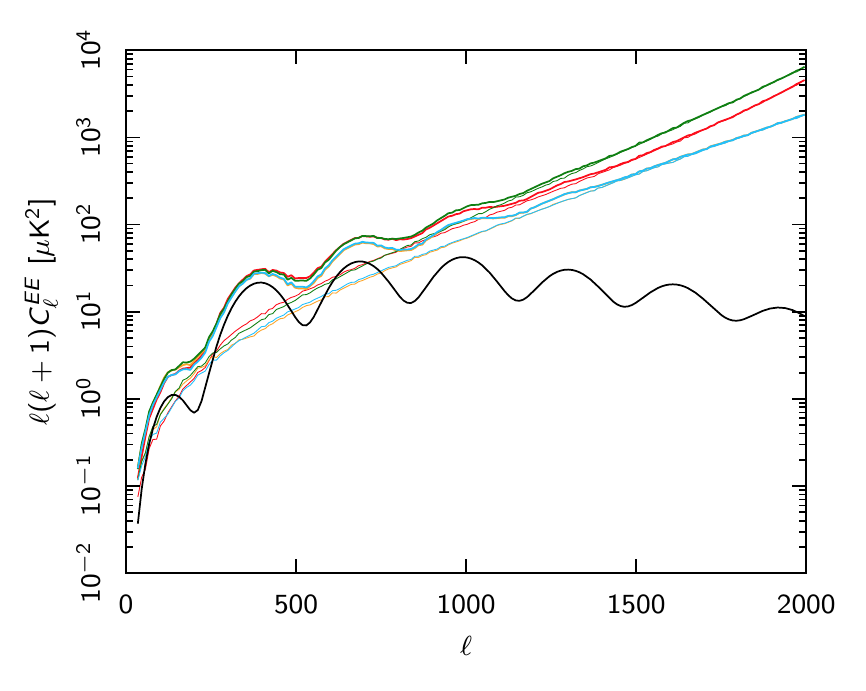}
  \end{center}
  \caption{Power spectra of the the foreground-cleaned CMB maps.
    \textit{Left}: $TT$ power spectra evaluated using the
    \texttt{UT78} mask.  \textit{Right}: $EE$ power spectra evaluated
    using the \texttt{UP78} mask.  The thick lines show the spectra of
    the half-mission half-sum maps containing signal and noise.  The
    thin lines show the spectra of the half-mission half-difference
    maps, which given an estimate of the noise. The black line shows
    the \Planck\ 2015 best-fit CMB spectrum for comparison.}
  \label{fig:dx11_spectra_raw}
  \vskip 6mm
  \begin{center}
    \includegraphics[width=\columnwidth]{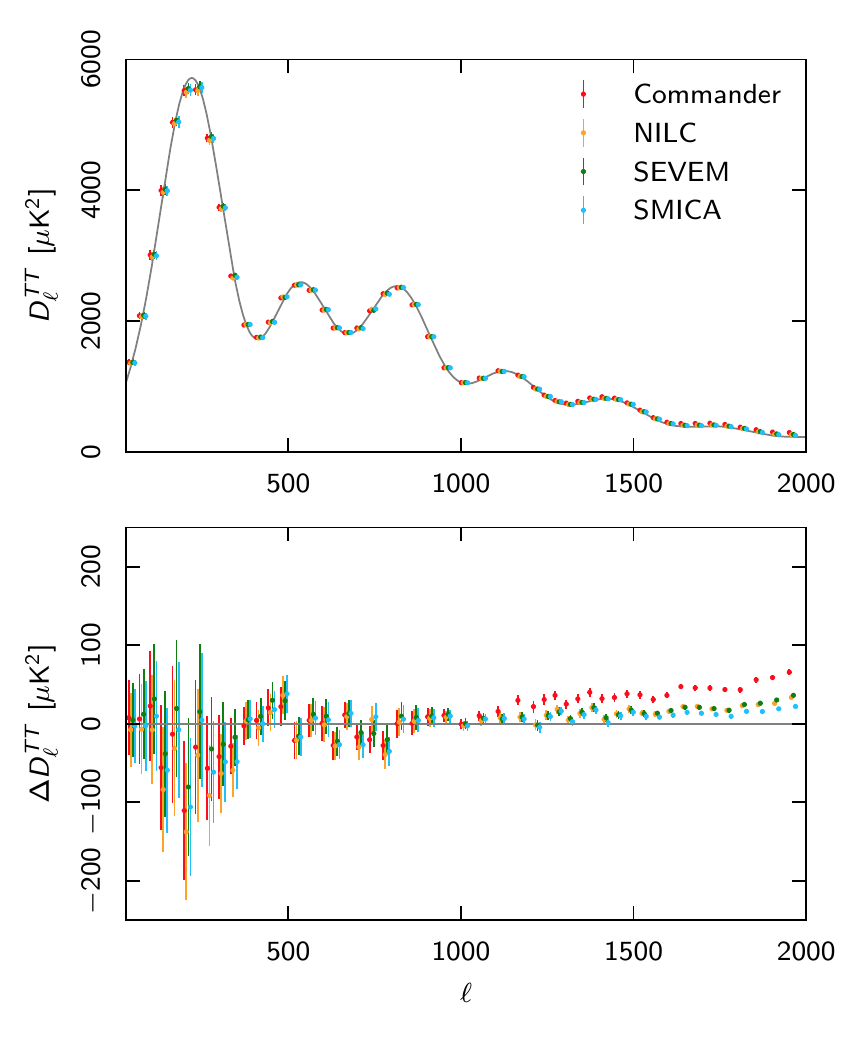}
    \includegraphics[width=\columnwidth]{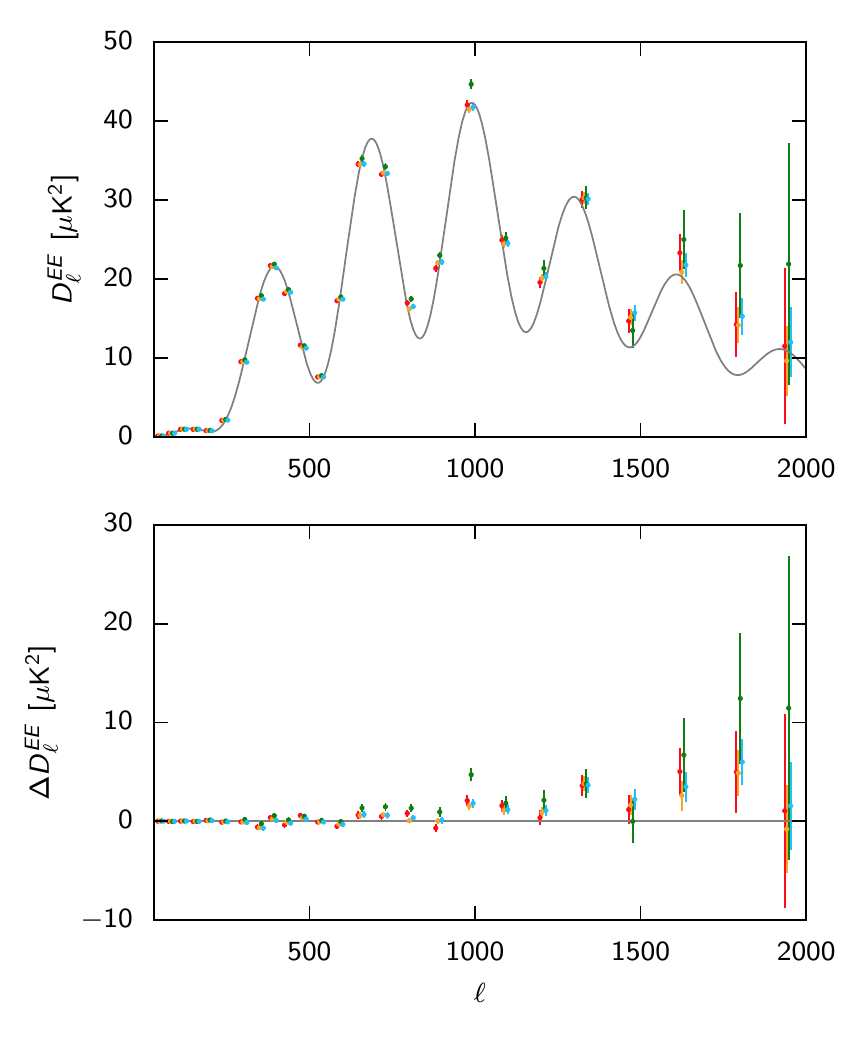}
  \end{center}
  \caption{CMB $TT$ (\emph{left}) and $EE$ (\emph{right}) power
    spectra for each of the four foreground-cleaned maps. Top panels
    show raw bandpowers with no subtraction of extragalactic
    foregrounds; the grey lines show the best-fit \LCDM\ model from
    the \Planck\ 2015 likelihood. The bottom panels show residual
    bandpowers after subtracting the best-fit \LCDM\ model showing the
    residual extragalactic foreground contribution.  The foregrounds
    are modelled and marginalized over when estimating parameters, see
    Figure~\ref{fig:dx11_params_TT_EE}.}
  \label{fig:dx11_spectra}
\end{figure*}

\begin{figure*}
  \begin{center}
    \includegraphics[width=\textwidth]{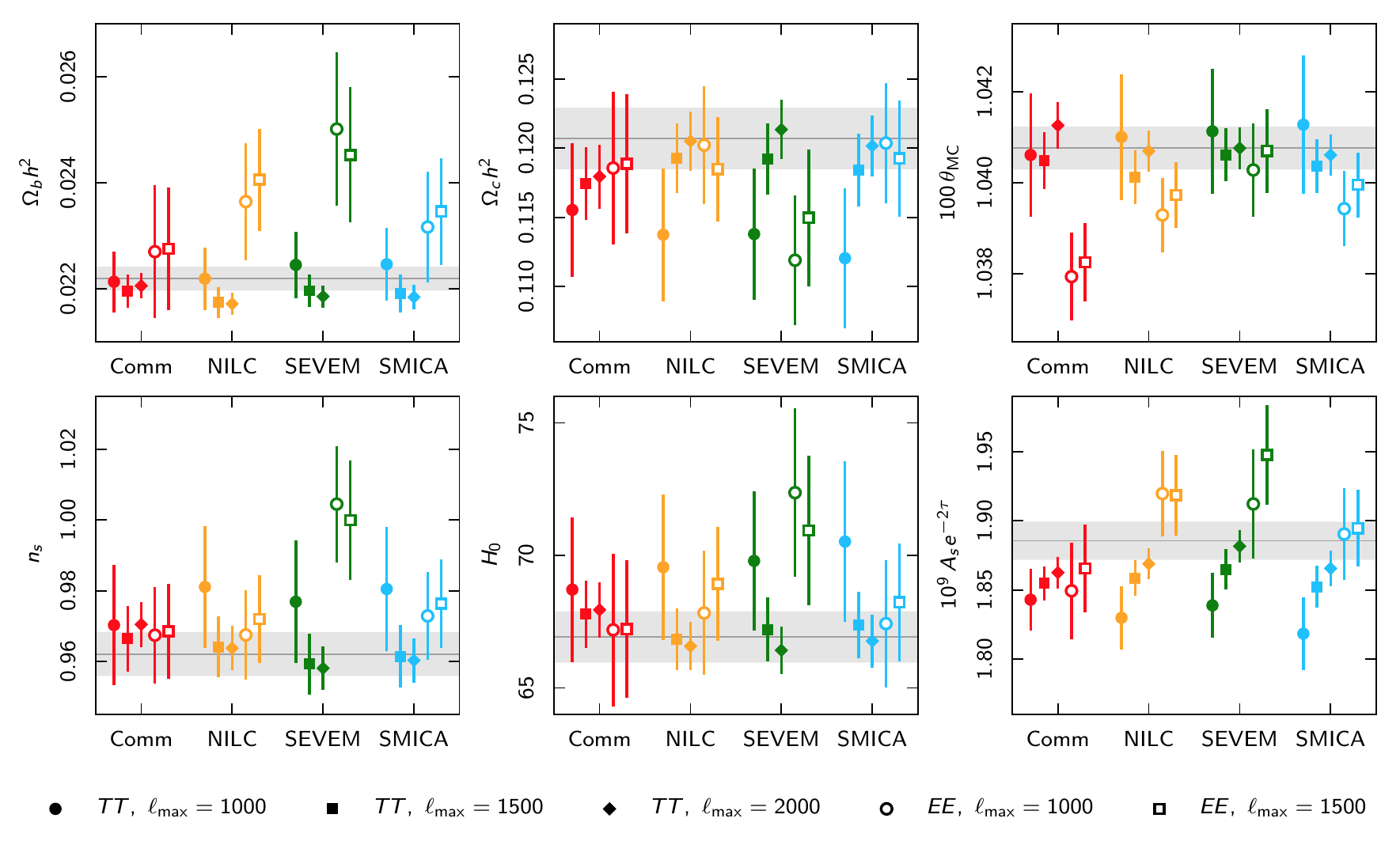}
  \end{center}
  \caption{Comparison of cosmological parameters estimated from the
    $TT$ and $EE$ spectra computed from the foreground-cleaned CMB
    maps. Within each group, the three left-most points show results
    for $TT$ with $\ell_{\mathrm{max}} = 1000$, $1500$, and $2000$ the
    two right-most points show results for $EE$ with
    $\ell_{\mathrm{max}} = 1000$ and $1500$. For comparison, we also
    show the corresponding parameters obtained with the \Planck\ 2015
    likelihood including multipoles up to $\ell_{\mathrm{max}} = 2500$
    as the horizontal line surrounded by a grey band giving the
    uncertainties.  The foreground model used for the cleaned CMB maps
    is the method-tailored full-sky model from the FFP8 simulations.}
    \label{fig:dx11_params_TT_EE}
\end{figure*}

Figure~\ref{fig:dx11_spectra_raw} shows the $TT$ and $EE$ power
spectra of the foreground-cleaned maps, applying the \texttt{UT78}
mask in temperature and the \texttt{UP78} mask in polarization.  The
power spectra of the half-mission half-sum (HMHS) data are shown as
thick lines; those of the half-mission half-difference (HMHD) data are
shown as thin lines. The HMHS spectra contain signal and noise,
whereas the HMHD spectra contain only noise and potential systematic
effects.  The variations in the temperature noise spectra at low to
intermediate multipoles are caused by the component separation methods
optimizing the trade-off between foreground signal and noise as a
function of scale.  At high mutipoles, the same spectra are smooth
because the relative weighting of the frequency channels is set by the
noise levels.  The breaks in the \commander\ noise spectra are caused
by the hybridization of maps at different resolutions to make the
final map.

Figure~\ref{fig:dx11_spectra} shows the power spectra of the signal in
the maps estimated using \XFaster\ \citep{rocha2010b,rocha2011} from
the HMHS and HMHD maps.  The top panels compare each of the four power
spectra derived from the component separation maps with the best-fit
\LCDM\ power spectrum derived from the \Planck\ likelihood including
multipoles up to $\ell = 2500$. The bottom panels show the spectrum
differences between the component separation methods and the best-fit
spectrum.  The bottom left panel shows significant differences between
the temperature spectra computed from the four raw component
separation maps. The \commander\ spectrum has significantly more
temperature power than the other three methods at $\ell \gtrsim 1000$,
with a clear break in amplitude between $\ell = 1000$ and 1200. As
described in Appendix~\ref{sec:commander}, \commander\ only uses
frequencies 217\GHz\ and above for multipoles above $\ell \sim 1000$;
the \commander\ hybridization approach achieves greater angular
resolution at the price of a higher point source contribution above
$\ell = 1000$, compared to the other three methods. On the other hand,
fitting foregrounds pixel-by-pixel ensures that point sources remain
localized in the final map, and the shape of the foreground spectrum
has the $D_{\ell} \propto \ell^2$ shape expected for a Poisson source
contribution.

\smica\ has the lowest high-$\ell$ temperature power excess from
extragalactic sources of all four methods. In this case, though, the
excess is almost flat between $\ell\approx1200$ and 1700, reflecting
the effective harmonic space frequency weighting as a function of
multipole, and the effective source contribution therefore has a more
complicated overall behaviour. To account at least partially for this
effect, we use the FFP8 simulations to construct an effective
extragalactic source template as a function of multipole for each
method individually, and marginalize over the corresponding amplitude
during parameter estimation. These templates, however, are only as
good as the inputs used for the simulations, and uncertainties in the
frequency dependency of the underlying true source populations
translate into an $\ell$-dependent error in the source templates for
\nilc\, \sevem, and \smica.  \commander\ is more robust against this
particular effect because it smooths all maps to a common angular
resolution prior to component separation, but, as seen in
Fig.~\ref{fig:dx11_spectra}, this robustness comes at the price of a
higher overall source amplitude.

The bottom right panel of Fig.~\ref{fig:dx11_spectra} shows the
corresponding information for the $EE$ spectrum. Once again, we see a
clear positive excess in these residual spectra. In the absence of a
detailed FFP8 model for this excess, we assume for now that it has a
shape $D_{\ell} \propto \ell^2$, and marginalize over its amplitude
during cosmological parameter estimation.

Cosmological parameters from the component separated maps in both
temperature and polarization are determined using \XFaster\ power
spectra, coupled to \texttt{CosmoMC} \citep{cosmomc} using a
correlated Gaussian likelihood.  Specifically, we include multipoles
between $\ell_{\textrm{min}} = 50$ and $\ell_{\textrm{max}}$, where
$\ell_{\textrm{max}} = 1000$, 1500, or 2000 for temperature, and 1000
or 1500 for polarization. We adopt a standard six-parameter
\LCDM\ model, and, since low-$\ell$ data are not used in the
likelihood, impose an informative Gaussian prior of $\tau = 0.07 \pm
0.006$.  As mentioned above, we construct foreground templates for
each CMB map by propagating the simulated full-sky FFP8 foreground
maps through the respective pipeline and estimating the resulting
power spectra normalized to some pivotal multipole.

The resulting cosmological parameters are summarized in
Fig.~\ref{fig:dx11_params_TT_EE} for both $TT$ (filled symbols) and
$EE$ (unfilled symbols). Starting with the temperature cases, we first
observe good overall internal agreement between the four component
separation methods, with almost all differences smaller than
$\sim\,$1$\,\sigma$ within each multipole band. Second, we also
observe good agreement with the best-fit \Planck\ 2015 \LCDM\ model
derived from the likelihood, as most of the differences are within
1$\,\sigma$. The notable exception is the power spectrum amplitude,
$A_{s}e^{-2\tau}$, which is systematically low at $\sim\,$2$\,\sigma$
for $\ell_{\textrm{max}} = 1000$ for all methods.

On a more detailed level, however, there is some evidence of internal
tensions at the 1--$2\,\sigma$ level, most clearly seen in
$\Omega_{\textrm{c}}h^2$ and $A_{\rm s} e^{-2\tau}$. For these two
parameters, there are almost $1\,\sigma$ shifts for \nilc, \sevem, and
\smica\ going from $\ell_{\textrm{max}} = 1000$ to 1500, and again
from $\ell_{\textrm{max}} = 1500$ to 2000. \commander\ appears to be
somewhat more robust with respect to multipole range than the other
three methods. Although some variation is indeed expected by
statistical variation alone, the combination of the shapes of the
power spectrum differences seen in Fig.~\ref{fig:dx11_spectra} and the
systematic parameter trends suggest systematic uncertainties at the
1--$2\,\sigma$ level due to extragalactic foreground modelling, as
discussed above. For this reason, we do not recommend using the
component separated maps for cosmological parameter estimation at this
time; for this purpose the \Planck\ likelihood method is preferred,
which shows much better stability with respect to multipole range
\citep{planck2014-a13}.

For polarization, the results are more ambiguous, with fluctuations
relative to the temperature prediction beyond $2\,\sigma$. Note,
however, that the cosmic variance contributions to the temperature and
polarization parameters are essentially independent, and the two
estimates are therefore only expected to agree statistically, not
point-by-point. Still, in particular \sevem\ appears to show evidence
of larger deviations than expected in polarization for several
parameters, and the $\theta$ parameter for \commander\ in polarization
shows some hints of tension with respect to the corresponding
temperature estimate.

Overall, however, we find good consistency between the four different
component separation methods in both temperature and polarization. For
temperature, the main outstanding issues are uncertainties in the
residual extragalactic foreground model at the
$\sim10\muK^2$ level for \nilc, \sevem, and \smica, and a
substantially larger effective point source amplitude in the
\commander\ map compared to the other three maps. The parameters
derived from polarization observations are generally in good agreement
with the corresponding temperature parameters, except for the outliers
noted above.

\section{Higher-order statistics}
\label{sec:higherorder}

We now consider the higher-order statistics of the CMB maps in the
form of 1-point statistics (variance, skewness, and kurtosis),
$N$-point correlation functions, and primordial non-Gaussianity
($f_{\textrm{NL}}$).  We focus in particular on the polarization maps
and their degree of consistency with the FFP8 simulations.  The
temperature results are described in \citet{planck2014-a18},
\citet{planck2014-a19}, and \citet{planck2014-a20}.

\subsection{$1$-point statistics}
\label{sec:onepointstatistics}

The variance, skewness, and kurtosis of the maps, and the
pre-processing steps needed to compute them, are described in detail
in \citet{planck2014-a18}, \citet{Monteserin2008}, and
\citet{Cruz2011}.  The procedure is to normalize the data, $d_{p}$, in
pixel $p$ by its expected rms, $\hat{d}_{p} = d_{p}/\sigma_{p}$.  The
rms is calculated from 1000 FFP8 realizations \citep{planck2014-a13},
for both CMB anisotropies and instrumental noise.  To the extent that
$\sigma_{p}^2$ represents an accurate description of the data
variance, and both the sky signal and the instrumental noise are
Gaussian-distributed quantities, $\hat{d}$ will be Gaussian
distributed with zero mean and unit variance.  In temperature, the
variance of the instrumental noise is subtracted in order to determine
the variance of the CMB signal in the data. Once the variance of the
CMB signal is estimated, it is used to extract the skewness and
kurtosis from the normalized map. This procedure is well established,
and provides a direct test for the presence of residual foregrounds
and of CMB Gaussianity.

The temperature analysis reveals that the \Planck\ 2015 CMB maps and
the FFP8 fiducial CMB maps are fully compatible with the Monte Carlo
simulations, therefore they can be used for further statistical
analyses.  For more details about the temperature results and the FFP8
validation analysis please refer to Appendix~\ref{sec:1pdf_ffp8}
and~\citet{planck2014-a18}.

For polarization, since the $Q$ and $U$ maps are not rotationally
invariant, we consider the polarized intensity $P=\sqrt{Q^2+U^2}$.
$P$ is not Gaussian-distributed with zero mean, and its skewness and
kurtosis are non-vanishing; however, by comparing the data with the
Monte Carlo ensemble, we can test for residual foregrounds and
quantify the performance of the component separation methods.  Noise
in the $P$ maps is complicated; rather than trying to remove the noise
contribution (as is done in the case of temperature), we compare the
$P$ maps with the Monte Carlo simulations of CMB and noise together.
Table~\ref{table:1pdf} gives the resulting lower-tail probabilities,
that is, the percentage of simulations that show a lower variance,
skewness, or kurtosis than the $P$ maps.  This is done at three
different resolutions ($\nside = 1024$, $256$, and $64$) for both the
data maps (using mask \texttt{UPB77}) and the fiducial FFP8 maps
(using mask \texttt{FFP8-UPA76}).

\begin{table}[th!]
\begingroup
\newdimen\tblskip \tblskip=5pt
\caption{Percentage of simulations showing a lower variance, skewness,
  or kurtosis than the $P$ maps, the ``lower-tail probability''.  The
  top half of the table shows results for the data, and the bottom
  half for the FFP8 fiducial map, using the \texttt{UPB77} and
  \texttt{FFP8-UPA76} masks, respectively.  In brackets, in the
  variance column, we report the excess variance of the data in
  percent with respect to the mean of the variance distribution of the
  Monte Carlo simulations.}
\label{table:1pdf}
\nointerlineskip
\vskip -3mm
\footnotesize
\setbox\tablebox=\vbox{
   \newdimen\digitwidth \setbox0=\hbox{\rm
   0} \digitwidth=\wd0 \catcode`*=\active \def*{\kern\digitwidth}
   \newdimen\signwidth 
   \setbox0=\hbox{+} 
   \signwidth=\wd0 
   \catcode`!=\active 
   \def!{\kern\signwidth}
\halign{ \hbox to 1.1in{#\leaderfil}\tabskip=2em&
         \hfil#\hfil&
         \hfil#\hfil&
         \hfil#\hfil\tabskip=0pt\cr
\noalign{\doubleline}
\omit&\multispan3\hfil\sc Lower-tail Probability\hfil\cr
\noalign{\vskip -3pt}
\omit&\multispan3\hrulefill\cr
\noalign{\vskip 2pt}
\omit\hfil\sc Map\hfil&\omit\hfil Variance \hfil&\omit\hfil Skewness \hfil&\omit\hfil Kurtosis \hfil\cr
\noalign{\vskip 3pt\hrule\vskip 5pt}
\omit\bf Data \hfil\cr
\noalign{\vskip 4pt}
\omit\hglue 1em \bf$N_{\rm side}=1024$ \hfil\cr
\noalign{\vskip 4pt}
\hglue 2em \commander& 100.0 (3.5)& 89.5& 45.9\cr
\hglue 2em \nilc&      100.0 (4.0)& 99.9& 95.9\cr
\hglue 2em \sevem&     100.0 (4.5)& 93.6& 85.1\cr
\hglue 2em \smica&     100.0 (3.2)& 73.0& 49.7\cr
\noalign{\vskip 4pt}
\omit\hglue 1em \bf$N_{\rm side}=256$ \hfil\cr
\noalign{\vskip 4pt}
\hglue 2em \commander& 100.0 (7.5)& 62.3& 44.5\cr
\hglue 2em \nilc&      100.0 (8.6)& 54.8& 39.4\cr
\hglue 2em \sevem&     100.0 (9.2)& 57.6& 52.8\cr
\hglue 2em \smica&     100.0 (6.9)& 33.0& 24.2\cr
\noalign{\vskip 4pt}
\omit\hglue 1em \bf$N_{\rm side}=64$ \hfil\cr
\noalign{\vskip 4pt}
\hglue 2em \commander& 100.0 (11.7)& *0.0& *0.1\cr
\hglue 2em \nilc&      100.0 (18.0)& 22.3& 22.1\cr
\hglue 2em \sevem&     100.0 (20.5)& 14.6& *6.6\cr
\hglue 2em \smica&     100.0 (16.5)& 26.4& 30.3\cr
\noalign{\vskip 4pt}
\omit\bf FFP8 \hfil\cr
\noalign{\vskip 4pt}
\omit\hglue 1em \bf$N_{\rm side}=1024$ \hfil\cr
\noalign{\vskip 4pt}
\hglue 2em \commander& *95.9&  77.0&  88.1\cr
\hglue 2em \nilc&      100.0&  90.1&  98.5\cr
\hglue 2em \sevem&     100.0&  64.6&  65.6\cr
\hglue 2em \smica&     100.0&  91.2&  99.4\cr
\noalign{\vskip 4pt}
\omit\hglue 1em \bf$N_{\rm side}=256$ \hfil\cr
\noalign{\vskip 4pt}
\hglue 2em \commander& *92.3&  72.5&  65.8\cr
\hglue 2em \nilc&      *82.5&  25.7&  23.1\cr
\hglue 2em \sevem&     *79.9&  17.1&  25.6\cr
\hglue 2em \smica&     *62.0&  62.3&  58.9\cr
\noalign{\vskip 4pt}
\omit\hglue 1em {\bf$N_{\rm side}=64$} \hfil\cr
\noalign{\vskip 4pt}
\hglue 2em \commander& *97.5&  89.8&  73.0\cr
\hglue 2em \nilc&      *90.9&  53.0&  27.5\cr
\hglue 2em \sevem&     *68.7&  48.2&  33.1\cr
\hglue 2em \smica&     *78.1&  66.8&  35.9\cr
\noalign{\vskip 5pt\hrule\vskip 3pt}}}
\endPlancktable
\endgroup
\end{table}

At lower resolutions, Table~\ref{table:1pdf} shows good agreement
between the fiducial CMB maps and the Monte Carlo simulations for all
methods.  At high resolution, skewness, and kurtosis are mostly in
agreement with the simulations, with \nilc\ and \smica\ showing
slightly high values for the kurtosis (lower tail probabilities of
98.5\,\% and 99.4\,\%, respectively).  However, there is excess
variance in the maps at high resolution, 2--3.5$\,\sigma$ away from
the mean of the simulations, with \commander\ deviating the least.

\begin{figure}[th!]
\begin{center}
  \includegraphics[width=\columnwidth]{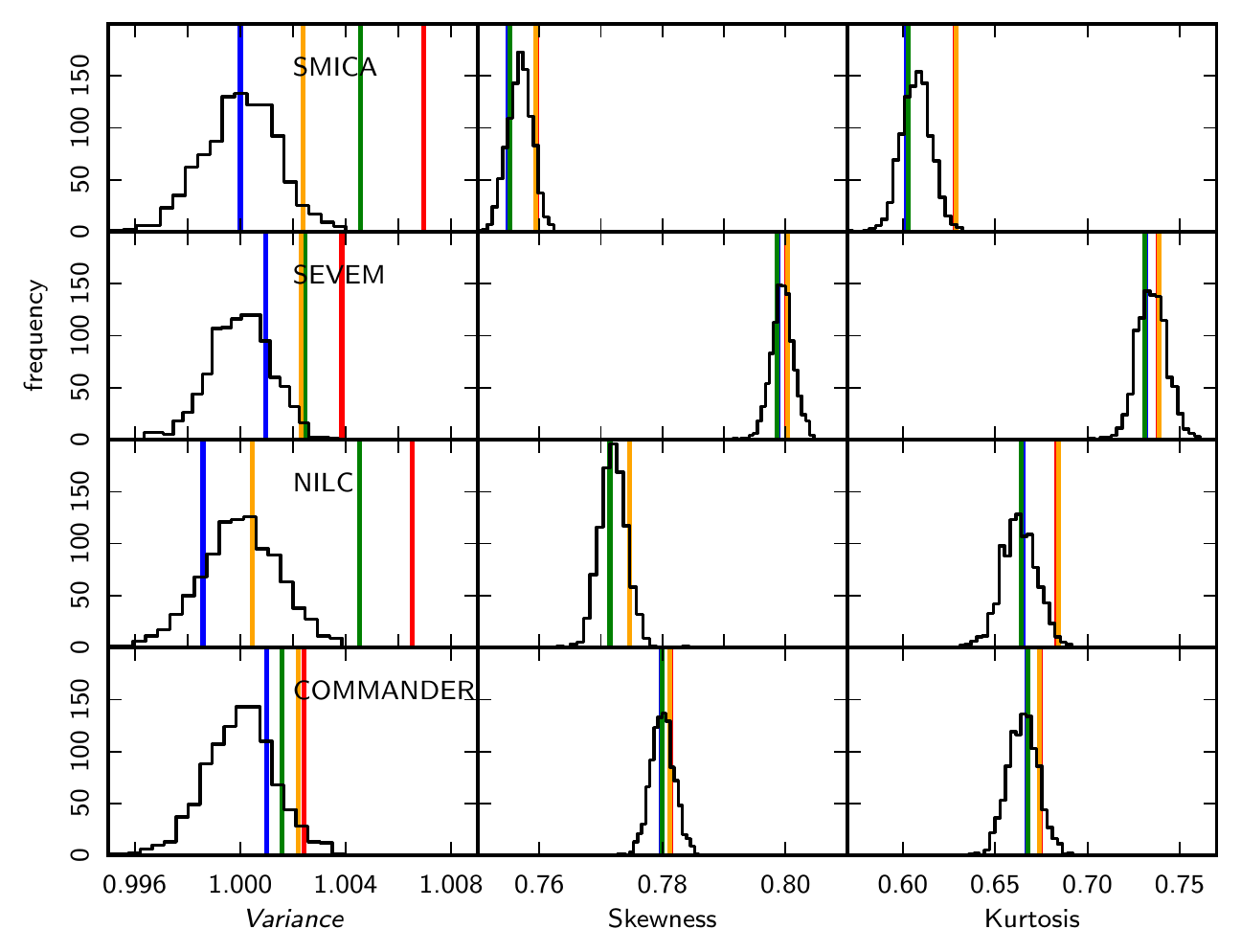}
\end{center}
\caption{Polarised intensity variance (\emph{left column}), skewness
  (\emph{middle column}), and kurtosis (\emph{right column}) evaluated
  from the FFP8 Monte Carlo simulations (histogram) and from
  components of the fiducial FFP8 map at $\nside = 1024$ outside the
  \texttt{FFP8-UPA76} mask.  The variance distributions have been
  normalized to the mean value of the Monte Carlo distributions for
  visualization purposes. Coloured vertical lines correspond to
  different combinations of components: the sum of CMB and noise is
  shown in blue; the sum of CMB, noise, and thermal dust is shown in
  green; the sum of CMB, noise, and radio point sources is shown in
  orange; the sum of CMB, noise, and all foregrounds is shown in red.}
\label{Fig:onepointsinglecomp}
\vskip12pt
\begin{center}
  \includegraphics[width=\columnwidth]{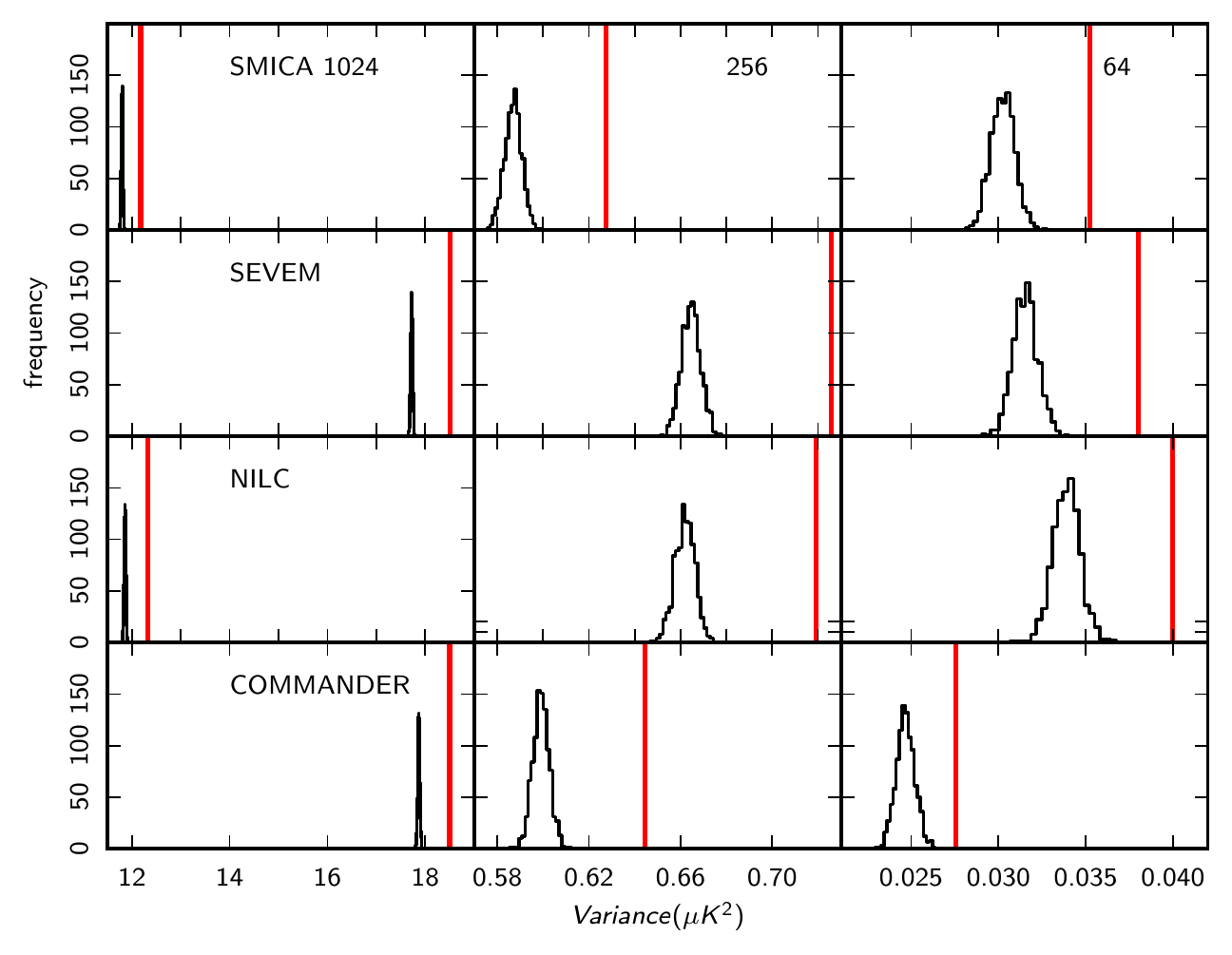}
\end{center}
\caption{Polarized intensity variance evaluated from the FFP8 Monte
  Carlo simulations (histogram) and from the \Planck\ 2015 maps
  (vertical red lines) outside the \texttt{UPB77} mask.  Columns from
  left to right show different resolutions ($\nside = 1024$, $256$,
  and $64$), while rows show results for the four component separation
  methods.  Unlike in Fig.~\ref{Fig:onepointsinglecomp}, the variance
  distributions are not normalized.}
\label{Fig:varianceDX11v2}
\end{figure}

Using the individual components of the FFP8 fiducial maps (CMB, noise,
and foregrounds), we are able to quantify the contributions to the
statistics of the high-resolution maps from each component separately.
Fig.~\ref{Fig:onepointsinglecomp} compares the variance (left column),
skewness (middle column), and kurtosis (right column) extracted from
the FFP8 Monte Carlo simulation $P$ maps (histograms) with those
extracted from the FFP8 fiducial realization of: the sum of CMB and
noise (blue); the sum of CMB, noise, and thermal dust (green); the sum
of CMB, noise, and radio point sources (orange); and the sum of CMB,
noise, and all foregrounds (red). The last defines the lower tail
probabilities in Table~\ref{table:1pdf}.  We have also investigated
other components, but found that their contributions are negligible.

We see that the only case that is fully compatible with the Monte
Carlo ensemble for all methods is that of CMB and noise (blue lines).
In particular, adding the thermal dust component (green lines)
increases the variance slightly outside the acceptable range for both
\nilc\ and \smica\, and the same effect becomes even stronger when
adding the rest of the foreground components for these
methods. \commander\ is only slightly affected by foregrounds by this
measure, and remains within the acceptable range even for the full
foreground model, while \sevem\ is an intermediate case.  For the
kurtosis, we find relatively high sensitivity to radio source
residuals (the orange and red lines coincide), but very low
sensitivity to diffuse foregrounds.  We conclude that the anomalous
statistics seen in analysis of the high-resolution component-separated
FFP8 CMB maps are due to the foreground components, which could
plausibly be caused by the additional complexity of the FFP8
foreground model with respect to the real sky.  The Monte Carlo
simulations are compatible with the CMB and noise components of the
fiducial map, as they should be by design.

Figure~\ref{Fig:varianceDX11v2} shows the variance of the Monte Carlo
simulations (from which the lower tail probabilities are summarized in
the top half of Table~\ref{table:1pdf}) compared with that of the
\Planck\ 2015 CMB $P$ maps.  The maps analysed are the sum of CMB and
noise signal.  Since the CMB is practically the same for all methods,
the different average values of the variance of the Monte Carlo
simulations are a reflection of somewhat higher noise on small angular
scales in the \sevem\ and \commander\ maps than in the \smica\ and
\nilc\ maps.  Looking at the data (red bars), we immediately see that
they do not match the simulations.  No simulation has a variance as
high as the data for any method at any resolution.  As already
mentioned, this discrepancy is due to an underestimation of the noise
in the FFP8 simulations~\citep{planck2014-a14}.  In the variance
column of Table~\ref{table:1pdf}, in parentheses, we report the excess
variance of the data in percent relative to the mean variance of the
Monte Carlo simulations.  The excess is 3--4\,\% at $\nside = 1024$,
increasing to 10--20$\,$\% at $\nside = 64$.  Moreover, we see that
there are some extreme values for both the skewness and kurtosis, in
particular at low resolution for \commander\ and at high resolution
for \nilc.  However, as long as the variance is inaccurate, it is
difficult to decide whether residual foregrounds, a true non-Gaussian
feature in the map, or noise underestimation is the cause.  For now,
we simply conclude that the simulations are inconsistent with the
data, and this effectively prevents studies of higher-order statistics
of this kind for the polarization maps.  For temperature, the same is
not true because of the much higher signal-to-noise ratio, although
care is warranted even then when probing into the noise-dominated
regime above $\ell\approx1500$-2000.

We also performed the 1-point analysis on the \Planck\ 2015
polarization maps using the \texttt{UP78} mask. In this case, the
values of the skewness and kurtosis for \smica, \nilc, and
\commander\ were significantly affected by the presence of point
sources. Conversely, the results for \sevem\ were consistent with the
expected distribution, because this method applies an inpainting
technique to remove the signal from the brightest point sources (see
Appendix~\ref{sec:sevem} for more details). This motivated the
construction of the \texttt{UPB77} mask, that excluded the brightest
point sources detected in polarization, which strongly alleviated this
problem for the other three methods as can be seen in the results of
Table~\ref{table:1pdf}. The same behaviour was also found for the FFP8
fiducial maps.

\subsection{$N$-point correlation functions}
\label{sec:npoint_correlation}

Real-space $N$-point correlation functions are a useful diagnostic of
the statistics of CMB maps complementary to harmonic analyses.  In
this section we describe their application to the \Planck\ 2015 CMB
polarization maps.  Results for the FFP8 CMB maps are given in
Section~\ref{sec:npoint_correlation_ffp8}. Details of their
application to temperature maps may be found
in~\citet{planck2014-a18}.

For observed fields $X$ measured in a fixed relative orientation on
the sky, the $N$-point correlation function is defined as
\begin{equation}
C_{N}(\theta_{1}, \ldots, \theta_{2N-3}) = \left\langle X(\vec{\hat{n}}_{1})\cdots
X(\vec{\hat{n}}_{N}) \right\rangle,
\label{eqn:npoint_def}
\end{equation}
where the unit vectors $\vec{\hat{n}}_1, \ldots,\vec{\hat{n}}_{N}$
span an $N$-point polygon on the sky. Assuming statistical isotropy,
$N$-point functions are functions only of the geometrical
configuration of the $N$-point polygon. In the case of the CMB, the
fields $X$ correspond to $\Delta T$ and two Stokes parameters $Q$ and
$U$ describing the linear polarization of the radiation in direction
$\vec{\hat{n}}$.  In standard CMB conventions, $Q$ and $U$ are defined
with respect to the local meridian of the spherical coordinate system
of choice.  However, $Q$ and $U$ form a spin-2 field and depend on a
rotational coordinate system transformation.  To obtain
coordinate-system-independent $N$-point correlation functions the
Stokes parameters are rotated with respect to a local coordinate
system defined by the centre of mass of the polygon (see
\citealp{gjerlow:2010} for details).  The Stokes parameters in this
new ``radial'' system are denoted by $Q_r$ and $U_r$.

Given rotationally invariant quantities $X \in \left\{ \Delta T, Q_r,
U_r \right\}$, the correlation functions are estimated by simple
product averages over all sets of $N$ pixels fulfilling the geometric
requirements set by $\theta_1, \ldots, \theta_{2N-3}$, which
characterize the shape and size of the polygon,
\begin{equation}
\hat{C}_{N}(\theta_{1}, \ldots, \theta_{2N-3}) 
= \frac{\sum_i \left(w_1^i \cdots w_N^i \right) \left( X_1^i \cdots
X_N^i \right) }{\sum_i w_1^i \cdots w_N^i} \ .
\label{eqn:npoint_estim}
\end{equation}
The pixel weights $w_1^i,\cdots,w_N^i$ are introduced to reduce noise
or mask boundary effects.  Masks set weights to 1 for included pixels
and 0 for excluded pixels.

The shapes of the polygons selected for the analysis are
pseudo-collapsed and equilateral configurations for the 3-point
function, and a rhombic configuration for the 4-point function
comprising two equilateral triangles sharing a common side.  The
4-point function is only computed in the analysis of temperature maps.
We use the same definition of pseudo-collapsed as in
\cite{eriksen2005}, that is, an isosceles triangle where the length of
the baseline falls within the second bin of the separation angles.
The length of the longer edges of the triangle, $\theta$, parametrizes
its size.  Analogously, in the case of the equilateral triangle and
rhombus, the size of the polygon is parametrized by the length of the
edge, $\theta$.  Note that these functions are chosen because of ease
of implementation, not because they are better suited for testing
Gaussianity than other configurations.  In the following, all results
refer to the connected 4-point function.

We analyse the high-pass filtered \Planck\ 2015 CMB maps at resolution
FWHM 160\arcm, $\nside = 64$. We used the downgraded version of the
\texttt{UP78} mask in the analysis.

\begin{figure*}[th!]
\begin{center}
\includegraphics[width=0.33\linewidth]{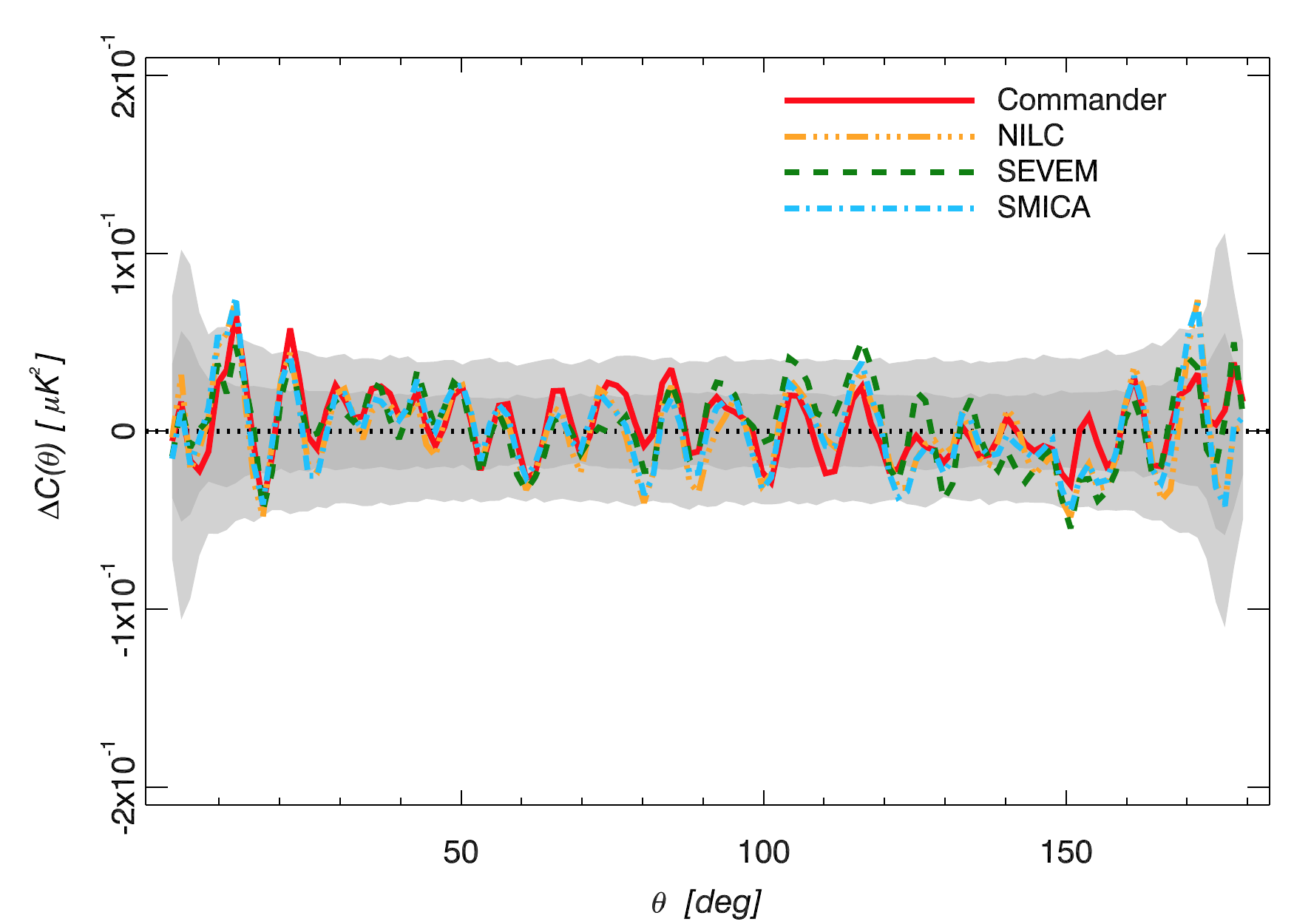}
\includegraphics[width=0.33\linewidth]{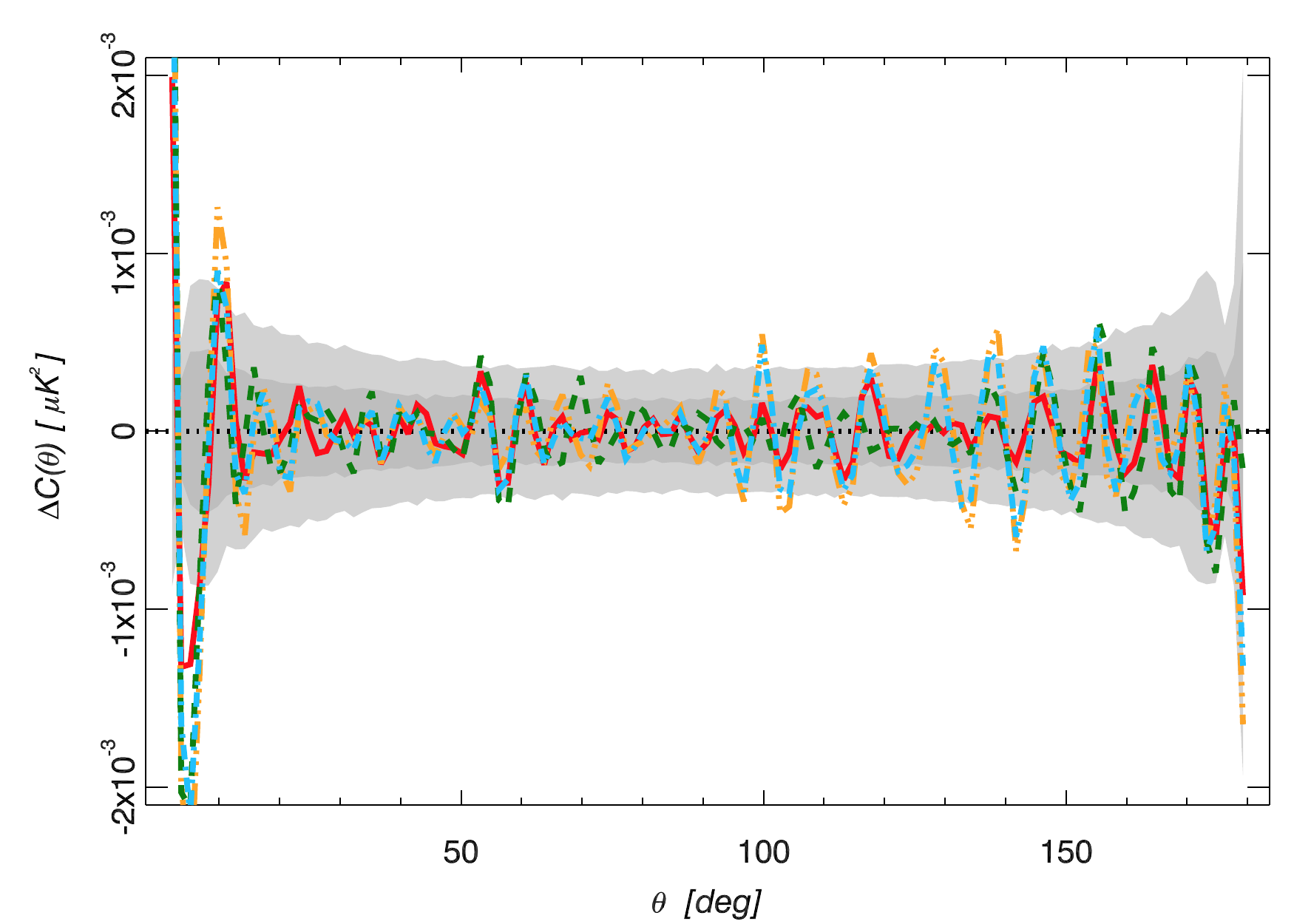}
\includegraphics[width=0.33\linewidth]{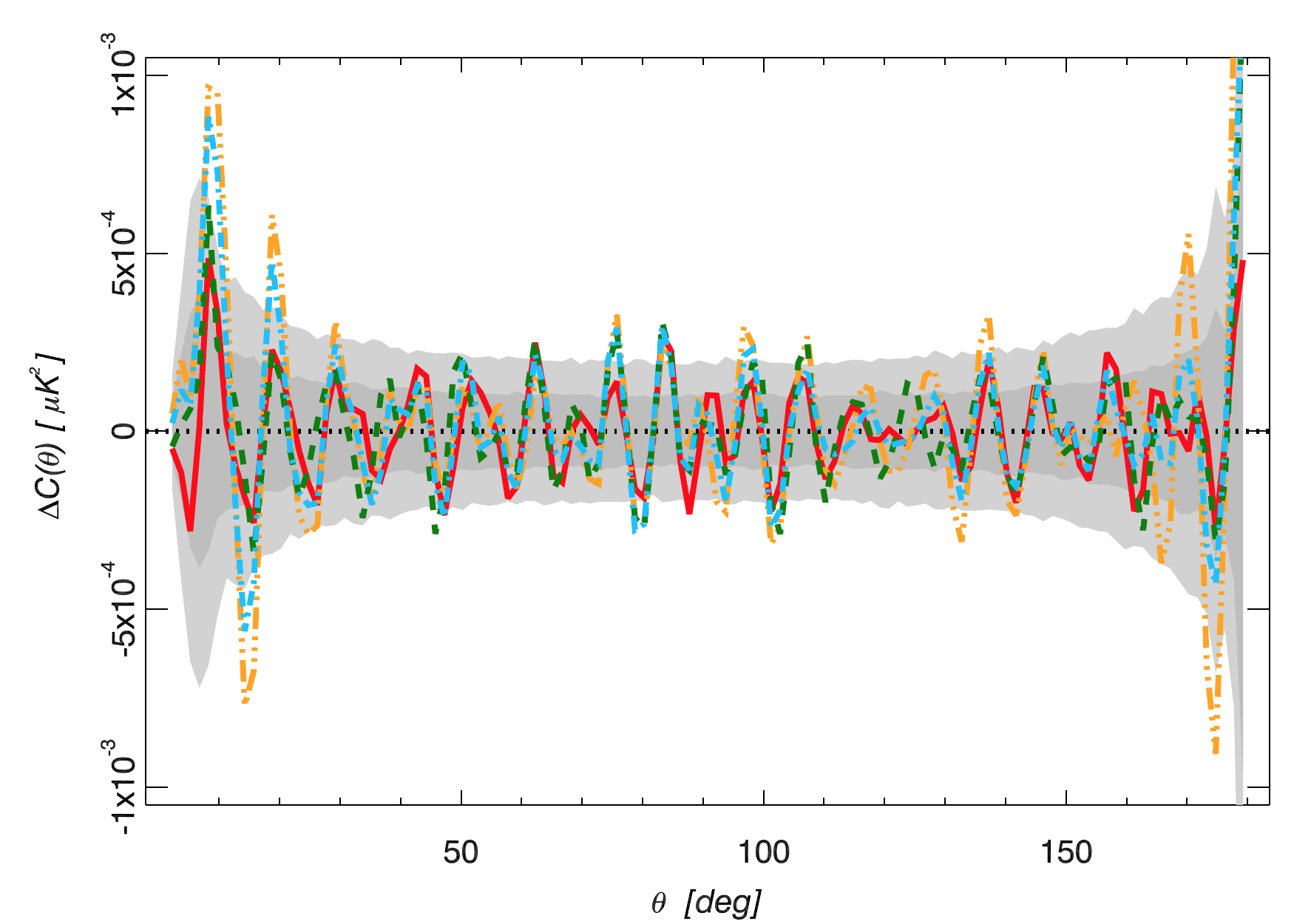}
\includegraphics[width=0.5\columnwidth]{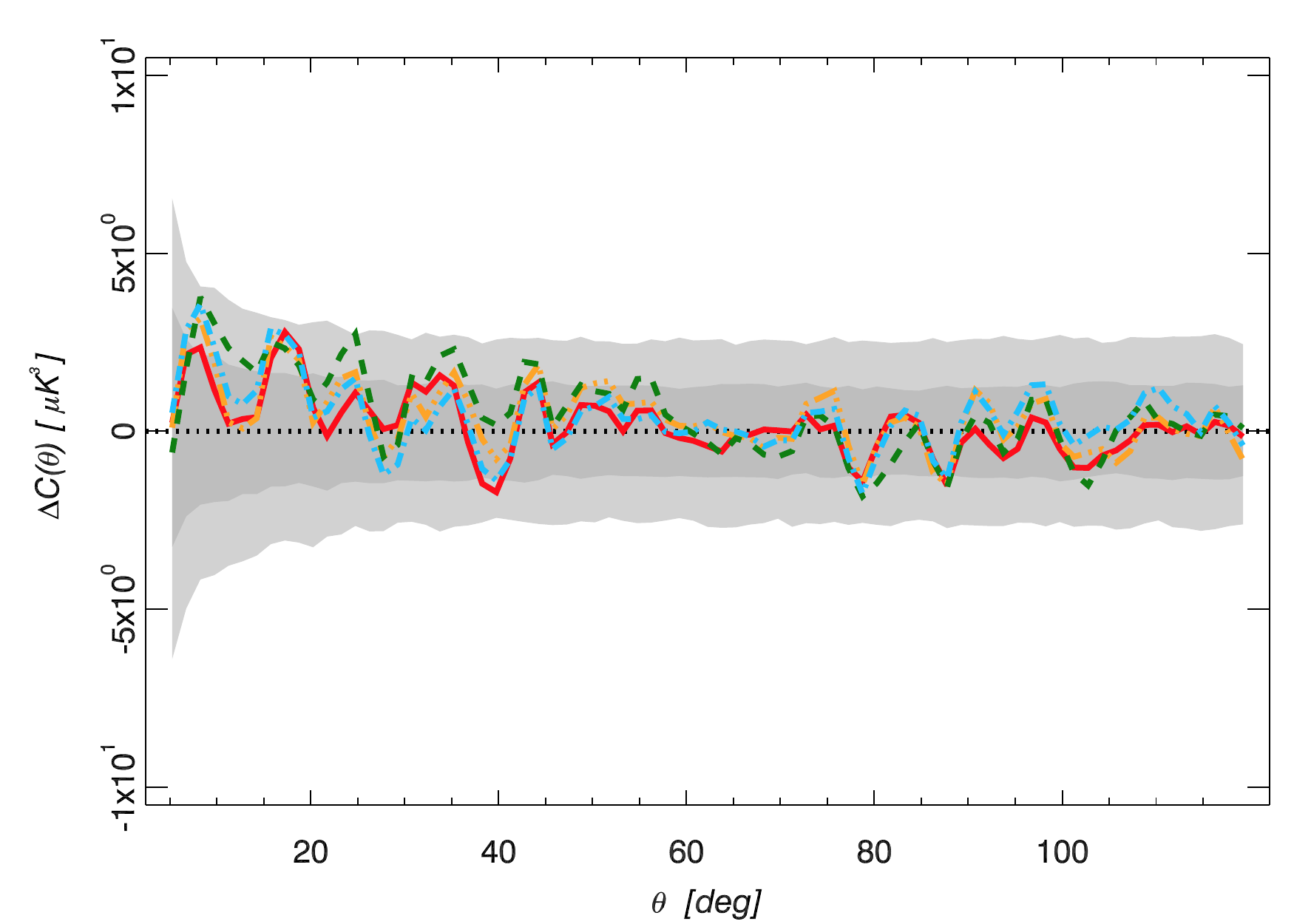}
\includegraphics[width=0.5\columnwidth]{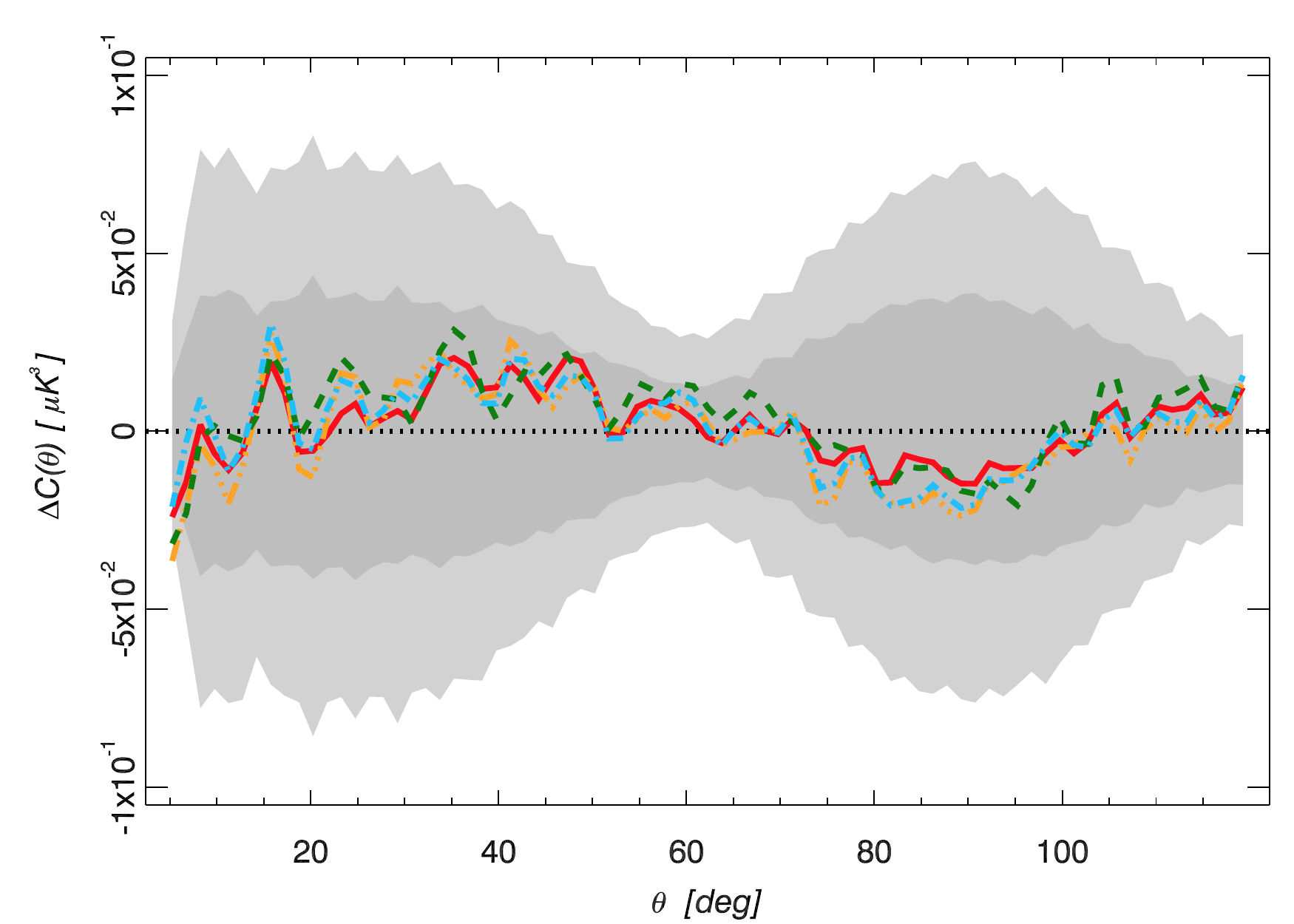}
\includegraphics[width=0.5\columnwidth]{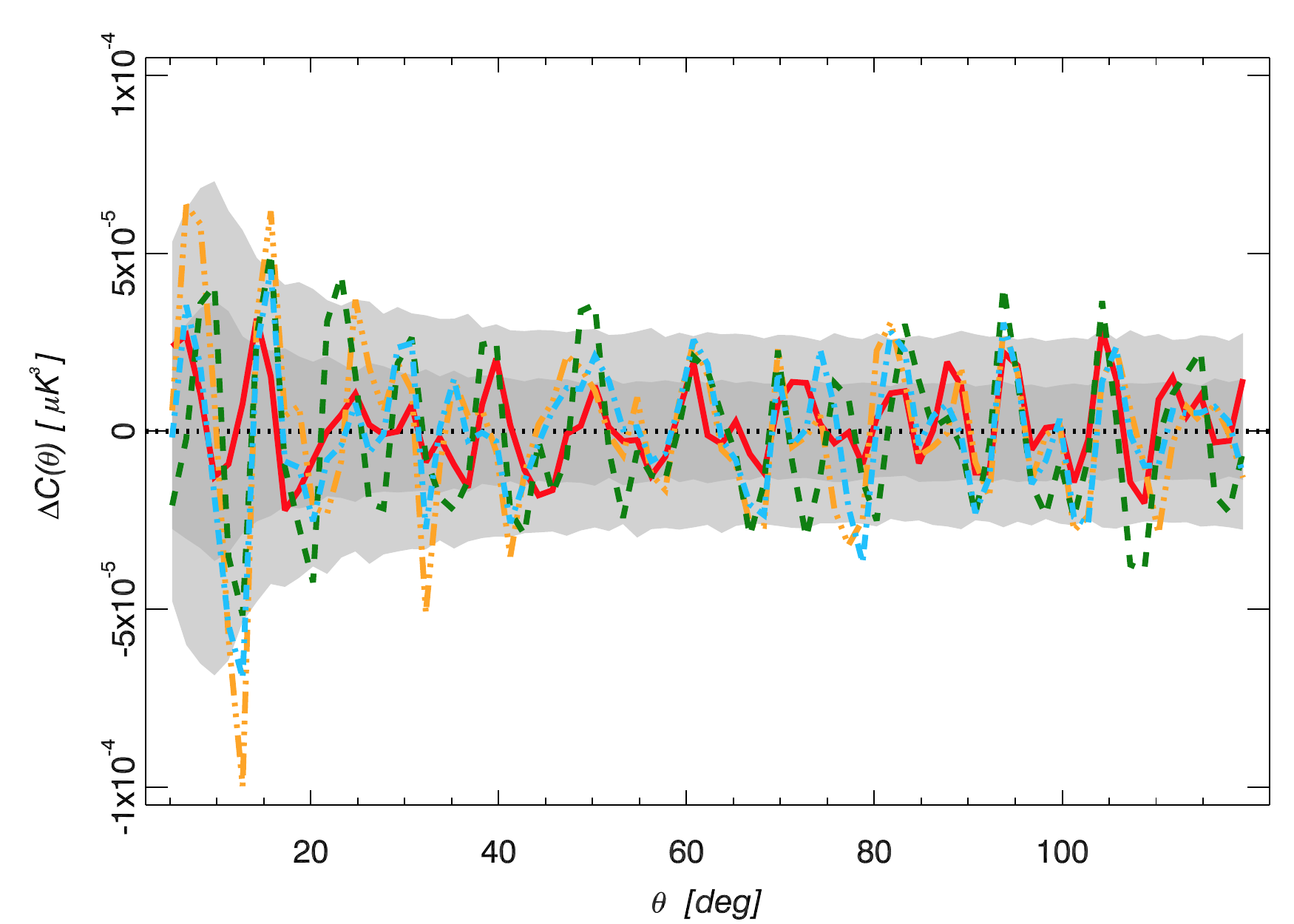}
\includegraphics[width=0.5\columnwidth]{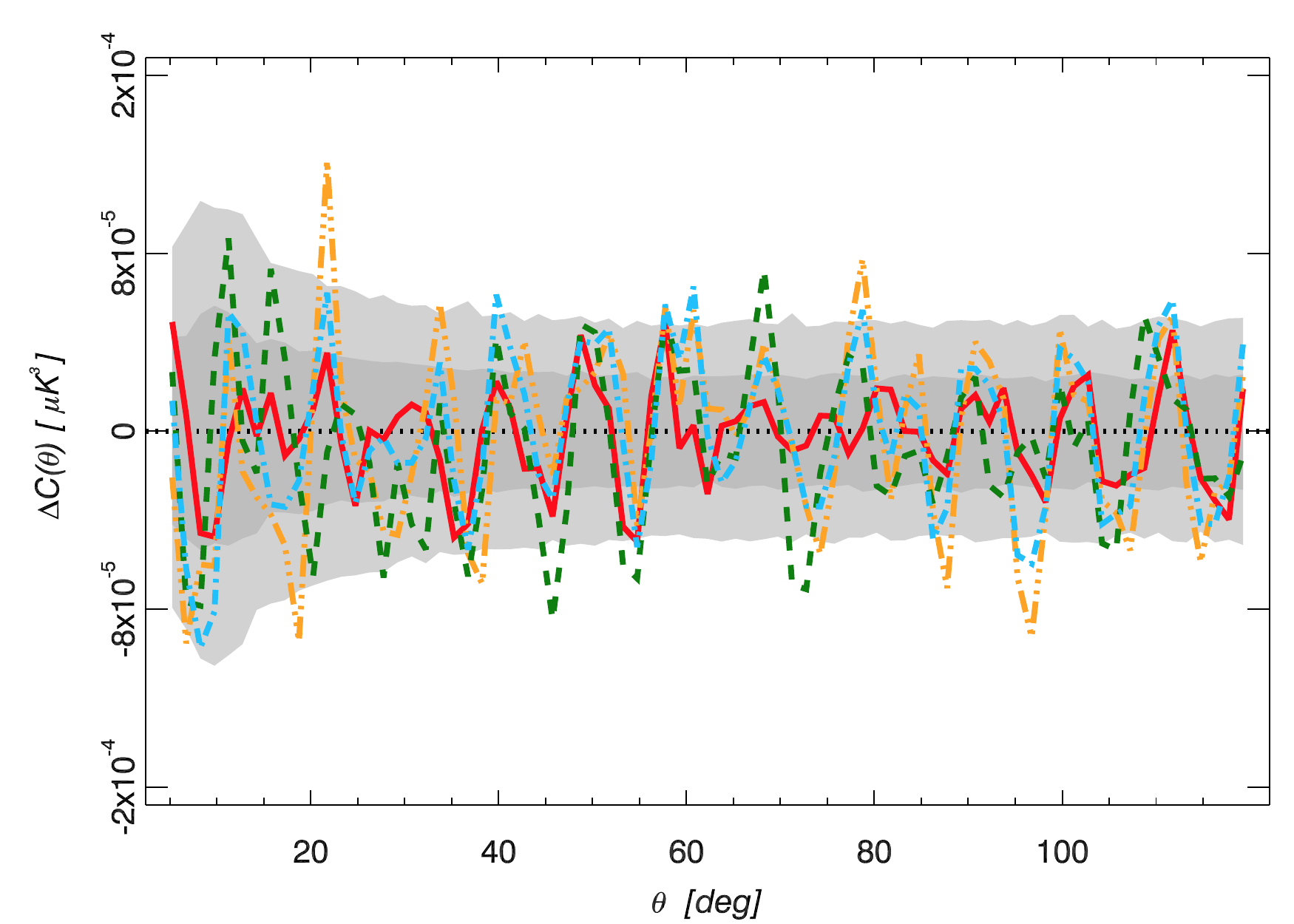}
\includegraphics[width=0.5\columnwidth]{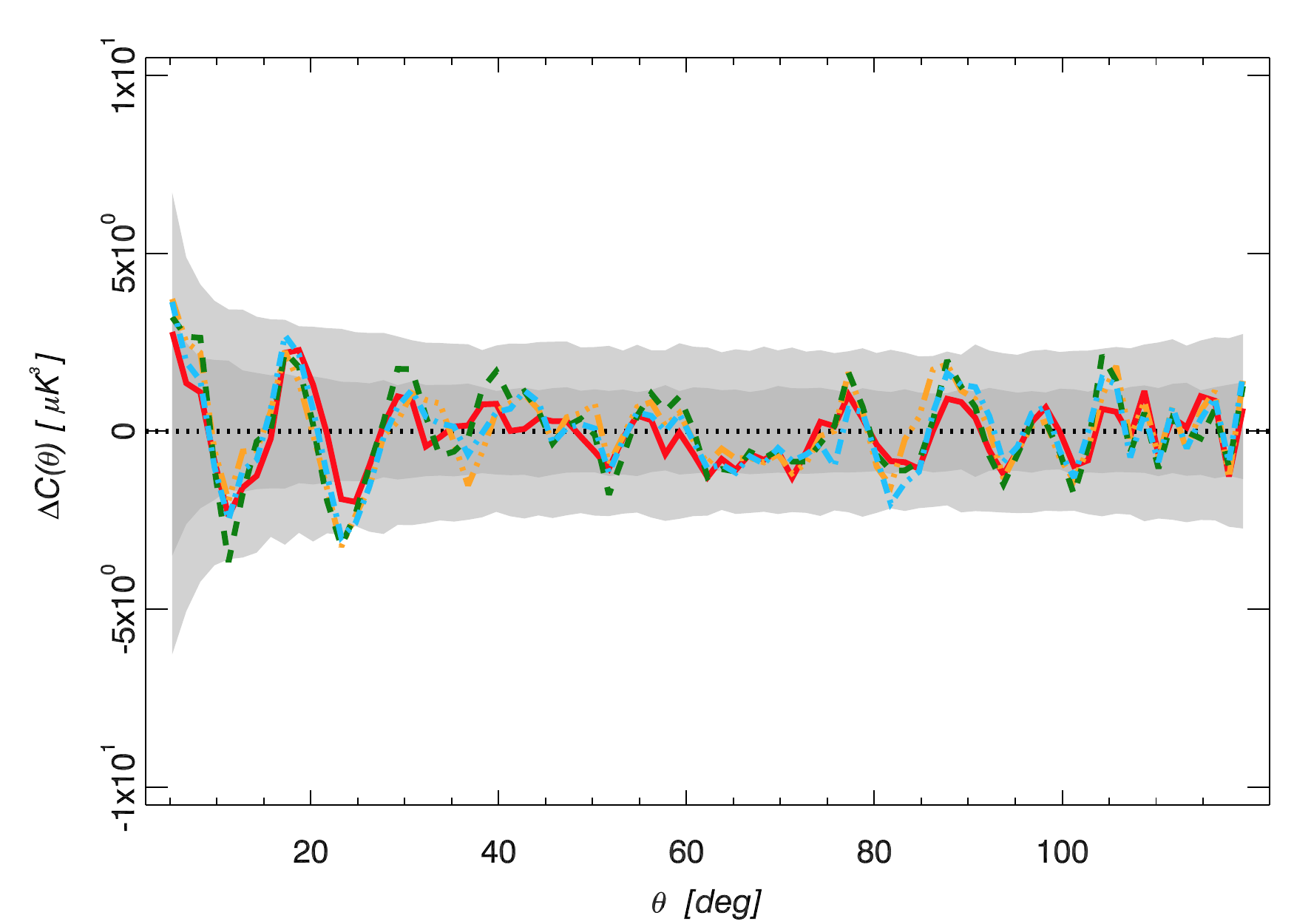}
\includegraphics[width=0.5\columnwidth]{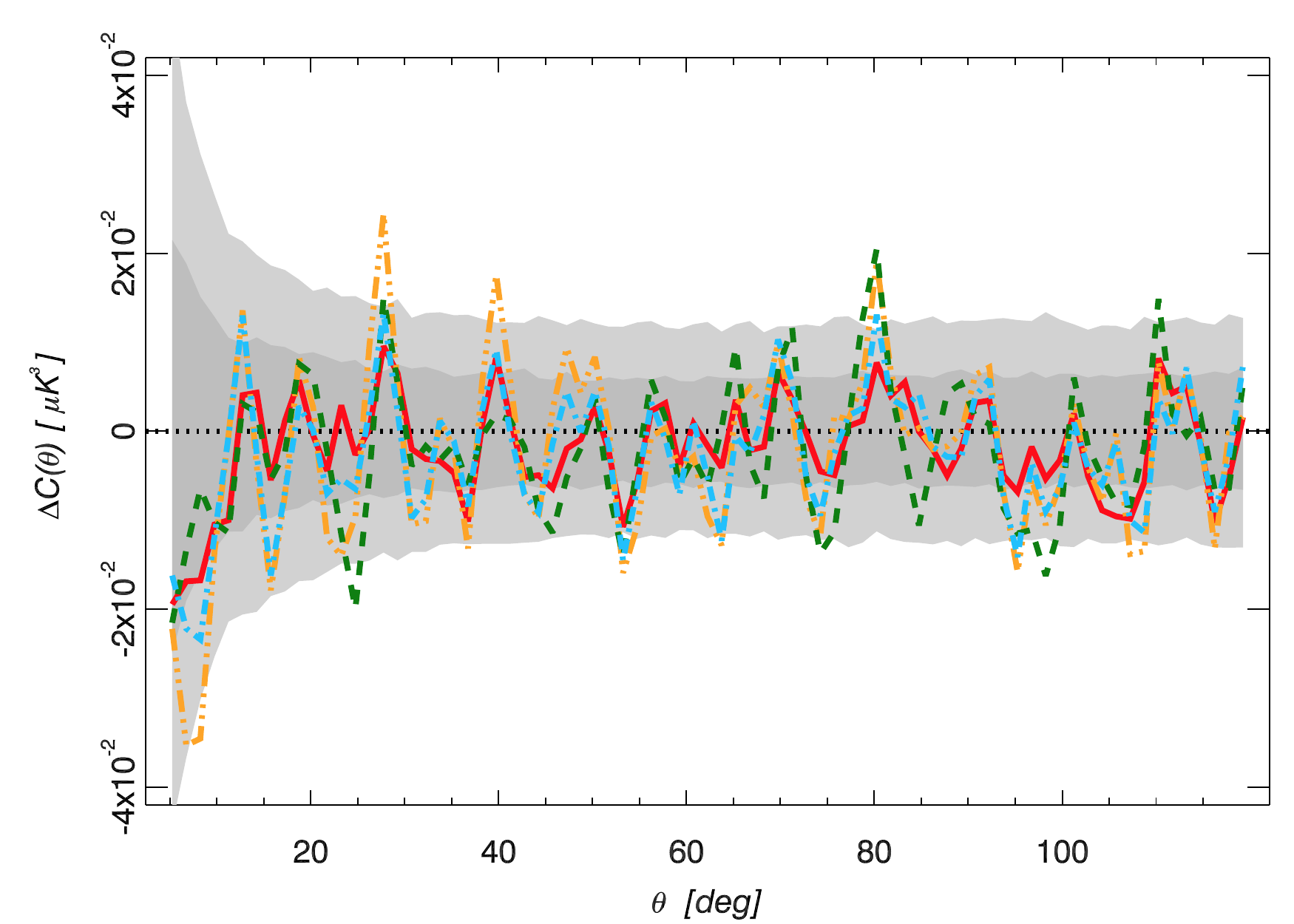}
\includegraphics[width=0.5\columnwidth]{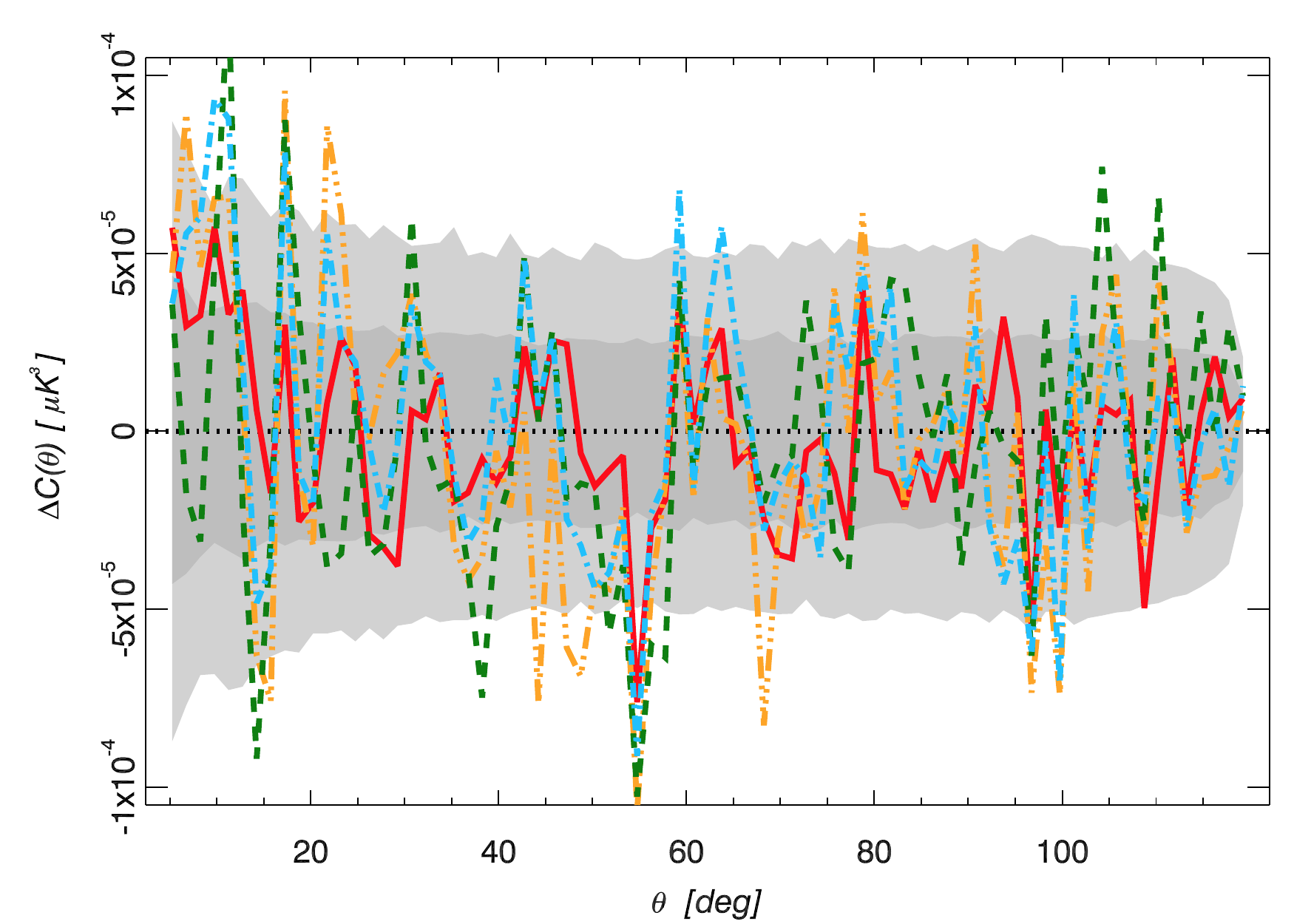}
\includegraphics[width=0.5\columnwidth]{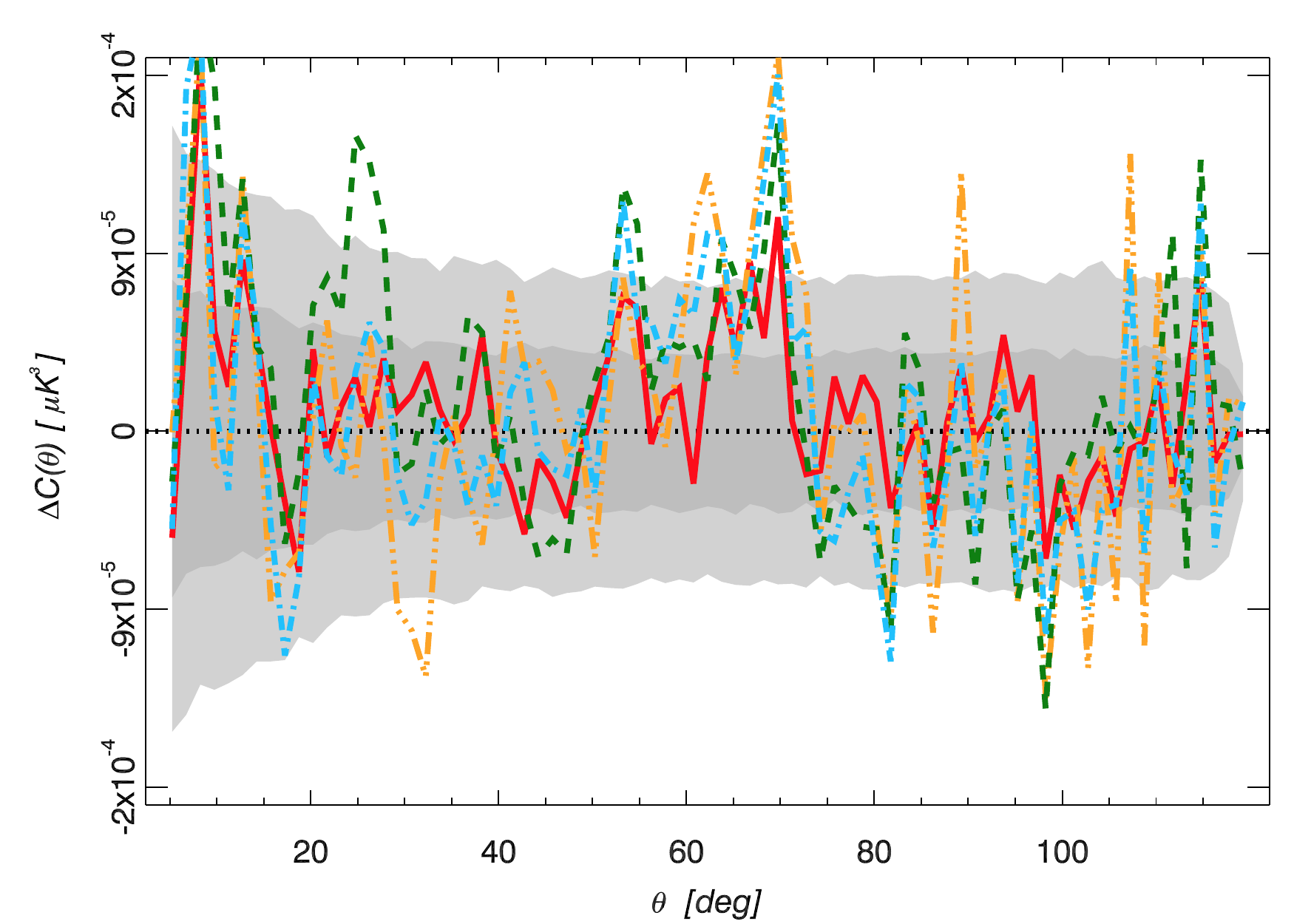}
\caption{The difference between the $N$-point functions for the
  high-pass filtered $N_{\rm side}=64$ \Planck\ 2015 CMB estimates and
  the corresponding means estimated from 1000 Monte Carlo
  simulations. The Stokes parameters $Q_r$ and $U_r$ were locally
  rotated so that the correlation functions are independent of
  coordinate frame. The first row shows results for the 2-point
  function, from left to right, $TQ_r$, $Q_rQ_r$, and $Q_rU_r$. The
  second row shows results for the pseudo-collapsed 3-point function,
  from left to right, $TTQ_r$, $TQ_rQ_r$, $Q_rQ_rU_r$, and
  $U_rU_rU_r$, and the third row shows results for the equialteral
  3-point function, from left to right, $TTQ_r$, $TQ_rQ_r$,
  $Q_rQ_rU_r$ and $U_rU_rU_r$. The red solid, orange dot dot
  dot-dashed, green dashed and blue dot-dashed lines correspond to the
  \commander, \nilc, \sevem, and \smica\ maps, respectively. The
  shaded dark and light grey regions indicate the 68\% and 95\%
  confidence regions, respectively, estimated using
  \smica\ simulations. See Sect.~\ref{sec:npoint_correlation} for the
  definition of the separation angle $\theta$.}
\label{fig:npt_data_hp}
\end{center}
\end{figure*}

The $N$-point functions are used to test the quality of the CMB
estimates derived from the \Planck\ data, and are shown in
Fig.~\ref{fig:npt_data_hp}.  We show the differences between the
$N$-point functions for the high-pass filtered CMB maps and the
corresponding mean values estimated from the 1000 Monte Carlo
simulations.  The probabilities of obtaining values of the $\chi^2$
statistic for the \Planck\ fiducial $\Lambda$CDM model at least as
large as for the CMB map are given in
Table~\ref{tab:prob_chisq_npt_data}.  The results of the analysis of
the temperature maps can be found in \citet{planck2014-a18}.

\begin{table}[th!] 
\begingroup
\newdimen\tblskip \tblskip=5pt
\caption{Probability-to-exceed (PTE) in percent for the $N$-point
  correlation function $\chi^2$ statistic applied to the \Planck\ 2015
  maps at $N_{\rm side} = 64$ for each of the four methods, as shown
  in Fig.~\ref{fig:npt_data_hp}. }
\label{tab:prob_chisq_npt_data}
\nointerlineskip
\vskip -5mm
\footnotesize
\setbox\tablebox=\vbox{
   \newdimen\digitwidth 
   \setbox0=\hbox{\rm 0} 
   \digitwidth=\wd0 
   \catcode`*=\active 
   \def*{\kern\digitwidth}
   \newdimen\signwidth 
   \setbox0=\hbox{+} 
   \signwidth=\wd0 
   \catcode`!=\active 
   \def!{\kern\signwidth}
\halign{ \hbox to 1.0in{#\leaderfil}\tabskip=1em&
         \hfil#\hfil\tabskip=1em&
         \hfil#\hfil\tabskip=2em&  
         \hfil#\hfil\tabskip=2em&
         \hfil#\hfil\tabskip=0pt\cr
\noalign{\doubleline}
\omit&\multispan4\hfil PTE [\%]\hfil\cr 
\noalign{\vskip -3pt}
\omit&\multispan4\hrulefill\cr
\noalign{\vskip 3pt}
\omit\hfil\sc  Function\hfil& \texttt{Commander}& \nilc& \sevem& \smica\cr
\noalign{\vskip 3pt\hrule\vskip 3pt}
\noalign{\vskip 4pt}
\multispan5 Two-point\hfil\cr
\noalign{\vskip 4pt}
\hglue 1em   $TQ_r$&    39.8&  *7.1&  11.0&  *6.4\cr
\hglue 1em   $TU_r$&    65.9&  39.4&  *0.5&  31.5\cr
\hglue 1em   $Q_rQ_r$&     *\llap{$<$}0.1&  *0.4&   *\llap{$<$}0.1&   *\llap{$<$}0.1\cr
\hglue 1em   $Q_rU_r$&     *\llap{$<$}0.1&  *0.4&   *\llap{$<$}0.1&   *\llap{$<$}0.1\cr
\hglue 1em   $U_rU_r$&     *\llap{$<$}0.1&  *0.4&   *\llap{$<$}0.1&   *\llap{$<$}0.1\cr
\noalign{\vskip 4pt}
\multispan5 Pseudo-collapsed three-point\hfil\cr
\noalign{\vskip 4pt}
\hglue 1em $TTQ_r$&     64.4&  61.2&  72.2&  72.1\cr
\hglue 1em  $TTU_r$&    44.2&  88.3&  69.6&  74.7\cr 
\hglue 1em  $TQ_rQ_r$&    21.9&  10.5&  12.7&  27.3\cr
\hglue 1em  $TQ_rU_r$&    *6.9&  *2.4&  *0.2&  *1.2\cr
\hglue 1em  $TU_rU_r$&    50.1&  11.1&  11.5&  *8.8\cr
\hglue 1em  $Q_rQ_rQ_r$&   *5.0&  *0.4&   *\llap{$<$}0.1&   *\llap{$<$}0.1\cr
\hglue 1em  $Q_rQ_rU_r$&   36.6&  *0.4&   *\llap{$<$}0.1&   *\llap{$<$}0.1\cr
\hglue 1em  $Q_rU_rU_r$&   *0.3&  *0.4&   *\llap{$<$}0.1&   *\llap{$<$}0.1\cr
\hglue 1em  $U_rU_rU_r$&   *1.4&  *0.5&  *0.2&  *1.6\cr
\noalign{\vskip 4pt}
\multispan5 Equilateral three-point\hfil\cr
\noalign{\vskip 4pt}
\hglue 1em  $TTQ_r$&     91.6&  75.1&  62.6&  80.8\cr
\hglue 1em  $TTU_r$&     50.1&  34.0&  63.5&  57.4\cr
\hglue 1em  $TQ_rQ_r$&     28.0&   6.0&  *7.1&  24.6\cr
\hglue 1em  $TQ_rU_r$&     21.9&   4.0&  12.6&  *7.7\cr
\hglue 1em  $TU_rU_r$&     22.1&  49.4&  20.4&  39.6\cr
\hglue 1em  $Q_rQ_rQ_r$&     *\llap{$<$}0.1&   0.4&   *\llap{$<$}0.1&   *\llap{$<$}0.1\cr
\hglue 1em  $Q_rQ_rU_r$&     *\llap{$<$}0.1&   0.4&   *\llap{$<$}0.1&   *\llap{$<$}0.1\cr
\hglue 1em  $Q_rU_rU_r$&     *\llap{$<$}0.1&   0.4&   *\llap{$<$}0.1&   *\llap{$<$}0.1\cr
\hglue 1em  $U_rU_rU_r$&     *\llap{$<$}0.1&   0.4&   *\llap{$<$}0.1&   *\llap{$<$}0.1\cr
\noalign{\vskip 3pt\hrule\vskip 3pt}}}
\endPlancktable
\endgroup
\end{table} 

The results for the data deviate significantly from the Monte Carlo
simulations for almost all $N$-point functions involving at least two
polarization fields.  The smallest deviation is seen for the
\commander\ map.  The $N$-point functions for Monte Carlo simulations
have smaller variance than for data.  By comparing HMHD maps with
Monte Carlo simulations of noise corresponding to the CMB estimates,
we find that the amplitude of the noise is underestimated by 18\,\%.
After adjusting the Monte Carlo simulations to compensate, the
$N$-point functions are more consistent with data.  It is difficult to
draw conclusions about the Gaussianity of the polarization maps from
these results, since the $N$-point functions themselves are used to
estimate the mismatch between the noise Monte Carlo simulations and
the data.

\subsection{Primordial non-Gaussianity}
\label{sec:fnl}

Primordial non-Gaussianity is often measured in terms of the
amplitude, $f_{\textrm{NL}}^{\textrm{local}}$, of the quadratic
corrections to the gravitational potential, as well as by means of the
three-point correlation function based on different triangle
configurations.  The results from these calculations for the
foreground-cleaned CMB maps are presented in \citet{planck2014-a19}.
Compared to the previous data release, we can now include both
temperature and polarization bispectra in the analysis.  We thus
consider bispectrum $f_{\textrm{NL}}$ estimates obtained from
temperature and polarization data only, as well as the full constraint
from all eight possible $TTT$, $TTE$, $EET$, and $EEE$ combinations.

Results obtained from application of the Komatsu-Spergel-Wandelt (KSW)
estimator to the CMB maps after subtraction of the lensing-ISW
correlation (see below) are listed in Table~\ref{tab:ng} for various
geometrical configurations that have been
considered~\citep{planck2014-a19}.  It is interesting to evaluate the
impact of polarization data on the estimation of $f_{\rm NL}$
measurements.  By considering only the temperature bispectrum for the
\smica\ map, we obtain $f_{\rm NL({\smica})}^{\rm local}=1.3 \pm 5.7$,
while the polarization alone yields $f_{\rm NL({\smica})}^{\rm
  local}=28.4 \pm 31.0$.  Uncertainties are evaluated by means of
Gaussian FFP8 simulations.  We find consistency between pipelines at
the $1\,\sigma$ level.  The results confirm and extend to polarization
the absence of evidence of non-Gaussianity of primordial origin
estimated through the bispectrum.  The performance of the methods is
also tested using Gaussian and non-Gaussian FFP8 simulations, showing
that \smica\ and \sevem\ give the results closest to the inputs in
both cases (see section 7.3 of \citealp{planck2014-a19}).

An interesting case to consider is that of the ISW-lensing bispectrum.
Unlike with primordial shapes, there is a specific prediction for the
expected amplitude of the ISW-lensing three-point signal in a given
cosmological model.  We can therefore use this shape to check whether
the expected level of NG is recovered, verifying that different
component separation methods do not either spuriously add or remove
any signal, at least in the squeezed limit where this shape is peaked.
This is indeed the case.  By normalizing the ISW-lensing shape in such
a way as to have an $f_{\textrm{NL}}$ amplitude of $1$ in the best-fit
model, we recover $f_{\textrm{NL({\smica)}}} = 0.85 \pm 0.2$ from the
full analysis including all bispectra, $f_{\textrm{NL({\smica)}}} =
0.6 \pm 0.3$ from temperature alone, and $f_{\textrm{NL({\smica)}}} =
4.7 \pm 6.0$ from polarization alone.

\begin{table}[th!]
\begingroup
\newdimen\tblskip \tblskip=5pt
\caption{Amplitude of primordial non-Gaussianity, $f_{\rm{NL}}$,
  estimated by the KSW estimator. See Table~10 in
  \citet{planck2014-a19} for full details.\label{tab:ng}}
\nointerlineskip
\vskip -5mm
%\footnotesize
\scriptsize
\setbox\tablebox=\vbox{
\newdimen\digitwidth
\setbox0=\hbox{\rm 0}
\digitwidth=\wd0
\catcode`*=\active
\def*{\kern\digitwidth}
\newdimen\signwidth
\setbox0=\hbox{+}
\signwidth=\wd0
\catcode`!=\active
\def!{\kern\signwidth}
\newdimen\decimalwidth
\setbox0=\hbox{.}
\decimalwidth=\wd0
\catcode`@=\active
\def@{\kern\signwidth}
\halign{ \hbox to 1.0in{#\leaderfil}\tabskip=1.5em&
    \hfil$#$\hfil\tabskip=1.5em&
    \hfil$#$\hfil\tabskip=1.5em&
    \hfil$#$\hfil\tabskip=1.5em&
    \hfil$#$\hfil\tabskip=0pt\cr
\noalign{\doubleline}
\omit&\multispan4\hfil $f_{\rm NL}$\hfil\cr 
\noalign{\vskip -3pt}
\omit&\multispan4\hrulefill\cr
\noalign{\vskip 3pt}
\omit\hfil\sc Type\hfil&\omit\hfil{\tt Commander}\hfil&\omit\hfil{\tt NILC}\hfil&\omit\hfil{\tt SEVEM}\hfil&\omit\hfil{\tt SMICA}\hfil\cr
\noalign{\vskip 4pt\hrule\vskip 8pt} 
T local&	    !**4\pm**6&  !**3\pm **6& !**4\pm **6& *!*3\pm **6\cr
T equilateral& *-20\pm*71& *-28\pm *69& **-2\pm *69& *-11\pm *70\cr
T orthogonal&  *-29\pm*35& *-45\pm *33& *-36\pm *33& *-34\pm *33\cr
\noalign{\vskip 6pt}
TE local&	    *!*4\pm**5& !**1\pm **5& **-3\pm **5& !**1\pm **5\cr
TE equilateral&*!14\pm*46& **-9\pm *44& !**8\pm *47& !**3\pm *43\cr
TE orthogonal& *-29\pm*22& *-25\pm *21& *-39\pm *23& *-25\pm *21\cr
\noalign{\vskip 6pt}
E local&	    *!33\pm*39& **-1\pm *33& !*60\pm *42& !*26\pm *32\cr
E equilateral& !327\pm165& !*75\pm 140& !292\pm 167& !144\pm 141\cr
E orthogonal&  *-52\pm*88& *-78\pm *76& -183\pm *91& -128\pm *72\cr
\noalign{\vskip 5pt\hrule\vskip 5pt}
}}
\endPlancktablewide
\endgroup
\end{table}

The constraint from polarization alone is, as expected, much looser
than the one from temperature alone.  However, polarization does not
have a negligible impact on the final combined measurement, mostly due
to contributions coming from $TTE$ configurations.  Adding
polarization reduces the final uncertainty by $\sim 30\,\%$, as well
as moving the recovered amplitude parameter closer to its expected
value of $1$.  The implications of these results in terms of the
physics of the early Universe, as well as the study of many additional
shapes, are discussed in \citet{planck2014-a19}, which explains the
algorithmic details and all results from the procedure summarized
here, and in \citet{planck2014-a24}, which discusses the implications
for inflationary physics.

\section{Gravitational lensing by large-scale structure}
\label{sec:lensing}

Gravitational lensing from intervening matter imprints a
non-Gaussian signature in the CMB temperature and polarization maps,
which in turn can be exploited to extract the gravitational potential
integrated along the line of sight back to the surface of last
scattering. \Planck\ accurately measures the lensing potential over
most of the sky \citep{planck2014-a17}. As it remaps the CMB
polarization, lensing partially transforms \emph{primordial} $E$~modes
into $B$~modes, resulting in a \emph{secondary} $B$~mode spectrum
peaking at around $\ell=1000$.  By forming a weighted product of the
$E$~modes from component separated CMB maps and the reconstructed
lensing potential, it is possible to generate a map of the expected
lensing-induced $B$~modes of CMB polarization. Cross-correlating this
lensing $B$-mode template to the total observed $B$~modes provides an
indirect measurement of the lensing $B$-mode power spectrum
\citep{planck2014-pip116, planck2014-a17}.  Lensing $B$-mode template
maps were synthesized for the four component separation methods
considered in this paper, using the Stokes parameter maps and the
lensing potential reconstruction from the intensity map as discussed
in \citet{planck2014-pip116}. A common mask was generated by combining
the union polarization mask and the lensing potential 80\,\% mask of
\citet{planck2014-pip116}. After apodization using a cosine function
over 3\deg, the resulting mask preserves an effective sky fraction of
about 60\,\%.

\begin{figure}
  \begin{center}
    \includegraphics[width=\columnwidth]{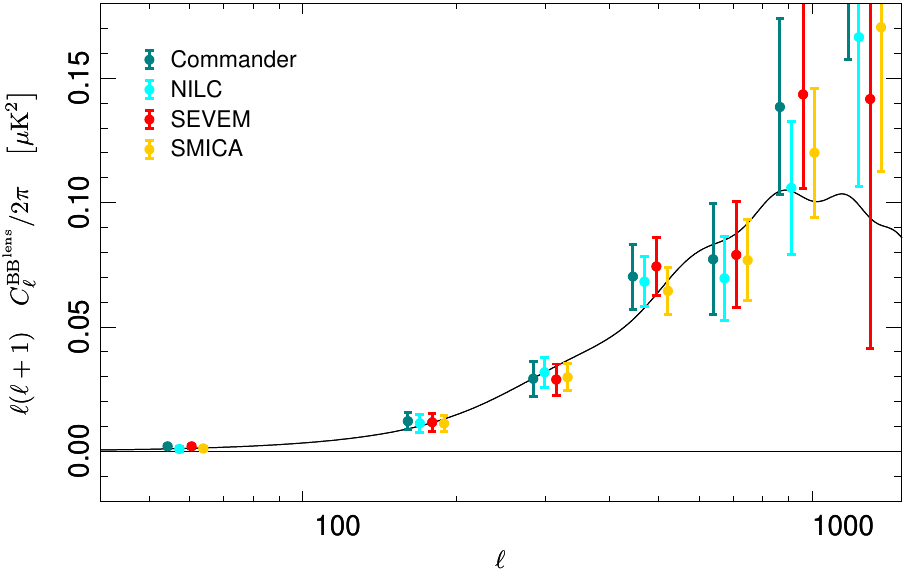}
  \end{center}
  \caption{Lensing-induced $B$-mode power spectra in the component-separated polarization CMB maps. The solid line represents  
    the best fit cosmology from the \Planck\ data release in 2015. Error
    bars were evaluated using a semi-analytical approximation
    validated over the FFP8 simulations as described in
    \citet{planck2014-pip116}.}
  \label{fig:lensing_b_modes}
\end{figure}

The lensing $B$-mode power spectra for \commander, \nilc, \sevem, and
\smica, which have been measured by cross-correlating with the
corresponding foreground cleaned polarization maps, are shown in
Fig.~\ref{fig:lensing_b_modes}.  The four CMB polarization solutions
lead to consistent lensing $B$-mode power spectrum measurements within
$1\,\sigma$ over the entire probed multipole range. In addition, the
$\chi^2$ relative to the 2015 \Planck\ \LCDM\ base model
\citep{planck2014-a15} of the lensing $B$-mode band powers in eight
multipole bins are 12.19, 8.99, 7.96, and 6.98, with corresponding
probability-to-exceed values of 14, 34, 44, and 54\,\% for \commander,
\nilc, \sevem, and \smica\, respectively, which indicate the agreement
of the \Planck\ lensing B-mode signal with theoretical expectations.
We measure amplitude fits with respect to the \Planck\ base model of
\begin{center}
\begin{tabular}{ll}
  $A_{B^{\mathrm{lens}}} = 1.03 \pm 0.11$ & (\commander),\\
  $A_{B^{\mathrm{lens}}} = 0.97 \pm 0.09$ & (\nilc),\\
  $A_{B^{\mathrm{lens}}} = 1.02 \pm 0.10$ & (\sevem),\\
  $A_{B^{\mathrm{lens}}} = 0.97 \pm 0.08$ & (\smica),
\end{tabular}
\end{center}
corresponding to 10, 11, 10, and 12$\,\sigma$ detections of lensing
$B$~modes for \commander, \nilc, \sevem, and \smica, respectively.

\section{Summary and recommendations}
\label{sec:discussion}

We now summarize the results, and provide a critical analysis of the
applicability of the derived maps for cosmological purposes.

Starting with the temperature case, we have shown that the four CMB
maps are in excellent agreement overall. The amplitudes of the
pairwise difference maps are smaller than $5\,\mu\textrm{K}$ over most
of the sky on large angular scales, and the high-$\ell$ power spectra
agree to $\sim\,$1$\,\sigma$. Correspondingly, the differences with
respect to the \Planck\ 2013 maps are typically smaller than
$10\,\mu\textrm{K}$ over most of the sky, and the morphology of the
differences is well understood in terms of improved treatment of
systematic errors in the 2015 analysis. We conclude that the
\Planck\ 2015 temperature maps provide a more accurate picture of the
CMB sky than the 2013 temperature maps, having both lower noise and
lower levels of systematics, and we expect these maps to find the same
cosmological applications as the previous generation of maps. However,
we emphasize that these maps are not cleaned of high-$\ell$
foregrounds, such as extragalactic point source or SZ
emission. Residuals lead to biases in cosmological parameters at
$2\,\sigma$ level, beyond $\ell\simeq 2000$.  Cosmological analyses
using small angular scales must therefore take care to marginalize
over such foregrounds as appropriate.

For polarization, the situation is significantly more complicated due
to two different problems. First, a low level of residual systematics
on large angular scales prevents a faithful CMB polarization
reconstruction on multipoles $\ell \lesssim 20$. These modes are
therefore removed by high-pass filtering in the current maps. Any
cosmological analysis of these maps must take into account the
corresponding transfer function in order to avoid biases. Second, due
to the current noise mismatch between the FFP8 simulations and the
data, we strongly caution against using the polarization maps provided
here for any cosmological analysis that depends sensitively on the
assumed noise level.  Nevertheless, the maps should prove useful for a
number of other applications that do not require detailed noise
simulations, for instance estimation of cross-spectra or
cross-correlations, or stacking analyses, and we therefore 
release the maps to the public despite these limitations.

Considering the four component separation methods in greater detail,
the results may be distinguished according to two
criteria, data selection and basis functions. While
\commander\ performs data selection at the detector (or detector set)
map level, rejecting potentially problematic maps, the other three
methods employ frequency channel maps, and thereby maximize the
signal-to-noise ratio. Likewise, while \commander\ and \sevem\ perform
their analyses in pixel space, \nilc\ and \smica\ perform all
operations in harmonic space. These distinctions can explain many of
the qualitative differences discussed in the previous sections.

We make the following recommendations regarding the use of the four
maps. First and foremost, we strongly recommend that any cosmological
analysis based on these maps consider all four maps in parallel in
order to assess the impact of specific choices of implementation and modelling. To be considered robust, no results should depend strongly on
the specifics of a given component separation algorithm. Considering
specific details, we generally consider \commander\ to be the
preferred solution on large and intermediate angular scales, due to
its somewhat lower large-scale effective polarization noise
(Table~\ref{tab:solution_parameters}), weaker cross-correlation with
the high-pass filtered WMAP K$-$Ka band synchrotron template
(Table~\ref{tab:cross_corr}), lower $N$-point correlation function
fluctuations (Fig.~\ref{fig:npt_data_hp} and
Table~\ref{tab:prob_chisq_npt_data}), and weaker cosmological
parameter dependence on $\ell_{\textrm{max}}$, suggesting less
internal tension between low, intermediate, and high multipoles
(Fig.~\ref{fig:dx11_params_TT_EE}). In addition, the method is able to
propagate uncertainties from the input maps to final products by means
of Monte Carlo samples drawn from the full posterior. For these
reasons, we adopt the \commander\ solution for the low-$\ell$
\Planck\ 2015 temperature likelihood \citep{planck2014-a13}.

However, at high multipoles the \commander\ solution exhibits a
significantly higher effective point source amplitude than the other
three maps, due to the exclusion of frequencies below 217$\,$GHz. For
temperature, the lowest residual high-$\ell$ foregrounds are instead
obtained by \smica, as shown in Fig.~\ref{fig:dx11_spectra}.  As a
result, as in 2013, we confirm our preference for the \smica\ map for
analyses that require full-resolution observations in temperature,
such as $f_{\textrm{NL}}$ (first three rows in Table~\ref{tab:ng} and
\citealp{planck2014-a19}) or lensing reconstruction
(Fig.~\ref{fig:lensing_b_modes} and \citealp{planck2014-a17}).
\sevem\ is also a very good choice for temperature, providing the map
with the lowest level of noise at a wide range of scales as well as a
smooth noise power spectrum, as measured by the HMHD maps (see
Table~\ref{tab:solution_parameters} and
Fig.~\ref{fig:dx11_spectra}). It also performs equally well as
\smica\ with regard to the estimation of $f_{\textrm{NL}}$.

In polarization, \nilc\ and \smica\ perform equivalently at high
multipoles (Fig.~\ref{fig:dx11_spectra}). The \nilc\ polarization maps
yield measurements of $f_{\textrm{NL}}$ which are most consistent with
zero (Table~\ref{tab:ng}).  The \nilc\ and \smica\ analyses also
provide an effective mapping of the weights of the
\Planck\ frequencies in the needlet/harmonic domains: a given weight
tends to be high if a given frequency channel, in a given band in the
harmonic domain, is relevant for foreground cleaning; on the other
hand, the higher the statistical noise at a given frequency, the lower
the associated weight. For \nilc\ and \smica, the weights for $T$,
$E$, and $B$ modes are shown in Figs.~\ref{fig:needlet-weights} and
\ref{fig:smica_filters}, respectively.  \sevem\ provides the most
stability with respect to the effect of bright point sources in
polarization, (see Section \ref{sec:onepointstatistics}). This is due
to the inpainting procedure applied to these sources, which
significantly reduces the effect of this contaminant. Therefore,
\sevem\ could be a suitable choice for those analyses in polarization
which cannot easily deal with the presence of point source holes in a
mask.

Finally, we note that the \sevem\ approach is unique in its ability to
provide independent CMB estimates in a number of frequency
channels. For analyses that benefit significantly from, or even
require, such information, \sevem\ is the only meaningful
choice. Specific examples include various isotropy estimators
\citep{planck2014-a18}, the integrated Sachs-Wolfe stacking analysis
\citep{planck2014-a26}, relativistic boosting
\citep{planck2013-pipaberration}, and Rayleigh scattering analyses
\citep{lewis2013}. 

\section{Conclusions}
\label{sec:conclusions}

We have presented four different foreground-reduced CMB maps in both
temperature and polarization derived from the \Planck\ 2015
observations. These maps are based on the full \Planck\ data,
including a total of 50 months of LFI observations and 29 months of
HFI observations. The temperature component of these maps represents
the most accurate description of the CMB intensity sky published to
date. In the polarization component, the characteristic
$E$-mode signal expected in a standard $\Lambda$CDM model is easily 
discernible over the full sky. Corresponding astrophysical
foreground products are described in \citet{planck2014-a12}. 

The CMB maps presented here are the direct result of the detailed
analyses of systematic errors described in \citet{planck2014-a07} and
\citet{planck2014-a09}, which led to an effective reduction of
systematic errors by almost two orders of magnitude in power on large
angular scales in polarization compared to 2013.  However, despite
these improvements, the polarization systematic errors in the
\Planck\ 2015 data set are not yet negligible in several frequencies
on the very largest scales.  Multipoles below $\ell\le20$ are
therefore suppressed by low-pass filtering in the current
component-separated CMB maps.

Additionally, as already noted in \citet{planck2014-a14}, we observe a
mismatch in the effective noise amplitude of 10--20$\,$\% when
comparing the latest generation of \Planck\ simulations (FFP8) with
the data. Considering that the current temperature sky maps are
signal-dominated up to $\ell\approx2000$, this noise mismatch is of
little practical importance for cosmological analyses based on
temperature observations except on the very smallest scales. As in
2013, the temperature maps presented here are therefore used for a
wide range of important applications, including large-scale
temperature likelihood estimation \citep{planck2014-a13},
gravitational lensing \citep{planck2014-a17}, studies of isotropy and
statistics \citep{planck2014-a18}, primordial non-Gaussianity
\citep{planck2014-a19}, and non-trivial cosmological topologies
\citep{planck2014-a20}. For high-$\ell$ power spectrum and likelihood
estimation, we recommend the cross-spectrum based methods described in
\citet{planck2014-a13}, primarily due to difficulties in establishing
sufficiently accurate models of unresolved extra-galactic high-$\ell$
foregrounds for the maps presented here. Cosmological parameters
derived by temperature power spectra from these maps have been
compared with the results of the \Planck\ 2015 likelihood analysis
\citep{planck2014-a13} and found to agree at the $1\sigma$ level.

For polarization, the noise mismatch is not negligible, and we
therefore do not yet recommend using the provided maps for
cosmological studies that require a highly accurate noise model.  The
polarization maps presented here may still be very useful for many
important cosmological applications, including cross-correlation and
cross-spectrum based analyses, and we therefore release the maps
despite the current noise mismatch. Analysis of the higher order
statistics of these maps has been performed within the current
framework of precision assessment and presented in this paper. The
bispectrum analysis, including $E$ mode polarization, applied to the
results of the four component separation methods, gives evidence of a
vanishing non-Gaussian signal for three geometrical configurations,
namely local, orthogonal, and equilateral.  The $B$-modes derived from
the current polarization maps have been cross-correlated with the
predicted lensing $B$-modes from the measured $E$ signal and the
lensing potential measured independently from temperature in
\citet{planck2014-a17}; the result is found to be in excellent
agreement among the four component separation methods, as well as with
the prediction of the \Planck\ best-fit cosmology.

On the basis of these encouraging results, intense work is on-going to
reduce the large-scale polarization systematics to negligible levels,
as well as to resolve the noise simulation mismatch, and good progress
is being made. Updated products will be published as soon as this work
has reached a successful completion.

\begin{acknowledgements}

The Planck Collaboration acknowledges the support of: ESA; CNES, and
CNRS/INSU-IN2P3-INP (France); ASI, CNR, and INAF (Italy); NASA and DoE
(USA); STFC and UKSA (UK); CSIC, MICINN, and JA (Spain); Tekes, AoF,
and CSC (Finland); DLR and MPG (Germany); CSA (Canada); DTU Space
(Denmark); SER/SSO (Switzerland); RCN (Norway); SFI (Ireland);
FCT/MCTES (Portugal); ERC and PRACE (EU).  A description of the Planck
Collaboration and a list of its members, indicating which technical or
scientific activities they have been involved in, can be found at
\texttt{http://www.cosmos.esa.int/web/planck/planck-collaboration}.
%\url{http://www.cosmos.esa.int/web/planck/planck-collaboration}.

Some of the results in this paper have been derived using the
\healpix\ \citep{gorski2005} package.

\end{acknowledgements}

% bibliography
\bibliographystyle{aat}
\bibliography{Planck_bib,cmb_compsep}

% appendices
\appendix
\section{Bayesian parametric fitting}
\label{sec:commander}

\commander\ \citep{eriksen2004,eriksen2008} fits a physical model to a
set of observations within a standard Bayesian parametric framework,
defined by a set of explicit physical parameters and priors.  The code
can be operated in two modes, either employing Gibbs sampling to map
out the full parameter posterior, or using iterative non-linear
searches to derive the maximum-likelihood solution; the only
implementational difference between the two is whether to sample from
or maximize the conditional posterior in each Gibbs step.  All maps
presented in this paper are derived in the maximum-likelihood mode,
and uncertainties are evaluated through simulations.

\commander\ forms the core of the \Planck\ 2015 foreground-targeted
diffuse component separation efforts, and the corresponding results
are described in full detail in \citet{planck2014-a12}.  In this
section we only summarize the most relevant steps for CMB-oriented
analysis.

\subsection{Intensity}
\label{sec:commander_intensity}

\begin{figure}
\begin{center}
\includegraphics[width=\columnwidth]{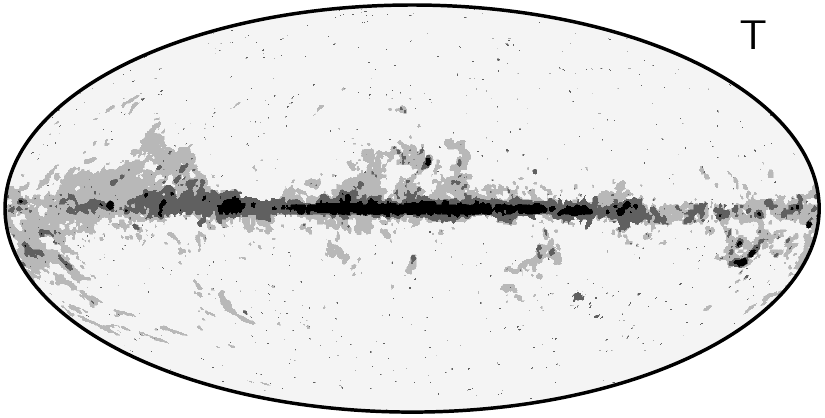}
\includegraphics[width=\columnwidth]{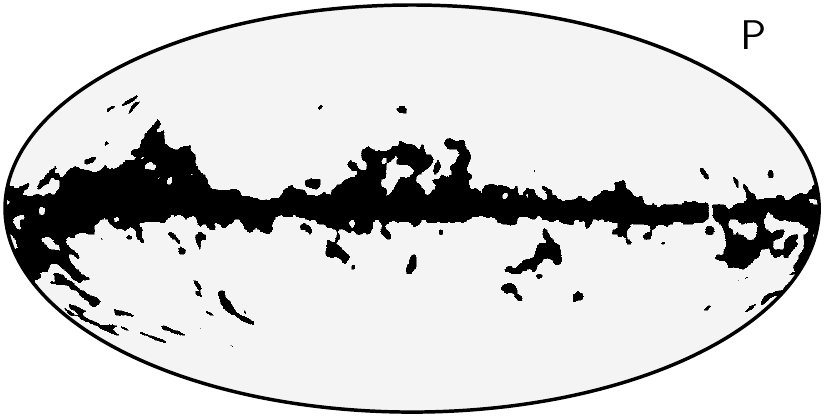}
\end{center}
\caption{\commander\ processing masks for temperature (\emph{top}) and
  polarization (\emph{bottom}).  For temperature, the different shades
  of grey correspond to different angular resolutions, ranging from
  $5\arcm$ (light grey) through $7.5\arcm$ (dark grey) to
  40\arcm\ FWHM (black).  For polarization, the same mask is used for
  both 10\arcm\ and 40\arcm\ FWHM resolution.}
\label{fig:comm_masks}
\end{figure}

In 2013, the \commander\ CMB temperature solution was derived using
only the seven lowest \Planck\ frequency maps between 30 and 353 GHz,
adopting a simple four-component signal model, including CMB, CO,
greybody thermal dust and a single power-law low-frequency component.
The only instrumental parameters included in the analysis were
monopole and dipole corrections.  In the current release,
significantly more data are included in the analysis, and the
astrophysical and instrument models have been expanded to account for
more effects.  Specifically, a total of 32 maps are considered in the
analysis, including 21 \Planck\ detector and detector set maps, 10
\emph{WMAP} differencing assembly maps, and a 408$\,$MHz low-frequency
survey map \citep{haslam1982}.  This wide frequency range allows us to
fit separately for synchrotron, free-free, spinning dust, thermal
dust, and CMB, as well individual CO transitions at 115, 230 and
353$\,$GHz, a common line emission component in the 94 and 100$\,$ GHz
channels (primarily HCN), and thermal Sunyaev Zeldovich emission for
the Coma and Virgo clusters.  On the instrumental side, we now fit for
both calibration and bandpass uncertainties, in addition to monopoles
and dipoles \citep{planck2014-a12}.

\subsection{Polarization}
\label{sec:commander_polarization}

The \commander\ CMB polarization map is derived in an analogous manner
to the temperature solution, but relying on \Planck\ observations
alone.  At low resolution, we derive a $40\arcm$ map from all
frequencies between 30 and 353$\,$GHz, including CMB, synchrotron and
thermal dust in the signal model.  At high resolution, we derive a
$10\arcm$ map from frequencies between 100 and 353 GHz, including only
CMB and thermal dust.

As discussed in \citet{planck2014-a12} several spectral index models
have been explored for polarized synchrotron and thermal dust
emission, including 1) spatially constant, 2) smoothed over large
angular scales, and 3) based on the temperature model.  The main
conclusion, however, is that the current polarization data have too
low signal-to-noise ratio for both synchrotron and thermal dust
indices to discriminate between the three models at a statistically
significant level.  Allowing such additional degrees of freedom only
increases the degeneracies between the various components without
improving the overall fit, and also leaves the solution sensitive to
large-scale residual systematics.  For now, we therefore adopt the
temperature-derived spectral parameters for both synchrotron and
thermal dust for the primary \commander\ CMB polarization map.

\subsection{Hybridization of multi-resolution sky maps}
\label{sec:hybridization}

As currently implemented through pixelized fits, the
\commander\ analysis requires uniform angular resolution across
frequencies in order to estimate spectral parameters correctly.  This
implies that all frequency maps must be smoothed to the resolution of
the lowest resolution channel before analysis.

\begin{figure}
\begin{center}
\includegraphics[width=\columnwidth]{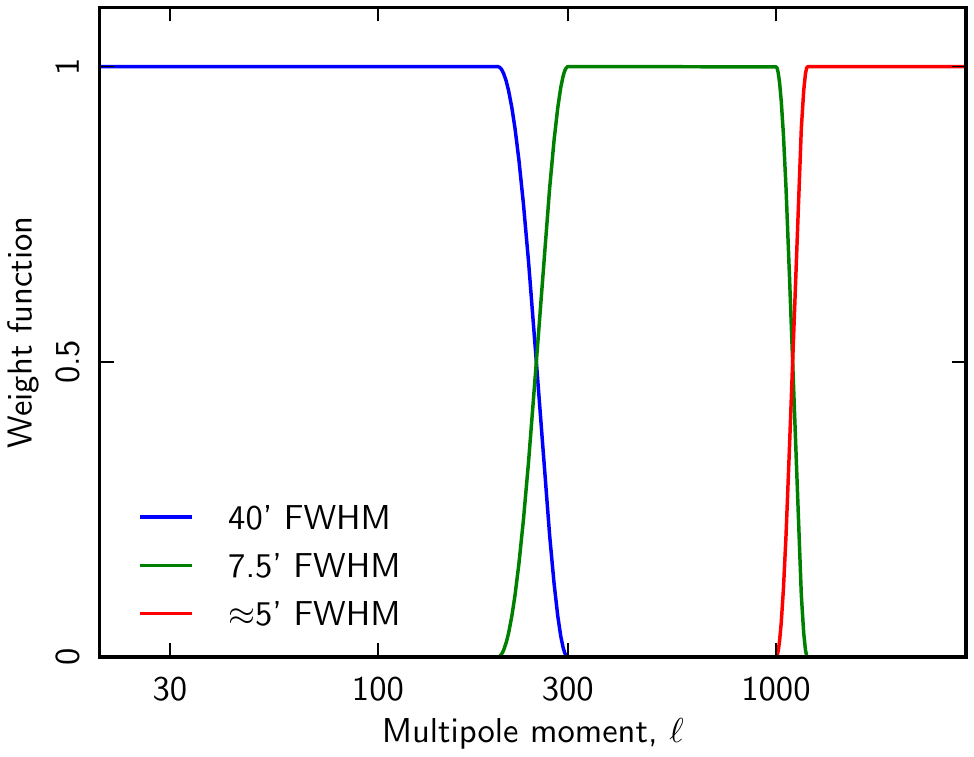}
\end{center}
\caption{Multipole moment weights used for multi-resolution
  hybridization in the \commander\ CMB map, as described by
  Eq.~\ref{eq:comm_hybrid}.}
\label{fig:hybridization}
\end{figure}

\begin{figure}
\begin{center}
\begin{tabular}{cc}
\includegraphics[width=0.45\columnwidth]{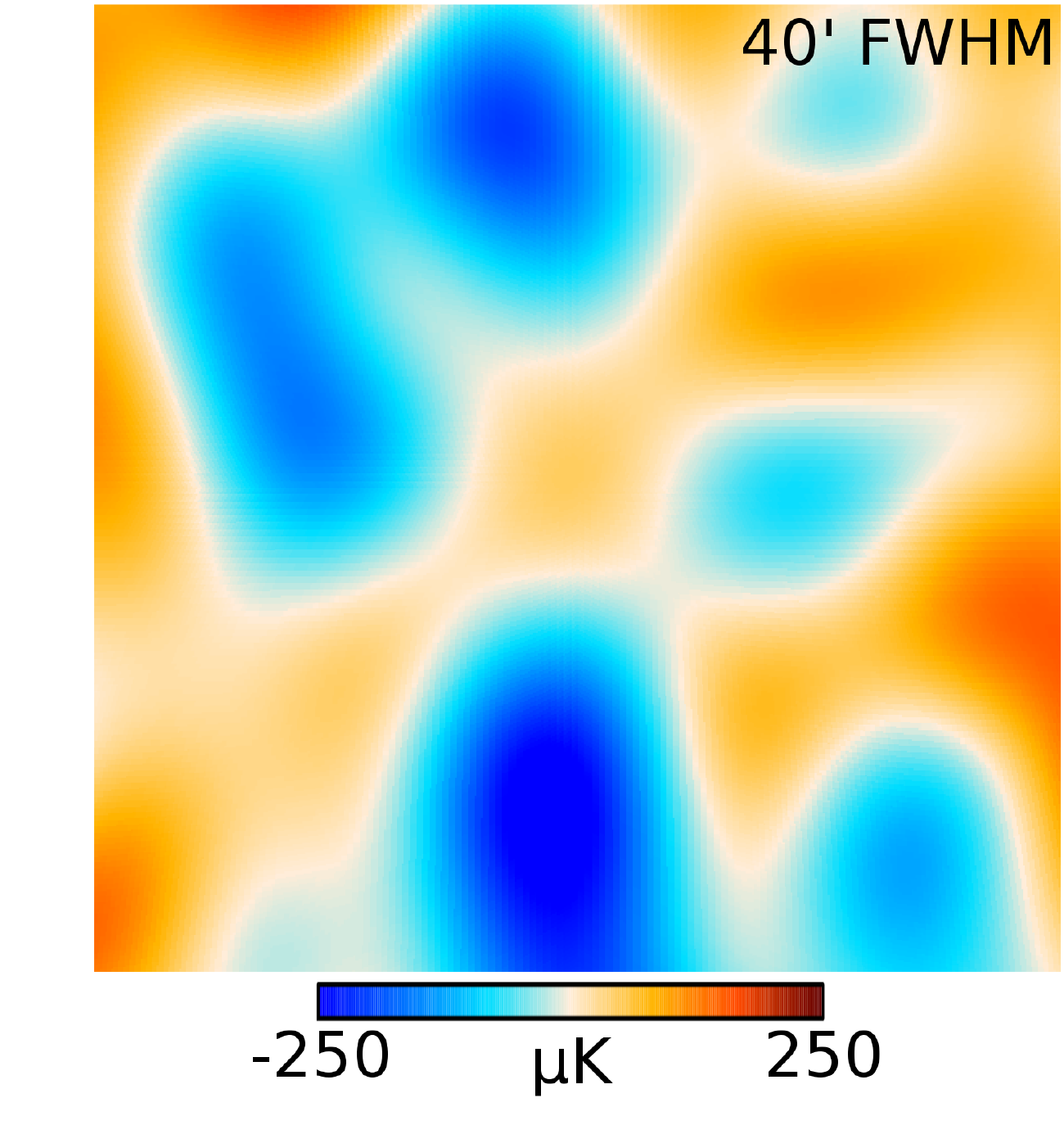}&
\includegraphics[width=0.45\columnwidth]{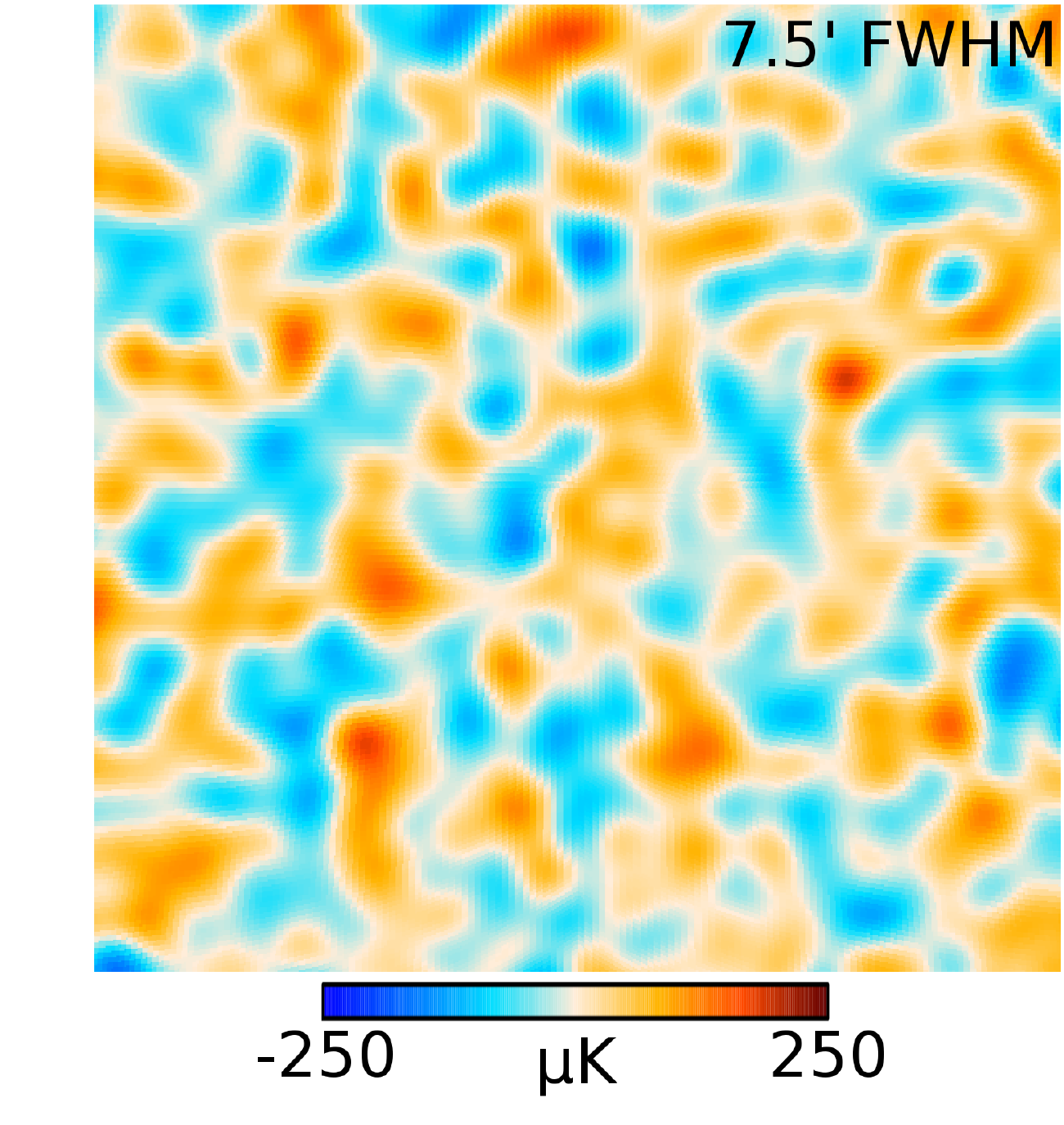}\\
\includegraphics[width=0.45\columnwidth]{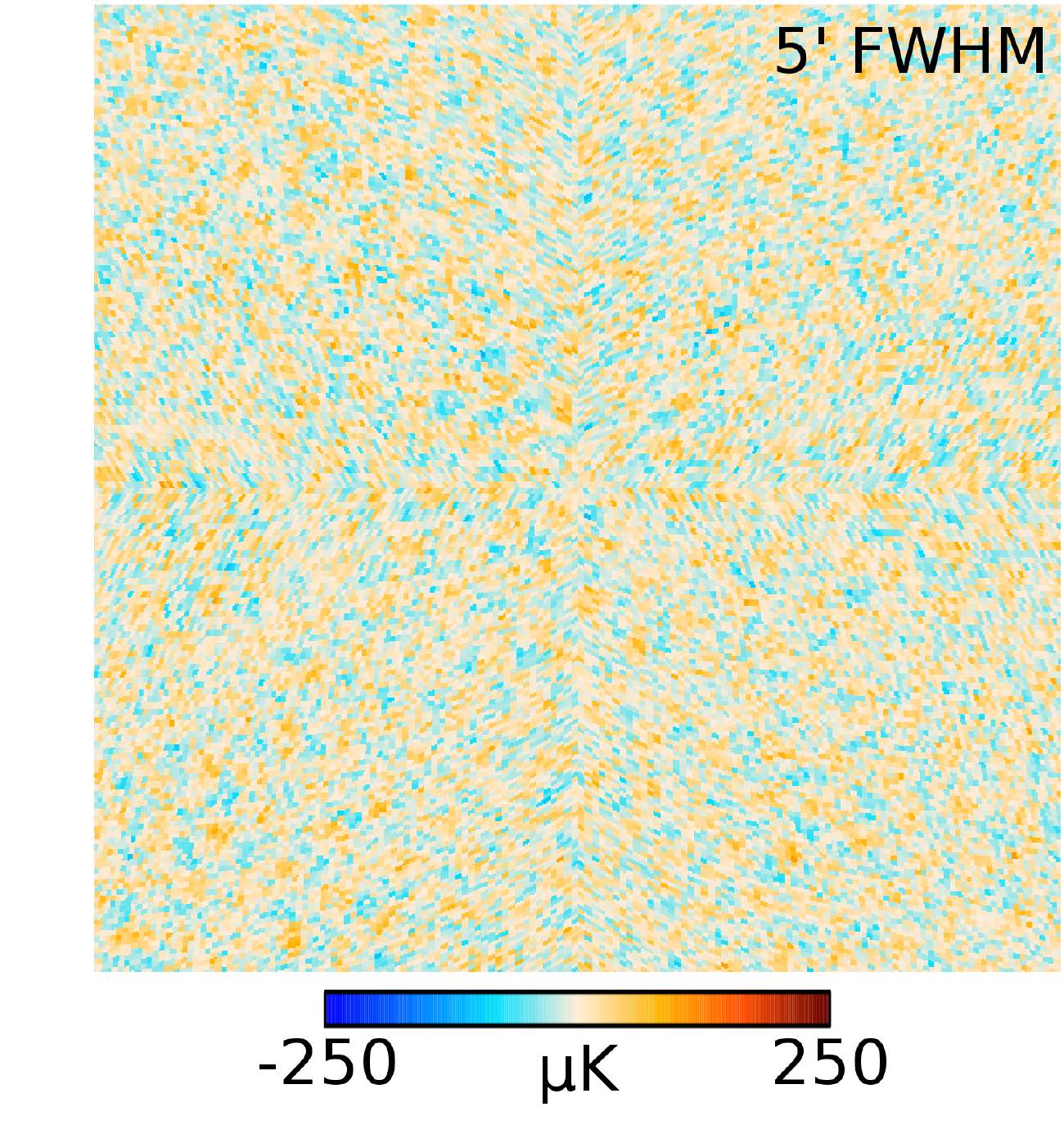}&
\includegraphics[width=0.45\columnwidth]{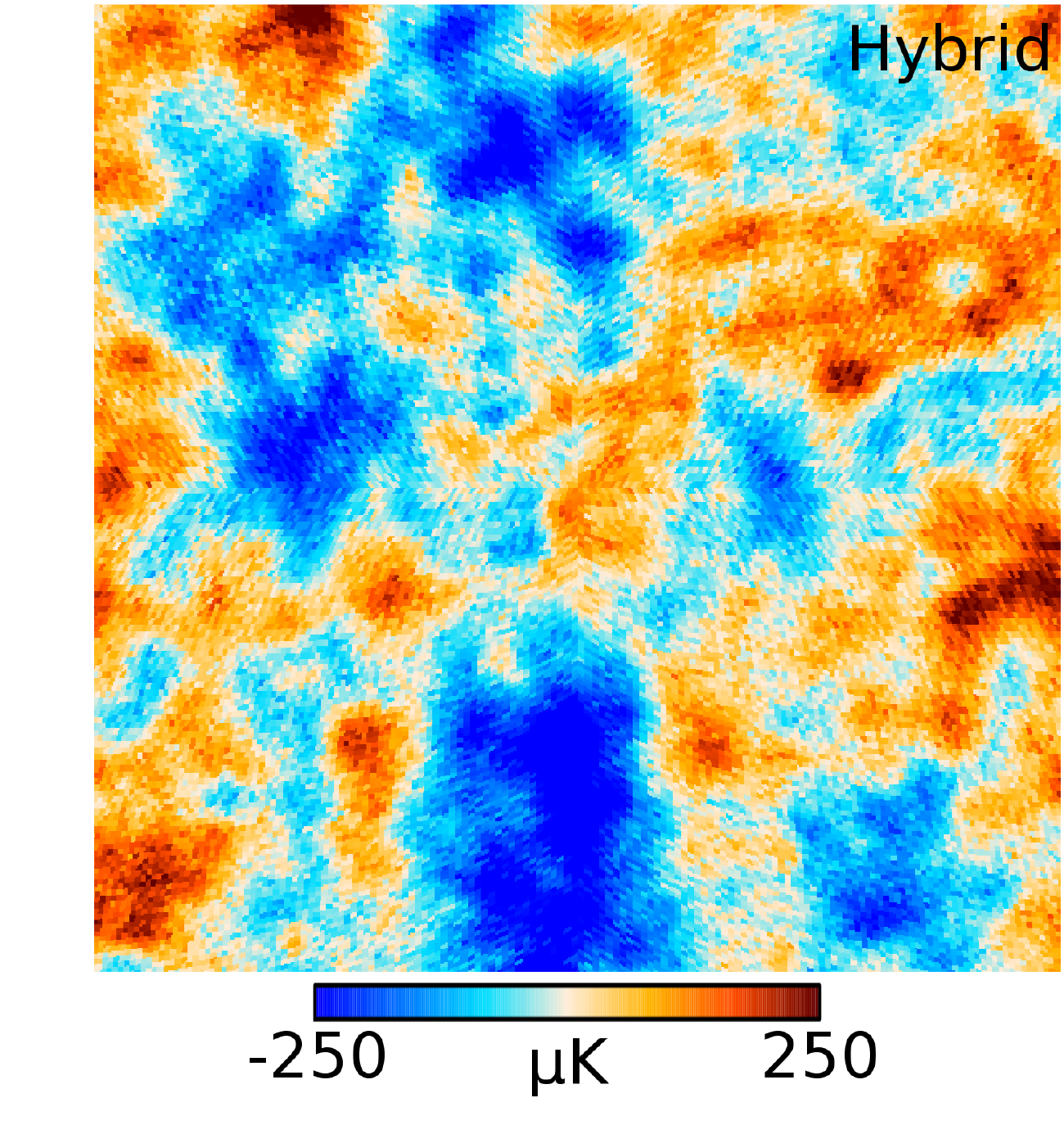}
\end{tabular}
\end{center}
\caption{$5\deg\times5\deg$ zoom-in of the multi-resolution
  contributions to the \commander\ hybrid CMB map from the $40\arcm$
  (\emph{top left}), $7\parcm5$ (\emph{top right}) and
  $\approx$$5\arcm$ (\emph{bottom left}) solutions, centered on the
  South Galactic Pole.  The hybrid map is shown in the bottom right
  panel.}
\label{fig:comm_hybrid_maps}
\end{figure}

In the 2013 \Planck\ release, this problem was partially solved by
first determining spectral parameters at low resolution using
\commander, as described above, and then solving for the component
amplitudes from full resolution data using a so-called \ruler\ step,
leading to the \commander-\ruler\ hybrid.  In the current release, we
adopt a \commander-only multi-stage approach, in which the system is
solved at four different angular resolutions, using different subsets
of the data, but each with internally coherent angular resolutions.
Explicitly, we first solve for the full parameter set at $1\deg$
resolution with temperature data only, combining \Planck\ observations
with external data (WMAP and the 408$\,$MHz Haslam observations; see
\citealt{planck2014-a12} for full details).  We then fix the global
parameters (monopoles, dipoles, calibration and bandpass corrections),
eliminate the external data, and solve again using only
\Planck\ observations at $40\arcm$ resolution, while at the same time
simplifying the low-frequency foreground model.  Next, we eliminate
all frequencies below 143$\,$GHz, and solve for CMB, CO and thermal
dust at $7\parcm5$ resolution, before finally eliminating also the
143$\,$GHz channel and the CO component, and solve only for CMB and
thermal dust at the native resolution of the 217$\,$GHz channel,
roughly $4\parcm8$.

Thus, a series of CMB estimates is established, ranging from low
resolution derived within a complete foreground model, to high
resolution derived within a greatly reduced foreground model.  The
basis for the low-$\ell$ likelihood is the lowest resolution solution,
implementing the most complete foreground model and exploiting both
\Planck\ and external data.  No further processing is required to
cover angular scales up to multipoles of $\ell\le 250$.

For high-resolution temperature CMB analysis, we hybridize the three
\Planck-only solutions, ranging between 40 and 5\arcm, into a single
map as follows.  We first define a mask for the low-resolution level
by simple $\chi^2$ thresholding.  For the second level, we also
threshold the corresponding $\chi^2$ map, but we additionally exclude
point sources, and we require that all pixels excluded by the lower
resolution maks are excluded by the higher resolution mask.  This is
repeated for the third and highest level, but in addition we also
exclude all pixels with a CO amplitude larger than 0.5\,K\,km/s (see
\citealp{planck2014-a12}), as CO is no longer included in the
foreground model.  The resulting masks admits 98.4, 95.0 and 82.4\,\%
of the sky, respectively, and are shown in the top panel of
Fig.~\ref{fig:comm_masks}.  For polarization, we construct a similar
mask from the product of the thresholded low-resolution $\chi^2$ and
CO maps, and the same mask is applied at both $40$ and $10\arcm$ FWHM.
The resulting mask admits 83.6\,\% of the sky, and is shown in the
bottom panel of Fig.~\ref{fig:comm_masks}.

Next, at each level the corresponding masked regions are replaced with
a constrained Gaussian realization \citep{eriksen2004}, as drawn from
$P(a_{\textrm{cmb}}|C_{\ell},\mathbf{d})$, before co-adding the three
maps in harmonic space using the following cosine apodization scheme,
\begin{equation}
\begin{array}{rcl}
a_{\ell m}^{\textrm{fullres}} &=&
(1-w_{\ell}^{40})\frac{p_{\ell}^{5}b_{\ell}^{5}}{p_{\ell}^{40}b_{\ell}^{40}}a_{\ell
  m}^{40} + \\
& & \quad\quad w_{\ell}^{40}(1-w_{\ell}^{7.5})\frac{p_{\ell}^{5}b_{\ell}^{5}}{p_{\ell}^{7.5}b_{\ell}^{7.5}}a_{\ell
  m}^{\textrm{7.5}} + w_{\ell}^{7.5} a_{\ell m}^{\textrm{5}}\nonumber,
\end{array}
\label{eq:comm_hybrid}
\end{equation}
adopting cosine apodization weights given by
\begin{eqnarray}
 w_{\ell}^{40} &=&
\left\{\begin{array}{ll}
1 &\textrm{for}\quad \ell \le 200\\
\frac{1}{2}[1-\cos(\pi\frac{300-\ell}{300-200})] & \textrm{for}\quad 200 < \ell \le 300
\end{array}\right.\\
w_{\ell}^{7.5} &=&
\left\{\begin{array}{ll}
1 &\textrm{for}\quad \ell \le 1000\\
\frac{1}{2}[1-\cos(\pi\frac{1200-\ell}{1200-1000})] & \textrm{for}\quad 1000 < \ell \le 1200
\end{array}\right.,
\end{eqnarray}
as illustrated in Fig.~\ref{fig:hybridization}.
Figure~\ref{fig:comm_hybrid_maps} shows the individual contributions
from the various resolutions, as well as the final sum.

The result is a single full-resolution CMB map with variable effective
sky fraction as a function of multipole.  For analyses employing only
multipoles below $\ell \le 200$, a total of 98.4\,\% of the sky
corresponds to direct CMB measurements, with the rest being filled
with a proper Gaussian constrained realization.  Therefore, for very
low multipoles, for which the partially missing modes can be nearly
exactly reproduced by the high-latitude information, the full sky is
in practice available for analysis.  For slightly smaller scales,
98\,\% of the sky is available for direct analysis.  Correspondingly,
below $\ell \le 1000$ a total of 95.0\,\% of the sky represents direct
measurements, while at higher multipoles, only 82\,\% of the sky
corresponds to direct measurements.  When using the 2015
\commander\ CMB map for cosmological analysis, it is therefore
important to consider which angular scales are relevant before
choosing the appropriate mask.  However, using the most conservative
mask is always safe at any angular scale.

Analogous processing is performed for the \commander\ polarization
map, but employing only two resolution levels, namely 40\arcm\ and
$10\arcm$ FWHM.  Cosine hybridization is performed between $\ell=200$
and 300, similar to the lowest two levels in the temperature analysis.

\section{Internal linear combination in needlet space}
\label{sec:nilc}

The goal of \nilc\ is to estimate the CMB from multifrequency
observations while minimizing the contamination from Galactic and
extragalactic foregrounds, and instrumental noise.  The method makes
a linear combination of the data from the input maps with minimum
variance on a frame of spherical wavelets called needlets
\citep{narcowich06localizedtight}.  Due to their unique properties,
needlets allow localized filtering in both pixel space and harmonic
space.  Localization in pixel space allows the weights of the linear
combination to adapt to local conditions of foreground contamination
and noise, whereas localization in harmonic space allows the method to
favour foreground rejection on large scales and noise rejection on
small scales.  Needlets permit the weights to vary smoothly on large
scales and rapidly on small scales, which is not possible by cutting
the sky in zones prior to processing \citep{2009A&A...493..835D}.

The \nilc\ pipeline \citep{2012MNRAS.419.1163B, 2013MNRAS.435...18B}
is applicable to scalar fields on the sphere, hence we work
separately on maps of temperature and the $E$ and $B$ modes of
polarization.  The decomposition of input polarization maps into $E$
and $B$ is done on the full sky.  At the end, the CMB $Q$ and $U$ maps
are reconstructed from the $E$ and $B$ maps.

\begin{figure}[th!]
  \centering
  \includegraphics[width=\linewidth]{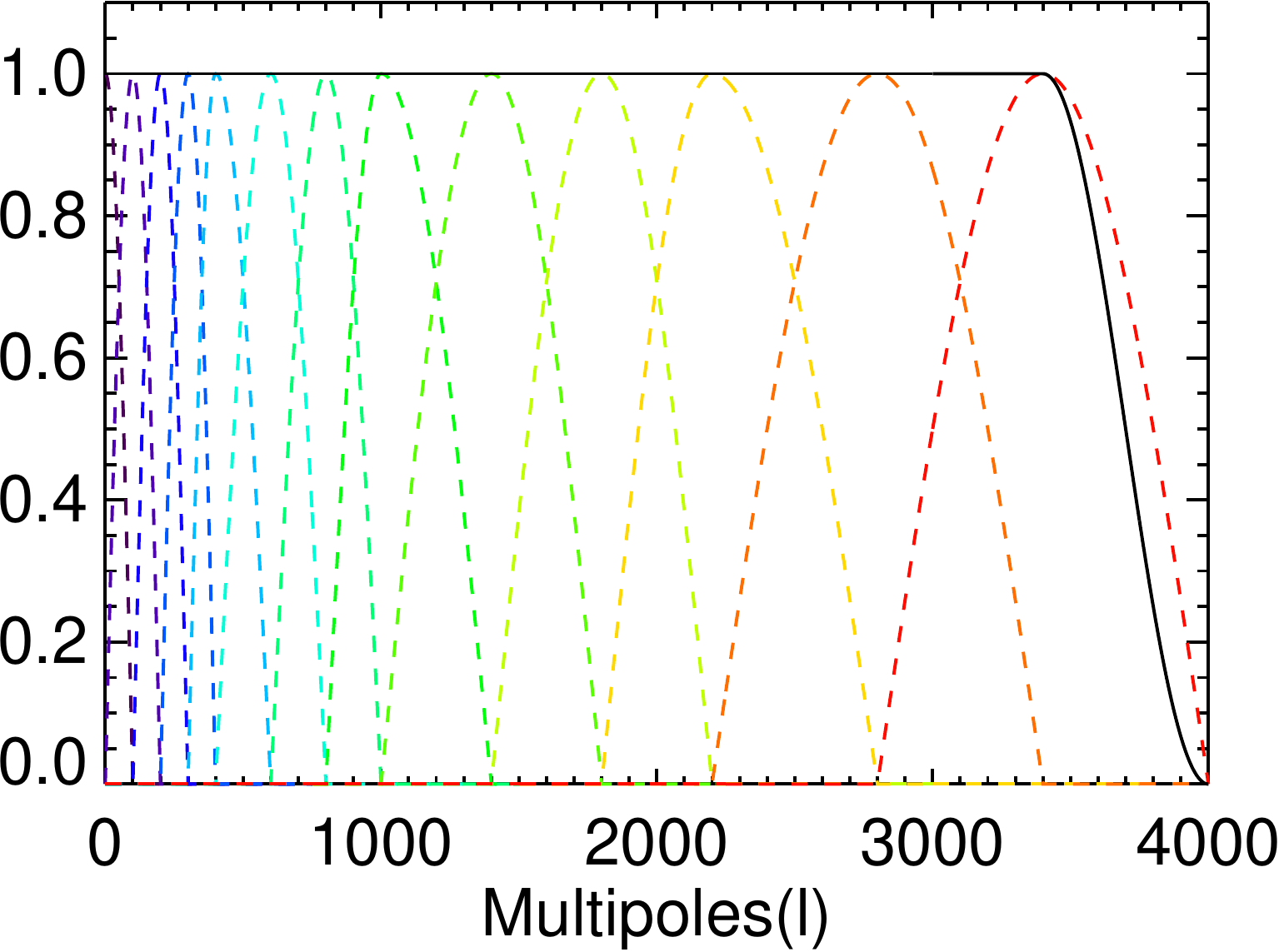}
  \caption{Needlet bands used in the analysis.  The solid black line
    shows the normalization of the needlet bands, that is, the total
    filter applied to the original map after needlet decomposition and
    synthesis of the output map from needlet coefficients.}
  \label{fig:needlet-bands}
\end{figure}

Prior to applying \nilc, all of the input maps are convolved or
deconvolved in harmonic space to a common resolution corresponding to
a Gaussian beam of 5\arcm\ FWHM.  Each map is then
decomposed into a set of needlet coefficients. For each scale $j$,
needlet coefficients of a given map are stored in the form of a single
\healpix\ map. The filters $h^{j}_{l}$ used to compute filtered maps
are shaped as follows
\begin{eqnarray*} 
h^{j}_{l} = \left\{
\begin{array}{rl} 
\cos\left[\left(\frac{l^{j}_{\mathrm{peak}}-l}{l^{j}_{\mathrm{peak}}-l^{j}_{\mathrm{min}}}\right)
\frac{\pi}{2}\right]& \mathrm{for } l^{j}_{\mathrm{min}} \le l < l^{j}_{\mathrm{peak}},\\ 
\\
1\hspace{0.5in} & \mathrm{for } l = l_{\mathrm{peak}},\\
\\
\cos\left[\left(\frac{l-l^{j}_{\mathrm{peak}}}{l^{j}_{\mathrm{max}}-l^{j}_{\mathrm{peak}}}\right)
\frac{\pi}{2}\right]& \mathrm{for } l^{j}_{\mathrm{peak}} < l \le l^{j}_{\mathrm{max}}. 
\end{array} \right. 
\end{eqnarray*}
For each scale $j$, the filter has compact support between the
multipoles $l^{j}_{\mathrm{min}}$ and $l^{j}_{\mathrm{max}}$ with a
peak at $l^{j}_{\mathrm{peak}}$ (see table \ref{tab:needlet-bands} and
figure \ref{fig:needlet-bands}). The needlet coefficients are computed
from these filtered maps on \healpix\ pixels with $\nside$ equal to
the smallest power of $2$ larger than $l^{j}_{\mathrm{max}}/2$.

\begin{table}[th!]
\begingroup \newdimen\tblskip \tblskip=5pt
  \caption{List of needlet bands used in the \nilc\ analysis.}
  \label{tab:needlet-bands} 
\vskip -4mm
\footnotesize
\setbox\tablebox=\vbox{
\newdimen\digitwidth
\setbox0=\hbox{\rm 0}
\digitwidth=\wd0
\catcode`*=\active
\def*{\kern\digitwidth}
\newdimen\signwidth
\setbox0=\hbox{+}
\signwidth=\wd0
\catcode`!=\active
\def!{\kern\signwidth}
\newdimen\decimalwidth
\setbox0=\hbox{.}
\decimalwidth=\wd0
\catcode`@=\active
\def@{\kern\signwidth}
\halign{ \hbox to 0.9in{#\leaderfil}\tabskip=2em&
    \hfil#\hfil&
    \hfil#\hfil&
    \hfil#\hfil&
    \hfil#\hfil\tabskip=0pt\cr
\noalign{\doubleline}
\omit\hfil Band\hfil&$\ell_{\rm min}$&$\ell_{\rm peak}$&$\ell_{\rm max}$&$N_{\rm side}$\cr
\noalign{\vskip 4pt\hrule\vskip 5pt}
$j=1$&            ***0& ***0& *100& **64\cr
\hglue 1.77em 2&  ***0& *100& *200& *128\cr
\hglue 1.77em 3&  *100& *200& *300& *256\cr
\hglue 1.77em 4&  *200& *300& *400& *256\cr
\hglue 1.77em 5&  *300& *400& *600& *512\cr
\hglue 1.77em 6&  *400& *600& *800& *512\cr
\hglue 1.77em 7&  *600& *800& 1000& *512\cr
\hglue 1.77em 8&  *800& 1000& 1400& 1024\cr
\hglue 1.77em 9&  1000& 1400& 1800& 1024\cr
\hglue 1.34em 10& 1400& 1800& 2200& 2048\cr
\hglue 1.34em 11& 1800& 2200& 2800& 2048\cr
\hglue 1.34em 12& 2200& 2800& 3400& 2048\cr
\hglue 1.34em 13& 2800& 3400& 4000& 2048\cr
\noalign{\vskip 5pt\hrule\vskip 4pt}
}}
\endPlancktable
\endgroup
\end{table}

\begin{figure}
  \centering
  \includegraphics[width=\columnwidth]{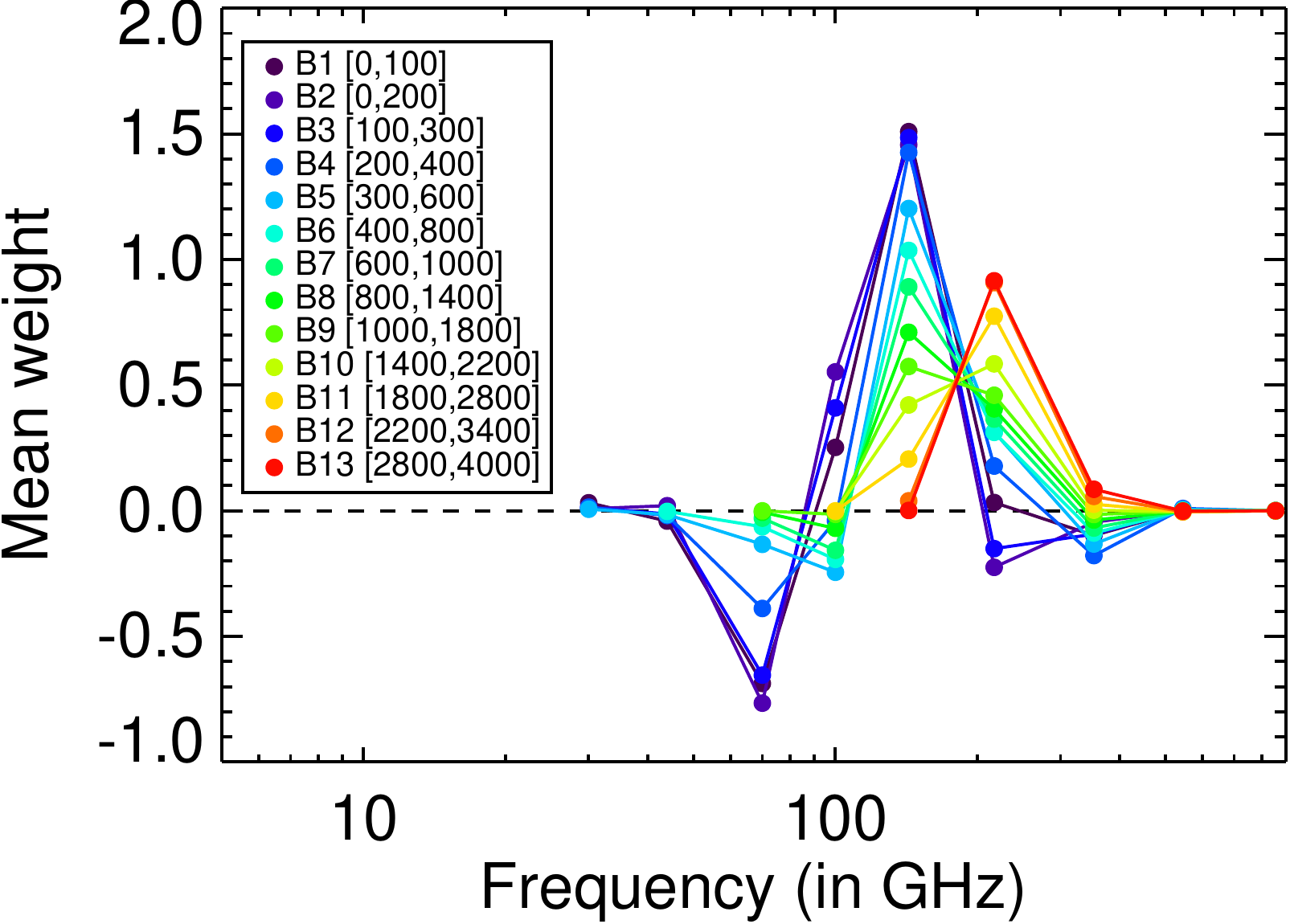}
  \vskip 4mm
  \includegraphics[width=\columnwidth]{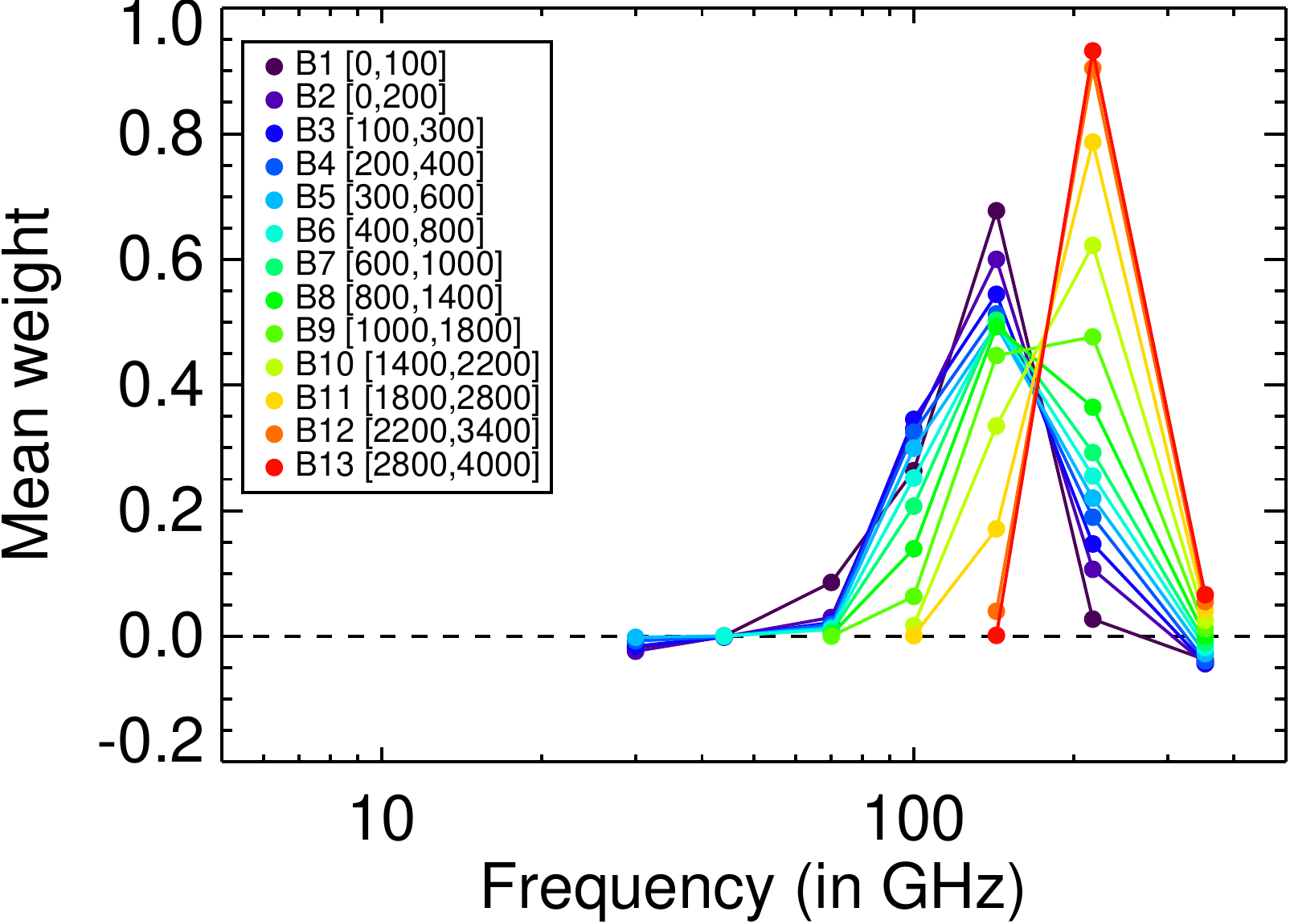}
  \vskip 4mm
  \includegraphics[width=\columnwidth]{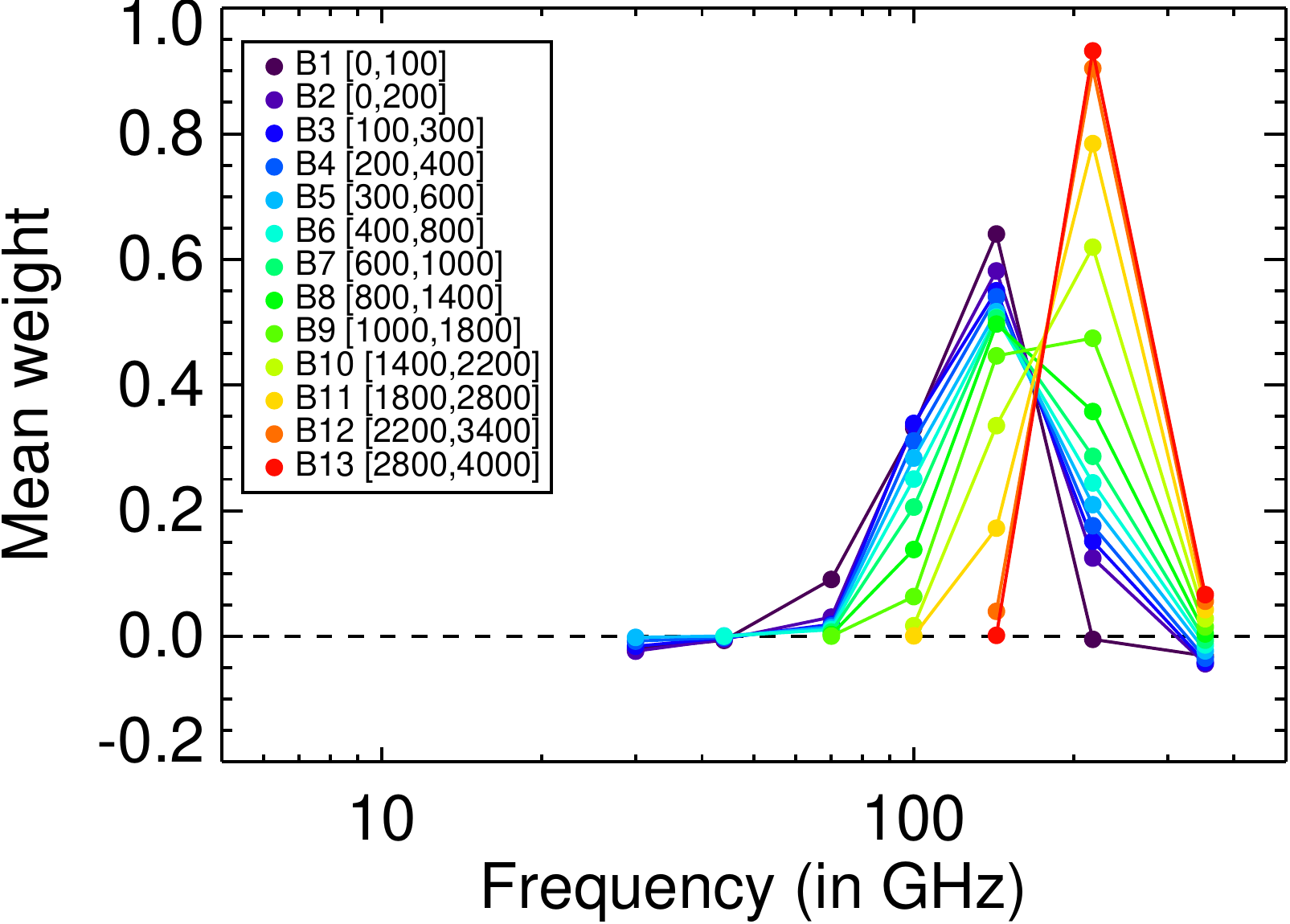}
  \caption{Full-sky average of needlet weights for different frequency
    channels and needlet bands. From top to bottom, the panels show
    results for temperature, $E$, and $B$ modes.}
  \label{fig:needlet-weights}
\end{figure}

In order to show the contribution of the various frequency channels to
the final CMB map at different needlet bands, we compute the full sky
average of needlet weights for each frequency channel and needlet
band.  Figure~\ref{fig:needlet-weights} shows that the most of
contribution to the reconstructed CMB maps comes from the 143\,GHz and
217\,GHz channels.  In the low $\ell$ needlet bands, the contribution
from 143\,GHz is large compared to that from 217\,GHz.  However, due
to better angular resolution, the 217\,GHz channel contributes more
than the 143\,GHz channel in the highest $\ell$ needlet bands. In the
intermediate needlet bands, the contributions from these two channels
are comparable.  It is interesting to note the contribution from the
LFI in the lowest $\ell$ bands. In intensity, the 70\,GHz channel
serves mostly for foreground removal, while for polarization it
contributes positively to the CMB solution very similarly between $E$
and $B$.  We stress again that the lowest $\ell$ modes have been
filtered out in the results presented here, and therefore the low
$\ell$ results will require further investigation.

\subsection{Masking}
\label{sec:nilc_masking}

The confidence masks for \nilc\ for intensity and polarization have
been generated following a procedure similar to that used by \smica,
but adopting a different parameterization.

\begin{figure}
\begin{center}
\includegraphics[width=\columnwidth]{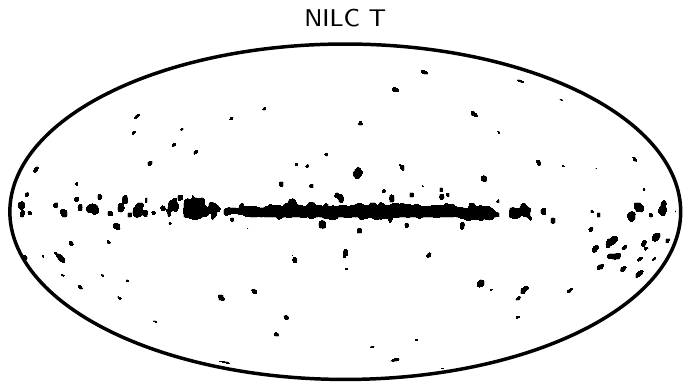}
\includegraphics[width=\columnwidth]{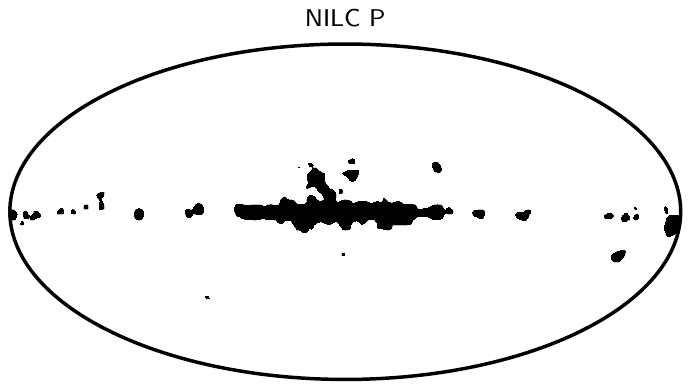}
\end{center}
\caption{\nilc\ masks for temperature (\emph{top}) and polarization
  (\emph{bottom}).}
\label{fig:dx11_masks_nilc}
\end{figure}

For intensity, the \nilc\ CMB map is filtered through a spectral window
\begin{equation}
f(l) = \exp\left[-\displaystyle\frac{1}{2}\left(\frac{l-1700}{200}\right)^2\right]\ . 
\label{eq:nilc_filtering_for_mask}
\end{equation}
The result is then squared and smoothed with a Gaussian circular beam
with FWHM 120\arcm.  The variance map obtained in this way is then
corrected for the noise contribution by subtracting the variance map
for the noise obtained in the same way from the \nilc\ HRHD map. The
confidence map is obtained by thresholding the noise-corrected
variance map at 73.5\muK$^2$.  For polarization, the polarized
intensity $P=\sqrt{Q^{2}+U^{2}}$ is obtained from the \nilc\ outputs,
and is filtered through a circularly symmetric Gaussian window
function of FWHM 30\arcm.  The result is then squared and smoothed
with a Gaussian of FWHM 210\arcm. The resulting variance map is
corrected for the noise contribution by following the same procedure
used for intensity.  The confidence map is obtained by thresholding at
6.75\muK$^2$.  The resulting masks are shown in
Fig.~\ref{fig:dx11_masks_nilc}.

\section{Template fitting}
\label{sec:sevem}

The \sevem\ method \citep{2008A&A...491..597L,2012MNRAS.420.2162F}
aims to produce clean CMB maps for several frequency channels by using
internal template fitting.  The templates are constructed from the
\Planck\ data, typically as the subtraction of two close
\Planck\ frequency channels to remove the CMB signal. The resolution
of the maps are equalized before subtraction. The cleaning is achieved
simply by subtracting a linear combination of the templates
$t_j(\vec{x})$ from the data, with coefficients $\alpha_j$ obtained by
minimizing the variance outside a given mask,
\begin{equation}
T_c(\vec{x},\nu)=d(\vec{x},\nu)- \sum_{j=1}^{n_t} \alpha_j t_j(\vec{x}).
\label{eq:sevem_basic_formula}
\end{equation}
where $n_t$ is the number of templates used, and $T_c(\vec{x},\nu)$
and $d(\vec{x},\nu)$ correspond to the cleaned and raw maps at
frequency $\nu$, respectively.  The same expression applies for $T$,
$Q$, or $U$.

The cleaned frequency maps are then combined in harmonic space, taking
into account the noise level, resolution, and, for temperature, a
rough estimate of the foreground residuals of each cleaned channel, to
produce a final CMB map at the required resolution.

\subsection{Implementation for temperature}

For temperature, we followed a similar procedure to that for the
\Planck\ 2013 release: the 100, 143, and 217\GHz\ maps are cleaned
using four templates constructed from the six remaining frequency
channels. A few differences have been implemented in the current
pipeline with respect to the previous work: the use of a single
coefficient over the whole sky for each template (instead of defining
two regions); the use of inpainting to reduce contamination from
sources; the use of the 857\GHz\ channel as a template (instead of
857$-$545).  Note that the other three templates (30$-$44, 44$-$70,
and 353$-$217) are the same as for the previous release.

The six frequency channels used to construct templates (30 to
70\GHz\ and 353 to 857\GHz) are inpainted at the position of sources
detected by the Mexican hat wavelet (MHW) algorithm
\citep{planck2014-a35}.  The size of the holes to be inpainted is
determined by taking into account the beam size of the channel as well
as the flux density of each source.  We do a simple diffuse
inpainting, which fills one pixel with the mean value of the
neighbouring pixels in an iterative way. To avoid inconsistencies when
subtracting two channels, each map is inpainted on the sources
detected in both that map and on the other map used to construct the
template. For example, to construct the (30$-$44) template, both maps
are inpainted in the positions of the sources detected at 30 and
44\GHz. This reduces significantly the contamination from compact
sources in the templates.

Once they have been inpainted, the maps are smoothed to a common
resolution and then subtracted. To construct the first three
templates, the first channel in the subtraction is smoothed with the
beam of the second map and vice versa. For the 857\GHz\ template, we
simply filter the map with the 545\GHz\ beam (this is done for
comparison with the 857$-$545 template from the 2013 pipeline, where
the 857\GHz\ map was smoothed by the 545\GHz\ beam.)  Note that the
coefficients used to multiply this template are typically $\sim
10^{-4}$, so the level of CMB signal introduced by this template in
the final cleaned map is negligible.  We take advantage of this to
drop subtraction of the 545\GHz\ map, as was done in 2013 nominally to
remove the CMB signal.  This simplifies the method and also reduces
the noise in this template.

The 100, 143, and 217\GHz\ maps are then cleaned by subtracting a
linear combination of the four templates. The coefficients of the
linear combination (Table~\ref{table:sevem_coef_T}) are obtained by
minimizing the variance outside an analysis mask. The main difference
with respect to the 2013 release is that we have used the same
coefficients for the whole sky (instead of dividing it in two
regions), since this simplifies the procedure without affecting the
quality of the reconstruction (other than on those pixels very close
to the Galactic centre that need to be masked in any case). This
analysis mask covers the 1\,\% of the sky with the brightest emission,
as well as sources detected at all frequency channels. Once the maps
are cleaned, each is inpainted on the source positions detected at
that (raw) channel.  Then, the MHW algorithm is run again, now on the
cleaned maps. A relatively small number of new sources are found and
are also inpainted at each channel. The resolution of the cleaned map
is the same as that of the raw map.  Note that no assumptions about
the noise or foregrounds are made in order to construct the
single-frequency cleaned maps.

\begin{table}[th!]
\begingroup \newdimen\tblskip \tblskip=5pt
  \caption{Linear coefficients, $\alpha_j$, of the templates used to
    clean individual frequency maps with \sevem\ for temperature.}
\label{table:sevem_coef_T}
\nointerlineskip
\vskip -3mm
\footnotesize
\setbox\tablebox=\vbox{
\newdimen\digitwidth
\setbox0=\hbox{\rm 0}
\digitwidth=\wd0
\catcode`*=\active
\def*{\kern\digitwidth}
\newdimen\signwidth
\setbox0=\hbox{+}
\signwidth=\wd0
\catcode`!=\active
\def!{\kern\signwidth}
\newdimen\expsignwidth
\setbox0=\hbox{$^{-}$}
\expsignwidth=\wd0
\catcode`@=\active
\def@{\kern\expsignwidth}
\halign{\hbox to 0.85in{#\leaderfil}\tabskip=1.5em&
     \hfil#\hfil&
     \hfil#\hfil&
     \hfil#\hfil\tabskip=0pt\cr
\noalign{\doubleline}
\omit&\multispan3\hfil Coefficients $\alpha_j$\hfil\cr
\noalign{\vskip -3pt}
\omit&\multispan3\hrulefill\cr
\noalign{\vskip 3pt}
\omit\hfil Template\hfil& 100\GHz& 143\GHz& 217\GHz\cr
\noalign{\vskip 3pt\hrule\vskip 5pt}
   *30$-$44* &$-6.38 \times 10^{-2}$& !$2.84\times 10^{-2}$&$-1.41\times 10^{-1}$\cr  
   *44$-$70* &!$3.53 \times 10^{-1}$& !$1.33\times 10^{-1}$&!$3.82 \times 10^{-1}$\cr 
   545$-$353 &!$4.34\times 10^{-3}$&$!6.56\times 10^{-3}$&!$1.79\times 10^{-2}$\cr 
      857    &$-3.63\times 10^{-5}$&$-5.66\times 10^{-5}$&$-1.18\times 10^{-4}$\cr 
\noalign{\vskip 5pt\hrule\vskip 3pt}}}
\endPlancktablewide
\endgroup
\end{table} 

Finally, the \sevem\ CMB map is constructed by combining the cleaned
and inpainted 143 and 217\GHz\ maps. In the combination, the maps are
weighted taking into account the noise, resolution, and a rough
estimation (obtained from realistic simulations) of the foreground
residuals at each map. The resolution of this map corresponds to a
Gaussian beam of FWHM 5\arcm\ and HEALPix resolution $\nside = 2048$
with a maximum multipole $\ell_{\mathrm{max}} = 4000$.

The same procedure, including the full inpainting process of point
sources, is applied when running the pipeline on FFP8 simulations.

\subsection{Implementation for polarization}

To clean the polarization maps, a procedure similar to the one used
for the temperature maps is applied to the $Q$ and $U$ maps
independently. However, given that narrower frequency coverage is
available for polarization, the templates and maps to be cleaned are
different. In particular, we clean the 70, 100, and 143\GHz\ maps
using three templates.  Since the signal-to-noise ratio is lower, at
100 and 143\GHz\ the clean maps are produced at $\nside = 1024$, with
resolution corresponding to a Gaussian beam of FWHM 10\arcm\ and a
maximum multipole $\ell_{max} = 3071$.  At 70\GHz, the map is produced
at its native resolution.

The first step of the pipeline is to inpaint the positions of the
sources detected using the MHW algorithm in those channels that will
be used to construct templates. As for the temperature case, for a
given template constructed as the difference of two frequency
channels, the inpainting is performed for all of the sources detected
in any of the channels. Note that the inpainting is performed in the
frequency maps at their native resolution.

These inpainted maps are then used to construct a total of four
templates. To trace the synchrotron emission, we construct a 30$-$44
template. For the dust emission, the following templates are used:
353$-$217 (smoothed to 10\arcm\ resolution), 217$-$143 (used to clean
70 and 100\GHz) and 217$-$100 (to clean 143\GHz). These last two
templates are constructed at 1\degr\ resolution, since an additional
smoothing becomes necessary in order to increase the signal-to-noise
ratio of the template. Otherwise, the estimated coefficients are
driven by the noise and the cleaning is less efficient. Since fewer
frequency channels are available in polarization, it becomes necessary
to use the maps to be cleaned as part of one of the
templates. Therefore the 100\GHz\ map is used to clean the
143\GHz\ frequency channel and vice versa, making the clean maps less
independent than in the temperature case.

These templates are then used to clean the non-inpainted 70 (at its
native resolution), 100 (at 10\arcm\ resolution), and 143\GHz\ maps
(also at 10\arcm). The corresponding linear coefficients (listed in
Tables~\ref{table:sevem_coef_Q} and \ref{table:sevem_coef_U}) are
estimated independently for $Q$ and $U$ by minimizing the variance of
the clean maps outside a mask that covers compact sources and the
3\,\% of sky with the brightest Galactic emission. Once the maps have
been cleaned, inpainting of the sources detected at each map is
carried out. The size of the holes to be inpainted takes into account
the additional smoothing of the 100 and 143\GHz\ maps. As for
temperature, the same inpainting processing is applied to the point
sources positions when running the pipeline on the FFP8
simulations. The 100 and 143\GHz\ clean maps are then combined in
harmonic space, using a full-sky $E$ and $B$ decomposition, to produce
the final CMB maps for the $Q$ and $U$ components with a Gaussian beam
of FWHM 10\arcm\ and \healpix\ resolution $\nside = 1024$. Each map is
weighted taking into account its noise level at each multipole.

\begin{table}[th!]
\begingroup
\newdimen\tblskip \tblskip=5pt
  \caption{Linear coefficients, $\alpha_j$, of the templates used to
    clean individual frequency maps with \sevem\ for $Q$.}
\label{table:sevem_coef_Q}
\nointerlineskip
\vskip -3mm
\footnotesize
\setbox\tablebox=\vbox{
\newdimen\digitwidth
\setbox0=\hbox{\rm 0}
\digitwidth=\wd0
\catcode`*=\active
\def*{\kern\digitwidth}
\newdimen\signwidth
\setbox0=\hbox{+}
\signwidth=\wd0
\catcode`!=\active
\def!{\kern\signwidth}
\newdimen\expsignwidth
\setbox0=\hbox{$^{-}$}
\expsignwidth=\wd0
\catcode`@=\active
\def@{\kern\expsignwidth}
\halign{\hbox to 0.85in{#\leaderfil}\tabskip=1.5em&
     \hfil#\hfil&
     \hfil#\hfil&
     \hfil#\hfil\tabskip=0pt\cr
\noalign{\doubleline}
\omit&\multispan3\hfil Coefficients $\alpha_j$\hfil\cr
\noalign{\vskip -3pt}
\omit&\multispan3\hrulefill\cr
\noalign{\vskip 3pt}
\omit\hfil Template\hfil& 70\GHz& 100\GHz& 143\GHz\cr
\noalign{\vskip 3pt\hrule\vskip 5pt}
   *30$-$44* &$2.66\times 10^{-2}$&$1.00\times 10^{-2}$&$5.72\times 10^{-3}$\cr  
   217$-$143 &$8.01\times 10^{-2}$&$1.24\times 10^{-1}$&\dots\cr 
   217$-$100 &\dots &\dots &$1.78\times10^{-1}$\cr 
   353$-$217  &$4.26\times 10^{-3}$&$1.14\times 10^{-2}$&$2.50\times 10^{-2}$\cr 
\noalign{\vskip 5pt\hrule\vskip 3pt}}}
\endPlancktablewide
\endgroup
\end{table} 

\begin{table}[th!]
\begingroup
\newdimen\tblskip \tblskip=5pt
  \caption{Linear coefficients, $\alpha_j$, of the templates used to
    clean individual frequency maps with \sevem\ for $U$.}
\label{table:sevem_coef_U}
\nointerlineskip
\vskip -3mm
\footnotesize
\setbox\tablebox=\vbox{
\newdimen\digitwidth
\setbox0=\hbox{\rm 0}
\digitwidth=\wd0
\catcode`*=\active
\def*{\kern\digitwidth}
\newdimen\signwidth
\setbox0=\hbox{+}
\signwidth=\wd0
\catcode`!=\active
\def!{\kern\signwidth}
\newdimen\expsignwidth
\setbox0=\hbox{$^{-}$}
\expsignwidth=\wd0
\catcode`@=\active
\def@{\kern\expsignwidth}
\halign{\hbox to 0.85in{#\leaderfil}\tabskip=1.5em&
     \hfil#\hfil&
     \hfil#\hfil&
     \hfil#\hfil\tabskip=0pt\cr
\noalign{\doubleline}
\omit&\multispan3\hfil Coefficients $\alpha_j$\hfil\cr
\noalign{\vskip -3pt}
\omit&\multispan3\hrulefill\cr
\noalign{\vskip 3pt}
\omit\hfil Template\hfil& 70\GHz& 100\GHz& 143\GHz\cr
\noalign{\vskip 3pt\hrule\vskip 5pt}
   *30$-$44* &$2.98\times 10^{-2}$&$1.47\times 10^{-2}$&$7.97\times 10^{-3}$\cr  
   217$-$143 &$4.08\times 10^{-2}$&$1.53\times 10^{-1}$&\dots\cr 
   217$-$100&\dots&\dots&$1.66\times10^{-1}$\cr 
   353$-$217  &$6.76\times 10^{-3}$&$0.66\times 10^{-2}$&$2.41\times 10^{-2}$\cr 
\noalign{\vskip 5pt\hrule\vskip 3pt}}}
\endPlancktable
\endgroup
\end{table} 

Before applying the post-processing high-pass filter to the cleaned
$Q$ and $U$ polarization maps, we inpaint the region with the
brightest Galactic residuals (5\,\% of the sky) with the same simple
algorithm used for point source holes. This is done to avoid
introducing ringing around the Galactic centre when the maps are
filtered.

In addition, $E$ and $B$ modes maps are constructed from the clean $Q$
and $U$ maps. In this case, prior to performing the decomposition, the
region of the $Q$ and $U$ maps defined by the \sevem\ confidence mask
is filled with a Gaussian-constrained realization \citep{eriksen2004}.

\subsection{Masks}
\label{sec:sevem_masking}

In temperature, the \sevem\ confidence mask is produced by
thresholding differences obtained between different CMB
reconstructions. In particular, we construct the difference map
between the clean 217 and 143\GHz\ maps at a resolution of FWHM
30\arcm\ and $\nside = 256$. The brightest pixels of this map (and
their direct neighbours) are successively removed from the CMB
combined map (at the same resolution) and the dispersion of the CMB
combined map calculated. If a sufficient number of pixels is removed,
the dispersion of the CMB map goes down and reaches a plateau,
indicating that convergence has been achieved. The removed pixels
constitute the mask. This mask is then smoothed with a Gaussian beam
of 1\degr\ to avoid sharp edges, and upgraded to full resolution. The
same procedure is repeated for other two difference maps: the clean
143$-$100 map and the difference of two clean CMB combined maps, whose
linear coefficients have been obtained by minimizing the variance
outside two different masks. Finally, the three masks are multiplied
in order to produce our final confidence mask, which leaves a suitable
sky fraction of approximately 85\,\%.  Residual monopoles and dipoles
outside this mask are subtracted for the single frequency and combined
cleaned maps.

\begin{figure}[th!]
\begin{center}
\includegraphics[width=\columnwidth]{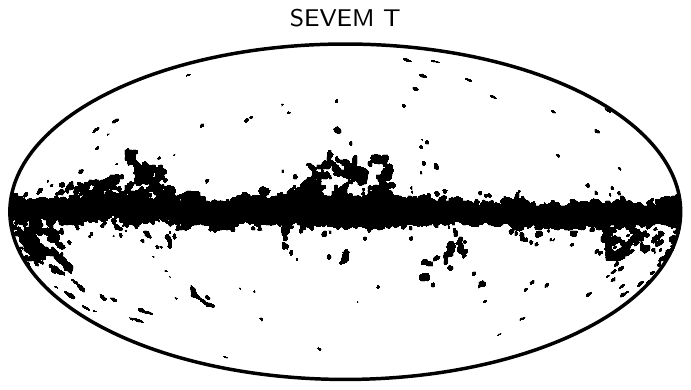}
\includegraphics[width=\columnwidth]{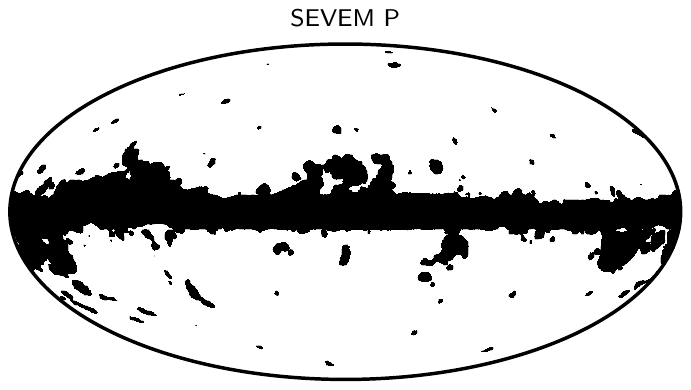}
a\end{center}
\caption{\sevem\ masks in temperature (\emph{top}) and polarization
  (\emph{bottom}).}
\label{fig:dx11_masks_sevem}
\end{figure}

For polarization, the clean combined map is downgraded to a resolution
of FWHM 90\arcm\ and $\nside = 128$.  The dispersion at each pixel is
estimated from a circle centered in the considered pixel. Those pixels
with a dispersion above a given threshold are included in the mask,
which is then smoothed with a Gaussian beam of FWHM 90\arcm\ to avoid
sharp edges and upgraded to the required $\nside$. Finally, this mask
is multiplied by a mask customized to cover the CO emission, in order
to discard those pixels contaminated by this foreground component due
to the bandpass leakage. An additional 1\,\% of the sky is added to
the final mask to remove those pixels most affected by the high-pass
filtering that is subsequently applied to the cleaned maps. The final
mask allows for a useful fraction of the sky of approximately 80\,\%.
The 2015 \sevem\ masks are shown in Fig.~\ref{fig:dx11_masks_sevem}.

\section{Spectral matching}
\label{sec:smica}

\def\lm{{\ell m}}
\def\adj{^\dagger}
\def\minv{^{-1}}
\def\nchan{N_{\mathrm{chan}}}

\newcommand\bivec[2]{\left[\begin{array}{c} #1 \\ #2 \end{array}\right]}
\newcommand\bimat[4]{\left[\begin{array}{cc} #1 & #2 \\ #3 & #4 \end{array}\right]}

Spectral Matching Independent Component Analysis (\smica) is a
semi-blind component separation algorithm which operates in harmonic
space.  CMB maps are synthesized from spherical harmonic coefficients,
$s_\lm$, obtained as weighted linear combinations of the coefficients
of $\nchan$ input maps,
\begin{equation}
  \label{eq:smica_synth}
  s_\lm = \mathbf{w}_{\ell}\adj \mathbf{x}_\lm\,,
\end{equation}
where $\mathbf{x}_\lm$ is the $\nchan\times1$ vector of the spherical
harmonic coefficients of the input maps and $\mathbf{w}_{\ell}$ is the
$\nchan\times1$ vector of weights.  The spectral weights used to
produce the \smica\ maps are designed to minimize the total foreground
and noise contamination at each multipole, under the constraint that
the resulting map has a well defined effective beam window function,
that of a Gaussian beam with 5\arcm\ FWHM.

In theory, such weights are given by
\begin{equation}
  \label{eq:ilcweights}
  \mathbf{w}_{\ell} = \frac
  {\mathbf{R}_{\ell}\minv \mathbf{a}}
  {\mathbf{a}\adj\mathbf{R}_{\ell}\minv \mathbf{a}}\,,
\end{equation}
where the $\nchan\times 1$ vector $\mathbf{a}$ is the frequency
spectrum of the CMB and the $\nchan\times\nchan$ spectral covariance
matrix $\mathbf{R}_{\ell}$ contains in its $(i,j)$ entry the
cross-power spectrum at multipole $\ell$ between the input frequency
maps~$i$ and~$j$, at 5\arcm\ resolution.  In practice, the spectral
covariance matrices must be estimated from the data.  The natural
sample estimate
\begin{equation}
  \label{eq:sampleSCM}
  \widehat{\mathbf{R}}_{\ell}=\frac1{2\ell+1}\sum_m\mathbf{x}_\lm\mathbf{x}_\lm\adj
\end{equation}
can be used, possibly after some binning, to replace
$\mathbf{R}_{\ell}$ in the weight formula~(\ref{eq:ilcweights}).  This
works well at high $\ell$ because a large number of modes are averaged
in (\ref{eq:sampleSCM}), so this estimate has low variance.  At
larger scales, however, it is necessary to constrain the spectral
covariance matrices in order to get reliable estimates.  For that
purpose, \smica\ uses a semi-parametric model of these matrices.

The CMB, the foregrounds, and the noise are independent processes, so
the spectral covariance matrices, after beam correction, can be
decomposed into
\begin{equation}\label{eq:smicamodelsplit}
  \mathbf{R}_{\ell} = \mathbf{R}^\mathrm{cmb}_{\ell} +
  \mathbf{R}^\mathrm{fgd}_{\ell} + \mathbf{R}^\mathrm{noise}_{\ell} .
\end{equation}
We assume that the noise is uncorrelated between frequency channels,
therefore the noise contribution $\mathbf{R}^\mathrm{noise}_{\ell}$ is
diagonal.  The signal parts are modelled as
\begin{equation}\label{eq:smicamodel}
  \mathbf{R}^\mathrm{cmb}_{\ell}  = \mathbf{a}\mathbf{a}\adj C_{\ell} ,
  \qquad
  \mathbf{R}^\mathrm{fgd}_{\ell}  =  \mathbf{F}\mathbf{P}_{\ell}\mathbf{F}\adj ,
\end{equation}
where $C_{\ell}$ is the CMB power spectrum, matrix $\mathbf{F}$ is
$\nchan\times\ d$, and $\mathbf{P}_{\ell}$ is a $d\times d$ positive
definite matrix.  A \smica\ fit consists of fitting the
model~(\ref{eq:smicamodelsplit}) and (\ref{eq:smicamodel}) to
empirical covariance matrices $\widehat{\mathbf{R}}_{\ell}$ by
minimizing the spectral matching criterion
\begin{displaymath}
  \sum_\ell (2\ell+1) \left[
    \mathrm{trace}(\widehat{\mathbf{R}}_{\ell}\,
    \mathbf{R}\inv_{\ell}) + \ln\det \mathbf{R}_{\ell} \right].
\end{displaymath}
The fitted values of $\mathbf{R}_{\ell}$ can be seen as regularized
versions of their empirical counterparts
$\widehat{\mathbf{R}}_{\ell}$, to be used in
equation~(\ref{eq:ilcweights}).  If no constraints are imposed on
matrices $\mathbf{F}$ and $\mathbf{P}_{\ell}$, except that the latter
is positive definite, the foreground model is equivalent to $d$ sky
templates with arbitrary frequency spectra (represented by the columns
of $\mathbf{F}$) and arbitrary power spectra and correlations between
them (represented by $\mathbf{P}_{\ell}$).  Ultimately, the foreground
contribution is controlled by a single parameter, the rank $d$ of the
foreground model.

More details regarding the principles of \smica\ can be found
in~\citet{cardoso2008}.  We now describe its extension to polarization
data.

There are several options to extend \smica\ to polarization.  We
choose to obtain the CMB $Q$ and $U$ maps through a joint processing
of the $E$ and $B$ modes.  The CMB $U$ and $Q$ maps are synthesized
from their $E$ and $B$ modes, $s_\lm^E$ and $s_\lm^B$, obtained as
linear combinations
\begin{equation}
  \label{eq:bivecSE}
  \bivec{s_\lm^E}{s_\lm^B} = \mathbf{W}_{\ell}\adj
  \bivec{\mathbf{x}_\lm^E}{\mathbf{x}_\lm^B}\,,
\end{equation}
where the $\nchan\times1$ vectors $\mathbf{x}_\lm^{E}$ and
$\mathbf{x}_\lm^{B}$ are the spherical harmonic coefficients of the
input maps and $\mathbf{W}_{\ell}$ is a $2\nchan\times2$ matrix
of weights.  The optimal weights are obtained as a simple
generalization of the single-field case,
\begin{equation}
  \label{eq:ilcbidi}
  \mathbf{W}_{\ell} = \left(\mathbf{A}\adj
  \mathbf{R}_{\ell}\minv\mathbf{A}\right)\minv \mathbf{A}\adj
  \mathbf{R}_{\ell}\minv \quad\mathrm{with}\quad \mathbf{A} =
  \bimat{\mathbf{a}} 0 0 {\mathbf{a}}\,,
\end{equation}
where $\mathbf{R}_{\ell}$ now is a $2\nchan\times2\nchan$ covariance
matrix. The weights defined by~(\ref{eq:ilcbidi}) can be safely
obtained at high $\ell$ by using the sample covariance matrices,
possibly after binning them.  At low multipoles, some regularization
via modelling is in order, as for the temperature analysis.  We use a
natural extension of temperature~(\ref{eq:smicamodel}), in the form
\begin{eqnarray}
  \label{eq:psmica_mod_cmb}
  \mathbf{R}^\mathrm{cmb}_{\ell} &=&
  \mathbf{A}\bimat{C^{EE}_{\ell}}{C^{EB}_{\ell}}{C^{EB}_{\ell}}{C^{EE}_{\ell}}\mathbf{A}\adj,\\
  \label{eq:psmica_mod_fgd}
  \mathbf{R}^\mathrm{fgd}_{\ell} &=&
  \bimat{\mathbf{F}_E}00{\mathbf{F}_B}
  \mathbf{P}_{\ell}
  \bimat{\mathbf{F}_E}00{\mathbf{F}_B}\adj ,
\end{eqnarray}
where $\mathbf{F}_E$ and $\mathbf{F}_B$ are $\nchan\times d$ matrices
and $\mathbf{P}_{\ell}$ is a $2d\times 2d$ matrix. For this release,
no constraints are imposed on those matrices, except that each
$\mathbf{P}_{\ell}$ must be positive definite.  Hence, as in
temperature, \smica\ fits a foreground model representing $d$
polarized templates with arbitrary frequency spectra, power spectra,
and correlation.

\subsection{Implementation for temperature}

The production of the CMB temperature map is mostly unchanged from 2013 (see \citealt{planck2013-p06} for details).  Here, we recall the 3-step fitting strategy adopted in
2013 and mention the changes made for this release.

\begin{itemize}

\item First fit: recalibration.  A preliminary and independent
  \smica\ fit is done with power spectra estimated over a clean
  fraction of the sky, including $\mathbf{a}$ as a free parameter.
  This can be understood as a recalibration procedure; the estimated
  value of $\mathbf{a}$ is kept fixed in the following steps. The
  number of foreground templates, $d$, is now 5, whereas it was 4 in
  2013. The 30\GHz\ channel is not recalibrated, unlike in 2013.

\item Second fit: foreground emission.  The foreground emission,
  matrix $\mathbf{F}$, is estimated in a second \smica\ fit with
  $\mathbf{a}$ kept fixed at the value found in the first step.
  Spectra are estimated over a large fraction of the sky ($\fsky =
  97\,\%$) and the fit is made over the range $4\leq\ell\leq150$.  For
  the data, this step is the same as in 2013.  For FFP8, foreground
  emission is captured using 7 templates, whereas 6 templates were
  used for the FFP6 simulations in 2013.

\item Third fit: power spectra.  The frequency spectra captured by
  vector $\mathbf{a}$ and matrix $\mathbf{F}$ are kept fixed at the
  values found in steps 1 and 2; we fit only the signal power
  spectra $C_{\ell}$ and $\mathbf{P}_{\ell}$ and the frequency
  channel noise power spectra.  Spectral covariance matrices are
  computed over 97\,\% of the sky and the fit includes all multipoles
  up to $\ell=1500$.  This step is the same as in 2013.

\end{itemize}

\begin{figure}[th!]
  \begin{center}
    \includegraphics[width=\columnwidth]{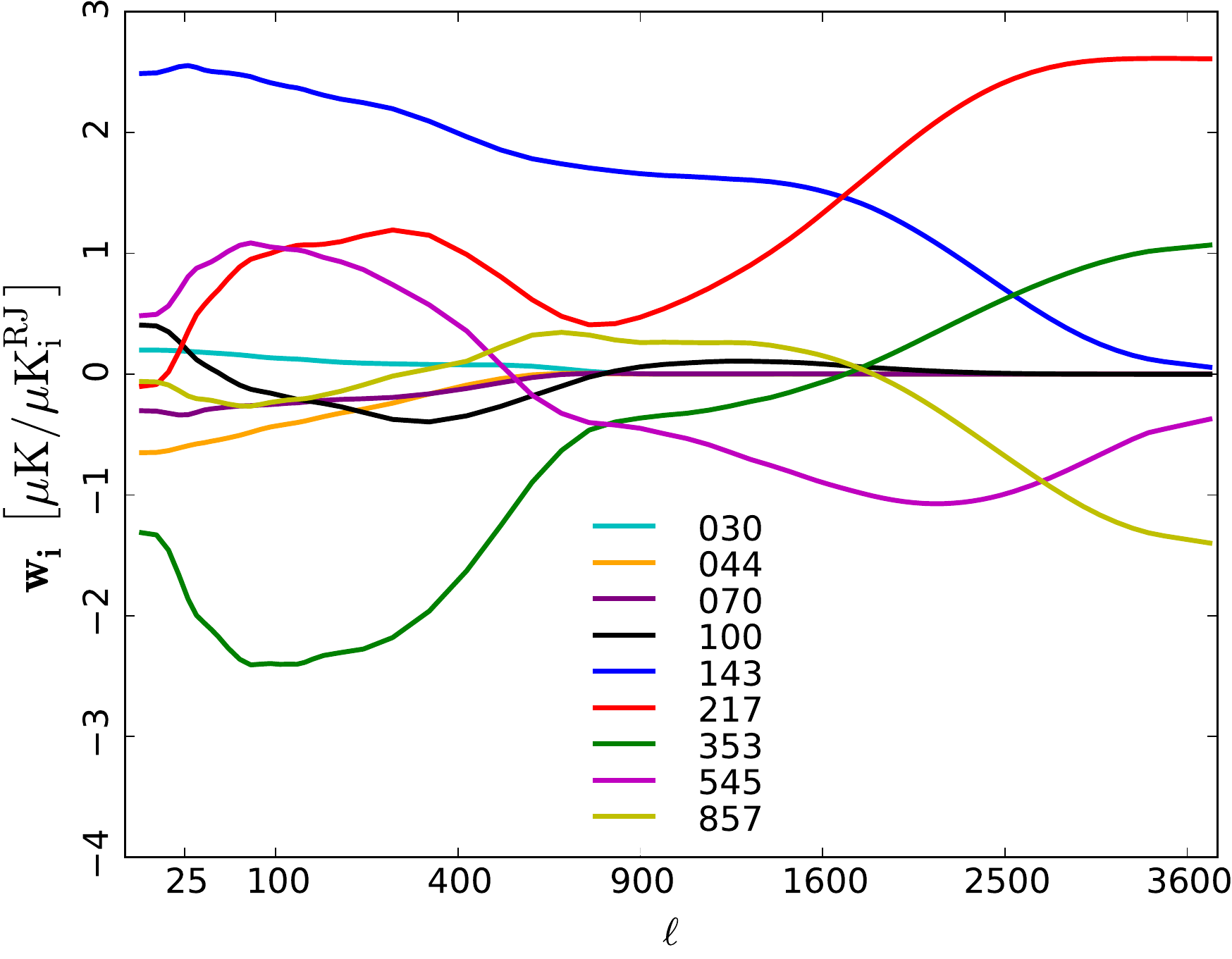}\\
    \includegraphics[width=\columnwidth]{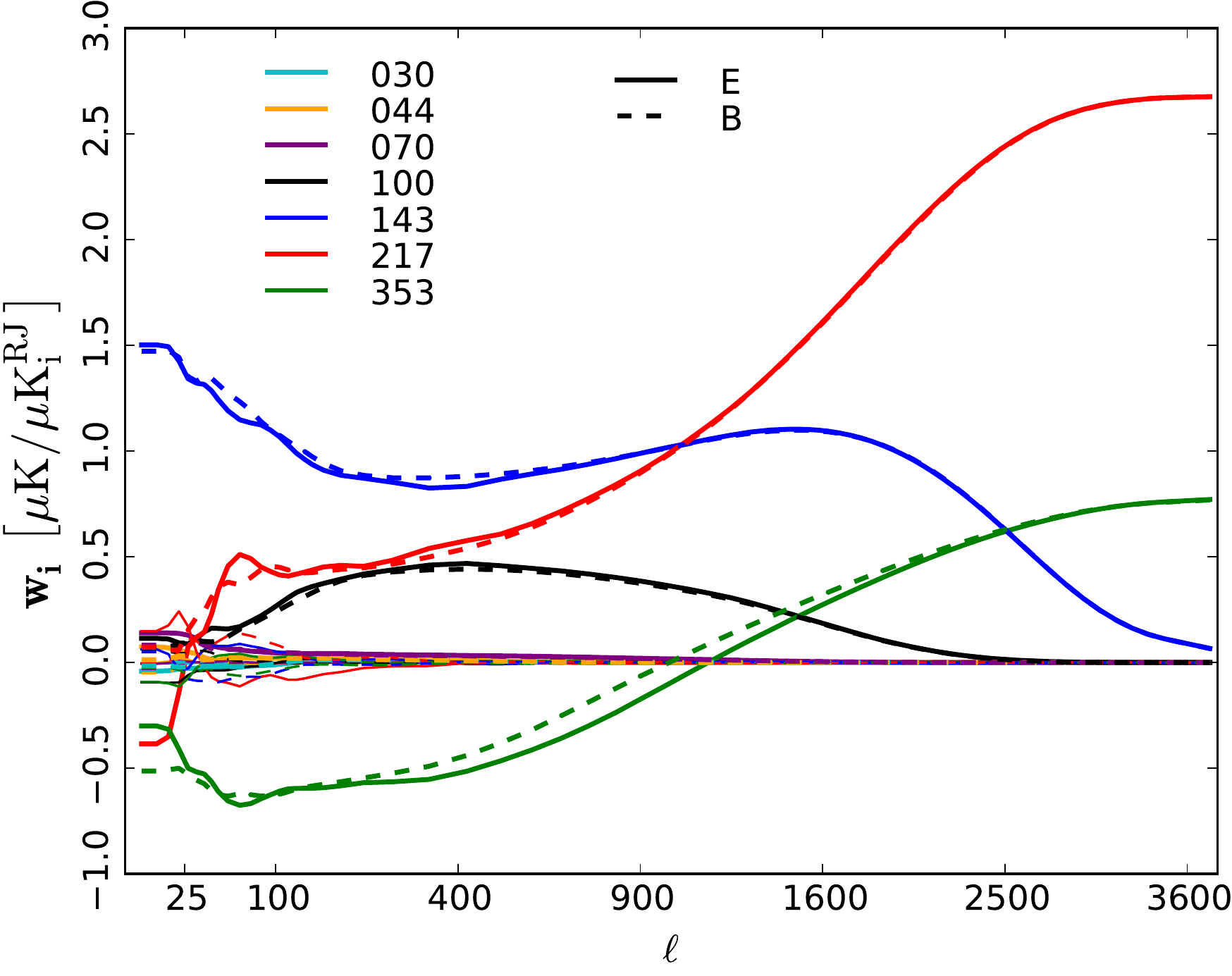}    
  \end{center}
  \caption{\smica\ weights for temperature (\emph{top}) and
    polarization (\emph{bottom}). For readibility, the values are
    shown for input maps in units of antenna temperature.  The plot
    goes up to $\ell\sim3600$, but the output maps are synthesized
    uses all multipoles up to $\ell=4000$.  For polarization, the
    thick solid lines show the contribution of input $E$ modes to the
    CMB $E$ modes and the thick dashed lines show the same for the $B$
    modes.  The thin lines, all close to zero, show
    ``cross-contributions'' of input $E$ modes to the CMB $B$ modes
    and vice versa.}
\label{fig:smica_filters}
\end{figure}

The weights determined for the temperature maps are shown at the top
of Fig.~\ref{fig:smica_filters}.  It should be noted that the CMB map
is synthesized from spherical harmonic coefficients which have been
set explicitly to zero for $\ell = 0$ and $\ell = 1$.  Therefore, the
\smica\ CMB map has no monopole or dipole components over the full
sky.

\subsection{Implementation for polarization}

All \Planck\ polarized channels are used to produce the polarized CMB
maps.  The process is easier in polarization than it is in temperature
because less precision is required due to the lower signal-to-noise
ratio and also because the foregrounds appear to have a simpler
structure.  In particular, we do not preprocess the frequency maps by
subtracting or masking point sources as is done to the temperature
maps.

The \smica\ fit in polarization is conducted with the same parameters
as in temperature, but with two differences.  First, the recalibration
step is omitted: we use the CMB frequency spectrum (vector
$\mathbf{a}$) determined from temperature maps.  Second, the
foreground model comprises 6 polarized templates (as for temperature)
but the matrices $\mathbf{F}_E$ and $\mathbf{F}_B$ are fitted in the
second step over the multipole range $4\leq\ell\leq 50$. The weights
determined for the polarization data are shown at the bottom of
Fig.~\ref{fig:smica_filters}.

In order to mitigate spectral leakage from $E$ to $B$ and from the
Galactic plane onto other regions of the sky, we do not compute
spherical transforms directly from masked $Q$ and $U$ maps.  Instead,
for each pair of input $Q$ and $U$ maps, we first produce full-sky $E$
and $B$ maps to which an apodized Galactic mask is applied.  It is
from those masked $E$ and $B$ maps that spherical harmonic
coefficients are computed for the estimation of the spectra and
cross-spectra going into $\widehat{\mathbf{R}}_{\ell}$ and for the
synthesis of the final CMB map.

\subsection{Masks}
\label{sec:smica_masking}

\begin{figure}[th!]
  \begin{center}
    \includegraphics[width=\columnwidth]{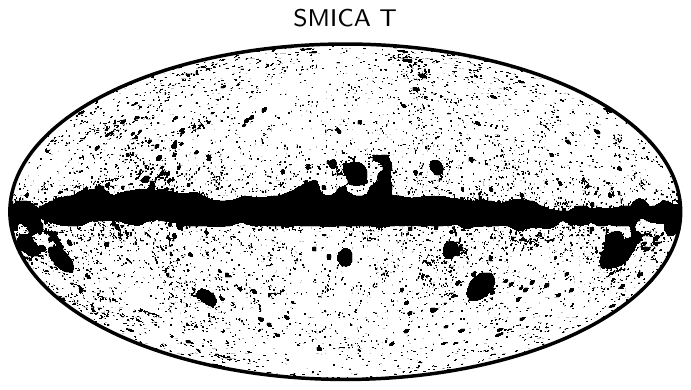}
    \includegraphics[width=\columnwidth]{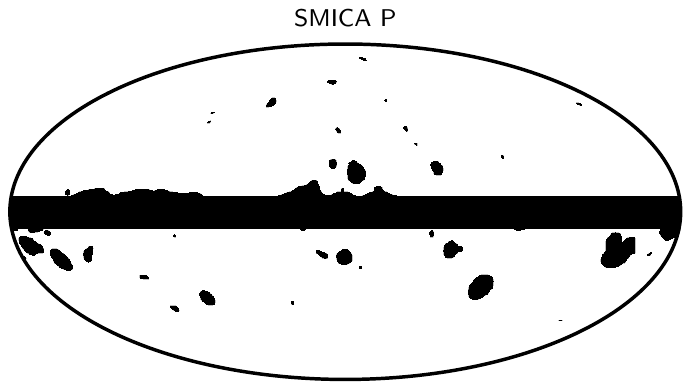}
  \end{center}
  \caption{\smica\ masks in temperature (\emph{top}) and polarization
    (\emph{bottom}).}
  \label{fig:dx11_masks_smica}
\end{figure}

Confidence masks, shown in Fig.~\ref{fig:dx11_masks_smica}, are built
using the procedure described in~\citet{planck2013-p06}, with the
following changes.  For temperature, we apply a bandpass filter to
the CMB map that is then squared and smoothed at 3\pdeg5 (it was
2\degr\ in 2013).  The confidence mask is obtained by thresholding the
resulting map of local power.  The threshold is determined by visual
inspection to be 50\muK$^2$ (it was 70\muK$^2$ in 2013).  The
resulting mask is then enlarged by multiplication by a Galactic mask
covering 10\,\% of the sky.  For polarization, the confidence mask is
obtained by a procedure similar to temperature, but the CMB $Q$ and
$U$ maps are low-pass filtered (rather than bandpassed) using a
Gaussian beam with 30\arcmin\ FWHM.  They are then squared and
smoothed to 3\pdeg5 resolution.  Any area where the resulting $P =
\sqrt{Q^2+U^2}$ map is above 5\muK$^2$ is excluded from the
confidence mask.  In addition to that, we exclude pixels which are
less than 7\deg\ away from the Galactic equator.

\section{FFP8 Simulations}
\label{sec:ffp8}

In this appendix we provide a compendium of analyses evaluated from
the FFP8 data set, corresponding directly to those performed on the
\Planck\ 2015 data in the main text.

\subsection{CMB map differences}
\label{sec:ffp8_cmb_map_differences}

\begin{figure*}[th!]
\begin{center}
\begin{tabular}{cc}
\includegraphics[width=\columnwidth]{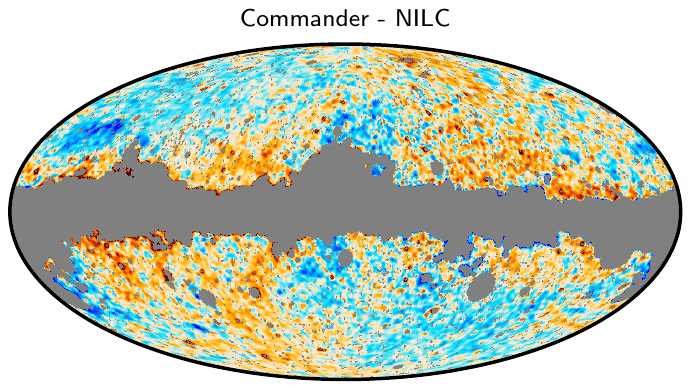}&
\includegraphics[width=\columnwidth]{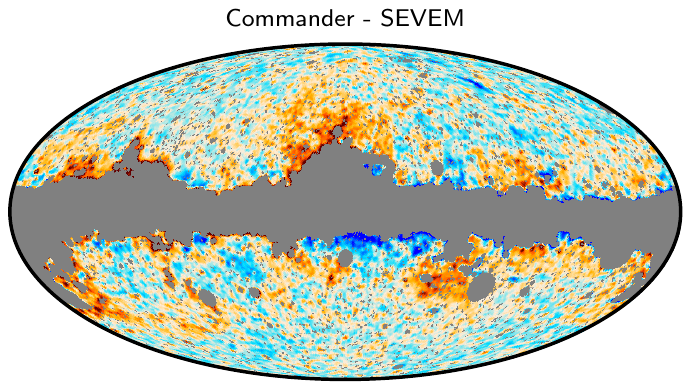}\\
\includegraphics[width=\columnwidth]{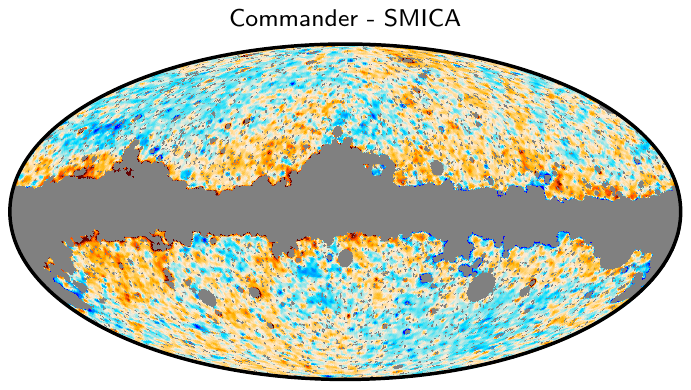}&
\includegraphics[width=\columnwidth]{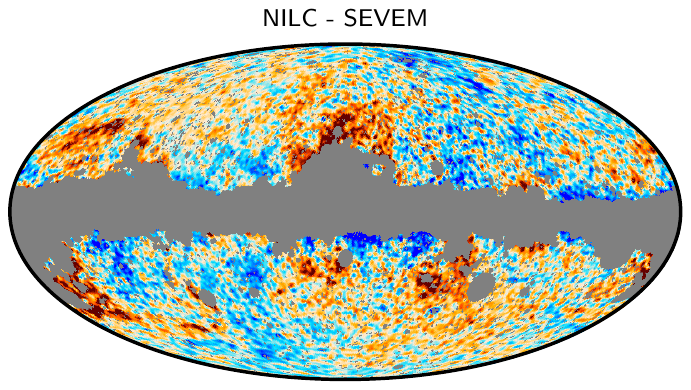}\\
\includegraphics[width=\columnwidth]{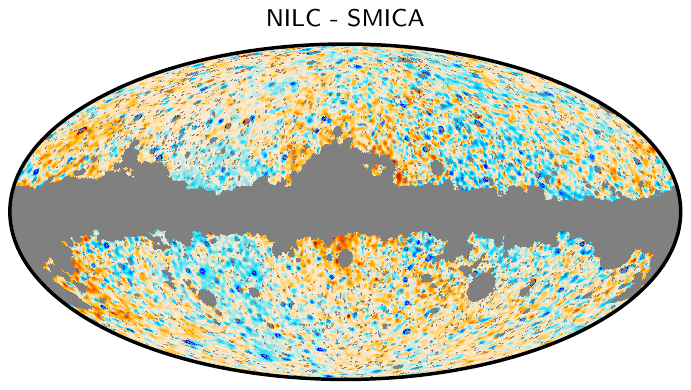}&
\includegraphics[width=\columnwidth]{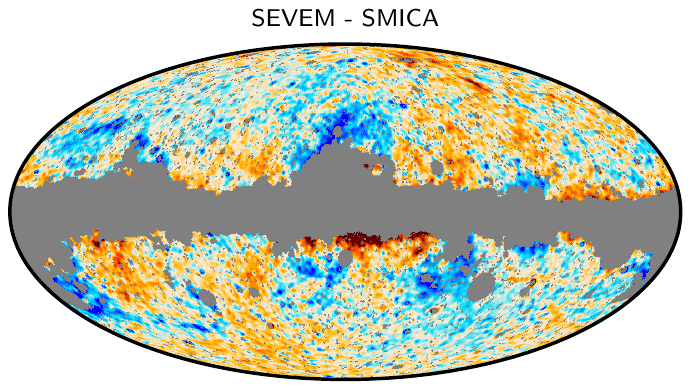}\\
\multicolumn{2}{c}{\includegraphics[height=1cm]{figs/colourbar_7_5uK.pdf}}
\end{tabular}
\end{center}
\caption{Pairwise difference maps between CMB temperature maps
  obtained on FFP8 simulations. Prior to differencing, the maps have
  been smoothed to 80 arcminutes FWHM and downgraded to N$_{side}$ =
  128.}
\label{fig:ffp8_diff_I}
\end{figure*}

\begin{figure*}[th!]
\begin{center}
\begin{tabular}{cc}
\includegraphics[width=\columnwidth]{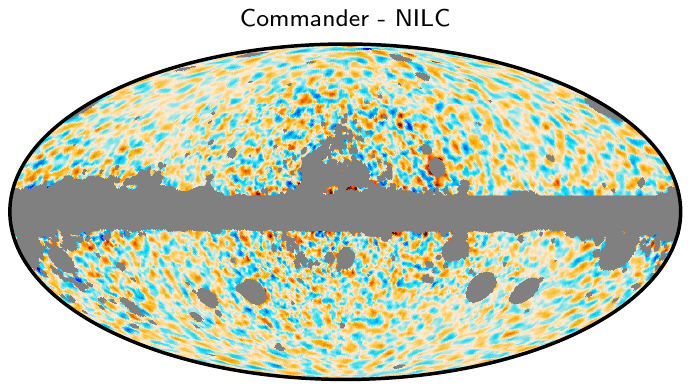}&
\includegraphics[width=\columnwidth]{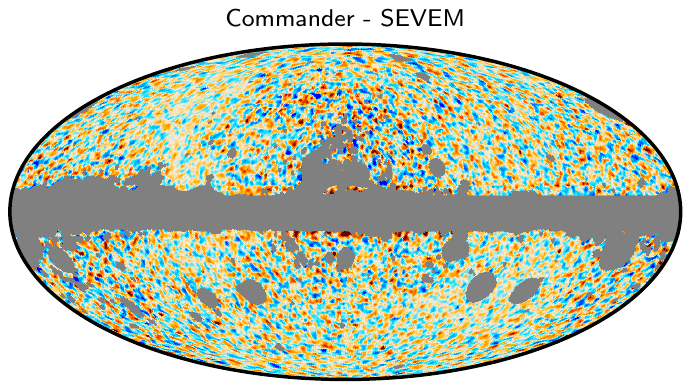}\\
\includegraphics[width=\columnwidth]{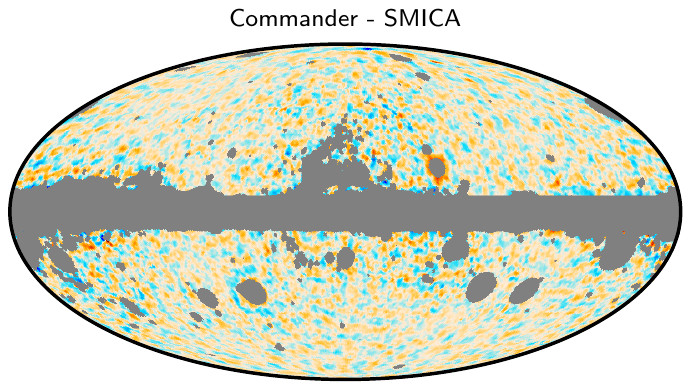}&
\includegraphics[width=\columnwidth]{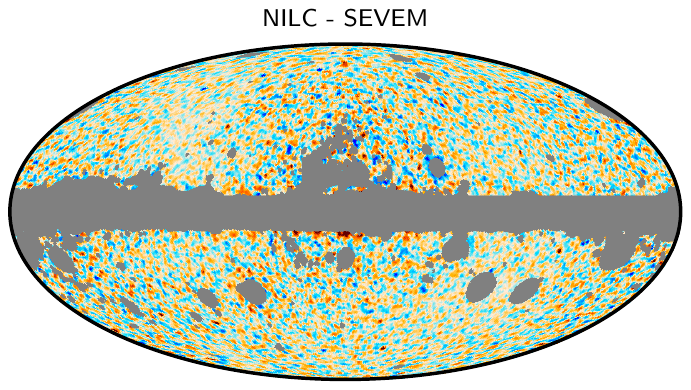}\\
\includegraphics[width=\columnwidth]{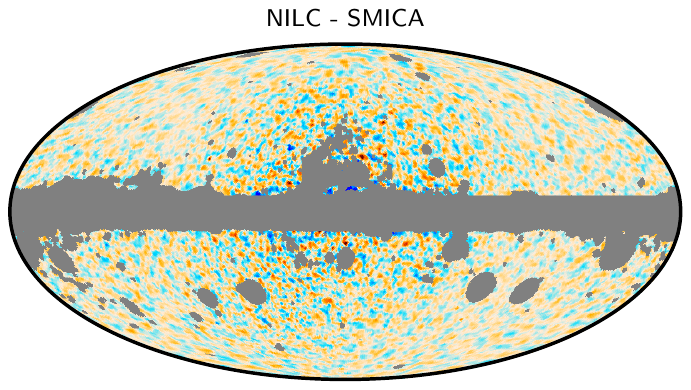}&
\includegraphics[width=\columnwidth]{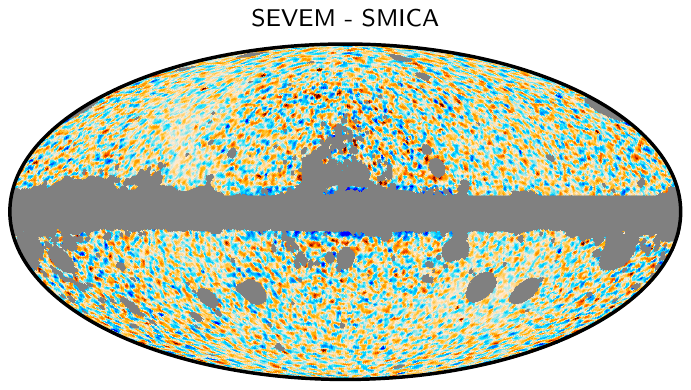}\\
\multicolumn{2}{c}{\includegraphics[height=1cm]{figs/colourbar_1uK.pdf}}
\end{tabular}
\end{center}
\caption{Pairwise difference maps between $Q$ maps obtained on FFP8 siulations. Smoothing and degrading as in 
Fig.~\ref{fig:dx11_diff_I}.}
\label{fig:ffp8_diff_Q}
\end{figure*}

\begin{figure*}[th!]
\begin{center}
\begin{tabular}{cc}
\includegraphics[width=\columnwidth]{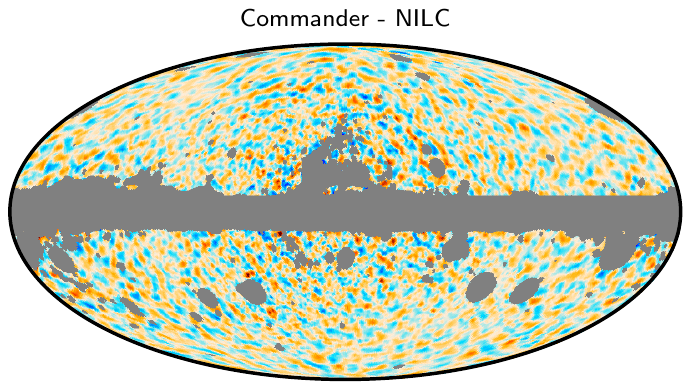}&
\includegraphics[width=\columnwidth]{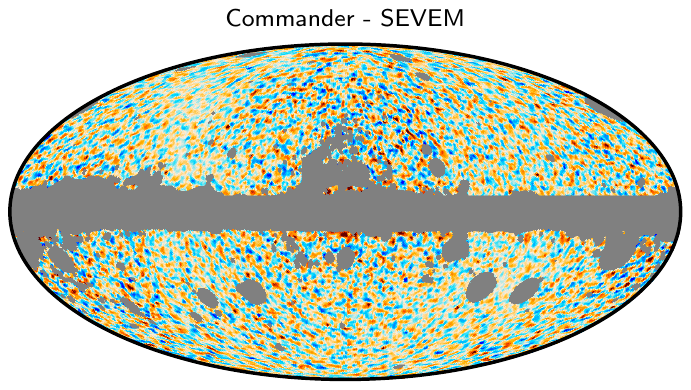}\\
\includegraphics[width=\columnwidth]{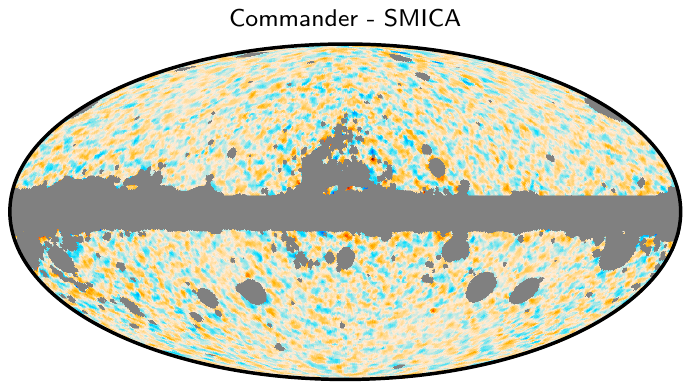}&
\includegraphics[width=\columnwidth]{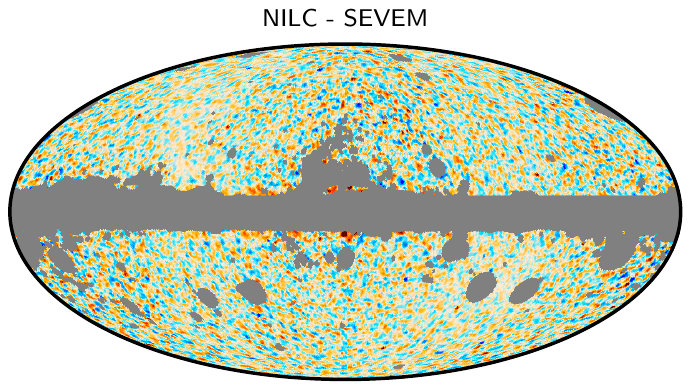}\\
\includegraphics[width=\columnwidth]{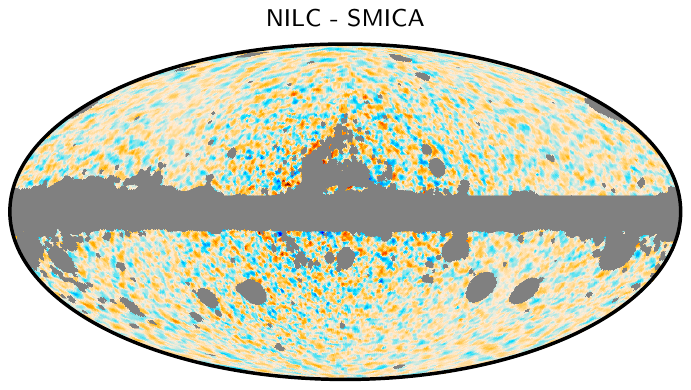}&
\includegraphics[width=\columnwidth]{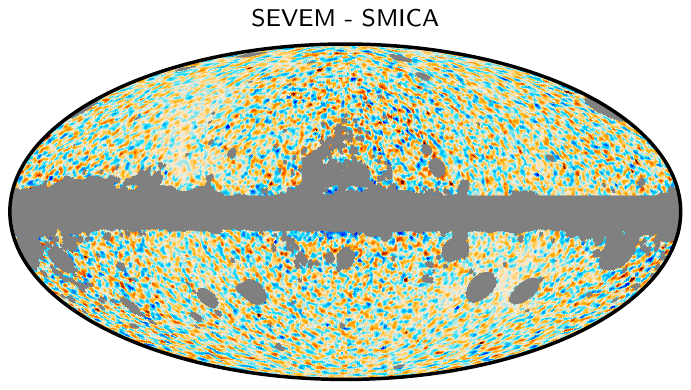}\\
\multicolumn{2}{c}{\includegraphics[height=1cm]{figs/colourbar_1uK.pdf}}
\end{tabular}
\end{center}
\caption{Pairwise difference maps between $U$ maps obtained on FFP8 siulations. Smoothing and degrading as in 
Figs.~\ref{fig:dx11_diff_I},\ref{fig:dx11_diff_Q}.}
\label{fig:ffp8_diff_U}
\end{figure*}

Figures~\ref{fig:dx11_diff_I}, \ref{fig:dx11_diff_Q} and
\ref{fig:dx11_diff_U} show pairwise difference maps between any of the
\commander, \nilc, \sevem, and \smica\ maps derived from the
\Planck\ 2015 data for each of the three Stokes parameters, $I$, $Q$,
and $U$. In Figs.~\ref{fig:ffp8_diff_I}--\ref{fig:ffp8_diff_U} we show
the same evaluated from the FFP8 simulation set.

Overall, the relative differences are of similar magnitude in the
simulated data set as in the data, although they have somewhat
different dominant morphology. The high-latitude differences are in
general somewhat weaker in FFP8 than in the data, while the
low-latitude differences are somewhat stronger. This is primarily due
to the rather complicated foreground model adopted for thermal dust
emission in the FFP8 simulations (see \citealt{planck2014-a14} for
details). In short, thermal dust emission is modelled in the FFP8
simulations by a sum of two greybody components, one of which has a
temperature-dependent spectral index, $\beta(T_{\textrm{d}})$.  This
was motivated by the results presented by \citet{planck2013-XIV};
however, as shown by the updated analysis in \citet{planck2014-a12},
there is no evidence for steepening from the Planck data alone. Only
when the IRAS 100$\,\mu$m data are included in the fit is any such
effect seen. At such high frequencies the thermal dust physics is far
more complicated, and the overall calibration problem more
difficult. For the present discussion, it is sufficient to note that
the thermal dust model adopted for the FFP8 simulations is
significantly more complicated in terms of frequency dependence than
the observed sky, and includes significant spectral curvature below
353\,GHz that is not seen in the actual sky. None of the component
separation codes accounts for this additional curvature, and this
results in the strong features near the Galactic plane seen in some of
the pairwise difference maps shown in Fig.~\ref{fig:dx11_diff_I}.

\subsection{1-point statistics of the total intensity FFP8 maps}
\label{sec:1pdf_ffp8}

Next, we consider the 1-point statistics of the FFP8 fiducial map in
temperature, using the {\tt FFP8-UT74} mask (Table~\ref{Table:1pdfFFP8}). These statistics were defined in
Sect.~\ref{sec:onepointstatistics}, and corresponding polarization
results were shown in Table~\ref{table:1pdf}. We see that all the
methods are in good agreement with the Monte Carlo simulations at all
resolutions considered, and the differences among codes are small.

\begin{table}[th!]
\begingroup
\newdimen\tblskip \tblskip=5pt
\caption{Lower tail probability in percent for the variance, skewness,
  and kurtosis for the FFP8 total intensity analysis at three
  different resolutions. The results have been obtained with the {\tt
    FFP8-UT74} mask.}
\label{Table:1pdfFFP8}
\nointerlineskip
\vskip -3mm
\footnotesize
\setbox\tablebox=\vbox{
   \newdimen\digitwidth 
   \setbox0=\hbox{\rm 0} 
   \digitwidth=\wd0 
   \catcode`*=\active 
   \def*{\kern\digitwidth}
   \newdimen\signwidth 
   \setbox0=\hbox{+} 
   \signwidth=\wd0 
   \catcode`!=\active 
   \def!{\kern\signwidth}
\halign{ \hbox to 1.2in{#\leaderfil}\tabskip=2em&
         \hfil#\hfil&
         \hfil#\hfil&
         \hfil#\hfil\tabskip=0pt\cr                           % Template goes here.0.682790.96
\noalign{\doubleline}
\omit&\multispan3\hfil\sc Lower-tail Probability\hfil\cr
\noalign{\vskip -3pt}
\omit&\multispan3\hrulefill\cr
\noalign{\vskip 2pt}
\omit\hfil\sc Map\hfil&\omit\hfil Variance \hfil&\omit\hfil Skewness \hfil&\omit\hfil Kurtosis \hfil\cr
\noalign{\vskip 3pt\hrule\vskip 5pt}
\noalign{\vskip 4pt}
\omit \bf$N_{\rm side}=2048$ \hfil\cr
\noalign{\vskip 4pt}
\hglue 1em {\tt Commander}& 81.1&  52.8&  37.0\cr
\hglue 1em {\tt NILC}&      85.8&  47.8&  16.2\cr
\hglue 1em {\tt SEVEM}&     76.1&  44.7&  30.6\cr
\hglue 1em {\tt SMICA}&     82.4&  50.4&  38.2\cr
\noalign{\vskip 4pt}
\omit \bf$N_{\rm side}=256$ \hfil\cr
\noalign{\vskip 4pt}
\hglue 1em {\tt Commander}& 79.3&  57.5&  16.9\cr
\hglue 1em {\tt NILC}&      85.0&  54.7&  12.9\cr
\hglue 1em {\tt SEVEM}&     76.8&  55.5&  21.3\cr
\hglue 1em {\tt SMICA}&     80.8&  55.2&  25.2\cr
\noalign{\vskip 4pt}
\omit \bf$N_{\rm side}=64$ \hfil\cr
\noalign{\vskip 4pt}
\hglue 1em {\tt Commander}& 81.3&  74.0&  29.1\cr
\hglue 1em {\tt NILC}&      85.0&  71.1&  23.8\cr
\hglue 1em {\tt SEVEM}&     81.1&  72.5&  29.0\cr
\hglue 1em {\tt SMICA}&     83.2&  71.1&  22.8\cr
\noalign{\vskip 5pt\hrule\vskip 3pt}}}
\endPlancktable                    % ends one-column \halign%\endPlancktablewide                 % ends two-column \halign
\endgroup
\end{table}

\subsection{The real-space $N$-point correlation functions for FFP8 maps} 
\label{sec:npoint_correlation_ffp8}

Finally, we present the real-space $N$-point correlation functions
derived from the FFP8 simulations, complementing the analysis shown in
Sect.~\ref{sec:npoint_correlation}.  As for the data, we
analyse the high-pass filtered FFP8 CMB estimates downgraded to
\mbox{$N_{\rm side}=64$} and smoothed with a Gaussian kernel of
$160\arcm$ FWHM. For the polarization analysis we employ a low
resolution version of the common polarization mask, and for the
temperature analysis we employ a low resolution version of the common
temperature mask.

\begin{figure*}[th!]
\begin{center}
\includegraphics[width=0.5\columnwidth]{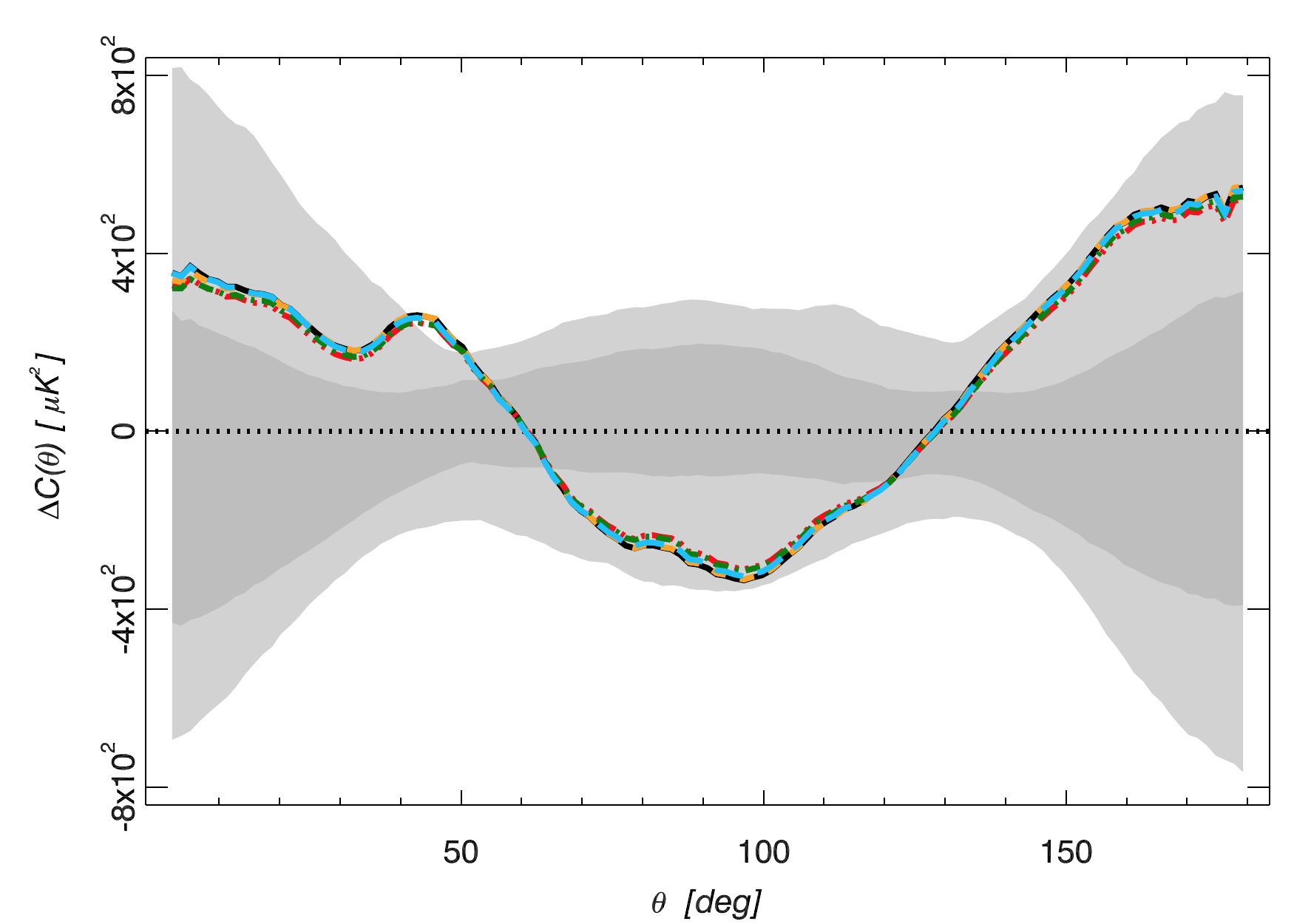}
\includegraphics[width=0.5\columnwidth]{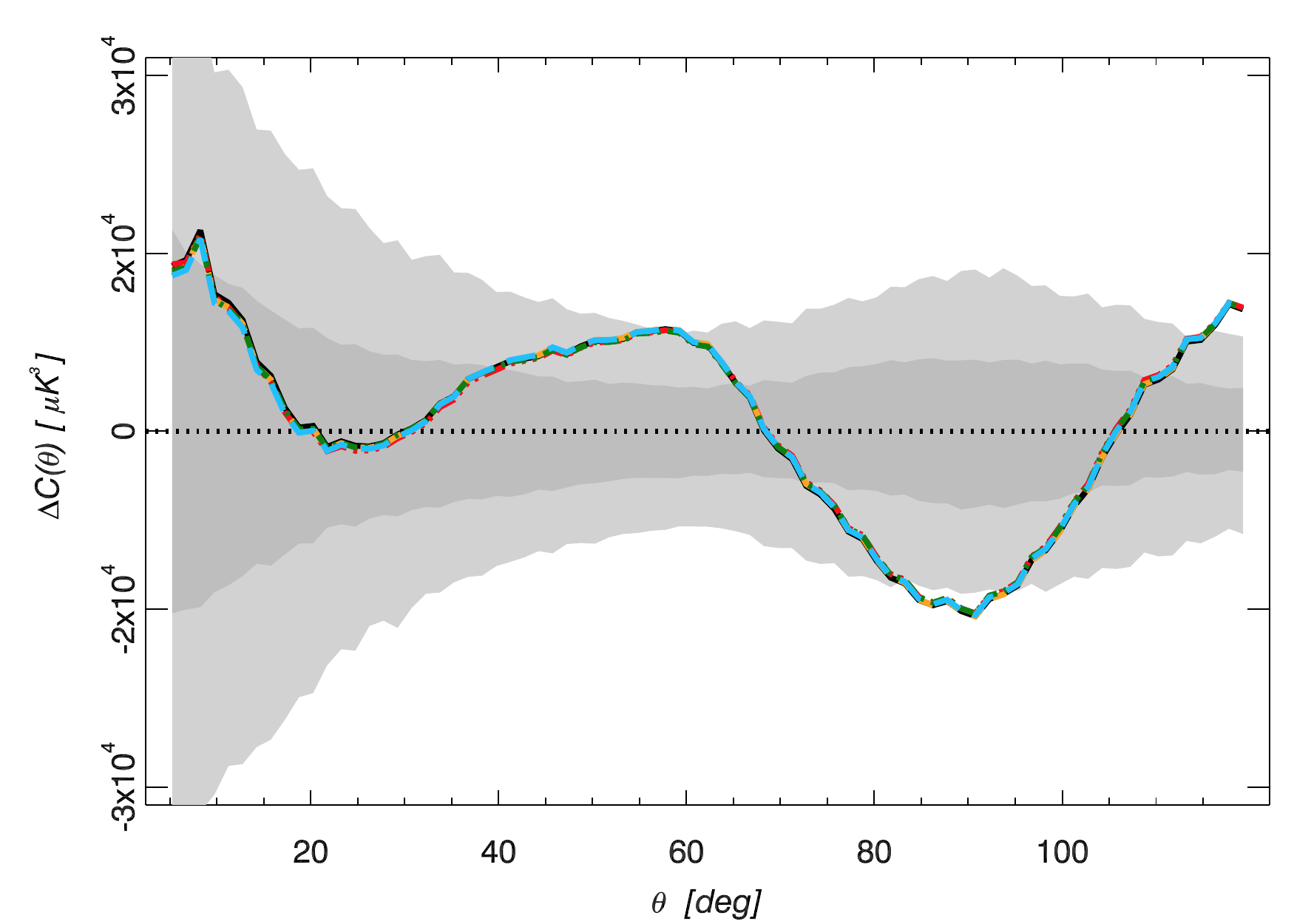}
\includegraphics[width=0.5\columnwidth]{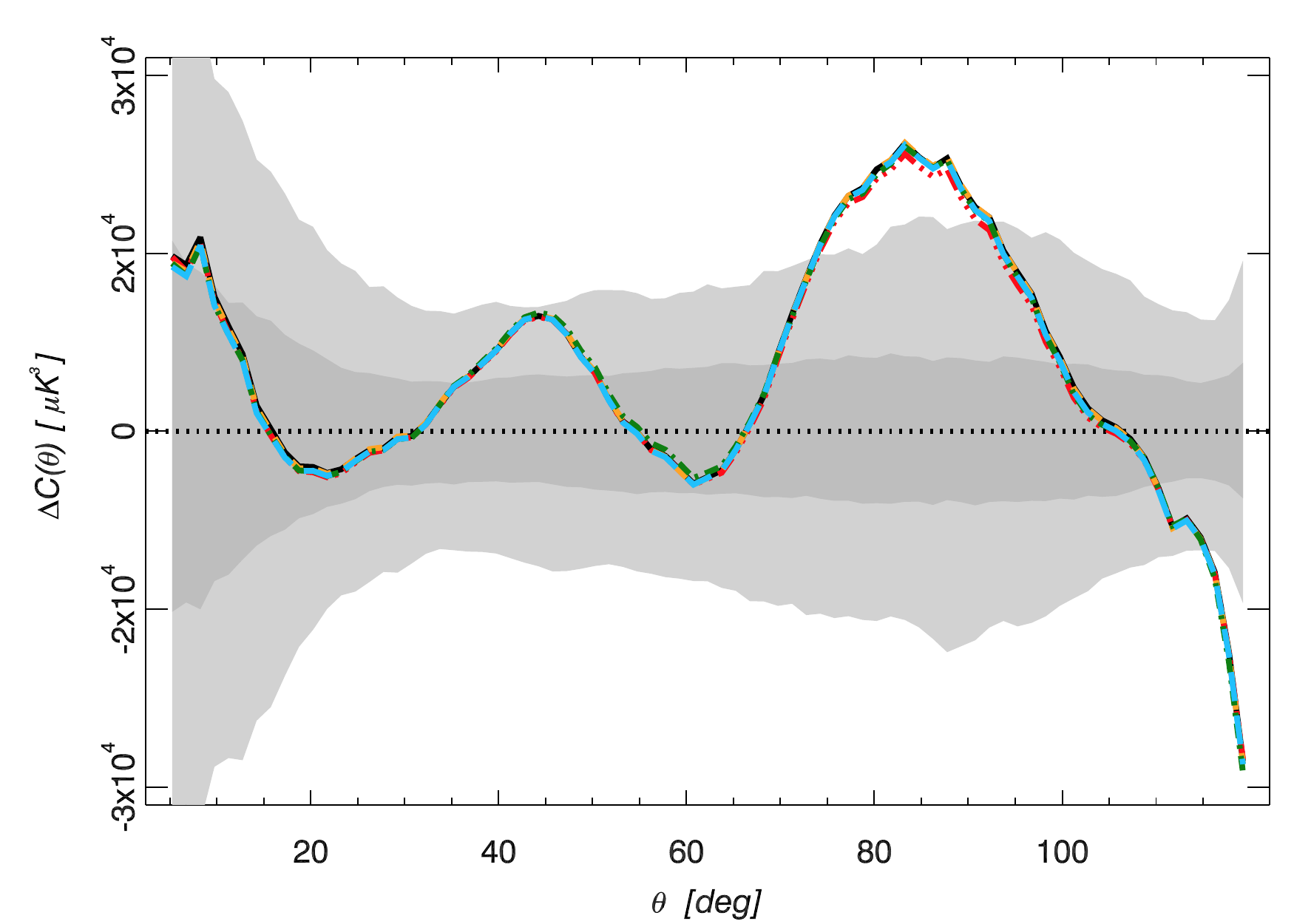}
\includegraphics[width=0.5\columnwidth]{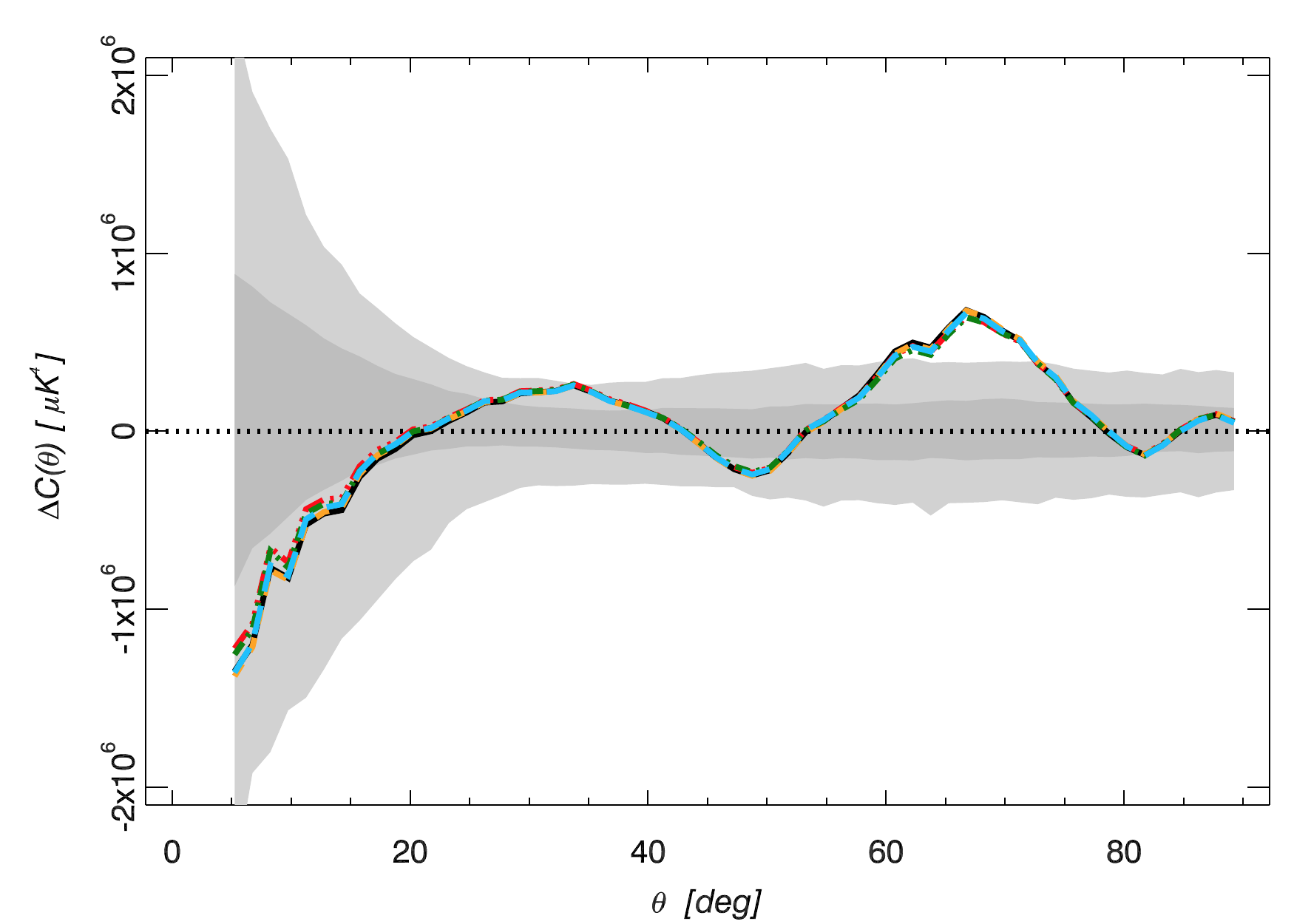}
\caption{The difference between the $N$-point functions and the
  corresponding means estimated from 1000 MC simulations.  From left
  to right, results for the 2-point, pseudo-collapsed 3-point,
  equilateral 3-point and connected rhombic 4-point functions for the
  $N_{\rm side}=64$ FFP8 CMB temperature estimates. The black solid,
  red dot dot dot-dashed, orange dashed, green dot-dashed, and blue
  long dashed lines correspond to the true, {\tt Commander}, {\tt
    NILC}, {\tt SEVEM}, and {\tt SMICA} maps, respectively. The true
  CMB map was analysed with added noise corresponding to the {\tt
    SMICA} component separation method. The shaded dark and light grey
  regions indicate the 68\% and 95\% confidence regions, respectively,
  estimated using \smica\ simulations. See
  Sect.~\ref{sec:npoint_correlation} for the definition of the
  separation angle $\theta$.}
\label{fig:npt_ffp8_temp}
\end{center}
\end{figure*}

\begin{figure*}[th!]
\begin{center}
\includegraphics[width=0.33\linewidth]{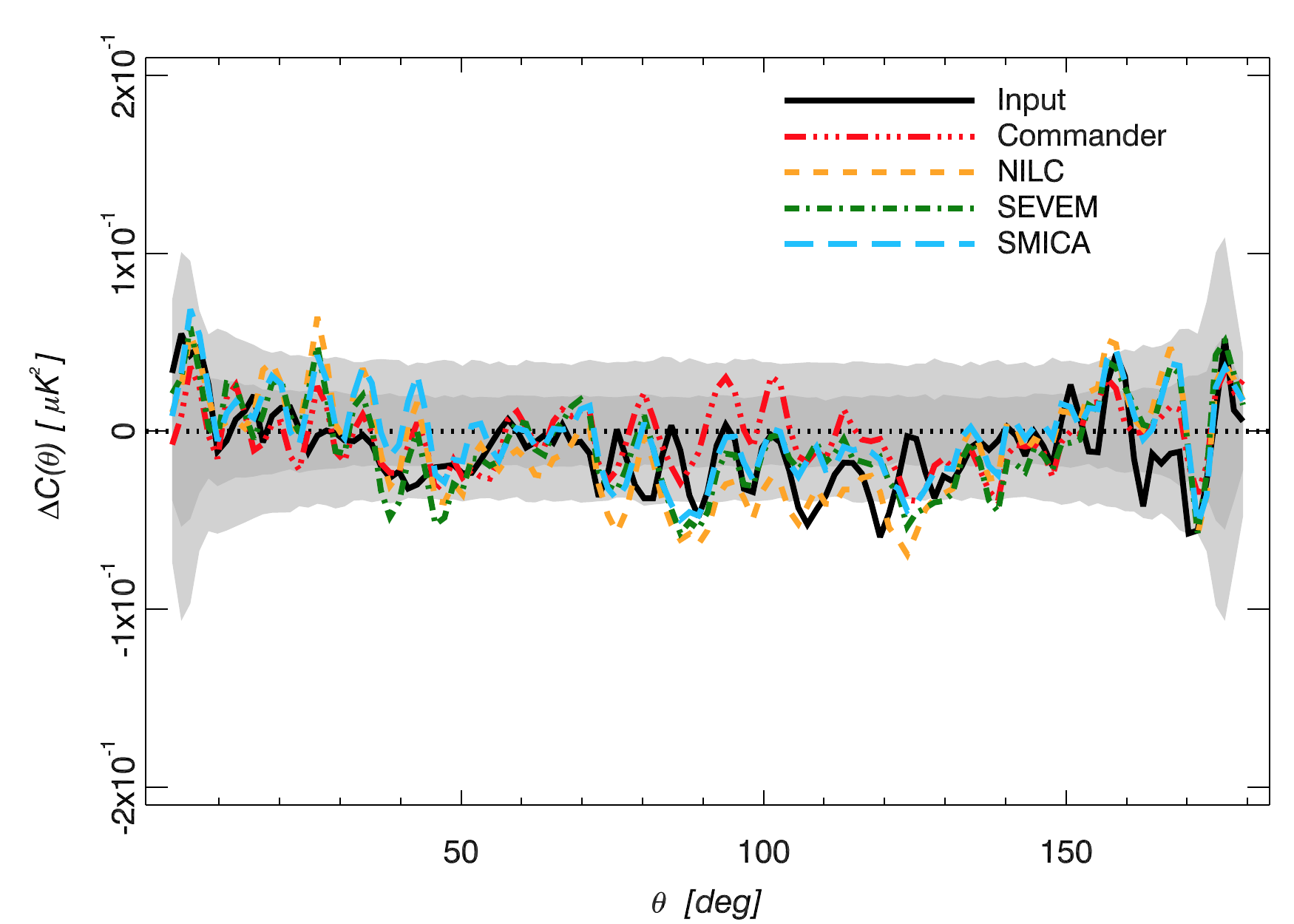}
\includegraphics[width=0.33\linewidth]{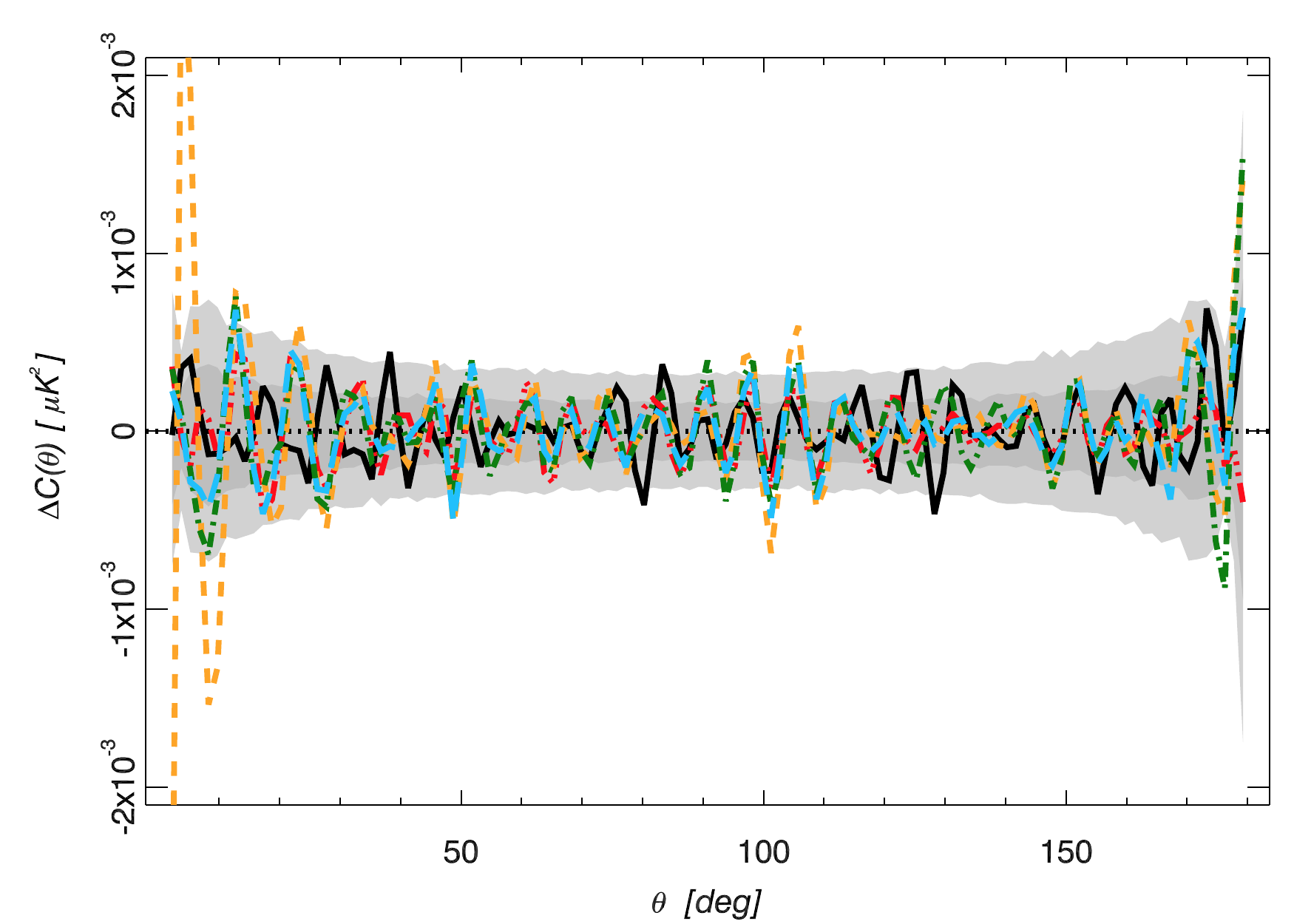}
\includegraphics[width=0.33\linewidth]{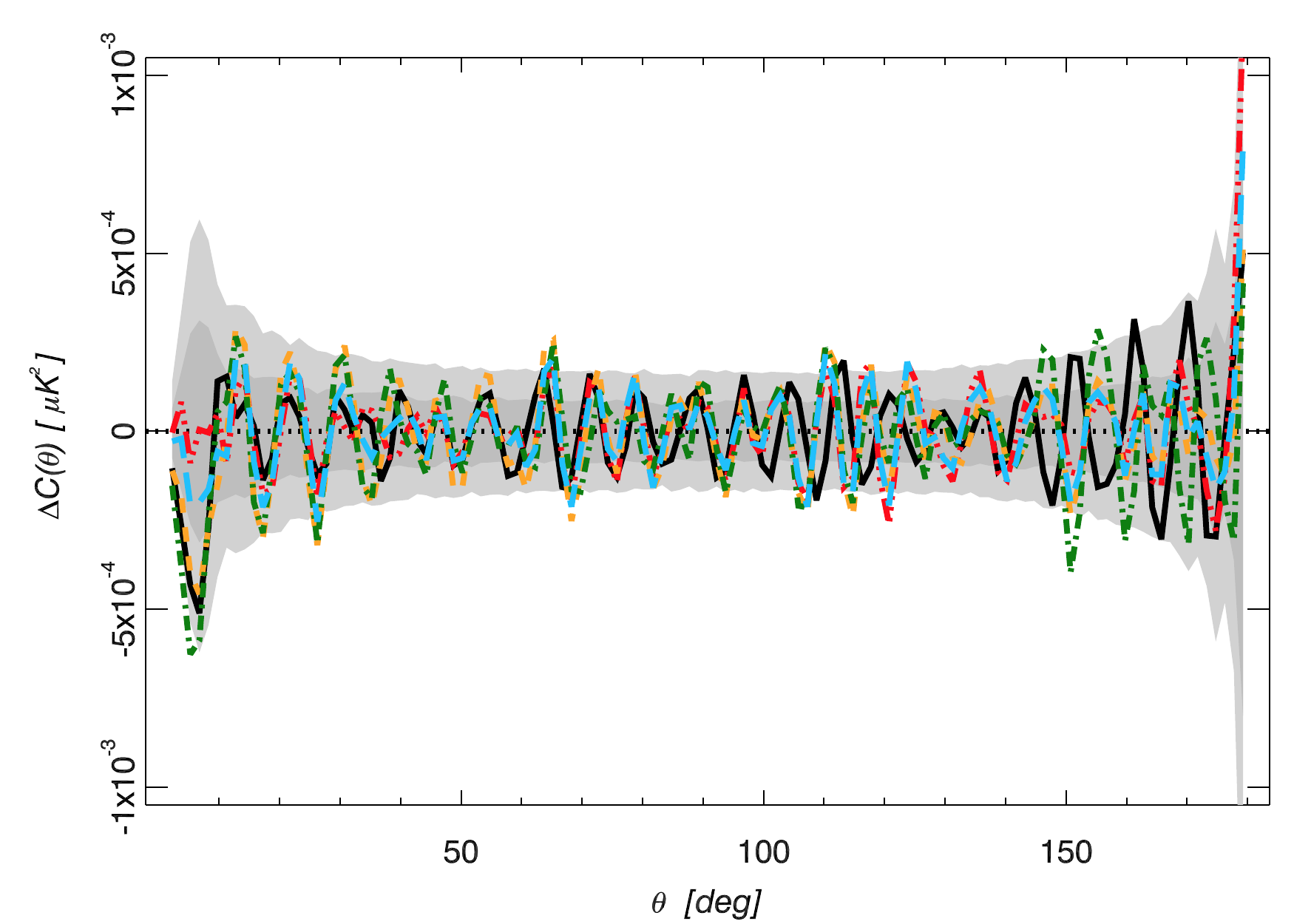}
\includegraphics[width=0.5\columnwidth]{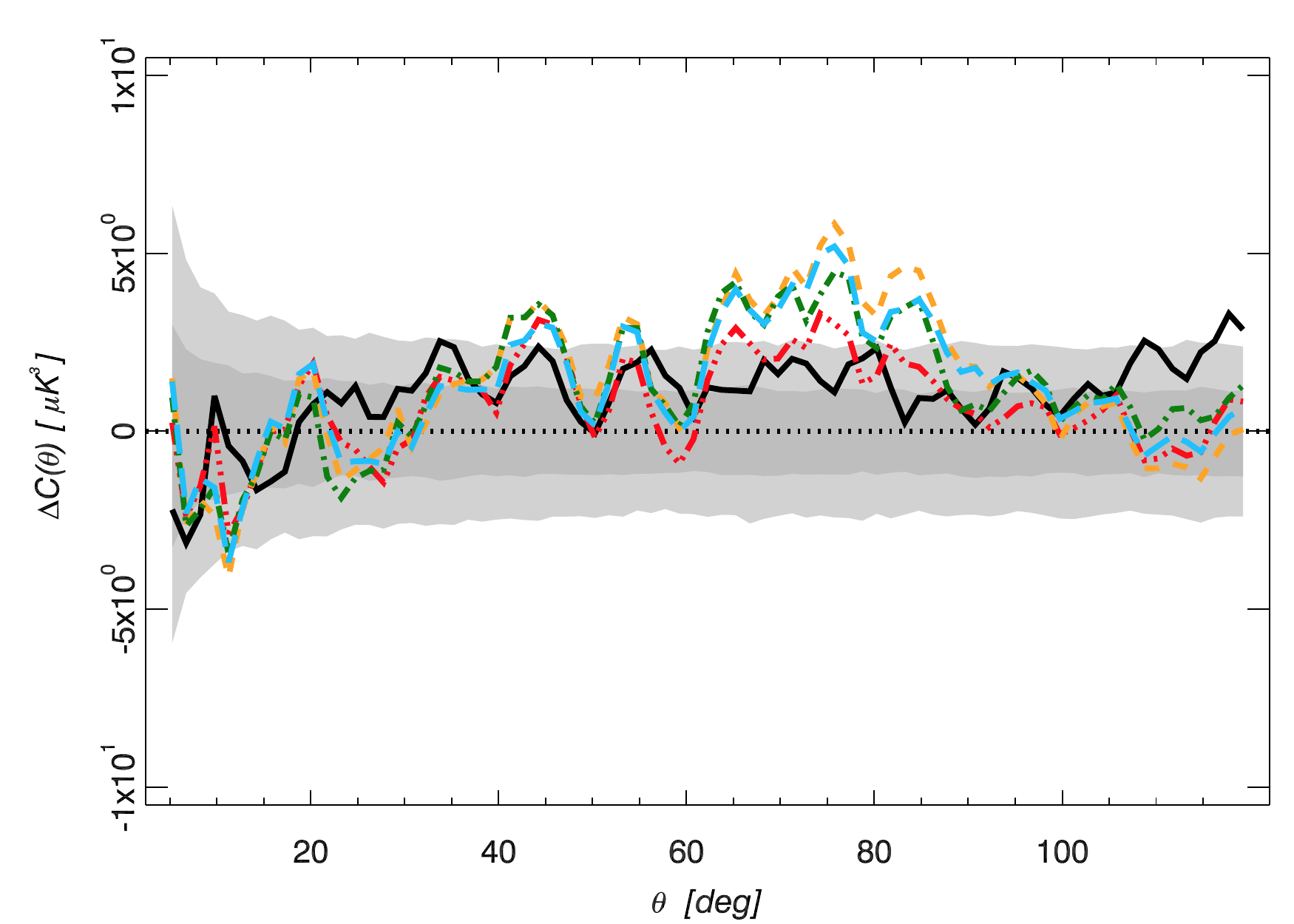}
\includegraphics[width=0.5\columnwidth]{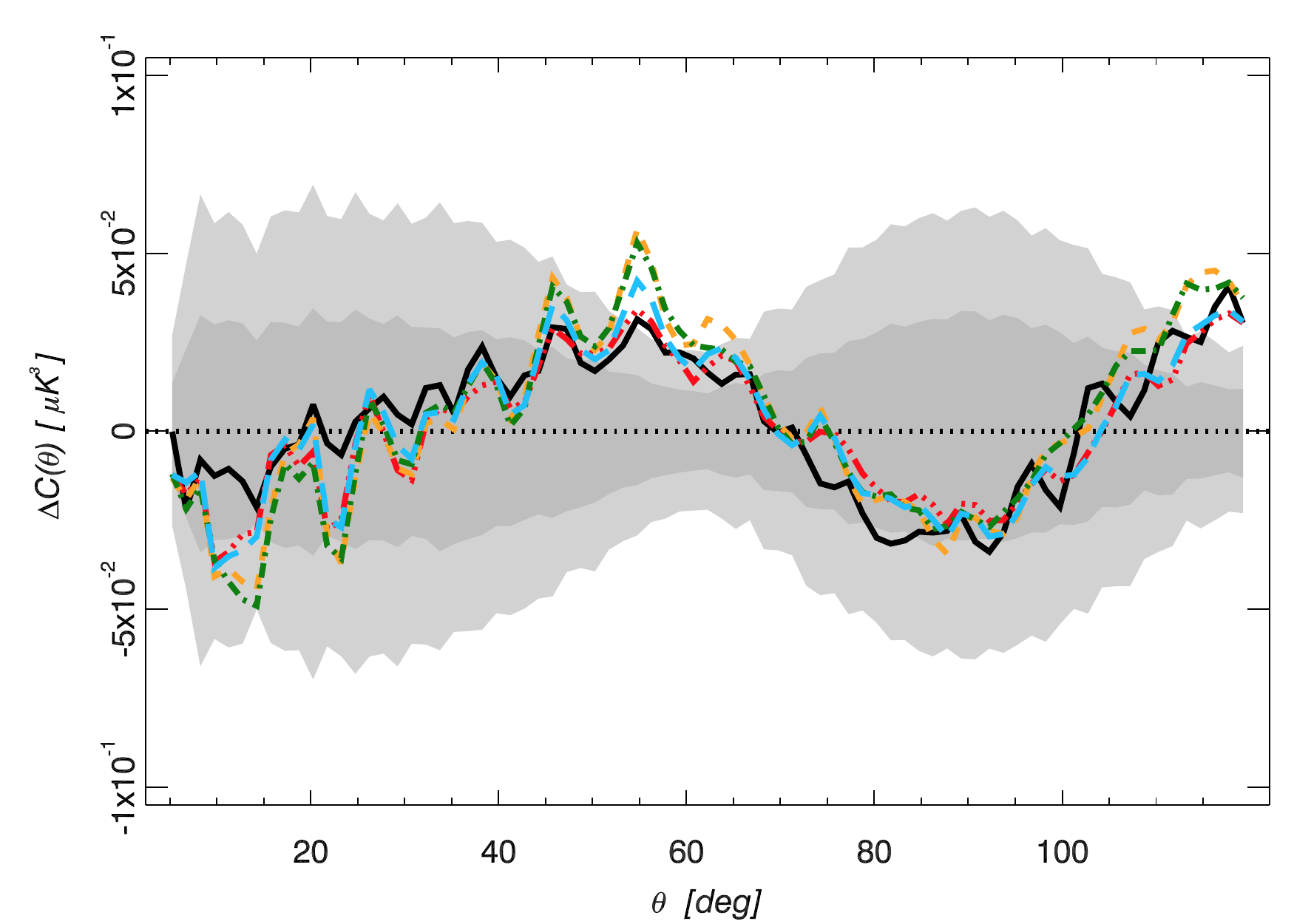}
\includegraphics[width=0.5\columnwidth]{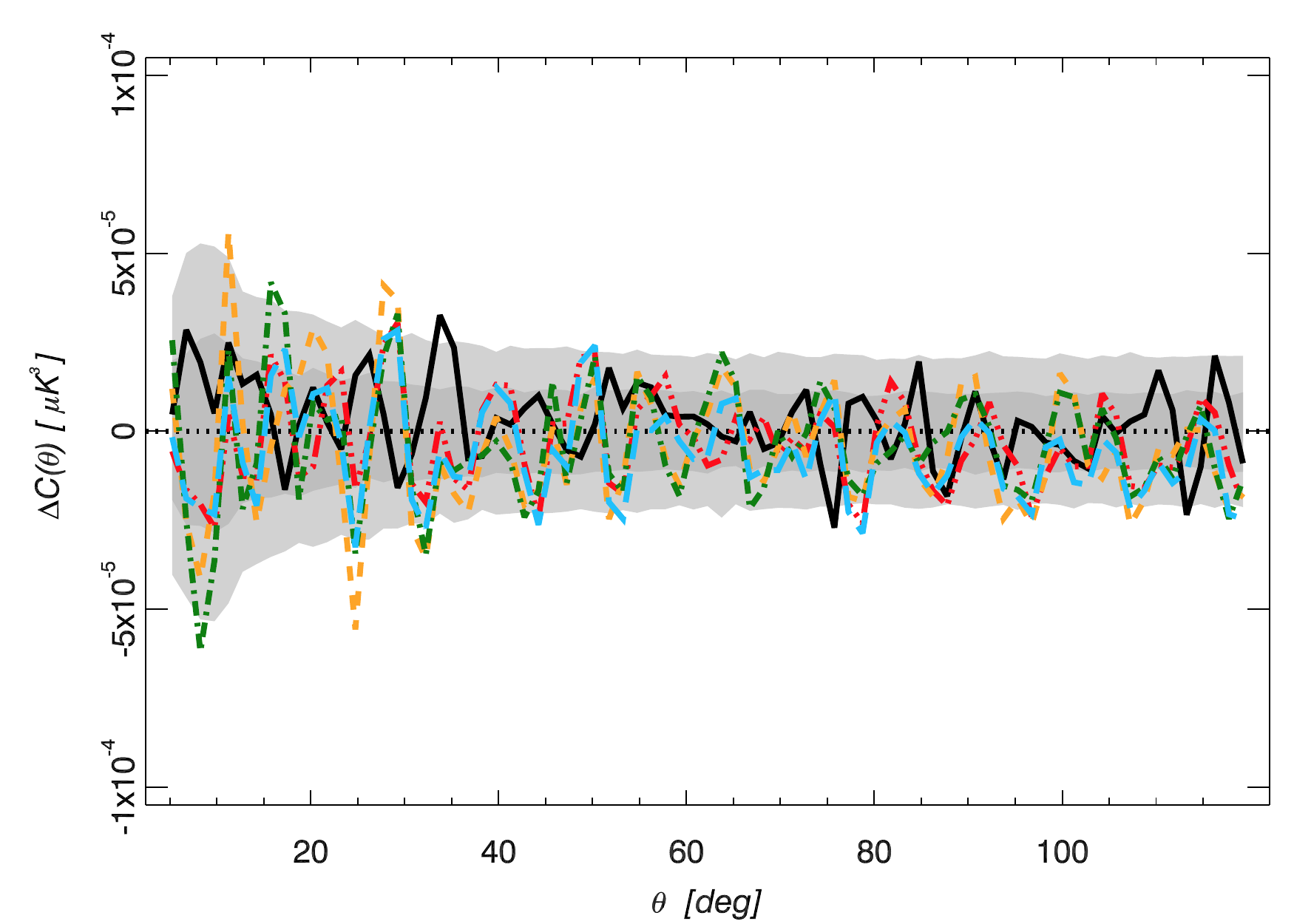}
\includegraphics[width=0.5\columnwidth]{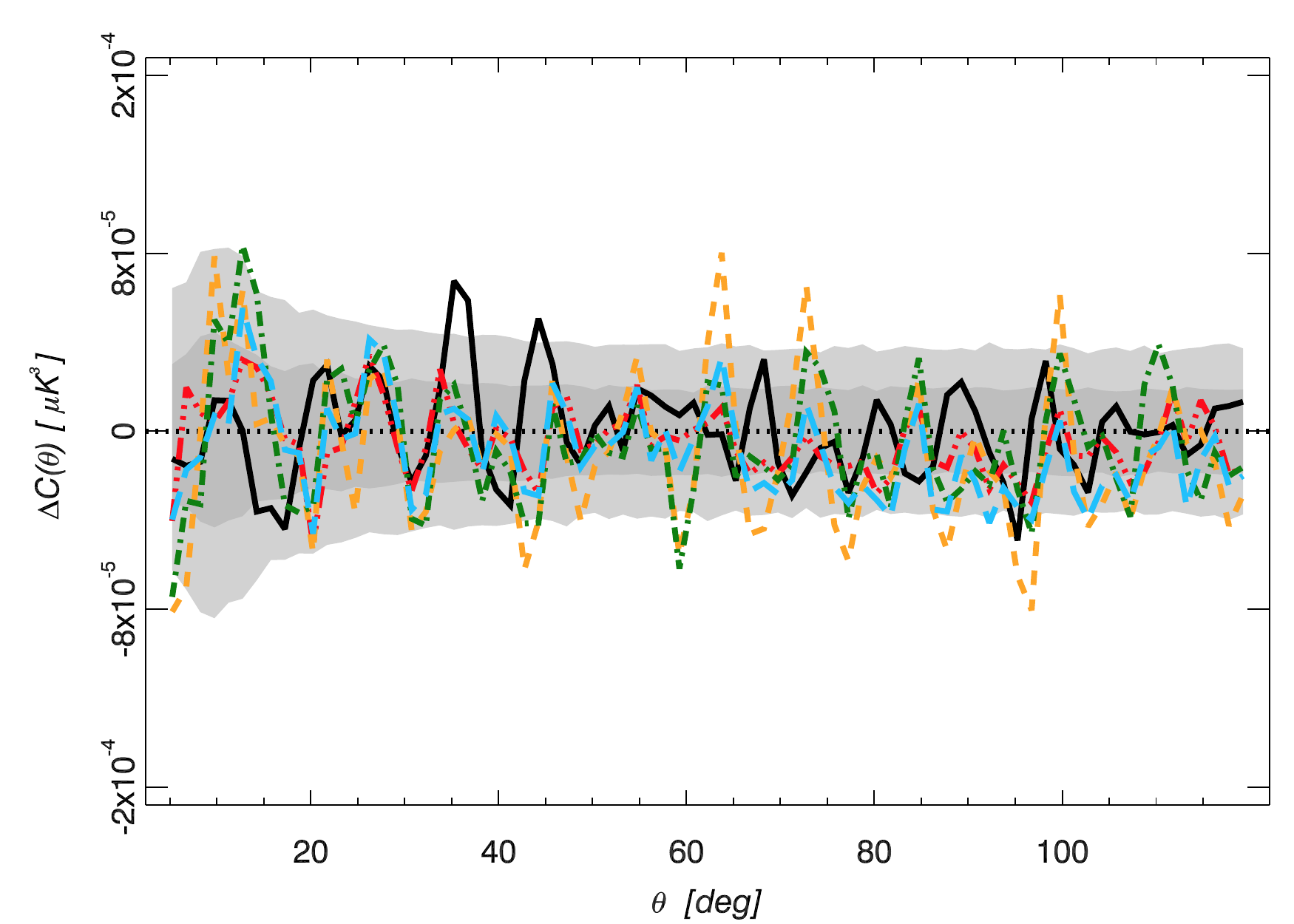}
\includegraphics[width=0.5\columnwidth]{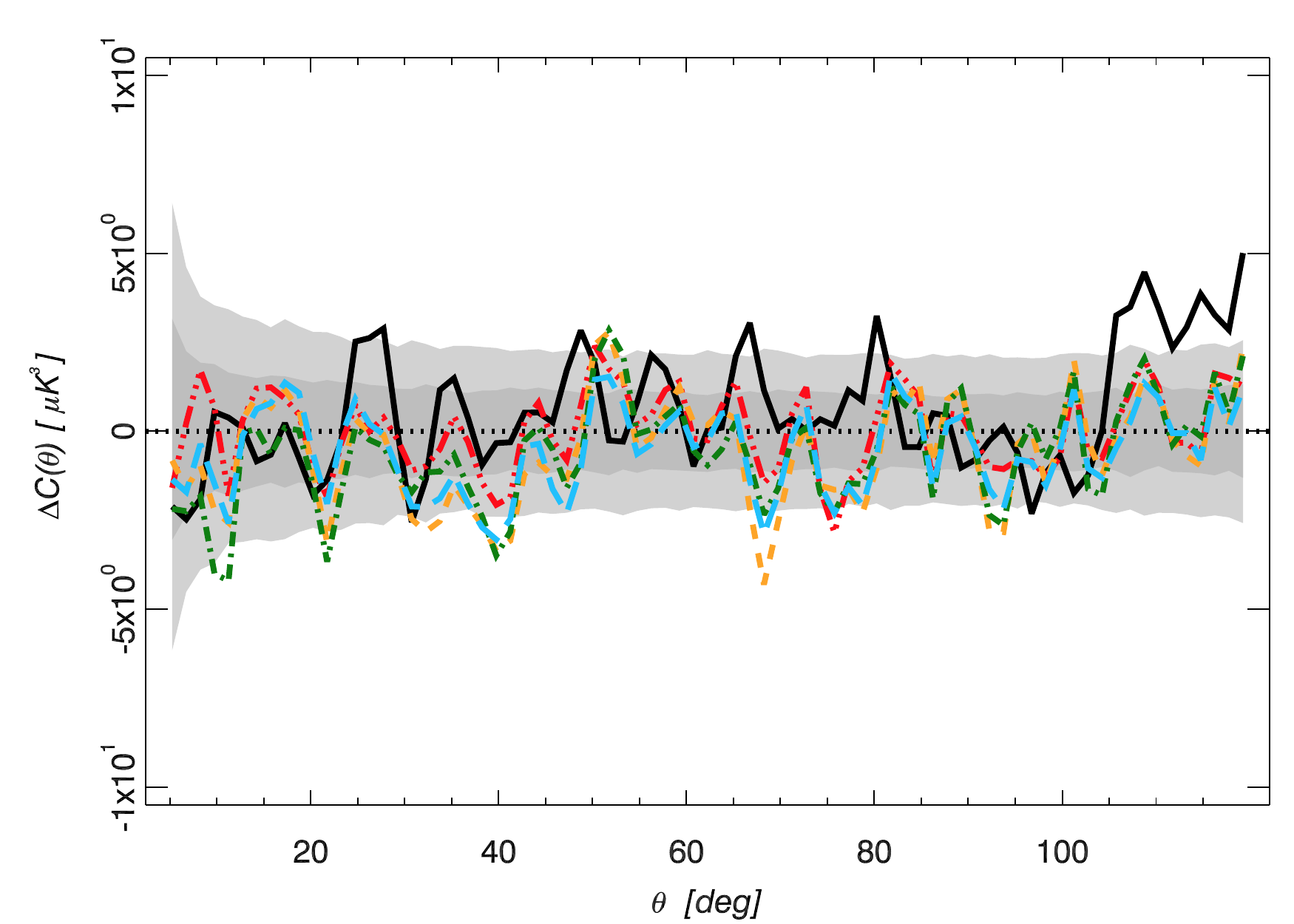}
\includegraphics[width=0.5\columnwidth]{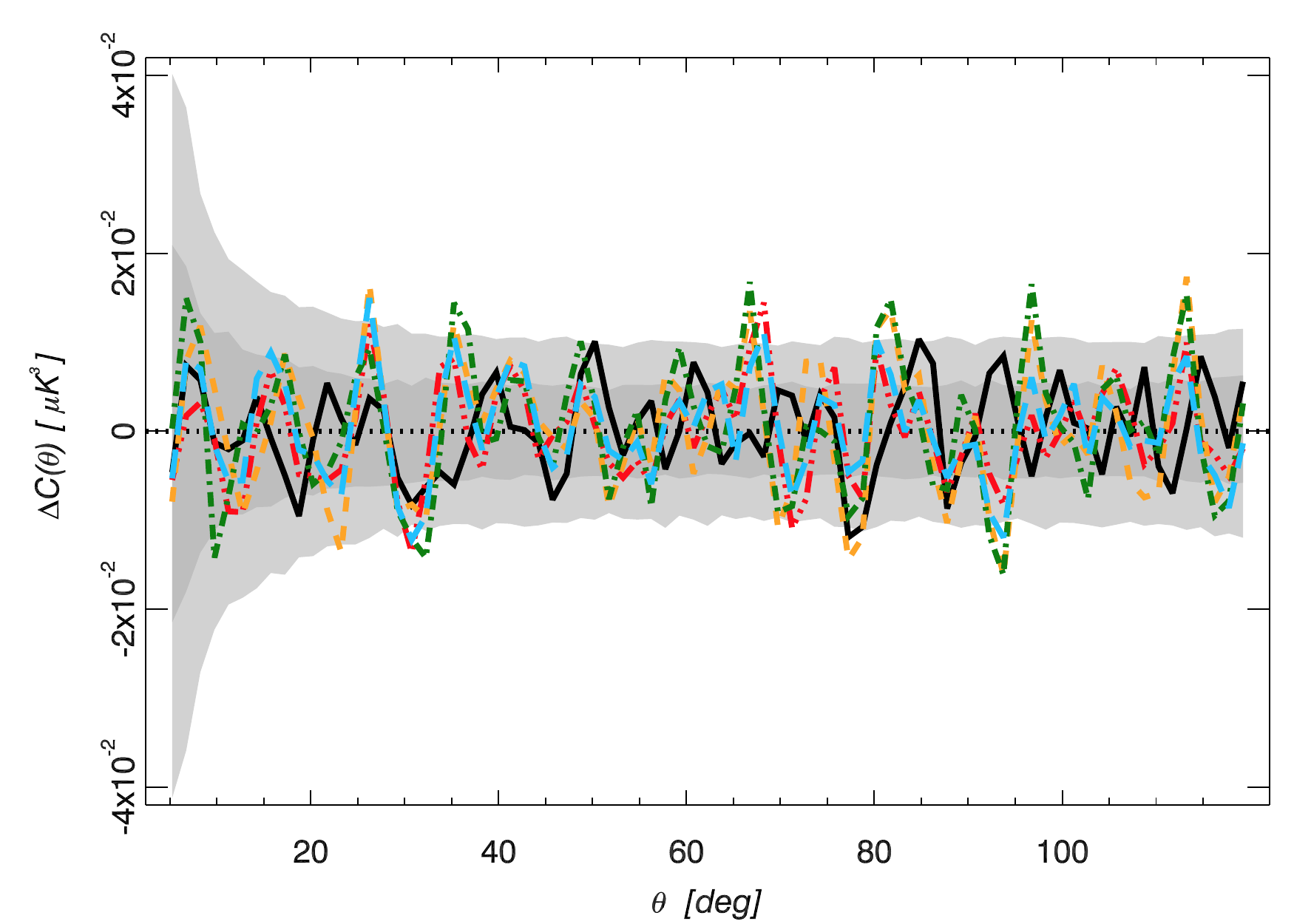}
\includegraphics[width=0.5\columnwidth]{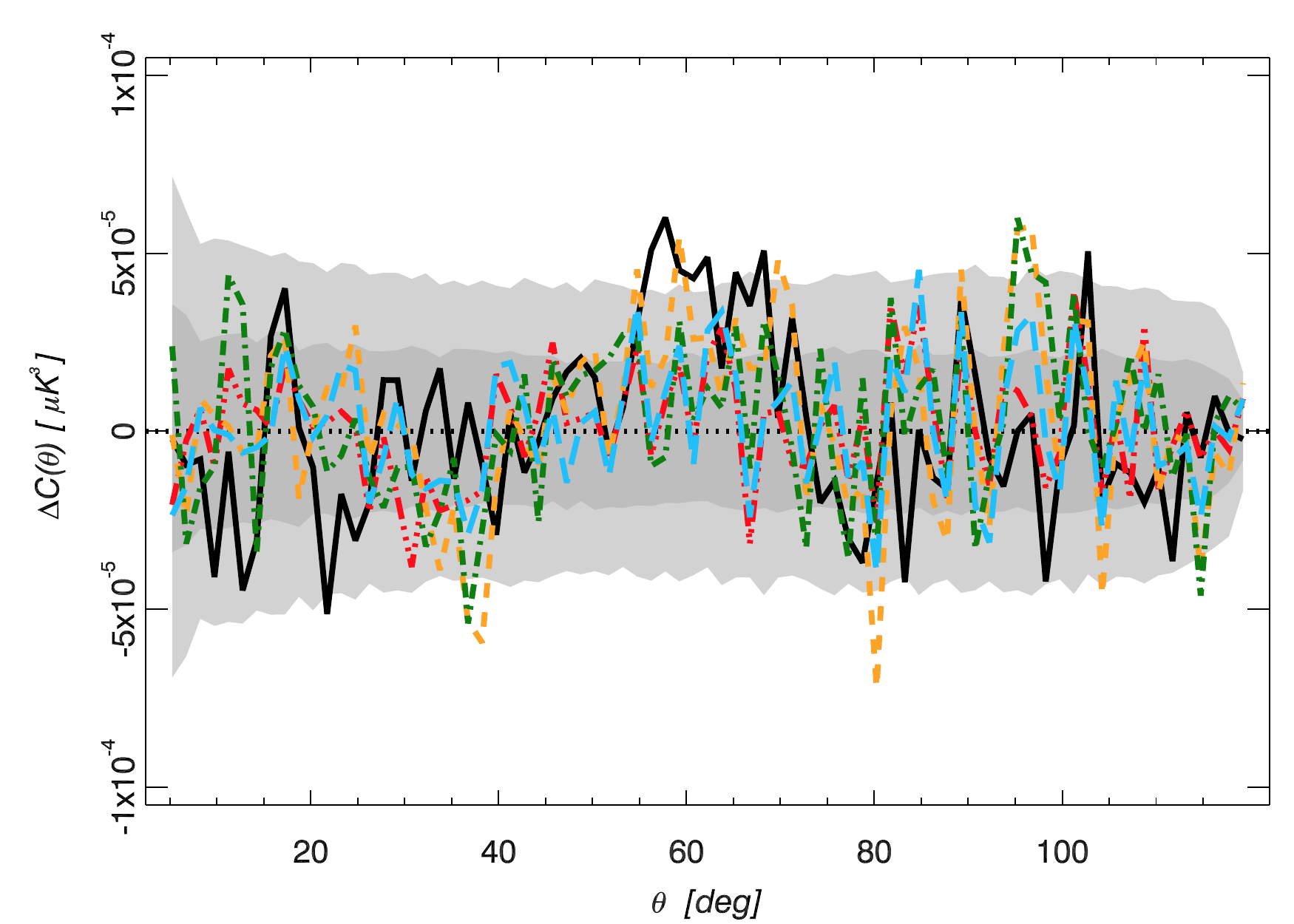}
\includegraphics[width=0.5\columnwidth]{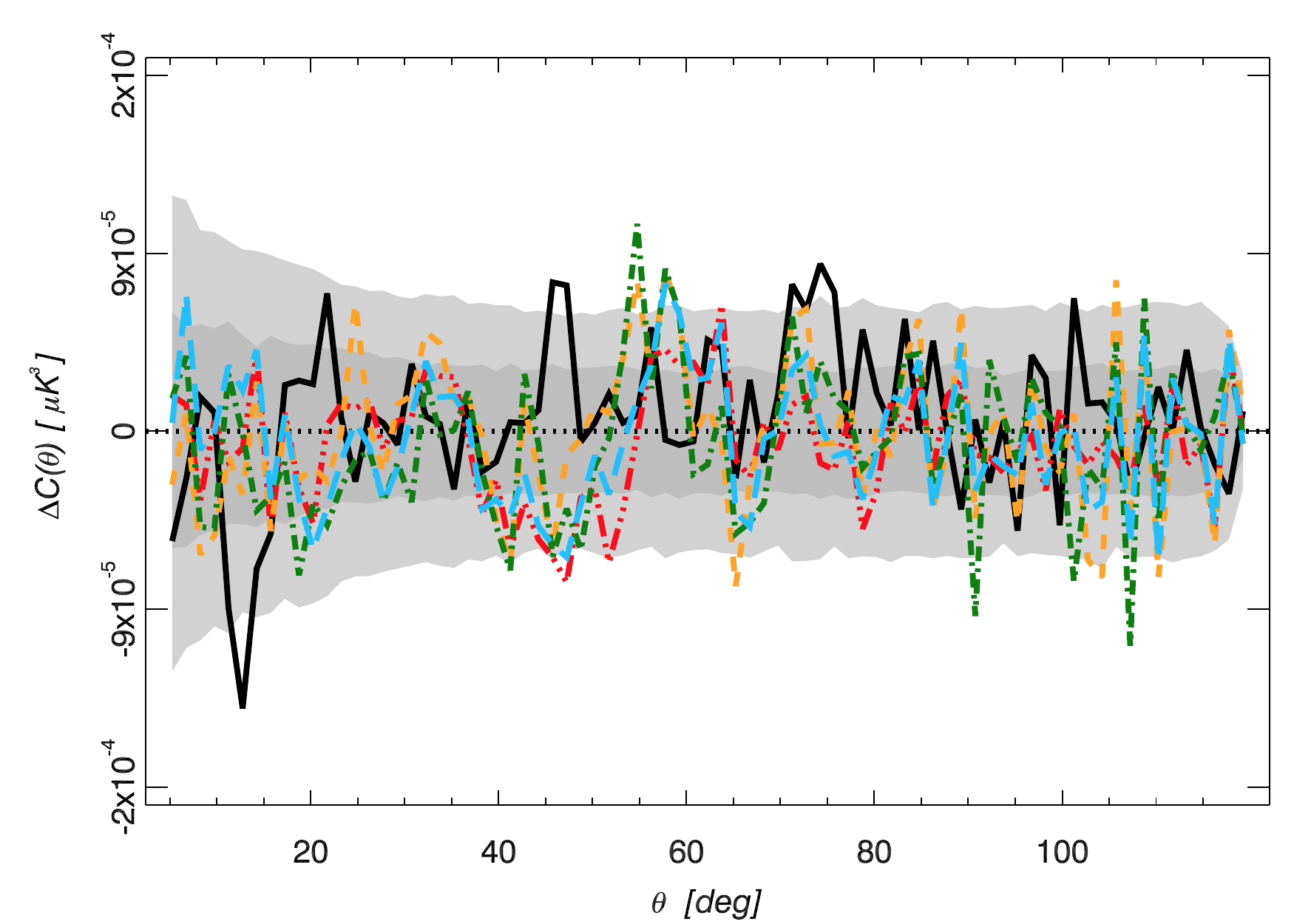}
\caption{The difference between the $N$-point functions for the
  high-pass filtered $N_{\rm side}=64$ FFP8 CMB estimates and the
  corresponding means estimated from 1000 MC simulations. The Stokes
  parameters $Q_r$ and $U_r$ were locally rotated so that the
  correlation functions are independent of coordinate frame. The first
  row shows results for the 2-point function, from left to right,
  $TQ_r$, $Q_rQ_r$, and $Q_rU_r$.  The second row shows results for
  the pseudo-collapsed 3-point function, from left to right, $TTQ_r$,
  $TQ_rQ_r$, $Q_rQ_rU_r$, and $U_rU_rU_r$, and the third row shows
  results for the equilateral 3-point function, from left to right,
  $TTQ_r$, $TQ_rQ_r$, $Q_rQ_rU_r$, and $U_rU_rU_r$. The black solid,
  red dot dot dot-dashed, orange dashed, green dot-dashed, and blue
  long dashed lines correspond to the true, {\tt Commander}, {\tt
    NILC}, {\tt SEVEM}, and {\tt SMICA} maps, respectively. The true
  CMB map was analysed with added noise corresponding to the {\tt
    SMICA} component separation method. The shaded dark and light grey
  regions indicate the 68\% and 95\% confidence regions, respectively,
  estimated using \smica\ simulations. See
  Sect.~\ref{sec:npoint_correlation} for the definition of the
  separation angle $\theta$.}
\label{fig:npt_ffp8_hp}
\end{center}
\end{figure*}

The $N$-point functions for the FFP8 simulations are shown in
Figs.~\ref{fig:npt_ffp8_temp} and \ref{fig:npt_ffp8_hp}, plotted in
terms of differences between the $N$-point functions for the high-pass
filtered fiducial FFP8 map and the mean value estimated from 1000
Monte Carlo simulations. The probabilities of obtaining larger values
for the $\chi^2$ statistic, compared to the \Planck\ fiducial
$\Lambda$CDM model, are tabulated in
Table~\ref{tab:prob_chisq_npt_ffp8}.

\begin{table}[th!]
\begingroup
\newdimen\tblskip \tblskip=5pt
\caption{Probability-to-exceed (PTE) in percent for the $N$-point
 correlation function $\chi^2$ statistic applied to the FFP8
 simulation at $N_{\rm side} = 64$ for each of the four CMB codes,
 as shown in Figs.~\ref{fig:npt_ffp8_temp} and \ref{fig:npt_ffp8_hp}. For
 reference, the second column lists the corresponding probabilities
 for the true input map with added noise corresponding to the {\tt
 SMICA} map.}
\label{tab:prob_chisq_npt_ffp8}
\nointerlineskip
\vskip -3mm
\setbox\tablebox=\vbox{
  \newdimen\digitwidth 
  \setbox0=\hbox{\rm 0} 
  \digitwidth=\wd0 
  \catcode`*=\active 
  \def*{\kern\digitwidth}
  \newdimen\signwidth 
  \setbox0=\hbox{+} 
  \signwidth=\wd0 
  \catcode`!=\active 
  \def!{\kern\signwidth}
\halign{ \hbox to 1.35in{#\leaderfil}\tabskip=1em&
        \hfil#\hfil&
        \hfil#\hfil&  
        \hfil#\hfil&
        \hfil#\hfil&
        \hfil#\hfil\tabskip=0pt\cr 
\noalign{\doubleline}
\omit&\multispan5\hfil PTE [\%]\hfil\cr 
\noalign{\vskip -3pt}
\omit&\multispan5\hrulefill\cr
\noalign{\vskip 3pt}
\omit \hfil Correlation function \hfil& {\tt Input}& {\tt Comm.}& {\tt NILC}& {\tt SEVEM}& {\tt SMICA} \cr
\noalign{\vskip 3pt\hrule\vskip 3pt}
\noalign{\vskip 4pt}
\multispan6 $N$-point; see Fig.~\ref{fig:npt_ffp8_temp}\hfil\cr
\noalign{\vskip 4pt}
\hglue 1em  2-pt                 &  30.4 &  33.4 &  40.4 &  33.3 & 30.7 \cr
\hglue 1em  Pseudo-coll.\ 3-pt &  11.1 &  13.4 &  14.0 &  13.5 &  11.7 \cr
\hglue 1em  Equil.\ 3-pt       &  10.3 &  11.0 &  12.2 &  11.0 &  10.1 \cr
\hglue 1em  Rhombic 4-pt      &  23.5 &  25.2 &  25.4 &  25.2 &  22.8 \cr
\noalign{\vskip 4pt}
\multispan6 Two-point; see Fig.~\ref{fig:npt_ffp8_hp}\hfil\cr
\noalign{\vskip 4pt}
\hglue 1em   $TQ_r$&    *7.8 &  18.5 &  67.5 &  *3.5 &  11.9 \cr
\hglue 1em   $TU_r$&    67.7 &  47.3 &  88.2 &  51.8 &  13.8 \cr
\hglue 1em   $Q_rQ_r$&  46.2 &  *6.1 &  33.3 &  *0.1 &  *3.6 \cr
\hglue 1em   $Q_rU_r$&  92.4 &  *0.2 &  94.5 &  *2.5 &  *0.1 \cr
\hglue 1em   $U_rU_r$&  75.8 &  39.5 &  58.0 &  *1.7 &  *3.8 \cr
\noalign{\vskip 4pt}
\multispan6 Pseudo-collapsed three-point; see Fig.~\ref{fig:npt_ffp8_hp}\hfil\cr
\noalign{\vskip 4pt}
\hglue 1em $TTQ_r$&     69.3 &  11.1 &  33.4 &  21.6 &  11.1 \cr
\hglue 1em  $TTU_r$&    *5.3 &  35.6 &  70.1 &  14.1 &  13.7 \cr 
\hglue 1em  $TQ_rQ_r$&   *4.4 &  11.0 &  70.1 &  24.5 &  14.7 \cr
\hglue 1em  $TQ_rU_r$&   39.6 &  18.3 &  98.5 &  65.5 &  58.0 \cr
\hglue 1em  $TU_rU_r$&    67.5 &  *2.6 &  55.5 &  *1.2 &  *0.9 \cr
\hglue 1em  $Q_rQ_rQ_r$&  35.3 &  *5.3 &  99.6 &  44.0 &  *6.1 \cr
\hglue 1em  $Q_rQ_rU_r$&  47.4 &  *0.2 &  99.7 &  41.7 &  *8.3 \cr
\hglue 1em  $Q_rU_rU_r$&  45.5 &  62.4 &  99.5 &  14.6 &  14.6 \cr
\hglue 1em  $U_rU_rU_r$&  55.5 &  60.5 &  98.7 &  75.7 &  22.4 \cr
\noalign{\vskip 4pt}
\multispan6 Equilateral three-point; see Fig.~\ref{fig:npt_ffp8_hp}\hfil\cr
\noalign{\vskip 4pt}
\hglue 1em  $TTQ_r$&     *1.6 &  24.4 &  50.0 &  16.6 &  18.7 \cr
\hglue 1em  $TTU_r$&     67.4 &  70.8 &  81.5 &  83.4 &  68.3 \cr
\hglue 1em  $TQ_rQ_r$&    84.2 &  14.5 &  98.0 &  56.3 &  55.0 \cr
\hglue 1em  $TQ_rU_r$&    20.3 &  54.1 &  95.6 &  54.2 &  30.0 \cr
\hglue 1em  $TU_rU_r$&    73.3 &  *2.2 &  82.4 &  22.2 &  *6.4 \cr
\hglue 1em  $Q_rQ_rQ_r$&   16.4 &  53.1 &  99.1 &  *6.5 &  75.9 \cr
\hglue 1em  $Q_rQ_rU_r$&   21.1 &  80.5 &  99.7 &  93.4 &  82.0 \cr
\hglue 1em  $Q_rU_rU_r$&   50.2 &  46.1 &  99.6 &  59.7 &  77.5 \cr
\hglue 1em  $U_rU_rU_r$&   *8.2 &  34.0 &  99.7 &  88.1 &  36.4 \cr
\noalign{\vskip 3pt\hrule\vskip 3pt}}}
\endPlancktable                    % ends one-column \halign%\endPlancktablewide                 % ends two-column \halign
\endgroup
\end{table}

Overall, there is good agreement between the recovered FFP8 CMB
estimates and the corresponding Monte Carlo ensembles. The main
outlier is \nilc, for which the fluctuations are larger than expected
for all three-point polarization $N$-point functions. 

\subsection{CMB Power spectra and residuals from individual components}
\label{sec:ffp8_power_spectra_recovery}

\begin{figure}[th!]
  \begin{center}
    \includegraphics[width=\columnwidth]{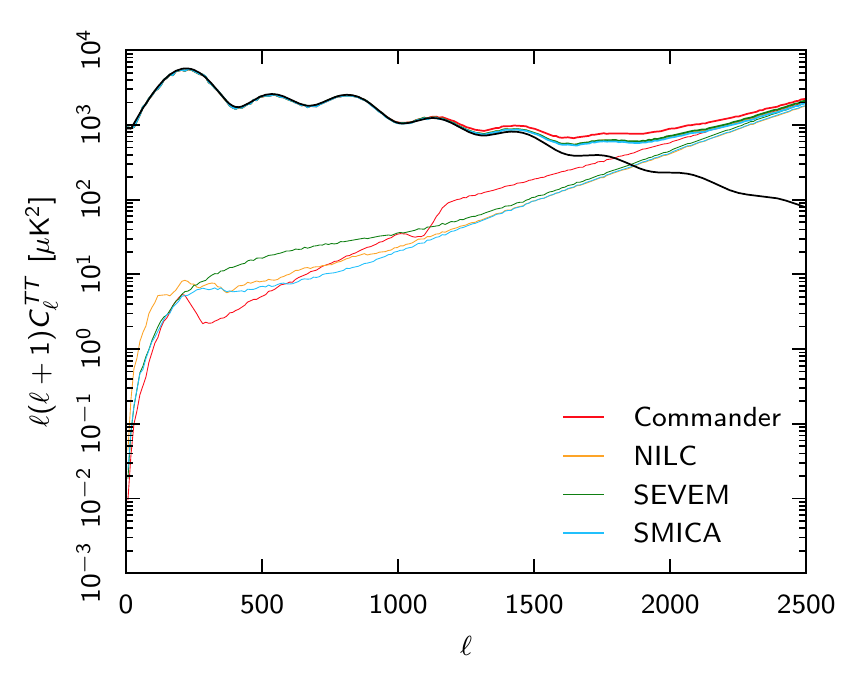}
    \includegraphics[width=\columnwidth]{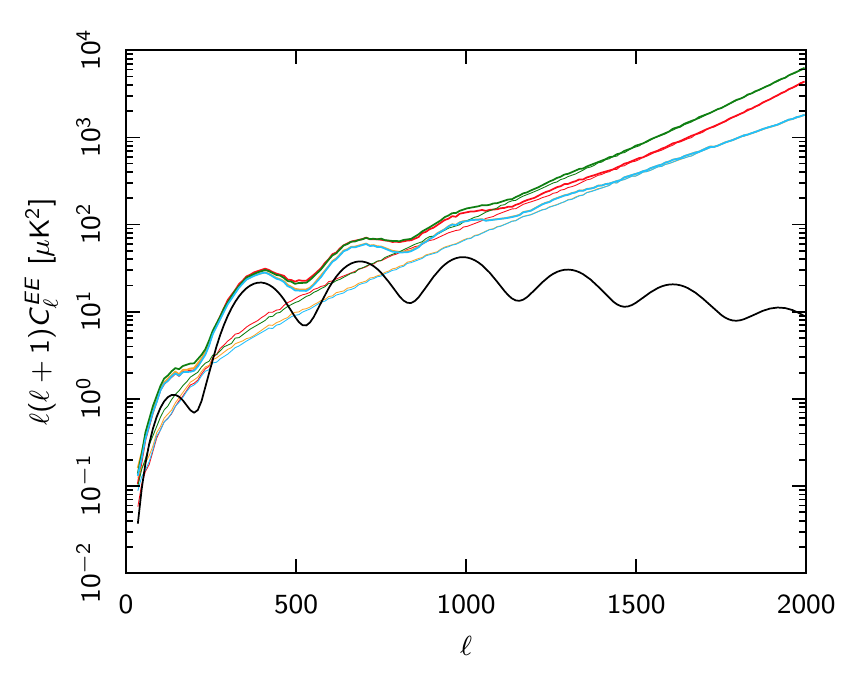}
  \end{center}
  \caption{Power spectra of the foreground cleaned CMB maps from FFP8 simulations.  \emph{Top:} $TT$ power spectra evaluated using the \texttt{FFP8-UT74} mask.  \emph{Bottom:} $EE$ power spectra evaluated using the \texttt{FFP8-UP76} mask.  Thick lines show the spectra of signal plus noise estimated from the half-mission half-sum maps; thin lines show the noise levels from half-mission half-difference maps. The black line shows the input spectrum.}
  \label{fig:ffp8_power_spectra}
\end{figure}

Here we discuss the recovery of the power spectrum in the FFP8
simulations, and assess the impact of foregrounds in the CMB solutions
presented in this paper. CMB angular power spectra from the foreground
cleaned CMB maps from the FFP8 are shown in
Fig.~\ref{fig:ffp8_power_spectra} compared to the input for $TT$,
$EE$.  The performance on simulations is comparable to that on the
data, as may be seen by comparison with Fig.~\ref{fig:dx11_spectra}.

\begin{figure}[th!]
  \begin{center}
    \includegraphics[width=0.95\columnwidth,trim = 0mm 4mm 0mm 6mm,clip]{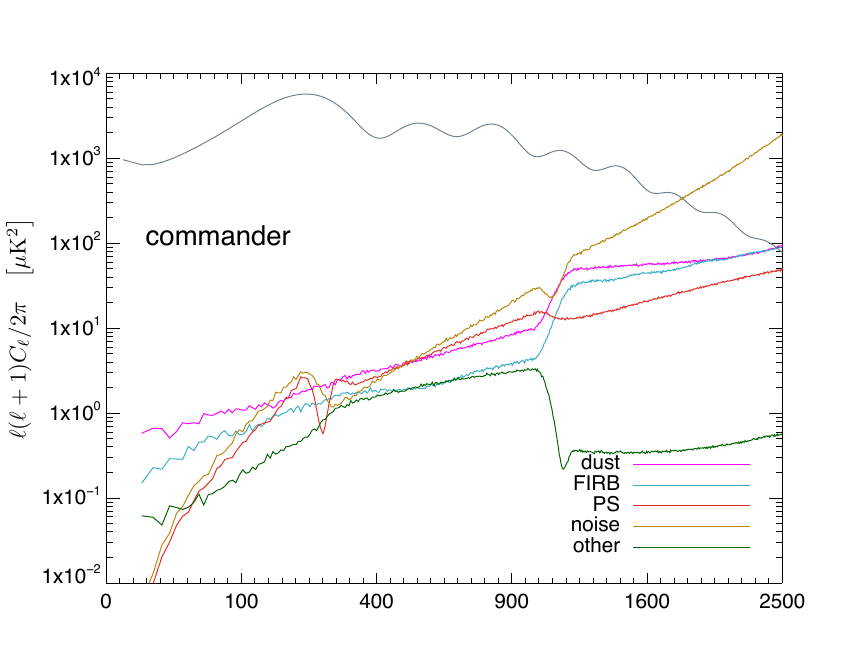}
    \includegraphics[width=0.95\columnwidth,trim = 0mm 4mm 0mm 6mm,clip]{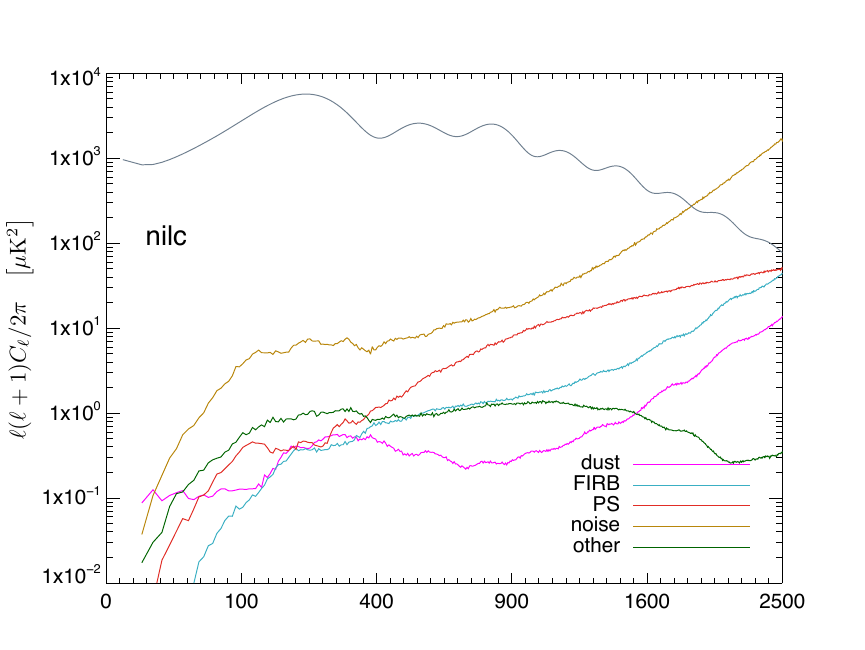}
    \includegraphics[width=0.95\columnwidth,trim = 0mm 4mm 0mm 6mm,clip]{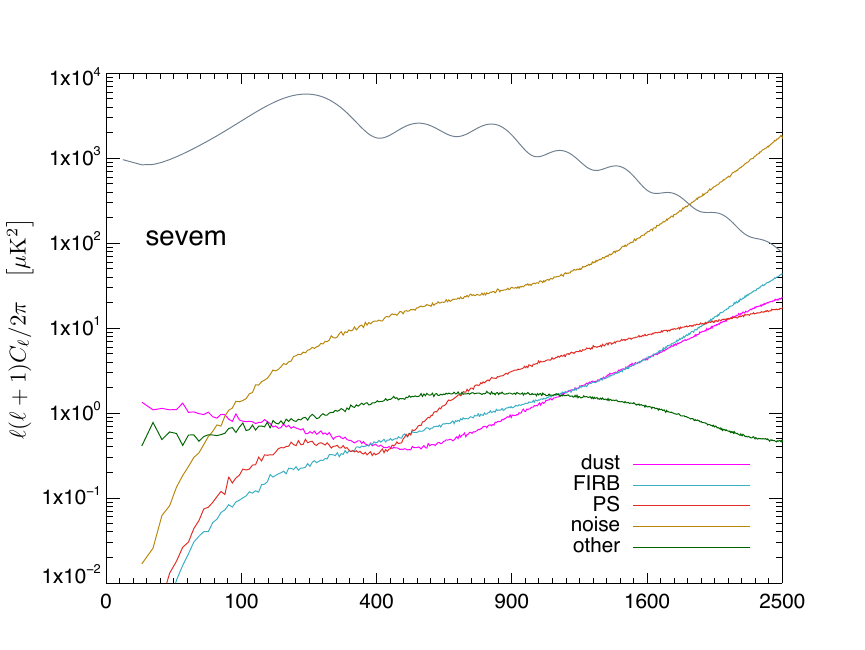}
    \includegraphics[width=0.95\columnwidth,trim = 0mm 1mm 0mm 6mm,clip]{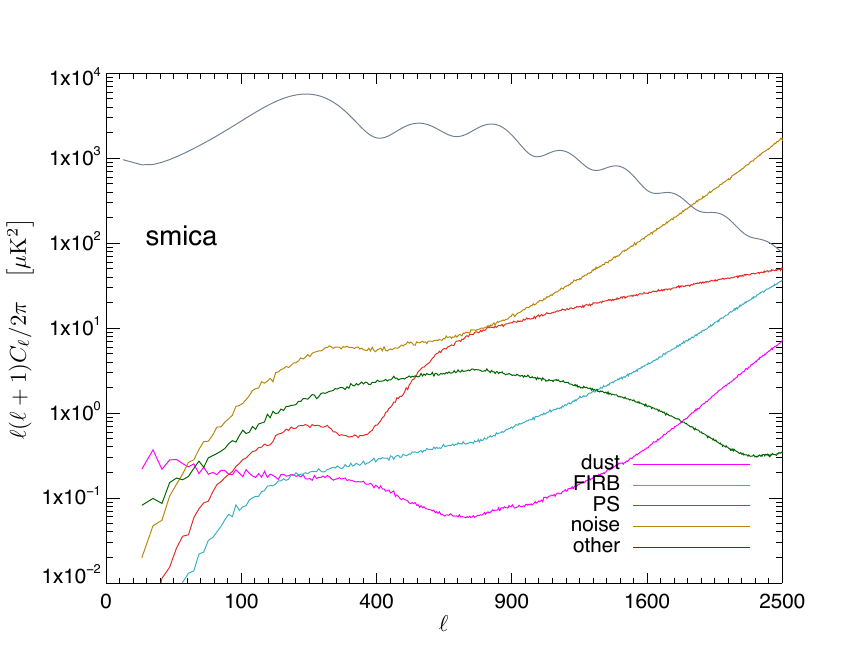}
  \end{center}
  \caption{$TT$ angular power spectra of residuals from the indicated
    FFP8 components in the \Planck\ 2015 CMB maps, compared with the
    predicted signal from the best fit cosmology. ``Other'' is the sum
    of CO, free-free, thermal and kinetic SZ, spinning dust, and
    synchrotron emission. The horizontal axis is linear in
    $\ell^{\,0.5}$.}
  \label{fig:ffp8_TT_spectra_from_individual_components}
\end{figure}

\begin{figure}[th!]
  \begin{center}
    \includegraphics[width=0.95\columnwidth,trim = 0mm 4mm 0mm 6mm,clip]{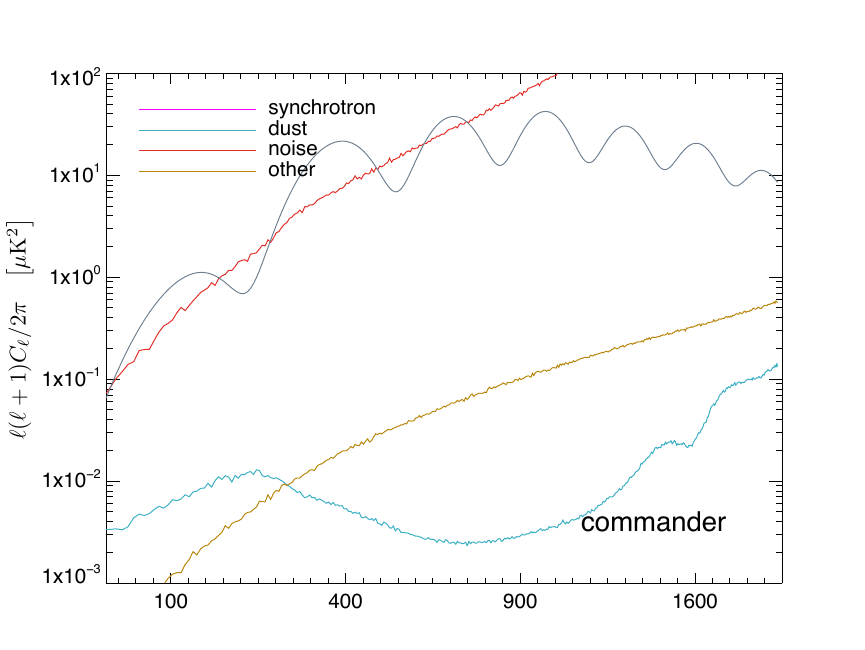}
    \includegraphics[width=0.95\columnwidth,trim = 0mm 4mm 0mm 6mm,clip]{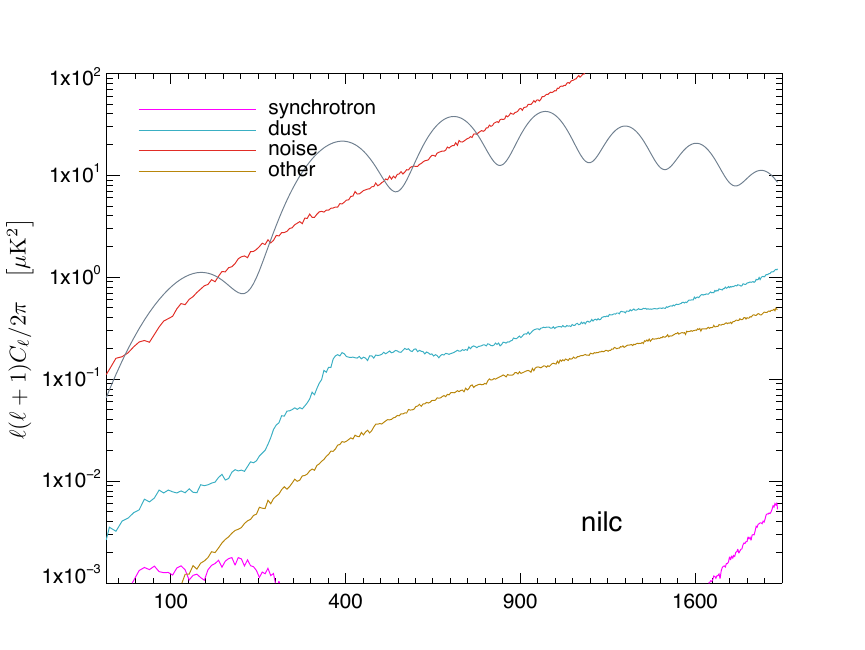}
    \includegraphics[width=0.95\columnwidth,trim = 0mm 4mm 0mm 6mm,clip]{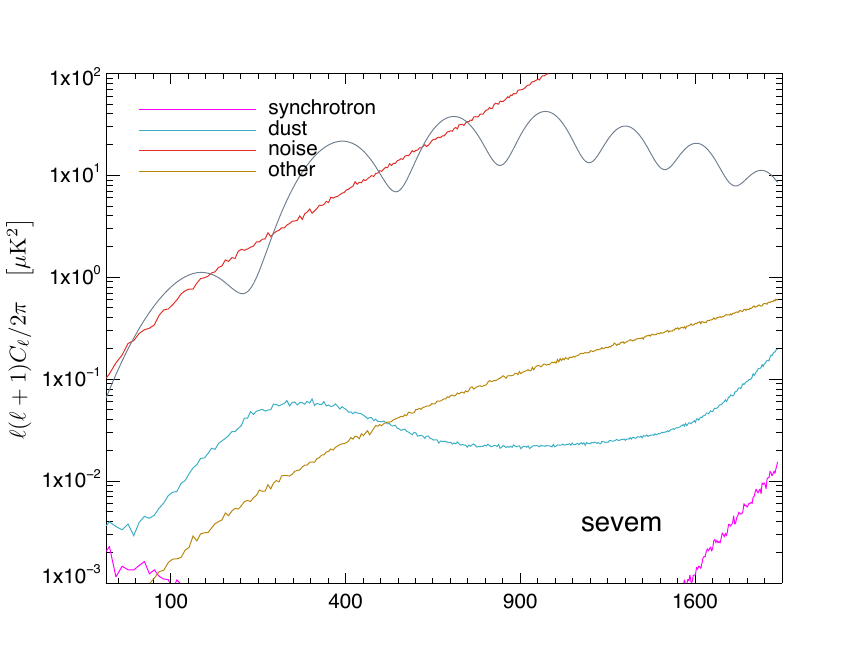}
    \includegraphics[width=0.95\columnwidth,trim = 0mm 1mm 0mm 6mm,clip]{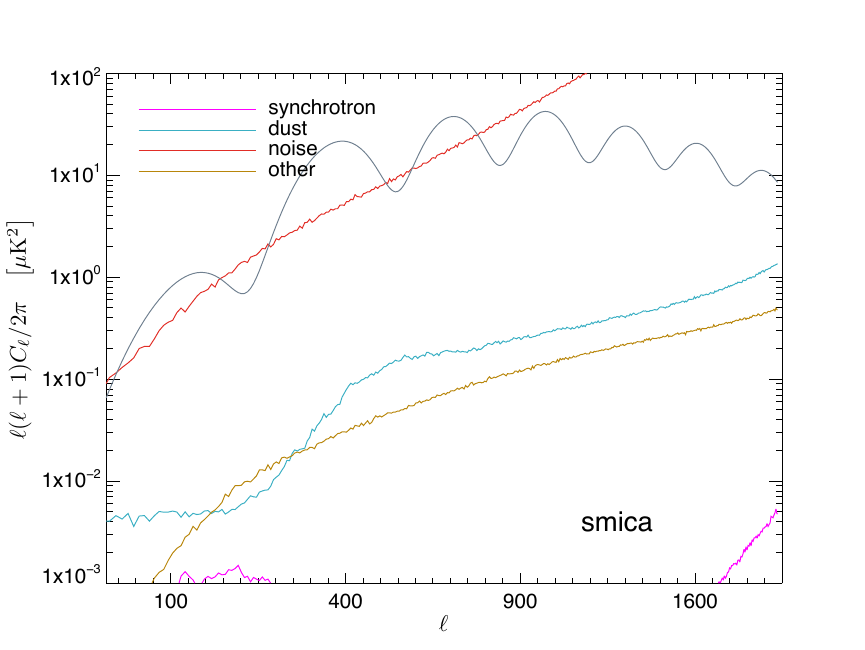}
  \end{center}
  \caption{$EE$ angular power spectra of residuals from the indicated
    FFP8 components in the \Planck\ 2015 CMB maps, compared with the
    predicted signal from the best fit cosmology. ``Other'' is the sum
    of CO, free-free, thermal and kinetic SZ, spinning dust,
    far-infrared background, and radio and infrared unresolved
    sources. The horizontal axis is linear in $\ell^{\,0.5}$.}
\label{fig:ffp8_EE_spectra_from_individual_components}
\end{figure}

Figures~\ref{fig:ffp8_TT_spectra_from_individual_components} and
\ref{fig:ffp8_EE_spectra_from_individual_components} show the residual
effect of combined individual components on the foreground-cleaned CMB
maps. These plots give an estimate of how the residuals of the
foreground components in the CMB solutions presented in this paper
compare in power and are distributed in $\ell$. The FFP8 simulated
skies of the individual components are propagated through the various
component separation pipelines, and their power spectra are calculated
using the {\tt FFP8-UT74} sky mask. The individual components are
those simulated in the FFP8 (Sec.~\ref{sec:ffp8_simulations} and
\citet{planck2014-a14}) and include in $TT$ dust, the far-infrared
background (FIRB), unresolved sources (PS), and all the remaining
components summed together (``other'').  In $EE$, they include
synchrotron, dust, noise, and remaining components. With the exception
of the lowest multipoles, the dominant contaminant to the $TT$ signal
is noise. For $EE$, the residuals from dust dominate over the other
foregrounds, but remain sub-dominant with respect to instrumental
noise at all multipoles.

\clearpage

\end{document}